\documentclass[a4paper,11pt]{article}
\pdfoutput=1 
\usepackage{jheppub}
\usepackage[T1]{fontenc} 
\usepackage[all]{xy}
\usepackage{rotating}
\usepackage{float}
\usepackage{tikz}
\usepackage{circuitikz}
\usepackage{tikz-network}
\usepackage{diagbox}
\usepackage{braket}
\usepackage{empheq}
\usepackage{pdflscape}
\usetikzlibrary{decorations.markings}
\usetikzlibrary{arrows.meta}
\usepackage[normalem]{ulem}
\tikzset{->-/.style={decoration={
  markings,
  mark=at position #1 with {\arrow{>}}},postaction={decorate}}}

\pgfarrowsdeclare{:}{:}{}{}

\let\a=\alpha \let\b=\beta \let\g=\gamma \let\d=\delta 
  \let\th=\theta  \let\k=\kappa
\let\l=\lambda \let\m=\mu    \let\r=\rho
\let\s=\sigma \let\t=\tau    
    \let\D=\Delta

\DeclareMathOperator{\Tr}{Tr}

\def\a{\alpha}
\def\b{\beta}

\def\CD{{\cal D}}

\def\CN{{\cal N}}

\def\Wm{\widetilde{m}}

\def\Wq{\widetilde{q}}
\def\Wr{\widetilde{r}}

\def\Wu{\widetilde{u}}
\def\Wv{\widetilde{v}}

\def\WN{\widetilde{N}}

\def\WQ{\widetilde{Q}}
\def\WR{\widetilde{R}}
\def\WS{\widetilde{S}}
\def\WT{\widetilde{T}}

\def\beq#1\eeq{\begin{align}#1\end{align}}

\makeatletter
\newcommand*{\rom}[1]{\expandafter\romannumeral #1}
\newcommand*\widefbox[1]{\fbox{\hspace{2em}#1\hspace{2em}}}
\makeatother

\preprint{
\begin{flushright}
    USTC-ICTS/PCFT-24-22\\
     KIAS-P24039
\end{flushright}}

\title{S-confinement of 3d Argyres--Douglas theories and the Seiberg-like duality with an adjoint matter}

\abstract{We propose an $\mathcal{N}=2$ preserving deformation that leads to the confining phase of the 3d reduction of the $D_p[SU(N)]$ Argyres--Douglas theories, referred to as $\mathbb{D}_p[SU(N)]$. This deformation incorporates monopole superpotential terms, which have recently played interesting roles in exploring possible RG fixed points of 3d supersymmetric gauge theories. Employing this confining phenomenon in 3d $\mathbb{D}_p[SU(N)]$ theories, we also propose a deconfined version of the Kim--Park duality, an IR duality for 3d $\mathcal{N}=2$ adjoint SQCDs, where an adjoint matter field is replaced by a linear quiver tail of $\mathbb{D}_p[SU(N)]$. Surprisingly, both the confinement of deformed $\mathbb{D}_p[SU(N)]$ and the deconfined Kim--Park duality can be proven only assuming some basic 3d $\mathcal{N}=2$ IR dualities. Finally, we propose a variant of the Kim–Park duality deformed by a single monopole superpotential term, which can also be derived using the same method.}

\author[a,b]{Chiung Hwang,}
\author[c]{Sungjoon Kim}

\affiliation[a]{Interdisciplinary Center for Theoretical Study, University of Science and Technology of China, Hefei, Anhui 230026, China}
\affiliation[b]{Peng Huanwu Center for Fundamental Theory, Hefei, Anhui 230026, China}
\affiliation[c]{Korea Institute for Advanced Study, 85 Hoegiro, Dongdaemun-Gu, Seoul 02455, Korea}

\emailAdd{chiung@ustc.edu.cn, sungjoon@kias.re.kr}

\begin{document} 
\maketitle
\flushbottom


\section{Introduction}

The strongly coupled nature of infrared (IR) dynamics of 3-dimensional gauge theories presents numerous intriguing phenomena. Particularly, IR dualities between different ultraviolet (UV) gauge theories are highly nontrivial and help us understand the non-perturbative aspects of 3d gauge theories, where monopole operators play crucial roles. To date, many examples of IR dualities have been proposed and utilized to study the IR dynamics of 3d supersymmetric gauge theories, ranging from the simplest Aharony duality \cite{Aharony:1997gp}, which is the 3d analog of the well-known Seiberg duality in 4d \cite{Seiberg:1994pq}, to more sophisticated dualities involving quiver gauge theories or matter fields in tensor representations of the gauge group.

Recently, many novel aspects of supersymmetric IR dualities with tensor matters in various dimensions have been discussed using the so-called (sequential) \emph{deconfinement} method, which allows us to replace the tensor matters by extra quiver gauge nodes \cite{Nii:2016jzi,Pasquetti:2019uop,Pasquetti:2019tix,Sacchi:2020pet,Benvenuti:2020gvy,Etxebarria:2021lmq,Benvenuti:2021nwt,Bajeot:2022kwt,Bottini:2022vpy,Amariti:2022wae,Bajeot:2022lah,Bajeot:2022wmu,Bajeot:2023gyl,Amariti:2023wts,Amariti:2024sde,Amariti:2024gco}.\footnote{See \cite{Berkooz:1995km,Pouliot:1995me,Luty:1996cg,Garcia-Etxebarria:2012ypj,Garcia-Etxebarria:2013tba} for earlier works on 4d $\mathcal N=1$ theories. This technique has also been discussed in the mathematics literature in the context of elliptic hypergeometric integrals. See \cite{Spiridonov:2009za,Spiridonov:2011hf,2003math......9252R,2003math......3205S,Albion:2023ndz}.} This technique turns out to be very useful, in particular, to find new examples of s-confining theories, which have IR dual theories described by Wess--Zumino models with the smooth superpotential along the entire vacuum moduli space \cite{Csaki:1996sm}. For example, in \cite{Bajeot:2023gyl}, novel s-confining phenomena of 4d $USp(2 N)$ theories with antisymmetric matters and those of 3d $USp/U$ theories with antisymmetric and adjoint matters, respectively, have been derived from known dualities using the sequential deconfinement and Higgsing; see \eqref{eq: Benvenuti conf} for the 3d $U(N)$ example.

In this paper, we extend such derivation of the s-confining phenomena of 3d $U(N)$ SQCDs with adjoint matters to a special class of quiver theories that are obtained as the 3d reduction \cite{Closset:2020afy,Giacomelli:2020ryy} of 4d $D_p[SU(N)]$ Argyres--Douglas theories \cite{Xie:2012hs,Cecotti:2012jx,Cecotti:2013lda}. We find that the 3d reduction of $D_p[SU(N)]$, which we call $\mathbb D_p[SU(N)]$, is confining once deformed by certain monopole superpotentials, which we prove using the deconfinement method and the Benini--Benvenuti--Pasquetti (BBP) duality \cite{Benini:2017dud}. The BBP duality is a cousin of the aforementioned Aharony duality that is obtained by deforming the Aharony duality by linear monopole superpotential terms. Or, in the other way around, one can obtain the Aharony duality from the BBP duality by mass deformation.

In general, such $\mathbb D_p[SU(N)]$ theories are given by linear quiver gauge theories consisting of $U/SU$ gauge nodes, where each node is either good or ugly, but not bad, in the sense of \cite{Gaiotto:2008ak}. If $\gcd(p,N) = 1$, every node is the unitary gauge group, and the corresponding quiver theory is given by
\begin{center}
\begin{equation}
\begin{tikzpicture}
  \tikzset{vertex/.style={circle,fill=white!25,minimum size=12pt,inner sep=2pt}}
  \tikzset{every loop/.style={}}
    \node at (-0.5,0) {$\mathbb{D}_p[SU(N)]\quad:\qquad$ };    
    \node[vertex] (m1) at (1.5,0) [shape=circle,draw=black] {$m_1$};
    \node[vertex] (m2) at (3,0) [shape=circle,draw=black] {$m_2$};
    \node[draw=none,fill=none] (ndots) at (4.5,0)  {$\cdots$};
    \node[vertex] (mp1) at (6,0) [shape=circle,draw=black] {\scriptsize $m_{p\!-\!1}$};
    \node[vertex] (Np) at (7.5,0) [shape=rectangle,draw=black,minimum height=2em, minimum width=2em] {$N$};

      \draw  (m1) -- (m2);
      \draw  (m2) -- (ndots);
      \draw  (ndots) -- (mp1);
      \draw (mp1) -- (Np);
\end{tikzpicture}
\end{equation}
\end{center}
where the gauge ranks $m_j$ are evaluated as $m_j = \lfloor j N / p \rfloor$. In this paper, we focus on the cases where the quiver is \emph{good}; namely, every node satisfies the condition
\begin{align}
m_{j-1}+m_{j+1}-2 m_j \geq 0
\end{align}
so that there is no decoupled monopole operator in the IR.
We show in appendix \ref{app: D_p[G] review} that this is true if and only if $N = \pm1 \mod p$, i.e.,
\begin{align}
N \; = \; p+1, \quad 2 p\pm1, \quad 3 p\pm1, \quad \dots \,.
\end{align}
The other cases, leading to quivers containing ugly nodes, will be discussed in a subsequent work \cite{ugly}.

Notice that our confining $\mathbb D_p[SU(N)]$ models are 3d reminiscent of the confinement-like behavior of the 4d $D_p[SU(2)] = (A_1,A_p)$ theories triggered by the deformation by the minimal Coulomb branch operator \cite{Xie:2021omd}, where the deformed non-Lagrangian $(A_1,A_p)$ theories are described by free chiral multiplets in the IR. This confining phenomenon is further generalized to the $D_p[SU(N)]$ theories in the context of an IR duality for 4d adjoint SQCDs \cite{Maruyoshi:2023mnv}, referred to as the Maruyoshi--Nardoni--Song (MNS) duality in this paper. In the same manner, we also use our confining $\mathbb D_p[SU(N)]$ models to derive a new IR duality for 3d adjoint SQCDs.

Furthermore, employing this result, we propose a \emph{deconfined} version of the Kim--Park duality, a Seiberg-like duality for the 3d $U(N)$ adjoint SQCD with the superpotential $W = \mathrm{Tr} \, X^{p+1}$ \cite{Kim:2013cma}, motivated by the 4d duality of the same type \cite{Kutasov:1995ve,Kutasov:1995np}. Surprisingly, this deconfined Kim--Park duality can also be proven simply using the Aharony duality, which, as mentioned, can be obtained from the BBP duality by mass deformation. Therefore, combined with the derivation of the confinement of $\mathbb D_p[SU(N)]$ previously explained, it provides a proof of the Kim--Park duality only assuming the BBP duality when $N = \pm1 \mod p$. Also interestingly, one can repeat the derivation of the deconfined Kim--Park duality but using the one-monopole BBP duality instead of the Aharony duality. This leads to another variant of the Kim--Park duality that is deformed by a \emph{single} monopole superpotential term. Although its existence has been widely believed, this variant has not been discussed in the literature, and our derivation shows it naturally descends from the BBP duality.

Our result sheds light on how to understand 3d Seiberg-like dualities with tensor matters in terms of more fundamental dualities without tensor matters. Indeed, this is parallel to the recent discovery that the 3d mirror symmetry and its $SL(2,\mathbb Z)$ cousins all originate from the basic Aharony duality \cite{Bottini:2021vms,Hwang:2021ulb,Comi:2022aqo,Giacomelli:2023zkk,Benvenuti:2023qtv,Giacomelli:2024laq}. While the mirror symmetry and the Seiberg-like duality with an adjoint matter discussed in this paper have completely different shapes, both of them are consequences of the simple Aharony duality, which strongly suggests that these two and other 3d IR dualities would be governed by the same microscopic mechanism encoded in the Aharony duality regardless of their details.

This paper is organized as follows. In section \ref{sec:confinement}, we propose the confining superpotential deformation of the good $\mathbb D_p[SU(N)]$ theories and provide the derivation of the confinement using the BBP duality. We also propose the 3d version of the 4d MNS duality for adjoint SQCDs employing these confining $\mathbb D_p[SU(N)]$ models. In section \ref{sec:deconfined KP}, utilizing the result of section \ref{sec:confinement}, we propose the deconfined version of the Kim--Park duality and its derivation from the Aharony duality. We also propose a version of the duality deformed by a single linear monopole term. In section \ref{sec:conclusion}, we conclude with some future directions.

There are also appendices. In appendix \ref{app: D_p[G] review}, we provide a brief review of the $D_p[SU(N)]$ Argyres--Douglas theories and their 3d reduction. In appendix \ref{app: MNS duality}, we provide a brief review of the confinement of $D_p[SU(N)]$ and the MNS duality. In appendix \ref{app: monopole map}, we discuss the duality map of monopole operators of linear quiver gauge theories under the Aharony duality and the BBP duality, the basic dualities used in this paper. In appendix \ref{app: monopole map under gauging}, we discuss the monopole map under the deconfinement of an adjoint field into a $\mathbb D_p[SU(N)]$ linear quiver tail. In appendix \ref{app: S_b^3 ptf}, we explain the integral identities of the $S^3_b$ supersymmetric partition functions for the basic dualities and how to read them from the quiver diagrams. We also provide an example of deriving the partition function identity for the deconfined Kim--Park duality we propose. In appendix \ref{app: rank reducing}, we derive a duality chain called the rank-reducing duality, which is employed to show the confinement of the deformed $\mathbb D_p[SU(N)]$ theories. Finally, in appendix \ref{app:moduli}, we examine the vacuum moduli spaces of the dual pair proposed in section \ref{sec:confinement} to determine the superpotential terms.

\paragraph{Note added.} While this manuscript was being finalized we learned about the work of \cite{Benvenuti:2024glr} which has a substantial overlap with our manuscript. We thank the authors of \cite{Benvenuti:2024glr} for sharing their draft with us and for coordinating the submission.
\\

\section{Confinement of 3d $\mathbb D_p[SU(N)]$ for $N = \pm1 \mod p$ and the 3d version of the Maruyoshi--Nardoni--Song duality}
\label{sec:confinement}

\subsection{Proposal}
The 3d reductions of 4d $\CN=2$ $D_p[SU(N)]$ Argyres--Douglas theories are given by 3d $\CN=4$ linear quiver gauge theories \cite{Closset:2020afy, Giacomelli:2020ryy}, which are denoted by $\mathbb{D}_p[SU(N)]$ in this paper. When two integers $(p,N)$ are coprime, the $\mathbb{D}_p[SU(N)]$ linear quiver theory, represented by the following quiver diagram:
\begin{equation}
\begin{tikzpicture}
  \tikzset{vertex/.style={circle,fill=white!25,minimum size=12pt,inner sep=2pt}}
  \tikzset{every loop/.style={}}
    \node at (-0.5,0) {$\mathbb{D}_p[SU(N)]\quad:\qquad$ };    
    \node[vertex] (m1) at (1.5,0) [shape=circle,draw=black] {$m_1$};
    \node[vertex] (m2) at (3,0) [shape=circle,draw=black] {$m_2$};
    \node[draw=none,fill=none] (ndots) at (4.5,0)  {$\cdots$};
    \node[vertex] (mp1) at (6,0) [shape=circle,draw=black] {\scriptsize $\!m_{p\texttt{-}1}\!$};
    \node[vertex] (Np) at (7.5,0) [shape=rectangle,draw=black,minimum height=2em, minimum width=2em] {$N$};

      \draw  (m1) -- (m2);
      \draw  (m2) -- (ndots);
      \draw  (ndots) -- (mp1);
      \draw (mp1) -- (Np);
\end{tikzpicture} \,,
\label{eq: 3d CD theory for gcd=1}
\end{equation}
only involves unitary groups $U(m_j)$ as its gauge nodes with the ranks $m_j = \lfloor j N / p \rfloor$, $j = 1,\dots,p-1$. This linear quiver is {\it good} if $N = \pm 1\mod p$ and {\it ugly} for others, whose detailed discussion is in appendix \ref{app: D_p[G] review}. In this paper, we focus on the good $\mathbb{D}_p[SU(N)]$ theories with $p, N > 1$ and propose a monopole superpotential deformation leading to the \emph{confinement} of $\mathbb{D}_p[SU(N)]$, whose IR dynamics is described by a Wess--Zumino (WZ) theory of a matrix of chiral fields, $X$, in the adjoint representation of the $U(N)$ flavor symmetry with the $A_p$-type superpotential $W = \Tr X^{p+1}$:\footnote{We borrow the classification of adjoint superpotentials from \cite{Intriligator:2003mi}:
\begin{align}
\begin{aligned}
& W_{\widehat O} = 0, \qquad W_{\widehat A} = \mathrm{Tr} Y^2, \qquad W_{\widehat D} = \mathrm{Tr} X Y^2, \qquad W_{\widehat E} = \mathrm{Tr} Y^3, \\
& W_{A_n} = \mathrm{Tr} \left(X^{n+1}+Y^2\right), \qquad W_{D_{n+2}} = \mathrm{Tr} \left(X^{n+1}+X Y^2\right), \\
& W_{E_6} = \mathrm{Tr} \left(Y^3+X^4\right), \quad W_{E_7} = \mathrm{Tr} \left(Y^3+Y X^3\right), \quad W_{E_8} = \mathrm{Tr} \left(Y^3+X^5\right).
\end{aligned}
\end{align}
}
\begin{equation}
\begin{tikzpicture}
  \tikzset{vertex/.style={circle,fill=white!25,minimum size=12pt,inner sep=2pt}}
  \tikzset{every loop/.style={}}
    \node[vertex] (N) at (0,0) [shape=rectangle,draw=black,minimum height=2em, minimum width=2em] {$N$};

    \draw[-to, min distance=1cm]  (N) edge [out=120, in=60] node {} (N);

    \node at (-4,0) {$\mathbb{D}_{p}[SU(N)]$};
    \node at (-1.8,0.2) {$\xrightarrow{\D W_{\text{conf}}^{(p),\pm}}$};
    
    \node at (0,1.2) {\tiny $X$};
    \node at (2,0) {$\quad,\quad W=\Tr X^{p+1}$};
\end{tikzpicture} \,.
    \label{eq: confining diagram}
\end{equation}
Notice that although the original flavor symmetry of $\mathbb D_p[SU(N)]$ is $SU(N)$, for later convenience, we promote it to $U(N)$ by coupling an extra singlet, which will be denoted by $\eta$. Consequently, the trace of the  flavor adjoint $X$ is non-vanishing and mapped to $\eta$ under the proposed confining duality. Together with $\eta$, we propose two types of {\it confining superpotential deformation} of $\mathbb D_p[SU(N)]$ labeled by superscripts $\pm$:
\begin{align}
   \D W_{\text{conf}}^{(p),\pm} = \eta^{p+1}+\eta \sum_{i=\a}^{p-1}\Tr \Phi^{(i)}
    +
    \sum_{i=\a}^{p-1} \hat{v}^{(i),\pm}
    +
     \hat{v}^{(\a,p\texttt{-}1),\mp}\;
     \sum_{i=\a}^{p-1}
     \hat{v}^{(i\texttt{-}\a\texttt{+}2,i),\mp}
    \label{eq: confining deformation}
\end{align}
where the gauge singlet $\eta$ couples to the traces of the adjoint fields $\Phi^{(i)}$, each belonging to the $\mathcal N=4$ vector multiplet of the $i$th gauge node in \eqref{eq: 3d CD theory for gcd=1}. We also turn on the power term of $\eta$, which is necessary to have the vacuum moduli space consistent with the dual WZ theory as discussed in appendix \ref{app:moduli}. $\a$ is the position of the first non-trivial gauge rank, i.e., $\a = \min[j]$ such that $m_j\neq 0$. Thus, $\alpha$ is 1 if $N > p$, and the only other value of $\alpha \neq 1$ beneath the {\it good} theory condition is $\alpha =2$ when $N=p-1$. The bare monopole operators having unit positive/negative charges under the $U(1)_T$ topological symmetries from the $l$-th gauge node through to the $r$-th gauge node are denoted by $\hat{v}^{(l,r),\pm}$, respectively. If $l=r$, a single index is used for simplicity: $\hat{v}^{(l,r),\pm} = \hat{v}^{(l),\pm} = \hat{v}^{(r),\pm}$. Also $\hat{v}^{(l,r),\pm} = 1$ if $l>r$. Therefore, \eqref{eq: confining deformation} becomes
\begin{align}
   \D W_{\text{conf}}^{(p),\pm} = \eta^{p+1}+\eta \sum_{i=1}^{p-1}\Tr \Phi^{(i)}
    +
    \sum_{i=1}^{p-1} \hat{v}^{(i),\pm}
    +
    \hat{v}^{(1,p\texttt{-}1),\mp}
\end{align}
if $N > p$ and
\begin{align}
   \D W_{\text{conf}}^{(p),\pm} = \eta^{p+1}+\eta \sum_{i=2}^{p-1}\Tr \Phi^{(i)}
    +
    \sum_{i=2}^{p-1} \hat{v}^{(i),\pm}
    +
     \hat{v}^{(2,p\texttt{-}1),\mp}\;
     \sum_{i=2}^{p-1}
     \hat{v}^{(i),\mp}
\end{align}
if $N = p-1$.
For example, consider $p=2$ and $N=2n+1$ for a positive integer $n$, whose confining deformation is simply given by
\begin{align}
    \D W_{\text{conf}}^{(p=2)}
    =
    \eta^{3}+\eta \Tr \Phi + \hat{v}^{+} + \hat{v}^{-}
\end{align}
where we have omitted the labeling of the gauge node since there's only a single of it. As a result, there is no distinction between $\D W_{\text{conf}}^{(p),+}$ and $\D W_{\text{conf}}^{(p),-}$, due to which the superscripts $\pm$ are also dropped. The s-confining duality is then given by
\begin{center}
\begin{equation}
\begin{tikzpicture}
  \tikzset{vertex/.style={circle,fill=white!25,minimum size=12pt,inner sep=2pt}}
  \tikzset{every loop/.style={}}
    \node[vertex] (n) at (-1,0) [shape=circle,draw=black,minimum size=2.5em] {$n$};
    \node[vertex] (2n+1) at (1,0) [shape=rectangle,draw=black,minimum height=2.5em, minimum width=2.5em] {$2n\!\!+\!\!1$};

    \draw[-to, min distance=1cm]  (n) edge [out=120, in=60] node {} (n);
    \draw[->-=.5] ([yshift= 2pt] n.east) to ([yshift= 2pt] 2n+1.west);
    \draw[->-=.5] ([yshift= -2pt] 2n+1.west) to ([yshift= -2pt] n.east);

    \node at (-1,1.4) {$\Phi$};
    \node at (0,0.3) {\tiny$Q$};
    \node at (0,-0.3) {\tiny$\WQ$};
    \node at (0,-1) {$W \!= \Tr \Phi Q \WQ  \!+\!\eta^{3}\!+\!\eta \Tr \Phi \!+ \hat{v}^{+} \!+ \hat{v}^{-}$};

    \node at (4,0.5) {Dual};
    \node at (4,0) {$\Longleftrightarrow$};

    \node[vertex] (N) at (7,0) [shape=rectangle,draw=black,minimum height=2.5em, minimum width=2.5em] {$2n\!\!+\!\!1$};

    \draw[-to, min distance=1cm]  (N) edge [out=120, in=60] node {} (N);

    \node at (7,1.4) {$X$};
    \node at (6.8,-1) {$\quad W=\Tr X^{3}$};
    
\end{tikzpicture}
\label{eq: Benvenuti conf}
\end{equation}
\end{center}
which was first observed in \cite{Bajeot:2023gyl}, up to the singlet $\eta$ and the trace part of $X$, which are included for later convenience. This shows s-confinement of $3d$ $\mathcal N=4$ $U(n)$ SQCD with $2n+1$ flavors into a WZ theory with the $A_2$-type superpotential. Our proposal \eqref{eq: confining diagram} is a generalization of this to the $A_p$-type s-confinement of $\mathbb D_p[SU(N)]$ with the good theory condition $N = \pm 1\mod p$.\footnote{The ugly cases will also be discussed in a subsequent paper \cite{ugly}.} In the next subsection, we provide a derivation of this proposal using some basic $3d$ $\mathcal N=2$ IR dualities.
\subsection{General derivation}
Our strategy for deriving the confinement \eqref{eq: confining diagram} is the mathematical induction: for a given integer $p>1$, we first check the confinement for some lowest values of $N$ satisfying the good theory condition $N = \pm 1 \mod p$, namely, $N=p-1$ and $N=p+1$. Then, we check the confinement of $\mathbb{D}_p[SU(N+p)]$ by assuming the confinement of $\mathbb{D}_p[SU(N)]$, which completes the derivation of the confinement of $\mathbb{D}_p[SU(N)]$ for all $N = \pm 1 \mod p$.
We employ two known 3d $\mathcal N=2$ IR dualities: the Aharony duality \cite{Aharony:1997gp} and the Benini--Benvenuti--Pasquetti (BBP) duality \cite{Benini:2017dud}, which provide a chain of dual theories resulting in the confinement of $\mathbb{D}_p[SU(N)]$. See appendix \ref{app: monopole map} for a brief review of those dualities. One should note that they are not independent dualities because the former can be derived from the latter by mass deformation. Each derivation step is presented with quiver diagrams with (normalized) real mass parameters, which include all the necessary information to specify the dual pair at each step. Hence, one can also read the partition function identities for each step from the given quiver diagrams; as an example, we explain how to write down the corresponding $S^3_b$ partition function identities \cite{Hama:2011ea} in appendix \ref{app: S_b^3 ptf}.

\begin{figure}[tbp]
\centering
\begin{tikzpicture}[thick,scale=1.1, every node/.style={scale=1.1}]
\tikzset{vertex/.style={circle,fill=white!25,minimum size=12pt,inner sep=2pt}}
  \tikzset{every loop/.style={}}
    \node[vertex] (AN) at (0.7,0) [shape=circle,draw=black,minimum size=2em] {$N$};
    \node[vertex] (AF) at (2.3,0) [shape=rectangle,draw=black,minimum height=2em, minimum width=2em] {$F$};

    \draw[-to, min distance=1cm]  (AN) edge [out=120, in=60] node {} (AN);
    \draw[->-=.5] ([yshift= 2pt] AN.east) to ([yshift= 2pt] AF.west);
    \draw[->-=.5] ([yshift= -2pt] AF.west) to ([yshift= -2pt] AN.east);

    \node at (0.7,1.2) {\scriptsize $2\D$};
    \node at (1.5,0.4) {\scriptsize $1\texttt{-}\D$};
    \node at (0.7,-0.7) {\scriptsize $F \D$};

    \node[vertex] (BN1) at (0,-2.5) [shape=circle,draw=black,minimum size=2em] {\scriptsize $N\texttt{-}1$};
    \node[vertex] (BN) at (1.5,-2.5) [shape=circle,draw=black,minimum height=2em, minimum width=2em] {$N$};
    \node[vertex] (BF) at (3,-2.5) [shape=rectangle,draw=black,minimum height=2em, minimum width=2em] {$F$};

    \draw[->-=.5] ([yshift= 2pt] BN1.east) to ([yshift= 2pt] BN.west);
    \draw[->-=.5] ([yshift= -2pt] BN.west) to ([yshift= -2pt] BN1.east);
    \draw[->-=.5] ([yshift= 2pt] BN.east) to ([yshift= 2pt] BF.west);
    \draw[->-=.5] ([yshift= -2pt] BF.west) to ([yshift= -2pt] BN.east);

    \node at (0.75,-2.5+0.4) {\scriptsize $\D$};
    \node at (2.2,-2.5+0.4) {\scriptsize $1\texttt{-}\D$};
    \node at (0,-2.5-0.7) {\scriptsize $N\D$};
    \node at (1.5,-2.5-0.7) {\scriptsize $(F\texttt{-}N\texttt{+}1)\D$};
    \node at (4.3,-2.5) {\scriptsize $+\{2N\D\}$};

    \node[vertex] (CN1) at (4,-5) [shape=circle,draw=black,minimum size=2em] {\scriptsize $N\texttt{-}1$};
    \node[vertex] (CF1) at (5.5,-5) [shape=circle,draw=black,minimum height=2em, minimum width=2em] {\scriptsize $F\texttt{-}1$};
    \node[vertex] (CF) at (7,-5) [shape=rectangle,draw=black,minimum height=2em, minimum width=2em] {$F$};

    \draw[-to, min distance=1cm]  (CN1) edge [out=120, in=60] node {} (CN1);
    \draw[-to, min distance=1cm]  (CF) edge [out=120, in=60] node {} (CF);
    \draw[->-=.5] ([yshift= 2pt] CN1.east) to ([yshift= 2pt] CF1.west);
    \draw[->-=.5] ([yshift= -2pt] CF1.west) to ([yshift= -2pt] CN1.east);
    \draw[->-=.5] ([yshift= 2pt] CF1.east) to ([yshift= 2pt] CF.west);
    \draw[->-=.5] ([yshift= -2pt] CF.west) to ([yshift= -2pt] CF1.east);

    \node at (4.75,-5+0.4) {\scriptsize $1\texttt{-}\D$};
    \node at (6.25,-5+0.4) {\scriptsize $\D$};
    \node at (4,-5+1.2) {\scriptsize $2\D$};
    \node at (7,-5+1.2) {\scriptsize $2\texttt{-}2\D$};
    \node at (4,-5-0.7) {\scriptsize $(F\texttt{+}1)\D$};
    \node at (5.5,-5-0.7) {\scriptsize $(N\texttt{-}F\texttt{-}1)\D$};
    \node at (9.3,-5) {\scriptsize $+\{2N\D,2(F\texttt{-}N\texttt{+}1)\D,0\}$};

    \node[vertex] (DN) at (8,-2.5) [shape=circle,draw=black,minimum size=2em] {\scriptsize $N\texttt{-}1$};
    \node[vertex] (DN1) at (9.5,-2.5) [shape=circle,draw=black,minimum height=2em, minimum width=2em] {\scriptsize $N\texttt{-}1$};
    \node[vertex] (DF) at (11,-2.5) [shape=rectangle,draw=black,minimum height=2em, minimum width=2em] {$F$};

    \draw[->-=.5] ([yshift= 2pt] DN.east) to ([yshift= 2pt] DN1.west);
    \draw[->-=.5] ([yshift= -2pt] DN1.west) to ([yshift= -2pt] DN.east);
    \draw[->-=.5] ([yshift= 2pt] DN1.east) to ([yshift= 2pt] DF.west);
    \draw[->-=.5] ([yshift= -2pt] DF.west) to ([yshift= -2pt] DN1.east);

    \node at (8.75,-2.5+0.4) {\scriptsize $\D$};
    \node at (10.2,-2.5+0.4) {\scriptsize $1\texttt{-}\D$};
    \node at (8,-2.5-0.7) {\tiny $1\texttt{+}(N\texttt{-}1)\D$};
    \node at (9.5,-2.5-0.7) {\tiny $(F\texttt{-}N\texttt{+}1)\D\texttt{-}1$};
    \node at (12.3,-2.5) {\scriptsize $+\{2N\D,0\}$};

    \node[vertex] (EN) at (8.2,0) [shape=circle,draw=black,minimum size=2em] {\scriptsize $N\texttt{-}1$};
    \node[vertex] (EF) at (9.8,0) [shape=rectangle,draw=black,minimum height=2em, minimum width=2em] {$F$};

    \draw[-to, min distance=1cm]  (EN) edge [out=120, in=60] node {} (EN);
    \draw[->-=.5] ([yshift= 2pt] EN.east) to ([yshift= 2pt] EF.west);
    \draw[->-=.5] ([yshift= -2pt] EF.west) to ([yshift= -2pt] EN.east);

    \node at (8.2,1.2) {\scriptsize $2\D$};
    \node at (9,0.4) {\scriptsize $1\texttt{-}\D$};
    \node at (8.2,-0.7) {\scriptsize $F \D$};
    \node at (11.7,0) {\scriptsize $+\{2N\D,2(1\texttt{-}N)\D\}$};

    \node at (1.5,-1.3) 
    {\begin{tikzpicture}
      \node [rotate=90] {\textcolor{black}{$\Longleftrightarrow$}};    
    \end{tikzpicture}
    };
    \node at (2.2,-1.3) {\scriptsize $\text{BBP}_1^+$};

    \node at (2.3,-4) 
    {\begin{tikzpicture}
      \node [rotate=135] {\textcolor{black}{$\Longleftrightarrow$}};    
    \end{tikzpicture}
    };
    \node at (1.3,-4) {\scriptsize $\text{Aharony}$};

    \node at (8.7,-4) 
    {\begin{tikzpicture}
      \node [rotate=45] {\textcolor{black}{$\Longleftrightarrow$}};    
    \end{tikzpicture}
    };
    \node at (9.5,-4) {\scriptsize $\text{BBP}_1^-$};

    \node at (9,-1.3) 
    {\begin{tikzpicture}
      \node [rotate=90] {\textcolor{black}{$\Longleftrightarrow$}};    
    \end{tikzpicture}
    };
    \node at (9.9,-1.3) {\scriptsize $\text{Aharony}$};

    \node at (5.3,0) 
    {\begin{tikzpicture}
      \node [rotate=0] {\Large \textcolor{black}{$\Longleftrightarrow$}};    
    \end{tikzpicture}
    };
    \node at (5.3,0.5) { $\text{Dual}$};
    
\end{tikzpicture}
\caption{\label{fig: SQCD confine} Duality moves from the $U(N)$ adjoint SQCD to the $U(N-1)$ adjoint SQCD for a specific value of the FI-parameter.}
\end{figure}
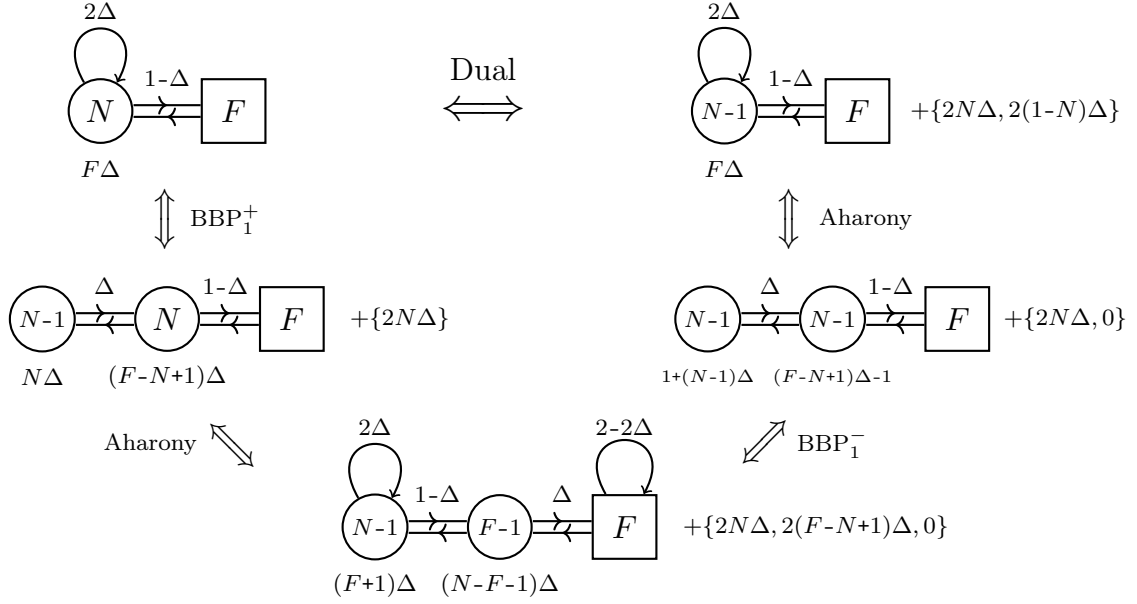

As an appetizer that will be useful in future steps, consider a chain of dualities described in Figure \ref{fig: SQCD confine}. The left-top theory is the $U(N)$ adjoint SQCD with $F$ flavors at a specific value of the FI parameter, $F\Delta$.\footnote{This value of the FI parameter implies a non-zero VEV of the highest dressed monopole operator $\hat V_{N-1}^-$ of negative $U(1)_T$ topological charge. This can be realized by superpotential $W = \xi\,\hat{V}_{N-1}^{-} + \xi$ with auxiliary singlet $\xi$, whose equation of motion requires $\langle\hat{V}_{N-1}^{-}\rangle \neq 0$.} The duality chain shows this left-top theory is dual to the right-top theory, the $U(N-1)$ adjoint SQCD with the same flavor and the same value of the FI parameter but with two extra singlets. Along the chain, any pair of singlets become massive and integrated out if their R-charges sum up to 2 as well as the other symmetry charges cancel, which is realized, e.g.,  in the $S^3_b$ partition function as the following relation:\footnote{This is due to the property of the double sine function
\begin{align}
s_b(z)s_b(-z) = 1\,.
\end{align}
}
\begin{align}
s_b\Big(\frac{iQ}{2}(1-x)\Big) s_b\Big(\frac{iQ}{2}(1-y)\Big) = 1
    \qquad \text{if}\quad x+y=2
\end{align}
where $x$ and $y$ are normalized real mass parameters defined by
\begin{align}
x = \frac{2}{iQ} \sum Q_i m_i+r
\end{align}
with the real masses $m_i$, their associated symmetry charges $Q_i$, and R-charge $r$. Namely, a (unordered) set of singlet fields can be represented by their normalized real mass parameters, satisfying the equivalence relation
\begin{align}
    \{x_1,\dots,x_{n-1},x_n\} \sim \{x_1,\dots,x_{n-2}\} \qquad \text{if} \quad x_{n-1}+x_n=2 \,.
\end{align}
Notice that the cost of reducing the rank by one is just introducing two singlets, and we can repeatedly employ the duality chain in Figure \ref{fig: SQCD confine} to obtain the WZ theory with a set of singlets, namely,
\begin{equation}
    \begin{tikzpicture}[thick,scale=1.1, every node/.style={scale=1.1}]
    \tikzset{vertex/.style={circle,fill=white!25,minimum size=12pt,inner sep=2pt}}
      \tikzset{every loop/.style={}}
        \node[vertex] (AN) at (0,0) [shape=circle,draw=black,minimum size=2em] {$N$};
        \node[vertex] (AF) at (1.5,0) [shape=rectangle,draw=black,minimum height=2em, minimum width=2em] {$F$};
    
        \draw[-to, min distance=1cm]  (AN) edge [out=120, in=60] node {} (AN);
        \draw[->-=.5] ([yshift= 2pt] AN.east) to ([yshift= 2pt] AF.west);
        \draw[->-=.5] ([yshift= -2pt] AF.west) to ([yshift= -2pt] AN.east);
    
        \node at (0,1.2) {\scriptsize $2\D$};
        \node at (0.75,0.4) {\scriptsize $1\texttt{-}\D$};
        \node at (0,-0.7) {\scriptsize $F \D$};

        \node at (3.5,0.4) {\text{Dual}};
        \node at (3.5,0) {\Large $\Longleftrightarrow$};

        \node at (7.5,0) {$+ \big\{\; 2j\D  \;,\; -2(j-1)\D \;\big\}_{j=1}^{N} $};
    \end{tikzpicture}
    \label{eq: SQCD confine}
\end{equation}
which shows the confinement of the $U(N)$ adjoint SQCD at a specific FI parameter. We will utilize this result in the following derivation.
\paragraph{Derivation for $N=p-1$\\}
\begin{figure}[tbp]
\centering
\begin{tikzpicture}[thick,scale=1.1, every node/.style={scale=1.1}]
\tikzset{vertex/.style={circle,fill=white!25,minimum size=12pt,inner sep=2pt}}
  \tikzset{every loop/.style={}}
    \node[vertex] (A1) at (0,0) [shape=circle,draw=black,minimum size=2em] {$1$};
    \node[vertex] (A2) at (1.2,0) [shape=circle,draw=black,minimum size=2em] {$2$};
    \node[vertex] (Acd) at (2.4,0)  {$\cdots$};
    \node[vertex] (Ap) at (3.6,0) [shape=circle,draw=black,minimum size=2em] { $\!p\texttt{-}2\!$};
    \node[vertex] (AF) at (4.8,0) [shape=rectangle,draw=black,minimum height=2em, minimum width=2em] {$\!p\texttt{-}1\!$};

    \draw[-to, min distance=0.8cm]  (A1) edge [out=120, in=60] node {} (A1);
    \draw[-to, min distance=0.8cm]  (A2) edge [out=120, in=60] node {} (A2);
    \draw[-to, min distance=0.8cm]  (Ap) edge [out=120, in=60] node {} (Ap);
    \draw[->-=.5] ([yshift= 2pt] A1.east) to ([yshift= 2pt] A2.west);
    \draw[->-=.5] ([yshift= -2pt] A2.west) to ([yshift= -2pt] A1.east);
    \draw[->-=.5] ([yshift= 2pt] A2.east) to ([yshift= 2pt] Acd.west);
    \draw[->-=.5] ([yshift= -2pt] Acd.west) to ([yshift= -2pt] A2.east);
    \draw[->-=.5] ([yshift= 2pt] Acd.east) to ([yshift= 2pt] Ap.west);
    \draw[->-=.5] ([yshift= -2pt] Ap.west) to ([yshift= -2pt] Acd.east);
    \draw[->-=.5] ([yshift= 2pt] Ap.east) to ([yshift= 2pt] AF.west);
    \draw[->-=.5] ([yshift= -2pt] AF.west) to ([yshift= -2pt] Ap.east);

    \node at (0.6,0.4) {\scriptsize $\D$};
    \node at (1.8,0.4) {\scriptsize $\D$};
    \node at (3,0.4) {\scriptsize $\D$};
    \node at (4.2,0.4) {\scriptsize $\D$};
    \node at (0,1.1) {\scriptsize $2\texttt{-}2\D$};
    \node at (1.2,1.1) {\scriptsize $2\texttt{-}2\D$};
    \node at (3.6,1.1) {\scriptsize $2\texttt{-}2\D$};
    \node at (5.1,0.8) {$+\{2\D\}$};

    \node at (0,-0.7) {\scriptsize $2\D$};
    \node at (1.2,-0.7) {\scriptsize $2\D$};
    \node at (3.6,-0.7) {\scriptsize $2\D$};

    \node[vertex] (B1) at (0+2.5,-3) [shape=circle,draw=black,minimum size=2em] {$j$};
    \node[vertex] (B2) at (1.2+2.5,-3) [shape=circle,draw=black,minimum size=2em] {\small $j\texttt{+}1$};
    \node[vertex] (Bcd) at (2.4+2.5,-3)  {$\cdots$};

    \draw[-to, min distance=0.8cm]  (B2) edge [out=120, in=60] node {} (B2);
    \draw[->-=.5] ([yshift= 2pt] B1.east) to ([yshift= 2pt] B2.west);
    \draw[->-=.5] ([yshift= -2pt] B2.west) to ([yshift= -2pt] B1.east);
    \draw[->-=.5] ([yshift= 2pt] B2.east) to ([yshift= 2pt] Bcd.west);
    \draw[->-=.5] ([yshift= -2pt] Bcd.west) to ([yshift= -2pt] B2.east);

    \node at (0.6+2.5,0.4-3) {\scriptsize $\D$};
    \node at (1.8+2.5,0.4-3) {\scriptsize $\D$};
    \node at (1.2+2.5,1.1-3) {\scriptsize $2\texttt{-}2\D$};

    \node at (3.4+2.5,0.4-3) {\scriptsize $\text{BBP}_1^+$};
    \node at (3.4+2.5,-3) {$\Longleftrightarrow$};
    \node at (3.4+2.5,-0.4-3) {\scriptsize$j\!=\!1,\!\cdots\!,p\texttt{-}2$};

    \node at (0+2.5,-0.7-3) {\scriptsize $(j\texttt{+}1)\D$};
    \node at (1.2+2.5,-0.7-3) {\scriptsize $2\D$};

    \node[vertex] (b2) at (4.8+2.5,-3) [shape=circle,draw=black,minimum size=2em] {\small $j\texttt{+}1$};
    \node[vertex] (bcd) at (6+2.5,-3)  {$\cdots$};

    \draw[->-=.5] ([yshift= 2pt] b2.east) to ([yshift= 2pt] bcd.west);
    \draw[->-=.5] ([yshift= -2pt] bcd.west) to ([yshift= -2pt] b2.east);

    \node at (5.4+2.5,0.4-3) {\scriptsize $\D$};
    \node at (4.8+2.5,-0.7-3) {\scriptsize $(j\texttt{+}2)\D$};
    \node at (5.7+2.5,0.8-3) {\scriptsize $+\{2\D(p\texttt{-}j)\}$};

    \node[vertex] (FF) at (11,0) [shape=rectangle,draw=black,minimum height=2em, minimum width=2em] { $\!p\texttt{-}1\!$};
    \draw[-to, min distance=0.8cm]  (FF) edge [out=120, in=60] node {} (FF);

    \node at (11,1.1) {\scriptsize $2\D$};


    \node at (8,-0.2) 
    {\begin{tikzpicture}
      \node [rotate=0] {\Large\textcolor{black}{$\Longleftrightarrow$}};    
    \end{tikzpicture}
    };
    \node at (8,0.2) {Dual};

    \node at (1,-1.7) 
    {\begin{tikzpicture}
      \node [rotate=-45] {\Large\textcolor{black}{$\Leftarrow\!\cdots$}};
     \end{tikzpicture}
    };

    \node at (10,-1.5) 
    {\begin{tikzpicture}
      \node [rotate=225] {\Large\textcolor{black}{$\Leftarrow\!\cdots$}};
     \end{tikzpicture}
    };
    
\end{tikzpicture}
\caption{\label{fig: N=p-1 confine} A diagram for the derivation of the confinement of the $\mathbb{D}_p[SU(p-1)]$ theory with a confining superpotential deformation $\D W_\text{conf}^{(p),+}$.}
\end{figure}
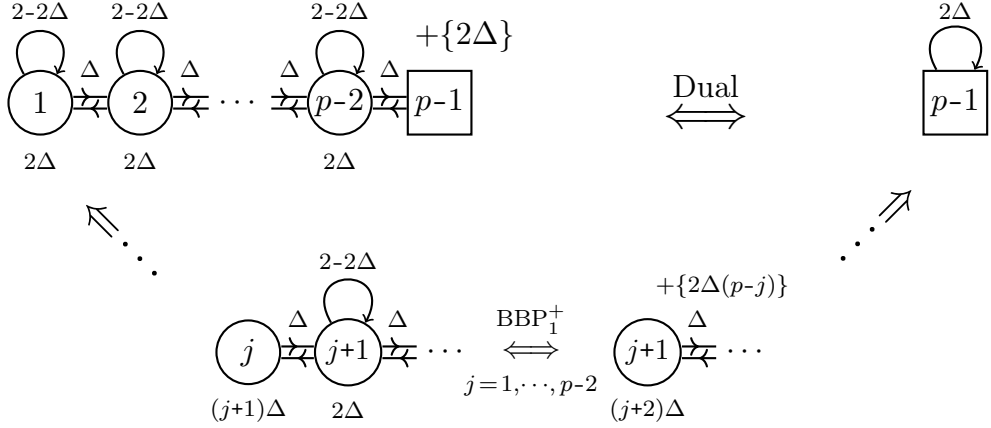
Consider $\mathbb{D}_p[SU(p-1)]$ for $p>2$,\footnote{$p = 1$ is excluded since $N = p-1$ is assumed to be larger than 1.} with the confining superpotential deformation $\D W_{\text{conf}}^{(p),+}$,\footnote{Derivation for the other deformation by $\D W_{\text{conf}}^{(p),-}$ is similar. The only difference is the sign of the FI parameter.}
\begin{align}
    \D W_\text{conf}^{(p),+} = 
    \eta^{p+1}+\eta \sum_{i=2}^{p-1} \Tr \Phi^{(i)} 
    +
    \sum_{i=2}^{p-1} \hat{v}^{(i),+}
    +
    \hat{v}^{(2,p\texttt{-}1),-}
    \sum_{i=2}^{p-1}
    \hat{v}^{(i),-} \,,
    \label{eq: N=p-1 superpotential}
\end{align}
which is merely the $\mathcal N=2$ deformed $T[SU(p-1)]$ theory whose normalized real masses and FI parameters are fixed as the left-top theory in Figure \ref{fig: N=p-1 confine} with $\D = \frac{1}{p+1}$. The singlet contribution $+\{2\D\}$ corresponds to $\eta$, which cancels that of the adjoint of the left-most $U(1)$ gauge node. Since there is no adjoint for this $U(1)$ node, we can apply the BBP duality with $W = V^+$, i.e., $\text{BBP}_1^+$,\footnote{There is also a BBP duality with two linear monopole superpotential terms, $W = V^++V^-$ \cite{Benini:2017dud}.} which removes the adjoint of the adjacent node. Hence, we can repeatedly apply the $\text{BBP}_1^+$ duality along the quiver, truncating all the gauge nodes from left to right as described in the second line of Figure \ref{fig: N=p-1 confine} for $j=1,\cdots,p-2$. While these steps generate $p-2$ singlet contributions, $+\{2\D(p-j)\}_{j=1}^{p-2}$, they exactly pair up to completely cancel each other. So, the resulting theory at the right-top corner has only a single $U(p-1)$ flavor adjoint chiral with R-charge $\frac{2}{p+1}$, which confirms the confinement of the $\mathbb{D}_p[SU(p-1)]$ theory with confining superpotential deformation \eqref{eq: N=p-1 superpotential}.
\paragraph{Derivation for $N=p+1$\\}
\begin{figure}[tbp]
\centering
\begin{tikzpicture}[thick,scale=1.1, every node/.style={scale=1.1}]
\tikzset{vertex/.style={circle,fill=white!25,minimum size=12pt,inner sep=2pt}}
  \tikzset{every loop/.style={}}
    \node[vertex] (A1) at (0,0) [shape=circle,draw=black,minimum size=2em] {$1$};
    \node[vertex] (A2) at (1.2,0) [shape=circle,draw=black,minimum size=2em] {$2$};
    \node[vertex] (Acd) at (2.4,0)  {$\cdots$};
    \node[vertex] (Ap2) at (3.6,0) [shape=circle,draw=black,minimum size=2em] { $\!p\texttt{-}2\!$};
    \node[vertex] (Ap1) at (4.8,0) [shape=circle,draw=black,minimum size=2em] { $\!p\texttt{-}1\!$};
    \node[vertex] (AF) at (6,0) [shape=rectangle,draw=black,minimum height=2em, minimum width=2em] { $\!p\texttt{+}1\!$};

    \draw[-to, min distance=0.8cm]  (A1) edge [out=120, in=60] node {} (A1);
    \draw[-to, min distance=0.8cm]  (A2) edge [out=120, in=60] node {} (A2);
    \draw[-to, min distance=0.8cm]  (Ap2) edge [out=120, in=60] node {} (Ap2);
    \draw[-to, min distance=0.8cm]  (Ap1) edge [out=120, in=60] node {} (Ap1);
    \draw[->-=.5] ([yshift= 2pt] A1.east) to ([yshift= 2pt] A2.west);
    \draw[->-=.5] ([yshift= -2pt] A2.west) to ([yshift= -2pt] A1.east);
    \draw[->-=.5] ([yshift= 2pt] A2.east) to ([yshift= 2pt] Acd.west);
    \draw[->-=.5] ([yshift= -2pt] Acd.west) to ([yshift= -2pt] A2.east);
    \draw[->-=.5] ([yshift= 2pt] Acd.east) to ([yshift= 2pt] Ap2.west);
    \draw[->-=.5] ([yshift= -2pt] Ap2.west) to ([yshift= -2pt] Acd.east);
    \draw[->-=.5] ([yshift= 2pt] Ap2.east) to ([yshift= 2pt] Ap1.west);
    \draw[->-=.5] ([yshift= -2pt] Ap1.west) to ([yshift= -2pt] Ap2.east);
    \draw[->-=.5] ([yshift= 2pt] Ap1.east) to ([yshift= 2pt] AF.west);
    \draw[->-=.5] ([yshift= -2pt] AF.west) to ([yshift= -2pt] Ap1.east);

    \node at (0.6,0.4) {\scriptsize $\D$};
    \node at (1.8,0.4) {\scriptsize $\D$};
    \node at (3,0.4) {\scriptsize $\D$};
    \node at (4.2,0.4) {\scriptsize $\D$};
    \node at (5.4,0.4) {\scriptsize $\D$};
    \node at (0,1.1) {\scriptsize $2\texttt{-}2\D$};
    \node at (1.2,1.1) {\scriptsize $2\texttt{-}2\D$};
    \node at (3.6,1.1) {\scriptsize $2\texttt{-}2\D$};
    \node at (4.8,1.1) {\scriptsize $2\texttt{-}2\D$};
    \node at (6.3,0.8) {$+\{2\D\}$};

    \node at (0,-0.7) {\scriptsize $2\D$};
    \node at (1.2,-0.7) {\scriptsize $2\D$};
    \node at (3.6,-0.7) {\scriptsize $2\D$};
    \node at (4.8,-0.7) {\scriptsize $(2\texttt{-}p)\D$};

    \node[vertex] (Cp2) at (5-3.5,-4) [shape=circle,draw=black,minimum size=2em] { $\!p\texttt{-}2\!$};
    \node[vertex] (Cp1) at (6.3-3.5,-4) [shape=circle,draw=black,minimum size=2em] { $\!p\texttt{-}1\!$};
    \node[vertex] (CF) at (7.6-3.5,-4) [shape=rectangle,draw=black,minimum height=2em, minimum width=2em] { $\!p\texttt{+}1\!$};

    \draw[-to, min distance=0.8cm]  (Cp1) edge [out=120, in=60] node {} (Cp1);
    \draw[->-=.5] ([yshift= 2pt] Cp2.east) to ([yshift= 2pt] Cp1.west);
    \draw[->-=.5] ([yshift= -2pt] Cp1.west) to ([yshift= -2pt] Cp2.east);
    \draw[->-=.5] ([yshift= 2pt] Cp1.east) to ([yshift= 2pt] CF.west);
    \draw[->-=.5] ([yshift= -2pt] CF.west) to ([yshift= -2pt] Cp1.east);

    \node at (5+0.65-3.5,0.4-4) {\scriptsize $\D$};
    \node at (6.3+0.65-3.5,0.4-4) {\scriptsize $\D$};
    \node at (6.3-3.5,1.1-4) {\scriptsize $2\texttt{-}2\D$};
    \node at (8.2-3.5,0.8-4) { $+\{2\texttt{-}2j\D\}_{j=2}^{p\texttt{-}2}$};

    \node at (5-3.5,-0.7-4) {\scriptsize $(p\texttt{-}1)\D$};
    \node at (6.3-3.5,-0.7-4) {\scriptsize $(2\texttt{-}p)\D$};

    \node[vertex] (Dp1) at (9.4-4.5,-6.5) [shape=circle,draw=black,minimum size=2em] { $\!p\texttt{-}1\!$};
    \node[vertex] (DF) at (10.6-4.5,-6.5) [shape=rectangle,draw=black,minimum height=2em, minimum width=2em] { $\!p\texttt{+}1\!$};

    \draw[->-=.5] ([yshift= 2pt] Dp1.east) to ([yshift= 2pt] DF.west);
    \draw[->-=.5] ([yshift= -2pt] DF.west) to ([yshift= -2pt] Dp1.east);

    \node at (10-4.5,0.4-6.5) {\scriptsize $\D$};
    \node at (10.7-4.5,1.2-6.5) {$+\{\overbrace{2\texttt{-}2j\D}^{\text{cancel}}\}_{j=2}^{p\texttt{-}1}$};
    \node at (9.4-4.5,-0.7-6.5) {\scriptsize $0$};

    \node[vertex] (F1) at (9.4,-4) [shape=circle,draw=black,minimum size=2em] { $1$};
    \node[vertex] (FF) at (10.7,-4) [shape=rectangle,draw=black,minimum height=2em, minimum width=2em] { $\!p\texttt{+}1\!$};

    \draw[-to, min distance=0.8cm]  (F1) edge [out=120, in=60] node {} (F1);
    \draw[-to, min distance=0.8cm]  (FF) edge [out=120, in=60] node {} (FF);
    \draw[->-=.5] ([yshift= 2pt] F1.east) to ([yshift= 2pt] FF.west);
    \draw[->-=.5] ([yshift= -2pt] FF.west) to ([yshift= -2pt] F1.east);

    \node at (9.4,1.1-4) {\scriptsize $2\D$};
    \node at (10,0.4-4) {\tiny $1\texttt{-}\D$};
    \node at (10.7,1.1-4) {\scriptsize $2\D$};
    \node at (10.7,1.5-4) {$+\{2,2\texttt{-}2\D\}$};
    \node at (9.4,-0.7-4) {\scriptsize $(p\texttt{+}1)\D$};

    \node[vertex] (EF) at (10,0) [shape=rectangle,draw=black,minimum height=2em, minimum width=2em] { $\!p\texttt{+}1\!$};
    \draw[-to, min distance=0.8cm]  (EF) edge [out=120, in=60] node {} (EF);

    \node at (10,1.1) {\scriptsize $2\D$};

    \node at (2,-1.7) 
    {\begin{tikzpicture}
      \node [rotate=90] {\textcolor{black}{$\Leftarrow \cdots \Rightarrow$}};    
    \end{tikzpicture}
    };
    \node at (1.3,-1.7) {\scriptsize $\text{BBP}_1^+$};

    \node at (3,-5.5) 
    {\begin{tikzpicture}
      \node [rotate=135] {\textcolor{black}{$\Longleftrightarrow$}};    
    \end{tikzpicture}
    };
    \node at (2,-5.6) {\scriptsize $\text{BBP}_1^+$};

    \node at (8.5,-5.5) 
    {\begin{tikzpicture}
      \node [rotate=45] {\textcolor{black}{$\Longleftrightarrow$}};    
    \end{tikzpicture}
    };
    \node at (9.5,-5.6) {\scriptsize $\text{BBP}_1^+$};

    \node at (10,-1.4) 
    {\begin{tikzpicture}
      \node [rotate=90] {\textcolor{black}{$\Longleftrightarrow$}};    
    \end{tikzpicture}
    };
    \node at (10.7,-1.4) {\scriptsize $\eqref{eq: SQCD confine}$};

    \node at (8,-0.2) 
    {\begin{tikzpicture}
      \node [rotate=0] {\Large\textcolor{black}{$\Longleftrightarrow$}};    
    \end{tikzpicture}
    };
    \node at (8,0.2) {Dual};
\end{tikzpicture}
\caption{\label{fig: N=p+1 confine} A diagram for the derivation of the confinement of the $\mathbb{D}_p[SU(p+1)]$ theory with a confining superpotential deformation $\D W_\text{conf}^{(p),+}$.}
\end{figure}
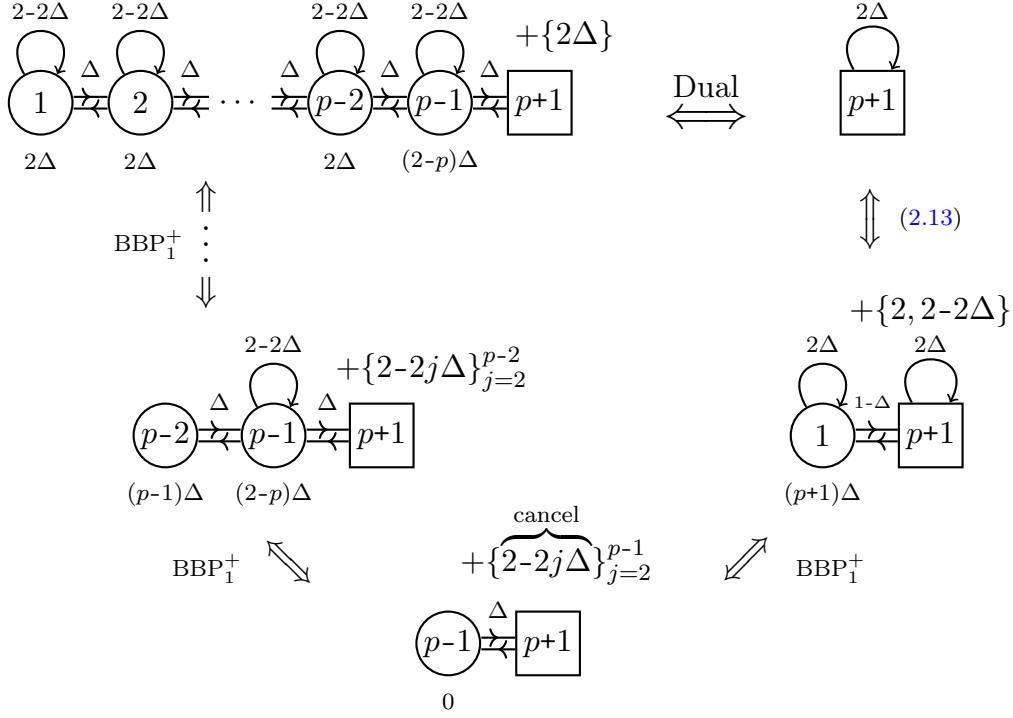
Now, consider the $\mathbb{D}_p[SU(p+1)]$ theory for $p>1$ with confining superpotential deformation $\D W_{\text{conf}}^{(p),+}$,
\begin{align}
    \D W_{\text{conf}}^{(p),+} = 
    \eta^{p+1}+\eta \sum_{i=1}^{p-1} \Tr \Phi^{(i)}
    + \sum_{i=1}^{p-1} \hat{v}^{(i),+}
    + \hat{v}^{(1,p\texttt{-}1),-} \,,
    \label{eq: N=p+1 confining potential}
\end{align}
which fixes the FI parameter of each gauge node as in the left-top theory in Figure \ref{fig: N=p+1 confine} with $\D = \frac{1}{p+1}$. In the same manner as the $N=p-1$ case,
the tail is truncated as we repeatedly apply the $\text{BBP}_1^+$ duality starting from the left-most gauge node to the right, obtaining the bottom theory, where all the generated singlet contributions completely cancel. Further applying the $\text{BBP}_1^+$ duality gives the right-middle theory, whose FI parameter satisfies the condition for the rank-reducing duality chain in \eqref{eq: SQCD confine}, which produces two singlet contributions $+\{0,2\D\}$, exactly canceling those of the right-middle theory. Hence, we obtain the resulting right-top theory, a WZ model of a single $U(p+1)$ flavor adjoint chiral of R-charge $2\D=\frac{2}{p+1}$, confirming the confinement of $\mathbb{D}_p[SU(p+1)]$ theory with confining superpotential \eqref{eq: N=p+1 confining potential}.
\paragraph{Derivation for $N \to N+p$\\}
\begin{figure}[tbp]
\centering
\begin{tikzpicture}[thick,scale=1.1, every node/.style={scale=1.1}]
\tikzset{vertex/.style={circle,fill=white!25,minimum size=12pt,inner sep=2pt}}
  \tikzset{every loop/.style={}}
    \node[vertex] (A1) at (0,0) [shape=circle,draw=black,minimum size=2em] {$m_1$};
    \node[vertex] (A2) at (1.2,0) [shape=circle,draw=black,minimum size=2em] {$m_2$};
    \node[vertex] (Acd) at (2.4,0)  {$\cdots$};
    \node[vertex] (Ap1) at (3.6,0) [shape=circle,draw=black,minimum size=2em] {\scriptsize $\!m_{p\texttt{-}1}\!$};
    \node[vertex] (AF) at (4.8,0) [shape=rectangle,draw=black,minimum height=2em, minimum width=2em] {\small $\!N\texttt{+}p\!$};

    \draw[-to, min distance=0.8cm]  (A1) edge [out=120, in=60] node {} (A1);
    \draw[-to, min distance=0.8cm]  (A2) edge [out=120, in=60] node {} (A2);
    \draw[-to, min distance=0.8cm]  (Ap1) edge [out=120, in=60] node {} (Ap1);
    \draw[->-=.5] ([yshift= 2pt] A1.east) to ([yshift= 2pt] A2.west);
    \draw[->-=.5] ([yshift= -2pt] A2.west) to ([yshift= -2pt] A1.east);
    \draw[->-=.5] ([yshift= 2pt] A2.east) to ([yshift= 2pt] Acd.west);
    \draw[->-=.5] ([yshift= -2pt] Acd.west) to ([yshift= -2pt] A2.east);
    \draw[->-=.5] ([yshift= 2pt] Acd.east) to ([yshift= 2pt] Ap1.west);
    \draw[->-=.5] ([yshift= -2pt] Ap1.west) to ([yshift= -2pt] Acd.east);
    \draw[->-=.5] ([yshift= 2pt] Ap1.east) to ([yshift= 2pt] AF.west);
    \draw[->-=.5] ([yshift= -2pt] AF.west) to ([yshift= -2pt] Ap1.east);

    \node at (0.6,0.4) {\scriptsize $\D$};
    \node at (1.8,0.4) {\scriptsize $\D$};
    \node at (3,0.4) {\scriptsize $\D$};
    \node at (4.2,0.4) {\scriptsize $\D$};
    \node at (0,1.1) {\scriptsize $2\texttt{-}2\D$};
    \node at (1.2,1.1) {\scriptsize $2\texttt{-}2\D$};
    \node at (3.6,1.1) {\scriptsize $2\texttt{-}2\D$};
    \node at (5.1,0.8) {$+\{2\D\}$};

    \node at (0,-0.7) {\scriptsize $\t_1$};
    \node at (1.2,-0.7) {\scriptsize $\t_2$};
    \node at (3.6,-0.7) {\scriptsize $\t_{p\texttt{-}1}$};

    \node[vertex] (B1) at (-0.7,-3.5) [shape=circle,draw=black,minimum size=2em] {\scriptsize$\!m_1\texttt{-}\!1\!$};
    \node[vertex] (B2) at (0.5,-3.5) [shape=circle,draw=black,minimum size=2em] {$m_1$};
    \node[vertex] (Bcd) at (1.7,-3.5) [shape=circle,draw=black,minimum size=2em] {$m_2$};
    \node[vertex] (Bp2) at (2.9,-3.5) {$\cdots$};
    \node[vertex] (Bp1) at (4.1,-3.5) [shape=circle,draw=black,minimum size=2em] {\scriptsize $\!m_{p\texttt{-}1}\!$};
    \node[vertex] (BF) at (5.3,-3.5) [shape=rectangle,draw=black,minimum height=2em, minimum width=2em] {\small $\!N\texttt{+}p\!$};

    \draw[-to, min distance=0.8cm]  (Bcd) edge [out=120, in=60] node {} (Bcd);
    \draw[-to, min distance=0.8cm]  (Bp1) edge [out=120, in=60] node {} (Bp1);
    \draw[->-=.5] ([yshift= 2pt] B1.east) to ([yshift= 2pt] B2.west);
    \draw[->-=.5] ([yshift= -2pt] B2.west) to ([yshift= -2pt] B1.east);
    \draw[->-=.5] ([yshift= 2pt] B2.east) to ([yshift= 2pt] Bcd.west);
    \draw[->-=.5] ([yshift= -2pt] Bcd.west) to ([yshift= -2pt] B2.east);
    \draw[->-=.5] ([yshift= 2pt] Bcd.east) to ([yshift= 2pt] Bp2.west);
    \draw[->-=.5] ([yshift= -2pt] Bp2.west) to ([yshift= -2pt] Bcd.east);
    \draw[->-=.5] ([yshift= 2pt] Bp2.east) to ([yshift= 2pt] Bp1.west);
    \draw[->-=.5] ([yshift= -2pt] Bp1.west) to ([yshift= -2pt] Bp2.east);
    \draw[->-=.5] ([yshift= 2pt] Bp1.east) to ([yshift= 2pt] BF.west);
    \draw[->-=.5] ([yshift= -2pt] BF.west) to ([yshift= -2pt] Bp1.east);

    \node at (-0.1,0.4-3.5) {\scriptsize $1\texttt{-}\D$};
    \node at (1.1,0.4-3.5) {\scriptsize $\D$};
    \node at (2.3,0.4-3.5) {\scriptsize $\D$};
    \node at (3.5,0.4-3.5) {\scriptsize $\D$};
    \node at (4.7,0.4-3.5) {\scriptsize $\D$};
    \node at (1.7,1.1-3.5) {\scriptsize $2\texttt{-}2\D$};
    \node at (4.1,1.1-3.5) {\scriptsize $2\texttt{-}2\D$};
    \node at (5.7,1.3-3.5) {$+\{2\D\}$};
    \node at (5.7,0.8-3.5) {$+\{2\texttt{-}2\k_0\}$};

    \node at (-0.7,-0.7-3.5) {\scriptsize $\s_1$};
    \node at (0.5,-0.7-3.5) {\scriptsize $\k_1$};
    \node at (1.7,-0.7-3.5) {\scriptsize $\t_2$};
    \node at (4.1,-0.7-3.5) {\scriptsize $\t_{p\texttt{-}1}$};

    \node[vertex] (C1) at (0+2.5,0-7.5) [shape=circle,draw=black,minimum size=2em] {$n_1$};
    \node[vertex] (Ccd) at (1.2+2.5,0-7.5) {$\cdots$};
    \node[vertex] (Cp1) at (2.4+2.5,0-7.5) [shape=circle,draw=black,minimum size=2em] {$\!n_{p\texttt{-}1}\!$};
    \node[vertex] (CN) at (3.6+2.5,0-7.5) [shape=circle,draw=black,minimum size=2em] {$N$};
    \node[vertex] (CF) at (4.8+2.5,0-7.5) [shape=rectangle,draw=black,minimum height=2em, minimum width=2em] {\small $\!N\texttt{+}p\!$};

    \draw[-to, min distance=0.8cm]  (C1) edge [out=120, in=60] node {} (C1);
    \draw[-to, min distance=0.8cm]  (Cp1) edge [out=120, in=60] node {} (Cp1);
    \draw[-to, min distance=0.8cm]  (CF) edge [out=120, in=60] node {} (CF);
    \draw[->-=.5] ([yshift= 2pt] C1.east) to ([yshift= 2pt] Ccd.west);
    \draw[->-=.5] ([yshift= -2pt] Ccd.west) to ([yshift= -2pt] C1.east);
    \draw[->-=.5] ([yshift= 2pt] Ccd.east) to ([yshift= 2pt] Cp1.west);
    \draw[->-=.5] ([yshift= -2pt] Cp1.west) to ([yshift= -2pt] Ccd.east);
    \draw[->-=.5] ([yshift= 2pt] Cp1.east) to ([yshift= 2pt] CN.west);
    \draw[->-=.5] ([yshift= -2pt] CN.west) to ([yshift= -2pt] Cp1.east);
    \draw[->-=.5] ([yshift= 2pt] CN.east) to ([yshift= 2pt] CF.west);
    \draw[->-=.5] ([yshift= -2pt] CF.west) to ([yshift= -2pt] CN.east);

    \node at (0.6+2.5,0.4-7.5) {\scriptsize $\D$};
    \node at (1.8+2.5,0.4-7.5) {\scriptsize $\D$};
    \node at (3+2.5,0.4-7.5) {\scriptsize $\D$};
    \node at (4.1+2.5,0.4-7.5) {\tiny $1\texttt{-}\D$};
    \node at (0+2.5,1.1-7.5) {\scriptsize $2\texttt{-}2\D$};
    \node at (2.4+2.5,1.1-7.5) {\scriptsize $2\texttt{-}2\D$};
    \node at (4.8+2.5,1.1-7.5) {\scriptsize $2\D$};
    \node at (6.3+2.5,0.5-7.5) {$+\{2\D\}$};
    \node at (6.5+2.5,0-7.5) {$+\{2\texttt{-}2\k_j\}_{j=0}^{p\texttt{-}1}$};

    \node at (0+2.5,-0.7-7.5) {\scriptsize $\t_1$};
    \node at (2.4+2.5,-0.7-7.5) {\scriptsize $\t_{p\texttt{-}1}$};
    \node at (3.6+2.5,-0.7-7.5) {\scriptsize $\s_p$};

    \node[vertex] (DN) at (9.2,-4) [shape=circle,draw=black,minimum size=2em] { $N$};
    \node[vertex] (DF) at (10.7,-4) [shape=rectangle,draw=black,minimum height=2em, minimum width=2em] {\small $\!N\texttt{+}p\!$};

    \draw[-to, min distance=0.8cm]  (DN) edge [out=120, in=60] node {} (DN);
    \draw[-to, min distance=0.8cm]  (DF) edge [out=120, in=60] node {} (DF);
    \draw[->-=.5] ([yshift= 2pt] DN.east) to ([yshift= 2pt] DF.west);
    \draw[->-=.5] ([yshift= -2pt] DF.west) to ([yshift= -2pt] DN.east);

    \node at (9.9,0.4-4) {\scriptsize $1\texttt{-}\D$};
    \node at (9.2,1.1-4) {\scriptsize $2\D$};
    \node at (10.7,1.1-4) {\scriptsize $2\D$};
    
    \node at (10.3,-2.4) {\small $+\{2\D(\!N\texttt{+}j)\}_{j=1}^{p}$};
    \node at (9.2,-0.7-4) {\scriptsize $(N\texttt{+}p)\D$};

    \node[vertex] (EF) at (10,0) [shape=rectangle,draw=black,minimum height=2em, minimum width=2em] {\small $\!N\texttt{+}p\!$};
    \draw[-to, min distance=0.8cm]  (EF) edge [out=120, in=60] node {} (EF);

    \node at (10,1.1) {\scriptsize $2\D$};

    \node at (0.5,-1.5) 
    {\begin{tikzpicture}
      \node [rotate=90] {\textcolor{black}{$\Longleftrightarrow$}};    
    \end{tikzpicture}
    };
    \node at (1.3-1.5,-1.5) {\scriptsize $\text{BBP}_1^+$};

    \node at (2.5,-5.3) 
    {\begin{tikzpicture}
      \node [rotate=135] {\textcolor{black}{$\Leftarrow \!\cdots\! \Rightarrow$}};    
    \end{tikzpicture}
    };
    \node at (1.5,-5.3) {\scriptsize $\text{BBP}_1^+$};
    \node at (3.5,-5.2) {\scriptsize $n_j \equiv m_j \texttt{-} j$};

    \node at (8,-5.5) 
    {\begin{tikzpicture}
      \node [rotate=45] {\textcolor{black}{$\Longleftrightarrow$}};    
    \end{tikzpicture}
    };
    \node at (10,-5.5) {\scriptsize $\mathbb{D}_p[SU(N)]$ confinement};

    \node at (10,-1.2) 
    {\begin{tikzpicture}
      \node [rotate=90] {\textcolor{black}{$\Longleftrightarrow$}};    
    \end{tikzpicture}
    };
    \node at (10.7,-1.2) {\scriptsize \eqref{eq: SQCD confine}};

    \node at (7.5,-0.2) 
    {\begin{tikzpicture}
      \node [rotate=0] {\Large\textcolor{black}{$\Longleftrightarrow$}};    
    \end{tikzpicture}
    };
    \node at (7.5,0.2) {Dual};
\end{tikzpicture}
\caption{\label{fig: N+p confine} A diagram for the derivation of the confinement of the $\mathbb{D}_p[SU(N+p)]$ theory with a confining superpotential deformation $\D W_\text{conf}^{(p),+}$.}
\end{figure}
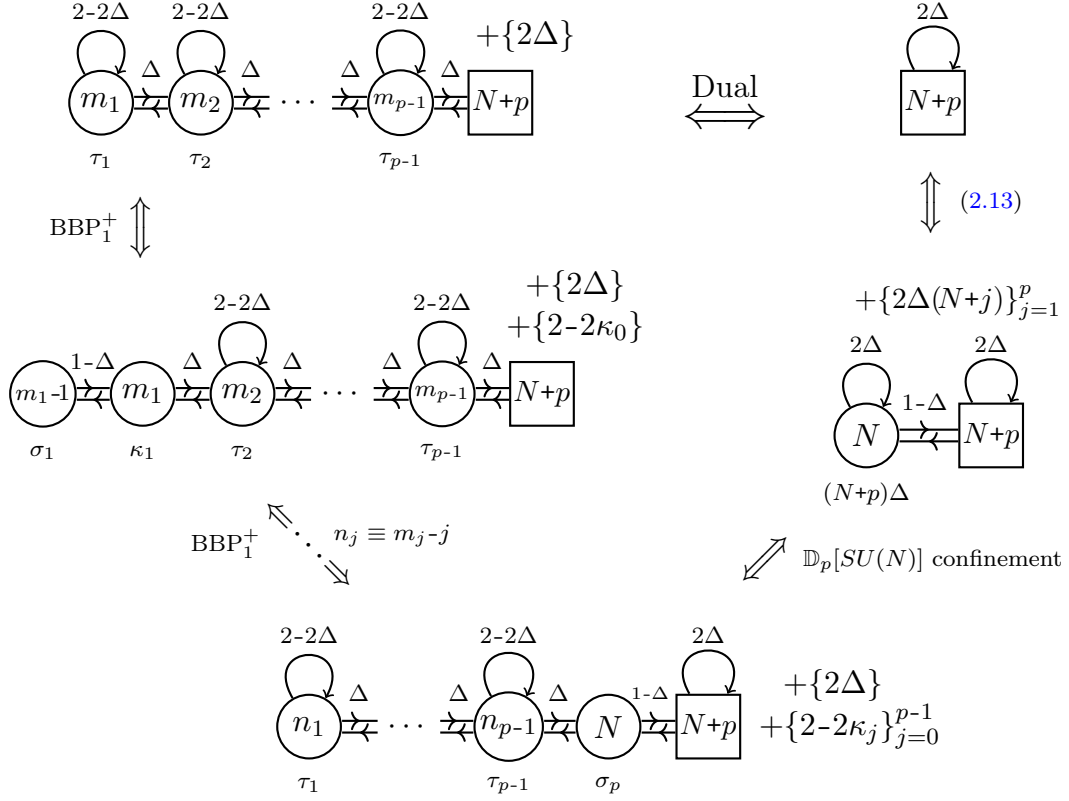
To complete the derivation, let us check the confinement of the $\mathbb{D}_p[SU(N+p)]$ theory assuming the confinement of $\mathbb{D}_p[SU(N)]$. Consider the left-top theory in Figure \ref{fig: N+p confine}, which is the $\mathbb{D}_p[SU(N+p)]$ theory with confining superpotential deformation $\D W_{\text{conf}}^{(p),+}$ that has the same form as \eqref{eq: N=p+1 confining potential}, which fixes the normalized FI parameters as $\t_j$,
\begin{align}
    \t_j = 2-(1-\D)(m_{j\texttt{-}1} - 2 m_j + m_{j\texttt{+}1} + 2 ) \,,\qquad
    j = 1,\cdots,p-1
\end{align}
with $\D = \frac{1}{p+1}$. The adjoint of the left-most gauge node can be deconfined into another gauge node of rank $m_1-1$ by applying the $\text{BBP}_1^+$ duality, producing the left-middle theory, whose second node $U(m_1)$ has no adjoint now. We can repeatedly apply $\text{BBP}_1^+$ from there to the $U(m_{p-1})$ node as follows:
\begin{equation}
\begin{tikzpicture}
  \tikzset{vertex/.style={circle,fill=white!25,minimum size=12pt,inner sep=2pt}}
  \tikzset{every loop/.style={}}
    \node[vertex] (n1) at (0,0) [shape=circle,draw=black,minimum size=2em] {$n_j$};
    \node[vertex] (n2) at (1.5,0) [shape=circle,draw=black,minimum height=2em, minimum width=2em] {$m_j$};
    \node[vertex] (n3) at (3,0) [shape=circle,draw=black,minimum height=2em, minimum width=2em] {\scriptsize $\!m_{j\texttt{+}1}\!$};

    \draw[-to, min distance=1cm]  (n3) edge [out=120, in=60] node {} (n3);
    \draw[->-=.5] ([yshift= 2pt] n1.east) to ([yshift= 2pt] n2.west);
    \draw[->-=.5] ([yshift= -2pt] n2.west) to ([yshift= -2pt] n1.east);
    \draw[->-=.5] ([yshift= 2pt] n2.east) to ([yshift= 2pt] n3.west);
    \draw[->-=.5] ([yshift= -2pt] n3.west) to ([yshift= -2pt] n2.east);

    \node at (-0.7,0) {$\cdots$};
    \node at (3.7,0) {$\cdots$};
    \node at (0.75,0.4) {\scriptsize $1\texttt{-}\D$};
    \node at (2.25,0.4) {\scriptsize $\D$};
    \node at (3,1.2) {\scriptsize $2\texttt{-}2\D$};
    \node at (0,-0.7) {\scriptsize $\s_j$};
    \node at (1.5,-0.7) {\scriptsize $\k_j$};
    \node at (3,-0.7) {\scriptsize $\t_{j\texttt{+}1}$};

    \node at (5.3,0.5) {\scriptsize $\text{BBP}_1^\pm$};
    \node at (5.3,0) {\Large$\Longleftrightarrow$};
    \node at (5.3,-0.5) {\scriptsize $j\!=\!1,\cdots,p\texttt{-}1$};

    \node[vertex] (N1) at (7.6,0) [shape=circle,draw=black,minimum size=2em] {$n_j$};
    \node[vertex] (N2) at (9.1,0) [shape=circle,draw=black,minimum height=2em, minimum width=2em] {\scriptsize $n_{j\texttt{+}1}\!$};
    \node[vertex] (N3) at (10.6,0) [shape=circle,draw=black,minimum height=2em, minimum width=2em] {\scriptsize $\!m_{j\texttt{+}1}\!$};

    \draw[-to, min distance=1cm]  (N1) edge [out=120, in=60] node {} (N1);
    \draw[->-=.5] ([yshift= 2pt] N1.east) to ([yshift= 2pt] N2.west);
    \draw[->-=.5] ([yshift= -2pt] N2.west) to ([yshift= -2pt] N1.east);
    \draw[->-=.5] ([yshift= 2pt] N2.east) to ([yshift= 2pt] N3.west);
    \draw[->-=.5] ([yshift= -2pt] N3.west) to ([yshift= -2pt] N2.east);

    \node at (6.9,0) {$\cdots$};
    \node at (11.3,0) {$\cdots$};
    \node at (7.6,1.2) {\scriptsize $2\texttt{-}2\D$};
    \node at (8.35,0.4) {\scriptsize $\D$};
    \node at (9.85,0.4) {\scriptsize $1\texttt{-}\D$};
    \node at (7.6,-0.7) {\scriptsize $\t_j$};
    \node at (9.1,-0.7) {\scriptsize $\s_{j\texttt{+}1}$};
    \node at (10.6,-0.7) {\scriptsize $\k_{j\texttt{+}1}$};

    \node at (11.2,0.8) {$+\{2\texttt{-}2\k_j\}$};
    
\end{tikzpicture}
\end{equation}
where we have defined $n_j \equiv m_j - j$ and the FI parameters $\s_j$ and $\k_j$ as
\begin{align}
    \s_j &= 1-j + (1-\D)(m_j - m_{j\texttt{-}1} + j - 1)
    \,,\quad j=1,\cdots,p \,,
    \nonumber\\
    \k_j &= j+1 + (1-\D)(m_j - m_{j\texttt{+}1} - j)
    \,,\qquad j=0,\cdots,p-1 \,.
\end{align}
Those repeated steps of $\text{BBP}_1^+$ bring us to the bottom theory with a set of decoupled chiral multiplets.
Note that the tail of gauge nodes of the bottom theory in Figure \ref{fig: N+p confine} is nothing but the $\mathbb{D}_p[SU(N)]$ theory with confining superpotential deformation $\D W_{\text{conf}}^{(p),+}$ since $n_j = \lfloor N\;j/p \rfloor$ and the FI parameters $\t_j$ are invariant under the change $m_j\to n_j$. By assuming its confinement, we would get the right-middle theory. Upon confining the $\mathbb{D}_p[SU(N)]$ tail into an adjoint, the FI parameter of the $U(N)$ gauge node is shifted as $\s_p \to \s_p + \D([N]-1)$ where $[N]:= N \mod p$. This shifting is necessary to properly map the Coulomb branch operators under the confinement of the $\mathbb{D}_p[SU(N)]$ tail; see appendix \ref{app: monopole map under gauging} for a detailed discussion. Surprisingly, this shifting tunes the FI parameter exactly as in the rank-reducing duality chain so that utilizing \eqref{eq: SQCD confine} results in the right-top WZ theory of an $U(N+p)$ flavor adjoint chiral of R-charge $2\D = \frac{2}{p+1}$, confirming the confinement of $\mathbb{D}_p[SU(N+p)]$ theory.

This completes the derivation of the confinement \eqref{eq: confining diagram} via mathematical induction. Note that we have only used the Aharony and one-monopole BBP dualities along the derivation, which are, as mentioned, not independent because the former can be derived from the latter by mass deformation \cite{Benini:2017dud}.
\subsection{Example: $p = 2$}
\label{sec: p=2 derivation example}
In this subsection, we present the $p=2$ case in more detail to explain how the operators and superpotentials are mapped along the duality chain.\footnote{As mentioned, the duality for $p=2$ is first proposed in \cite{Bajeot:2023gyl} by reducing a similar 4d duality for $USp(2 N)$ gauge theories to 3d. On the other hand, our derivation is carried out solely in 3d and also holds for higher $p$ although here it is demonstrated for $p=2$ for simplicity.} For this purpose, we use the maps of the monopole operators under the Aharony and BBP dualities acting on a linear quiver and under the deconfinement of an adjoint into a $\mathbb D_p[SU(N)]$ tail, which are explained in appendix \ref{app: monopole map} and \ref{app: monopole map under gauging}, respectively. 
The seed for the derivation of the $p=2$ confinement is the $\mathbb{D}_{2}[SU(3)]$ theory with the confining superpotential deformation $\D W_{\text{conf}}^{(p=2),+}$, whose confinement can easily be shown by the BBP duality with two linear monopole superpotential terms \cite{Benini:2017dud}:\footnote{One can also derive the confinement of the $p=2$ seed by using the Aharony and one-monopole BBP dualities as we did in the previous subsection. Here we just adopt the two-monopole BBP duality for a simpler discussion.}
\begin{center}
\begin{equation}
\begin{tikzpicture}
  \tikzset{vertex/.style={circle,fill=white!25,minimum size=8pt,inner sep=2pt}}
  \tikzset{every loop/.style={}}
    \node[vertex] (A1) at (0,0) [shape=circle,draw=black,minimum size=2em] {$1$};
    \node[vertex] (A3) at (1.5,0) [shape=rectangle,draw=black,minimum height=2em, minimum width=2em] {$3$};

    \draw[-to, min distance=1cm]  (A1) edge [out=120, in=60] node {} (A1);
    \draw[->-=.5] ([yshift= 2pt] A1.east) to ([yshift= 2pt] A3.west);
    \draw[->-=.5] ([yshift= -2pt] A3.west) to ([yshift= -2pt] A1.east);

    \node at (0,1.2) {\scriptsize $\Phi$};
    \node at (0.7,0.3) {\scriptsize $Q$};
    \node at (0.7,-0.3) {\scriptsize $\WQ$};
    \node at (0.7,-1) {\scriptsize $W \!=\! \Phi \Tr \WQ Q \!+\! \eta^{3}\!+\!\eta\, \Phi \!+\! \hat{V}^{+} \!+\! \hat{V}^{-}$};
    
    \node at (3,0.7) {\scriptsize Integrate};
    \node at (3,0.4) {\scriptsize out $\Phi$};
    \node at (3,0) {$\longrightarrow$};
    
    \node[vertex] (B1) at (4.5,0) [shape=circle,draw=black,minimum size=2em] {$1$};
    \node[vertex] (B3) at (1.5+4.5,0) [shape=rectangle,draw=black,minimum height=2em, minimum width=2em] {$3$};

    \draw[->-=.5] ([yshift= 2pt] B1.east) to ([yshift= 2pt] B3.west);
    \draw[->-=.5] ([yshift= -2pt] B3.west) to ([yshift= -2pt] B1.east);

    \node at (0.7+4.5,0.3) {\scriptsize $Q$};
    \node at (0.7+4.5,-0.3) {\scriptsize $\WQ$};
    \node at (0.75+4.5,-1) {\scriptsize $W = \hat{V}^{+} \!+\! \hat{V}^{-}+(\mathrm{Tr} \, \tilde Q Q)^3$};

    \node at (8,0.7) {\scriptsize BBP with};
    \node at (8,0.4) {\scriptsize two-monopole};
    \node at (8,0) {$\Longleftrightarrow$};
    
    \node[vertex] (C3) at (10,0) [shape=rectangle,draw=black,minimum height=2em, minimum width=2em] {$3$};

    \draw[-to, min distance=1cm]  (C3) edge [out=120, in=60] node {} (C3);

    \node at (10,1.2) {\scriptsize $M$};
    \node at (10,-1) {\scriptsize $W = \Tr M^3$};

\end{tikzpicture}
\end{equation}
\end{center}
where $\Tr$ here is the trace over the $SU(3)$ flavor group. $\Phi$ is a singlet coupled to another singlet $\eta$ so that both are massive and can be integrated out to result in the SQED with the superpotential of two linear monopole terms and a cubic term of the meson, whose dual description is given by the WZ theory with the cubic superpotential of the flavor adjoint chiral $M$. Note that without the cubic term $(\mathrm{Tr} \, \tilde Q Q)^3$, the SQED is dual to WZ with $W = \det_3 M$ \cite{Benini:2017dud}. In appendix \ref{app:moduli}, we argue that deformation by such a cubic term leads to the desired dual WZ with the cubic superpotential by comparing the moduli spaces of vacua.
\begin{figure}[tbp]
\centering
\begin{tikzpicture}[thick,scale=1.1, every node/.style={scale=1.1}]
\tikzset{vertex/.style={circle,fill=white!25,minimum size=12pt,inner sep=2pt}}
  \tikzset{every loop/.style={}}
    \node at (-0.5,1.5) {\scriptsize $(A)$};
    \node[vertex] (AN) at (0.7,0) [shape=circle,draw=black,minimum size=2em] {$\! n\texttt{+}1\!$};
    \node[vertex] (AF) at (2.3,0) [shape=rectangle,draw=black,minimum height=2em, minimum width=2em] {\scriptsize $\!2n\texttt{+}3\!$};

    \draw[-to, min distance=1cm]  (AN) edge [out=120, in=60] node {} (AN);
    \draw[->-=.5] ([yshift= 2pt] AN.east) to ([yshift= 2pt] AF.west);
    \draw[->-=.5] ([yshift= -2pt] AF.west) to ([yshift= -2pt] AN.east);

    \node at (0.7,1.2) {\scriptsize $\Phi$};
    \node at (1.5,0.4) {\scriptsize $Q$};
    \node at (1.5,-0.4) {\scriptsize $\WQ$};

    \node at (-0.5,1.2-3) {\scriptsize $(B)$};
    \node[vertex] (BN1) at (0,-3) [shape=circle,draw=black,minimum size=2em] {$n$};
    \node[vertex] (BN) at (1.5,-3) [shape=circle,draw=black,minimum height=2em, minimum width=2em] {$\! n\texttt{+}1\!$};
    \node[vertex] (BF) at (3,-3) [shape=rectangle,draw=black,minimum height=2em, minimum width=2em] {\scriptsize $\!2n\texttt{+}3\!$};

    \draw[->-=.5] ([yshift= 2pt] BN1.east) to ([yshift= 2pt] BN.west);
    \draw[->-=.5] ([yshift= -2pt] BN.west) to ([yshift= -2pt] BN1.east);
    \draw[->-=.5] ([yshift= 2pt] BN.east) to ([yshift= 2pt] BF.west);
    \draw[->-=.5] ([yshift= -2pt] BF.west) to ([yshift= -2pt] BN.east);

    \node at (0.75,-3+0.4) {\scriptsize $R$};
    \node at (0.75,-3-0.4) {\scriptsize $\WR$};
    \node at (2.2,-3+0.4) {\scriptsize $Q$};
    \node at (2.2,-3-0.4) {\scriptsize $\WQ$};

    \node at (5.2,1.5-5.5) {\scriptsize $(C)$};
    \node[vertex] (CN1) at (4,-5.5) [shape=circle,draw=black,minimum size=2em] {$n$};
    \node[vertex] (CF1) at (5.5,-5.5) [shape=circle,draw=black,minimum height=2em, minimum width=2em] {\scriptsize $\!2n\texttt{+}1\!$};
    \node[vertex] (CF) at (7,-5.5) [shape=rectangle,draw=black,minimum height=2em, minimum width=2em] {\scriptsize $\!2n\texttt{+}3\!$};

    \draw[-to, min distance=1cm]  (CN1) edge [out=120, in=60] node {} (CN1);
    \draw[-to, min distance=0.9cm]  (CF) edge [out=120, in=60] node {} (CF);
    \draw[->-=.5] ([yshift= 2pt] CN1.east) to ([yshift= 2pt] CF1.west);
    \draw[->-=.5] ([yshift= -2pt] CF1.west) to ([yshift= -2pt] CN1.east);
    \draw[->-=.5] ([yshift= 2pt] CF1.east) to ([yshift= 2pt] CF.west);
    \draw[->-=.5] ([yshift= -2pt] CF.west) to ([yshift= -2pt] CF1.east);

    \node at (4,-5.5+1.2) {\scriptsize $\phi$};
    \node at (4.75,-5.5+0.4) {\scriptsize $r$};
    \node at (4.75,-5.5-0.4) {\scriptsize $\Wr$};
    \node at (6.25,-5.5+0.4) {\scriptsize $q$};
    \node at (6.25,-5.5-0.4) {\scriptsize $\Wq$};
    \node at (7,-5.5+1.2) {\scriptsize $M$};

    \node at (7.2,1.5-3) {\scriptsize $(D)$};
    \node[vertex] (DN1) at (8.2,-3) [shape=circle,draw=black,minimum height=2em, minimum width=2em] {\scriptsize $\! 2n\texttt{+}1\!$};
    \node[vertex] (DF) at (9.8,-3) [shape=rectangle,draw=black,minimum height=2em, minimum width=2em] {\scriptsize $\!2n\texttt{+}3\!$};

    \draw[-to, min distance=1cm]  (DN1) edge [out=120, in=60] node {} (DN1);
    \draw[-to, min distance=0.9cm]  (DF) edge [out=120, in=60] node {} (DF);
    \draw[->-=.5] ([yshift= 2pt] DN1.east) to ([yshift= 2pt] DF.west);
    \draw[->-=.5] ([yshift= -2pt] DF.west) to ([yshift= -2pt] DN1.east);

    \node at (8.2,-3+1.2) {\scriptsize $X$};
    \node at (9,-3+0.4) {\scriptsize $q$};
    \node at (9,-3-0.4) {\scriptsize $\Wq$};
    \node at (9.8,-3+1.2) {\scriptsize $M$};

    \node at (7.7,1.5) {\scriptsize $(E)$};
    \node[vertex] (EF) at (9,0) [shape=rectangle,draw=black,minimum height=2em, minimum width=2em] {\scriptsize $\!2n\texttt{+}3\!$};

    \draw[-to, min distance=0.9cm]  (EF) edge [out=120, in=60] node {} (EF);

    \node at (9,1.2) {\scriptsize $M$};

    \node at (1.5,-1.5) 
    {\begin{tikzpicture}
      \node [rotate=90] {\textcolor{black}{$\Longleftrightarrow$}};    
    \end{tikzpicture}
    };
    \node at (0.8,-1.5) {\scriptsize $\text{BBP}_1^+$};

    \node at (2.3,-4.5) 
    {\begin{tikzpicture}
      \node [rotate=135] {\textcolor{black}{$\Longleftrightarrow$}};    
    \end{tikzpicture}
    };
    \node at (1.3,-4.5) {\scriptsize $\text{BBP}_1^+$};

    \node at (8.3,-4.5) 
    {\begin{tikzpicture}
      \node [rotate=45] {\textcolor{black}{$\Longleftrightarrow$}};    
    \end{tikzpicture}
    };
    \node at (9.7,-4.3) {\scriptsize $\mathbb{D}_{2}[SU(2n\texttt{+}1)]$};
    \node at (9.7,-4.7) {\scriptsize confinement };

    \node at (9,-1.2) 
    {\begin{tikzpicture}
      \node [rotate=90] {\textcolor{black}{$\Longleftrightarrow$}};    
    \end{tikzpicture}
    };
    \node at (9.9,-1.2) {\scriptsize $\text{Figure \ref{fig: p=2 DE step}}$};

    \node at (5.3,0) 
    {\begin{tikzpicture}
      \node [rotate=0] {\Large \textcolor{black}{$\Longleftrightarrow$}};    
    \end{tikzpicture}
    };
    \node at (5.3,0.5) { $\text{Dual}$};
    
\end{tikzpicture}
\caption{\label{fig: p=2 induction} The mathematical induction procedure for deriving the $p=2$ confinement.}
\end{figure}
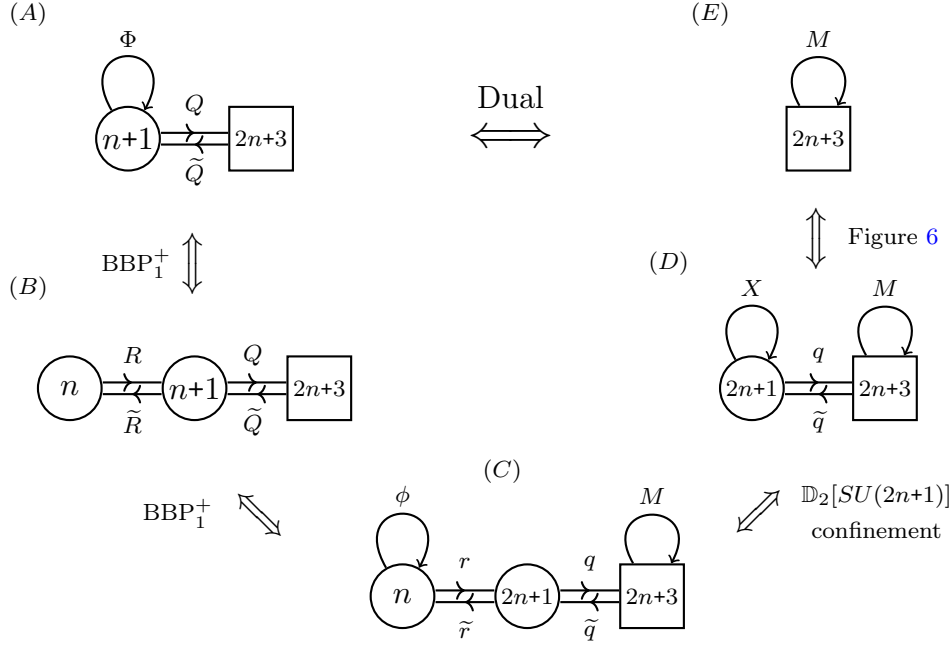
The procedure of the mathematical induction is depicted in Figure \ref{fig: p=2 induction}, where we start with the $\mathbb{D}_{2}[SU(2n\texttt{+}3)]$ theory $(n\geq 1)$ deformed by $\D W_{\text{conf}}^{(2),+}$; namely, the total superpotential of Theory A is
\begin{align}
\label{eq:W_A}
    W_A = \Tr \Phi Q \WQ + \eta^{3}+\eta\,\Tr \Phi + \hat V^{+} + \hat V^{-}
\end{align}
where $\Tr$ is now taken over the $U(n+1)$ gauge group, and the $SU(2n+3)$ flavor indices of $Q$ and $\tilde Q$ are assumed to be contracted.\footnote{This can easily be understood once regarding each field as a matrix.} $\hat V^{\pm}$ are the bare monopole operators charged under the $U(1)_T$ topological symmetry from the $U(n+1)$ gauge group.

The adjoint $\Phi$ can be deconfined into another $U(n)$ gauge node by applying the one-monopole BBP duality, producing Theory B with the superpotential
\begin{align}
    W_B = \Tr \WR R Q \WQ + \eta^{3}+\eta\,\Tr \WR R + \hat{v}^{(2),+} + \hat{v}^{(1,2),-} + \hat v^{(1),+} + \xi\, \hat{v}^{(1),-} \,,
\end{align}
where the chiral fields are mapped as $\Phi\to\WR R$, $\hat V^+ \to \hat v^{(2),+}$, $\hat V^-\to \hat v^{(1,2),-}$. The last two terms have arisen from the dual superpotential of the one-monopole BBP duality. Note that the singlet $\xi$ is mapped to the determinant of $\Phi$. Naively, $\det_{n\texttt{+}1} \Phi$ seems mapped to $\det_{n\texttt{+}1} \WR R$, which however classically vanishes in Theory B due to the rank condition $\text{rank}(\WR R) < n+1$. Instead, it is mapped to the new singlet $\xi$ having the same symmetry charges as $\det_{n\texttt{+}1} \Phi$. This can easily be seen by using the fact that the BBP dual of the $U(n)$ theory with $n-1$ flavors with a single linear monopole superpotential term is a WZ theory with superpotential $W = \xi\ \det \Phi$.

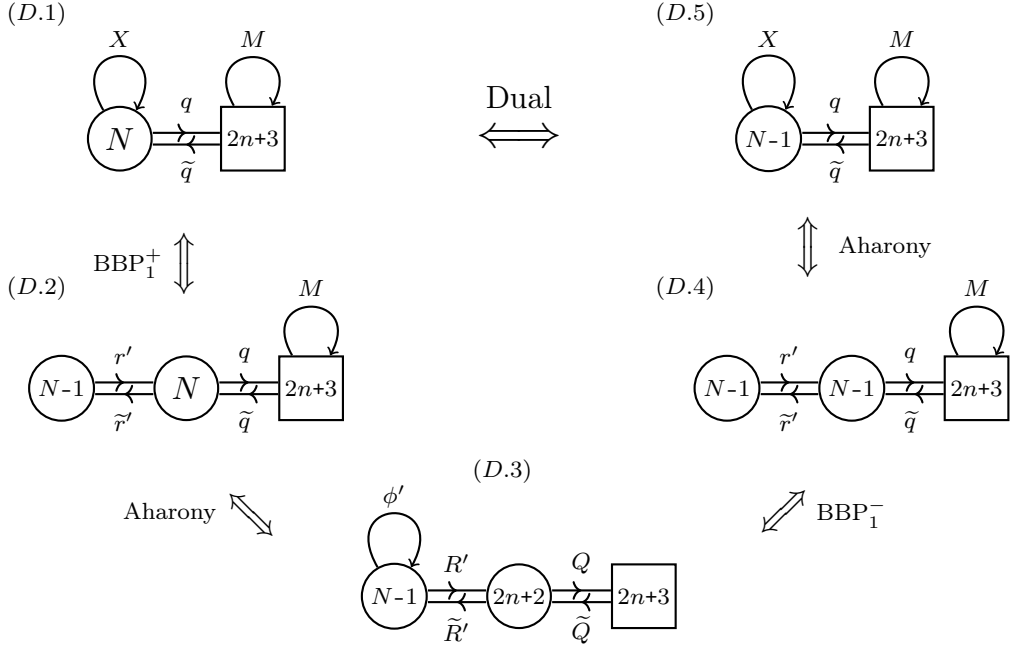
\begin{figure}[tbp]
\centering
\begin{tikzpicture}[thick,scale=1.1, every node/.style={scale=1.1}]
\tikzset{vertex/.style={circle,fill=white!25,minimum size=12pt,inner sep=2pt}}
  \tikzset{every loop/.style={}}
    \node at (-0.3,1.5) {\scriptsize $(D.1)$};
    \node[vertex] (AN) at (0.7,0) [shape=circle,draw=black,minimum size=2em] {$N$};
    \node[vertex] (AF) at (2.3,0) [shape=rectangle,draw=black,minimum height=2em, minimum width=2em] {\scriptsize $\! 2n\texttt{+}3 \!$};

    \draw[-to, min distance=1cm]  (AN) edge [out=120, in=60] node {} (AN);
    \draw[->-=.5] ([yshift= 2pt] AN.east) to ([yshift= 2pt] AF.west);
    \draw[->-=.5] ([yshift= -2pt] AF.west) to ([yshift= -2pt] AN.east);
    \draw[-to, min distance=0.9cm]  (AF) edge [out=120, in=60] node {} (AF);

    \node at (0.7,1.2) {\scriptsize $X$};
    \node at (1.5,0.4) {\scriptsize $q$};
    \node at (1.5,-0.4) {\scriptsize $\Wq$};
    \node at (2.3,1.2) {\scriptsize $M$};

    \node at (-0.3,1.2-3) {\scriptsize $(D.2)$};
    \node[vertex] (BN1) at (0,-3) [shape=circle,draw=black,minimum size=2em] {\scriptsize $N\texttt{-}1$};
    \node[vertex] (BN) at (1.5,-3) [shape=circle,draw=black,minimum height=2em, minimum width=2em] {$N$};
    \node[vertex] (BF) at (3,-3) [shape=rectangle,draw=black,minimum height=2em, minimum width=2em] {\scriptsize $\! 2n\texttt{+}3 \!$};

    \draw[->-=.5] ([yshift= 2pt] BN1.east) to ([yshift= 2pt] BN.west);
    \draw[->-=.5] ([yshift= -2pt] BN.west) to ([yshift= -2pt] BN1.east);
    \draw[->-=.5] ([yshift= 2pt] BN.east) to ([yshift= 2pt] BF.west);
    \draw[->-=.5] ([yshift= -2pt] BF.west) to ([yshift= -2pt] BN.east);
    \draw[-to, min distance=0.9cm]  (BF) edge [out=120, in=60] node {} (BF);

    \node at (0.75,-3+0.4) {\scriptsize $r'$};
    \node at (0.75,-3-0.4) {\scriptsize $\Wr'$};
    \node at (2.2,-3+0.4) {\scriptsize $q$};
    \node at (2.2,-3-0.4) {\scriptsize $\Wq$};
    \node at (3,-3+1.2) {\scriptsize $M$};

    \node at (5.3,1.5-5.5) {\scriptsize $(D.3)$};
    \node[vertex] (CN1) at (4,-5.5) [shape=circle,draw=black,minimum size=2em] {\scriptsize $N\texttt{-}1$};
    \node[vertex] (CF1) at (5.5,-5.5) [shape=circle,draw=black,minimum height=2em, minimum width=2em] {\scriptsize $\! 2n\texttt{+}2\!$};
    \node[vertex] (CF) at (7,-5.5) [shape=rectangle,draw=black,minimum height=2em, minimum width=2em] {\scriptsize $\! 2n\texttt{+}3 \!$};

    \draw[-to, min distance=1cm]  (CN1) edge [out=120, in=60] node {} (CN1);
    \draw[->-=.5] ([yshift= 2pt] CN1.east) to ([yshift= 2pt] CF1.west);
    \draw[->-=.5] ([yshift= -2pt] CF1.west) to ([yshift= -2pt] CN1.east);
    \draw[->-=.5] ([yshift= 2pt] CF1.east) to ([yshift= 2pt] CF.west);
    \draw[->-=.5] ([yshift= -2pt] CF.west) to ([yshift= -2pt] CF1.east);

    \node at (4,-5.5+1.2) {\scriptsize $\phi'$};
    \node at (4.75,-5.5+0.4) {\scriptsize $R'$};
    \node at (4.75,-5.5-0.4) {\scriptsize $\WR'$};
    \node at (6.25,-5.5+0.4) {\scriptsize $Q$};
    \node at (6.25,-5.5-0.4) {\scriptsize $\WQ$};

    \node at (7.5,1.2-3) {\scriptsize $(D.4)$};
    \node[vertex] (DN) at (8,-3) [shape=circle,draw=black,minimum size=2em] {\scriptsize $N\texttt{-}1$};
    \node[vertex] (DN1) at (9.5,-3) [shape=circle,draw=black,minimum height=2em, minimum width=2em] {\scriptsize $N\texttt{-}1$};
    \node[vertex] (DF) at (11,-3) [shape=rectangle,draw=black,minimum height=2em, minimum width=2em] {\scriptsize $\! 2n\texttt{+}3 \!$};

    \draw[->-=.5] ([yshift= 2pt] DN.east) to ([yshift= 2pt] DN1.west);
    \draw[->-=.5] ([yshift= -2pt] DN1.west) to ([yshift= -2pt] DN.east);
    \draw[->-=.5] ([yshift= 2pt] DN1.east) to ([yshift= 2pt] DF.west);
    \draw[->-=.5] ([yshift= -2pt] DF.west) to ([yshift= -2pt] DN1.east);
    \draw[-to, min distance=0.9cm]  (DF) edge [out=120, in=60] node {} (DF);

    \node at (8.75,-3+0.4) {\scriptsize $r'$};
    \node at (8.75,-3-0.4) {\scriptsize $\Wr'$};
    \node at (10.2,-3+0.4) {\scriptsize $q$};
    \node at (10.2,-3-0.4) {\scriptsize $\Wq$};
    \node at (11,-3+1.2) {\scriptsize $M$};

    \node at (7.5,1.5) {\scriptsize $(D.5)$};
    \node[vertex] (EN) at (8.5,0) [shape=circle,draw=black,minimum size=2em] {\scriptsize $N\texttt{-}1$};
    \node[vertex] (EF) at (10.1,0) [shape=rectangle,draw=black,minimum height=2em, minimum width=2em] {\scriptsize $\! 2n\texttt{+}3 \!$};

    \draw[-to, min distance=1cm]  (EN) edge [out=120, in=60] node {} (EN);
    \draw[->-=.5] ([yshift= 2pt] EN.east) to ([yshift= 2pt] EF.west);
    \draw[->-=.5] ([yshift= -2pt] EF.west) to ([yshift= -2pt] EN.east);
    \draw[-to, min distance=0.9cm]  (EF) edge [out=120, in=60] node {} (EF);

    \node at (8.5,1.2) {\scriptsize $X$};
    \node at (9.3,0.4) {\scriptsize $q$};
    \node at (9.3,-0.4) {\scriptsize $\Wq$};
    \node at (10.1,1.2) {\scriptsize $M$};

    \node at (1.5,-1.5) 
    {\begin{tikzpicture}
      \node [rotate=90] {\textcolor{black}{$\Longleftrightarrow$}};    
    \end{tikzpicture}
    };
    \node at (0.8,-1.5) {\scriptsize $\text{BBP}_1^+$};

    \node at (2.3,-4.5) 
    {\begin{tikzpicture}
      \node [rotate=135] {\textcolor{black}{$\Longleftrightarrow$}};    
    \end{tikzpicture}
    };
    \node at (1.3,-4.5) {\scriptsize $\text{Aharony}$};

    \node at (8.7,-4.5) 
    {\begin{tikzpicture}
      \node [rotate=45] {\textcolor{black}{$\Longleftrightarrow$}};    
    \end{tikzpicture}
    };
    \node at (9.5,-4.5) {\scriptsize $\text{BBP}_1^-$};

    \node at (9,-1.3) 
    {\begin{tikzpicture}
      \node [rotate=90] {\textcolor{black}{$\Longleftrightarrow$}};    
    \end{tikzpicture}
    };
    \node at (9.9,-1.3) {\scriptsize $\text{Aharony}$};

    \node at (5.5,0) 
    {\begin{tikzpicture}
      \node [rotate=0] {\Large \textcolor{black}{$\Longleftrightarrow$}};    
    \end{tikzpicture}
    };
    \node at (5.5,0.5) { $\text{Dual}$};
    
\end{tikzpicture}
\caption{\label{fig: p=2 DE step} Expansion for the $(D)$ $\to$ $(E)$\; step of the $p=2$ confinement derivation in Figure \ref{fig: p=2 induction}. The gauge rank decreases from $N=2n+1$ to $N=1$ as repeatedly applying this cycle, resulting in the confined Theory E.}
\end{figure}

Since $W_B$ has a linear monopole term $\hat v^{(2),+}$, we can further apply the one-monopole BBP duality to the $U(n\!+\!1)$ gauge node in Theory B, obtaining Theory C with the superpotential,
\begin{align}
    W_C = \Wr \phi r + \eta^{3}+\eta \Tr \phi + \hat v^{(1),+} + \hat v^{(1),-}
    + \Wq \Wr r q + M \Wq q + \hat v^{(2),+} + \widetilde{\xi}\, \hat v^{(2),-} + \xi\, \hat v^{(1,2),-} \,,
\end{align}
where the chiral fields are mapped as $R\WR \to \phi$, $\WQ Q \to M$, $\hat v^{(1),-} \to \hat v^{(1,2),-}$, $\hat v^{(1,2),-}\to \hat v^{(1),-}$, $\hat v^{(2),-}\to \widetilde{\xi}$ with $\widetilde{\xi}$ being a singlet. The positively charged monopole operators are trivially mapped, and the term $\Wq \Wr r q$ has arisen by integrating out massive flavors of the $U(n)$ gauge group. Note that the first five terms in $W_C$ are exactly the $p=2$ confining superpotential as well as the original superpotential of the $\mathbb D_2[SU(2 N+1)]$ tail of the quiver. Thus, assuming the confinement of $\mathbb{D}_2[SU(2n+1)]$, we arrive at Theory D with the superpotential
\begin{align}
    W_D = \Tr X^3 + \hat V_0^+ + \widetilde{\xi}\, \hat V_0^- + \xi\,\hat V_1^- + \Wq X q + M \Wq q \,,
\end{align}
where the chiral fields are mapped as $\Wr r \to X$, $\hat v^{(2),\pm} \to \hat V_0^\pm$, $\hat v^{(1,2),-}\to \hat V_1^-$ with $\hat V_i^\pm$ being the $i$-th dressed monopole operators of Theory D. To achieve Theory E, we should repeatedly use the rank-reducing duality chain depicted as in Figure \ref{fig: p=2 DE step}. From now on, for simplicity, we will focus on the terms involving monopole operators and dual singlets, which are crucial to determine which duality one can employ at each step. In appendix \ref{app:moduli}, we show how to fix the omitted terms by comparing the vacuum moduli spaces as a shortcut.
To follow the rank-reducing duality chain, let us start with Theory D.1 in Figure \ref{fig: p=2 DE step} with gauge rank $N=2n+1$. The superpotential is given by
\begin{align}
    W_{D.1} = \hat V_0^+ + \xi_1\, \hat V_0^- + \xi_2\,\hat V_1^-
\end{align}
where we have renamed the singlets as $\widetilde \xi \to \xi_1 $, $\xi\to \xi_2$. By deconfining $X$ into another gauge node via the one-monopole BBP duality, we get Theory D.2 with the superpotential
\begin{align}
\label{eq: N=2n+1 W_D2}
    W_{D.2} = \hat v^{(2),+} + \xi_1\,\hat v^{(1,2),-} + \xi_2\, \text{tr}(\Wr' r'\,\hat v^{(1,2),-}) + \hat v^{(1),+} + \xi_3 \, \hat v^{(1),-}
\end{align}
where the chiral fields are mapped as $X\to \Wr' r'$, $\hat V_0^+ \to \hat v^{(2),+}$, $\hat V_0^- \to \hat v^{(1,2),-}$, $\hat V_1^- \to \text{tr}(\Wr'r' \hat v^{(1,2),-})$. The last two terms have arisen from the dual superpotential of the BBP duality. $\text{tr}(\Wr' r'\,\hat v^{(1,2),-})$ is a monopole operator dressed by $\Wr' r'$ carrying negative unit charges under both gauge nodes . Note that it is an independent operator from the $\Tr \left(\Wr' r\right) \times \hat v^{(1,2),-}$. Applying the Aharony duality to the second gauge node of Theory D.2 gives Theory D.3 with the superpotential
\begin{align}
    W_{D.3} &=  v^+ + \xi_1\, \hat v^{(1),-} + \xi_2\, \hat v_1^{(1),-} 
     + \hat v^{(1,2),+} + \xi_3 \, \hat v^{(1,2),-} + v^+ \, \hat v^{(2),+} + v \, \hat v^{(2),-}
     \label{eq: N=2n+1 W_D3}
\end{align}
where the chiral fields are mapped as $r\Wr'\to \phi$, $\hat v^{(2),+}\to v^+$, $\hat v^{(2),-}\to v$, $\hat v^{(1),\pm} \to \hat v^{(1,2),\pm}$, $\hat v^{(1,2),\pm} \to \hat v^{(1),\pm}$ with $v^+$, $v$, the singlets arising from the Aharony duality, and $\hat v_1^{(1),-}$, a dressed monopole operator negatively charged under the first gauge node.

Since there is no $U(1)$ global symmetry that can be mixed with the R-symmetry along the RG flow, the conformal R-charges are exactly fixed by the superpotential, which requires the R-charge of $v$ to be 0. Therefore, one can redefine $v \rightarrow v -1$, which results in an extra term, $\hat v^{(2),-}$. Such a linear monopole term allows us to apply the one-monopole BBP duality and produce Theory D.4 with the superpotential
\begin{align}
    W_{D.4} 
    &=  \hat v^{(2),+} + \xi_1\, \hat v^{(1),-} + \xi_2 \, \hat v^{(1,2),-} + \hat v^{(1),+}
     + \xi_3 \,\text{tr}(r'\Wr'\,\hat v^{(1,2),-}) + v\, \hat v^{(2),-} + \hat v^{(2),-}
    \label{eq: N=2n+1 W_D4}
\end{align}
where the chiral fields are mapped as $\phi\to r'\Wr'$, $\WQ Q \to M$, $\hat v^{(1,2),+}\to \hat v^{(1),+}$, $\hat v^{(1,2),-}\to \text{tr}(r'\Wr'\,\hat v^{(1,2),-})$. The monopole term $\hat v^{(2),+}$ results from integrating out massive $v^+$. At this stage, it is important to notice \emph{quantum} relations arising among monopole operators. Let us first consider a theory without the confining deformation, which provides more refined information since $U(1)_A$ and $U(1)_T$ survive. Analyzing the superconformal index, we find the following quantum relations:
\begin{align}
    \hat v^{(2),+} \; &\sim \; \hat v^{(1),-}\,\text{tr}\big((r'\Wr')^{N-2}\,\hat v^{(1,2),+}\big) \,, \label{eq: monopole relation+} \\
    \hat v^{(2),-} \; &\sim \; \hat v^{(1),+}\,\text{tr}\big((r'\Wr')^{N-2}\,\hat v^{(1,2),-}\big) \,, \label{eq: monopole relation-} 
\end{align}
which are consistent with the quantum numbers of the operators as it should. Once we turn on the confining deformation, $U(1)_A$ and $U(1)_T$ are explicitly broken. Both sides are now charged under the R-symmetry only, with charge 2 as expected from the superpotential, and more other terms can be involved in the relations. For example, there can be mixing between the two relations. However, the right hand side of the first relation is actually trivial because $\hat v^{(1),-}$ vanishes due to the F-term condition. On the other hand, $\hat v^{(1),+}$ is still nontrivial and may appear as extra terms in the relations since it has R-charge 2 on its own. Here, we do not attempt to determine whether the relations are further quantum corrected by such extra terms or not. Instead, we have checked that the final result does not depend on these extra $\hat v^{(1),+}$ terms, and the original relations \eqref{eq: monopole relation+} and \eqref{eq: monopole relation-} are enough for the subsequent discussions; hence, for simplicity, we will stick to them without possible extra terms. Adopting these relations, we further confine the left-end tail of Theory D.4 into an adjoint chiral field $X$ by applying the Aharony duality, arriving at Theory D.5 with superpotential\footnote{Though the dressed monopole operators $\hat V_{N\texttt{-}2}^\pm$ may not be the chiral ring elements due to the truncation by the equations of motion for the adjoint chiral $X$, we still use the notation $\hat V_{N\texttt{-}2}^\pm$ formally for denoting some possible composite elements having the same quantum number since the only role of these dressed monopole operators is just fixing the FI parameter to a proper value for the rank-reducing duality chain.} 
\begin{align}
    W_{D.5} 
    &=  u\, v \, \hat V_{N\texttt{-}2}^- + u + u\, \hat V_{N\texttt{-}2}^- + \xi_2 \hat V_0^- + \xi_3 \hat V_1^- 
    \label{eq: N=2n+1 W_D5}
\end{align}
where the chiral fields are mapped as $\Wr' r' \to X$, $\hat v^{(1),+}\to u$, $\hat v^{(1,2),\pm} \to \hat V_0^\pm$, and we have also integrated out massive singlets.

Now, we finally employ the rank-reducing duality for the theory with superpotential \eqref{eq: N=2n+1 W_D5}, whose derivation is given in appendix \ref{app: rank reducing}. It is shown that this rank-reducing duality leads to the WZ theory of $(2n+3)^2$ free chiral fields. However, there are also additional interactions terms we have omitted for simplicity, which can be restored by comparing the vacuum moduli spaces as shown in appendix \ref{app:moduli}. As a result, the final theory obtained through the duality chain is the WZ theory with the superpotential
\begin{align}
W = \mathrm{Tr} X^3
\end{align}
as proposed. This completes the derivation of of the confinement of $\mathbb{D}_2[SU(2n\texttt{+}1)]$ for all positive integers $n$ via mathematical induction.

\subsection{The 3d version of the Maruyoshi--Nardoni--Song duality}
Recently, a new 4-dimensional IR duality has been proposed in \cite{Maruyoshi:2023mnv}, claiming a 4d $\CN=1$ SQCD with the $SU(N)$ gauge group coupled to the $D_p[SU(N)]$ theory of a pair of coprime integers $(p,N)$ via an $\CN=1$ vector multiplet is dual to the same SQCD with an extra adjoint field and the $A_p$-type superpotential. We will call this duality the \emph{Maruyoshi--Nardoni--Song} (MNS) duality; see appendix \ref{app: MNS duality} for a brief review. As an application of the confinement of 3d $\mathbb{D}_p[SU(N)]$, we propose a 3d version of the MNS duality that can be represented by Figure \ref{fig: 3d MNS}. On one side we have the deformed $\mathbb{D}_p[SU(N)]$ theory of $N=\pm 1\mod p$ whose flavor node is gauged by a 3d $\CN=2$ vector multiplet with extra $N_f$ flavors, while on the other side we have an adjoint SQCD with the $A_p$-type superpotential.\footnote{For later use in section \ref{sec:deconfined KP}, it is useful to define the $N=1$ case as well. In this case, the duality trivially holds because on the LHS, $m_j = 0$ for $j=1,\cdots,p-1$, whereas on the RHS, $X$ is simply a gauge singlet. The superpotentials also match upon the manifest duality map $\eta \leftrightarrow X$.}
\begin{figure}[tbp]
\centering
\begin{tikzpicture}
  \tikzset{vertex/.style={circle,fill=white!25,minimum size=12pt,inner sep=2pt}}
  \tikzset{every loop/.style={}}

    \node[vertex] (m1) at (0,0) [shape=circle,draw=black,minimum size=2em] {$m_1$};
    \node (cdot) at (1.2,0) {$\cdots$};
    \node[vertex] (mpm1) at (2.5,0) [shape=circle,draw=black,minimum size=2em] {\tiny$m_{p\!-\!1}$};
    \node[vertex] (NN) at (4,0) [shape=circle,draw=black,minimum size=2em] {$N$};
    \node[vertex] (NfNf) at (5.5,0) [shape=rectangle,draw=black,minimum height=2em, minimum width=2em] {$N_f$};

    \draw[-to, min distance=1cm]  (m1) edge [out=60, in=120] node {} (m1);
    \draw[-to, min distance=1cm]  (mpm1) edge [out=60, in=120] node {} (mpm1);
    \draw[->-=.5] ([yshift= 2pt] m1.east) to ([yshift= 2pt] cdot.west);
    \draw[->-=.5] ([yshift= -2pt] cdot.west) to ([yshift= -2pt] m1.east);
    \draw[->-=.5] ([yshift= 2pt] cdot.east) to ([yshift= 2pt] mpm1.west);
    \draw[->-=.5] ([yshift= -2pt] mpm1.west) to ([yshift= -2pt] cdot.east);
    \draw[->-=.5] ([yshift= 2pt] mpm1.east) to ([yshift= 2pt] NN.west);
    \draw[->-=.5] ([yshift= -2pt] NN.west) to ([yshift= -2pt] mpm1.east);
    \draw[->-=.5] ([yshift= 2pt] NN.east) to ([yshift= 2pt] NfNf.west);
    \draw[->-=.5] ([yshift= -2pt] NfNf.west) to ([yshift= -2pt] NN.east);

    \node at (0.6,0.3) {\tiny$Q^{(1)}$};
    \node at (0.6,-0.3) {\tiny$\WQ^{(1)}$};
    \node at (1.7,0.3) {\tiny$Q^{(p\!-\!2)}$};
    \node at (1.7,-0.3) {\tiny$\WQ^{(p\!-\!2)}$};
    \node at (3.3,0.3) {\tiny$Q^{(p\!-\!1)}$};
    \node at (3.3,-0.3) {\tiny$\WQ^{(p\!-\!1)}$};
    \node at (4.7,0.3) {\tiny$Q$};
    \node at (4.7,-0.3) {\tiny$\WQ$};
    \node at (0,1.2) {\tiny $\Phi^{(1)}$};
    \node at (2.5,1.2) {\tiny $\Phi^{(p\!-\!1)}$};

    \node at (3,-1) {\scriptsize$W = \sum_{i=\a}^{p-1} \WQ^{(i)} \Phi^{(i)} Q^{(i)} + \sum_{i=\a+1}^{p-1} Q^{(i-1)}\Phi^{(i)} \WQ^{(i-1)} + \D W_{\text{conf}}^{(p),\pm}$};

    \node at (8,0.5) {Dual};
    \node at (8,0.1) {$\Longleftrightarrow$};

    \node[vertex] (N) at (-2.5+13,0) [shape=circle,draw=black,minimum size=2em] {$N$};
    \node[vertex] (Nf) at (-1+13,0) [shape=rectangle,draw=black,minimum height=2em, minimum width=2em] {$N_f$};

    \draw[-to, min distance=1cm]  (N) edge [out=60, in=120] node {} (N);
    \draw[->-=.5] ([yshift= 2pt] N.east) to ([yshift= 2pt] Nf.west);
    \draw[->-=.5] ([yshift= -2pt] Nf.west) to ([yshift= -2pt] N.east);

    \node at (-2.5+13,1.2) {\tiny $X$};
    \node at (-1.8+13,0.3) {\tiny$Q$};
    \node at (-1.8+13,-0.3) {\tiny$\WQ$};
    \node at (-1.7+13,-1) {\scriptsize$W = \Tr X^{p+1}$};

\end{tikzpicture}
\caption{\label{fig: 3d MNS} 3d version of the MNS duality where the LHS theory is a $U(N)$ SQCD coupled to the $\mathbb{D}_p[SU(N)]$ theory of $N = \pm 1 \mod p$ with confining superpotential deformation $\D W_\text{conf}^{(p),\pm}$, while the RHS theory is an adjoint SQCD with the $A_p$-type superpotential. The letters in the quivers denote chiral multiplets. Here $\a$ is the position of the first non-trivial gauge node, i.e., $\a = \min[j]$ such that $m_j\neq 0$.}
\end{figure}
Even though the flavor node of $\mathbb{D}_p[SU(N)]$ is $SU(N)$, we can add $U(1)_G$ having a BF coupling with another $U(1)_T$ symmetry, which becomes the topological global symmetry when $U(1)_G$ and the original $SU(N)$ flavor symmetry are gauged together by a $U(N)$ vector multiplet. We also turn on deformation $\D W_{\text{conf}}^{(p),\pm}$, including monopole terms fixing FI parameters as follows:
\begin{align}
    \t_p = \t \pm \D (1-[N])
    \,,\qquad\text{for}\quad \D W_\text{conf}^{(p),\pm}
    \label{eq: mixing under gauging}
\end{align}
where $\t_p$ and $\t$ are the normalized FI parameters of the LHS and RHS theories respectively, and $\D = \frac{1}{p+1}$. This mixing is necessary to properly match the R-charge spectrum of the independent monopole operators on two sides, in other words, to match the Coulomb branch of the moduli space. We present a map of the chiral ring generators between two theories for the confining superpotential deformation $W_\text{conf}^{(p),+}$:
\begin{align}
    \tilde{Q} Q 
    &\qquad\longleftrightarrow\qquad
    \tilde{Q} Q
    \nonumber\\
    \tilde{Q} \tilde{Q}^{(p\texttt{-}1)} \!\cdots\! \tilde{Q}^{(p\texttt{-}i)}
    Q^{(p\texttt{-}i)} \!\cdots\! Q^{(p\texttt{-}1)} Q
    &\qquad\longleftrightarrow\qquad
    \tilde{Q} X^i Q
    \nonumber\\
    \tilde{Q}^{(p\texttt{-}1)} \cdots \tilde{Q}^{(p\texttt{-}i)}
    Q^{(p\texttt{-}i)} \cdots Q^{(p\texttt{-}1)}
    &\qquad\longleftrightarrow\qquad
    \Tr X^i
    \nonumber\\
    \hat{v}^{(j,p),+}
    &\qquad\longleftrightarrow\qquad
    \hat{V}_{[-j]}^+
    \nonumber\\
    \hat{v}^{(j,p),-}
    &\qquad\longleftrightarrow\qquad
    \hat{V}_{[j\texttt{-}1\texttt{+}N]}^-
    \qquad\qquad\qquad
\end{align}
where $i$ runs from $1$ to $p-\a$, and $j$ runs from $\a$ to $p$. The signs in the superscripts of the monopole operators are flipped as $\pm \to \mp$ if we choose the other confining superpotential deformation $W_\text{conf}^{(p),-}$. Recall that $\a=2$ when $N=p-1$, in which case, the chiral ring elements $\tilde Q X^{p-1} Q$, $\Tr X^{p-1}$, and $\hat V_{p-1}^{\pm}$ are absent as expected from the classical rank condition on the adjoint SQCD side. This classical truncation is reflected in the deconfined quiver theory as the triviality of the first gauge node, i.e., $m_1=0$.
\\

\section{The deconfined Kim--Park duality}
\label{sec:deconfined KP}

In this section, we propose a new duality for 3d quiver gauge theories that is equivalent to the Kim--Park (KP) duality \cite{Kim:2013cma}, a 3d Seiberg-like duality with an adjoint matter, upon the confinement of the 3d $\mathbb D_p[SU(N)]$ we have discussed in the previous section. Combined with such confinement, our duality provides a novel way of understanding Seiberg-like dualities with tensor matters. In particular, we will show that the deconfined KP duality actually descends from the Aharony duality; i.e., it can be \emph{proven} in a simple way once we assume the latter. Moreover, adopting the same method but using the one-monopole BBP duality instead of the Aharony duality, we also derive another variant of the KP duality deformed by a \emph{single} monopole superpotential term, which hasn't been discussed in the literature.

\begin{figure}[tbp]
\centering
\begin{tikzpicture}[thick,scale=0.95, every node/.style={scale=0.95}]
\tikzset{vertex/.style={circle,fill=white!25,minimum size=12pt,inner sep=2pt}}
  \tikzset{every loop/.style={}}
    \node[vertex] (eNc) at (-1,0) [shape=circle,draw=black,minimum size=2em] {$N_c$};
    \node[vertex] (eNf1) at (1,0.7) [shape=rectangle,draw=black,minimum height=2em, minimum width=2em] {$N_f$};
    \node[vertex] (eNf2) at (1,-0.7) [shape=rectangle,draw=black,minimum height=2em, minimum width=2em] {$N_f$};

    \draw[-to, min distance=1cm]  (eNc) edge [out=120, in=60] node {} (eNc);
    \draw[->-=.5] ([yshift= 2pt] eNc.east) to ([yshift= 0pt] eNf1.west);
    \draw[->-=.5] ([yshift= 0pt] eNf2.west) to ([yshift= -2pt] eNc.east);

    \node at (-1,1.3) {$X$};
    \node at (0,0.8) {$Q$};
    \node at (0,-0.8) {$\WQ$};

    \node at (4,0) {\Large$\Longleftrightarrow$};

    \node[vertex] (mNc) at (7,0) [shape=circle,draw=black,minimum size=2em] {$N_c'$};
    \node[vertex] (mNf1) at (9,0.7) [shape=rectangle,draw=black,minimum height=2em, minimum width=2em] {$N_f$};
    \node[vertex] (mNf2) at (9,-0.7) [shape=rectangle,draw=black,minimum height=2em, minimum width=2em] {$N_f$};

    \draw[-to, min distance=1cm]  (mNc) edge [out=120, in=60] node {} (mNc);
    \draw[->-=.5] ([yshift= 0pt] mNf1.west) to ([yshift= 2pt] mNc.east);
    \draw[->-=.5] ([yshift= -2pt] mNc.east) to ([yshift= 0pt] mNf2.west);
    \draw[->-=.7] ( [xshift= -5pt] mNf2.north) to ( [xshift= -5pt] mNf1.south);
    \draw[->-=.7] ( [xshift= 5pt] mNf2.north) to ( [xshift= 5pt] mNf1.south);

    \node at (7,1.2) {$x$};
    \node at (8,0.8) {$\Wq$};
    \node at (8,-0.8) {$q$};
    \node at (10.7,0) {$M_{i=0,\cdots,p\texttt{-}1}$};
    \node at (9-0.1,-0.1) {\tiny $\cdot$};
    \node at (9,-0.1) {\tiny $\cdot$};
    \node at (9+0.1,-0.1) {\tiny $\cdot$};
    \node at (8,-1.8) {$+\;V_{i=0,\cdots,p\texttt{-}1}^\pm$};
\end{tikzpicture}
\caption{\label{fig:KP} The quiver representation of the Kim--Park duality. The dual gauge rank $N_c'$ is given by $N_c' = p N_f-N_c$. $V_i^\pm$ on the dual side are gauge singlet fields coupled to the monopole operators of the dual theory.}
\end{figure}
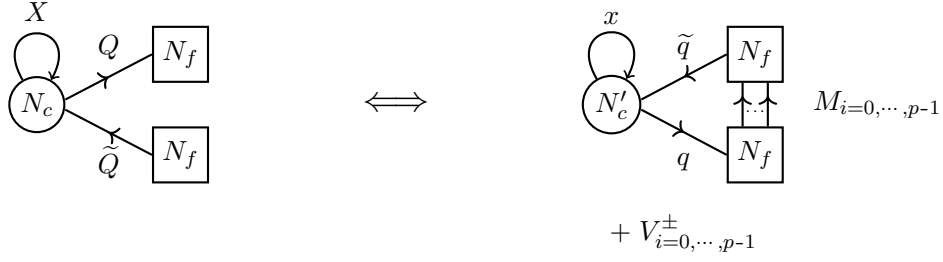

First, let us briefly review the Kim--Park duality. It is an IR duality between the following two theories \cite{Kim:2013cma}.
\begin{itemize}
\item The 3d $\mathcal N=2$ $U(N_c)$ gauge theory with $N_f$ pairs of fundamental and antifundamental fields and one adjoint $X$ having the superpotential
\begin{align}
W = \mathrm{Tr} \, X^{p+1} \,.
\end{align}
\item The 3d $\mathcal N=2$ $U(p N_f-N_c)$ gauge theory with $N_f$ pairs of fundamental and antifundamental $q$ and $\tilde q$, one adjoint $x$, $p N_f^2+2 p$ singlet $M_n$ and $V_n^\pm$, and the superpotential
\begin{align}
\label{eq:dual KP sup}
W = \mathrm{Tr} \, x^{p+1}+\sum_{i = 0}^{p-1} \left(M_{p-1-i} \, \tilde q x^i q+V_{p-1-i}^+ \, \hat v_i^++V_{p-1-i}^- \, \hat v_i^-\right) \,,
\end{align}
where $\hat v_n^\pm$ are the monopole operators with $U(1)_T$ charges $\mp 1$.
\end{itemize}
The corresponding quiver diagram is shown in Figure \ref{fig:KP}.\footnote{The charges of the elementary matter fields and the chiral monopole operators can be found in the original paper \cite{Kim:2013cma}.}

\begin{figure}[tbp]
\centering
\begin{tikzpicture}[thick,scale=0.65, every node/.style={scale=0.65}]
  \tikzset{vertex/.style={circle,fill=white!25,minimum size=12pt,inner sep=2pt}}
  \tikzset{every loop/.style={}}
    
    \node[vertex] (m1) at (0,0) [shape=circle,draw=black,minimum size=2em] {$m_1$};
    \node[vertex] (m2) at (1.5,0) [shape=circle,draw=black,minimum size=2em] {$m_2$};
    \node (ecdot) at (3,0) {$\cdots$};
    \node[vertex] (mp-1) at (4.5,0) [shape=circle,draw=black,minimum size=2em] {\tiny$m_{p\texttt{-}1}$};
    \node[vertex] (Nc) at (6,0) [shape=circle,draw=black,minimum size=2em] {$N_c$};
    \node[vertex] (eNf1) at (8,0.8) [shape=rectangle,draw=black,minimum height=2em, minimum width=2em] {$N_f$};
    \node[vertex] (eNf2) at (8,-0.8) [shape=rectangle,draw=black,minimum height=2em, minimum width=2em] {$N_f$};

    \draw[-to, min distance=1cm]  (m1) edge [out=120, in=60] node {} (m1);
    \draw[-to, min distance=1cm]  (m2) edge [out=120, in=60] node {} (m2);
    \draw[->-=.5] ([yshift= 2pt] m1.east) to ([yshift= 2pt] m2.west);
    \draw[->-=.5] ([yshift= -2pt] m2.west) to ([yshift= -2pt] m1.east);
    \draw[->-=.5] ([yshift= 2pt] m2.east) to ([yshift= 2pt] ecdot.west);
    \draw[->-=.5] ([yshift= -2pt] ecdot.west) to ([yshift= -2pt] m2.east);
    \draw[->-=.5] ([yshift= 2pt] ecdot.east) to ([yshift= 2pt] mp-1.west);
    \draw[->-=.5] ([yshift= -2pt] mp-1.west) to ([yshift= -2pt] ecdot.east);
    \draw[-to, min distance=1cm]  (mp-1) edge [out=120, in=60] node {} (mp-1);
    \draw[->-=.5] ([yshift= 2pt] mp-1.east) to ([yshift= 2pt] Nc.west);
    \draw[->-=.5] ([yshift= -2pt] Nc.west) to ([yshift= -2pt] mp-1.east);
    \draw[->-=.5] ([yshift= 2pt] Nc.east) to ([yshift= 0pt] eNf1.west);
    \draw[->-=.5] ([yshift= 0pt] eNf2.west) to ([yshift= -2pt] Nc.east);
    
    \node at (-0.5,2) {$(A)$};
    
    \node at (0,1.3) {$\Phi^{(1)}$};
    \node at (1.5,1.3) {$\Phi^{(2)}$};
    \node at (4.5,1.3) {$\Phi^{(p-1)}$};
    \node at (0.8,0.4) {\footnotesize$R^{(1)}$};
    \node at (0.8,-0.4) {\footnotesize$\WR^{(1)}$};
    \node at (2.3,0.4) {\footnotesize$R^{(2)}$};
    \node at (2.3,-0.4) {\footnotesize$\WR^{(2)}$};
    \node at (3.7,0.4) {\footnotesize$R^{(p\texttt{-}2)}$};
    \node at (3.7,-0.4) {\footnotesize$\WR^{(p\texttt{-}2)}$};
    \node at (5.3,0.4) {\footnotesize$R^{(p\texttt{-}1)}$};
    \node at (5.3,-0.4) {\footnotesize$\WR^{(p\texttt{-}1)}$};
    \node at (6.8,0.8) {$Q$};
    \node at (6.8,-0.8) {$\WQ$};

    \node at (10,0) {\Large $\Longleftrightarrow$};
    
    \node[vertex] (dm1) at (12,0) [shape=circle,draw=black,minimum size=2em] {$\Wm_1$};
    \node[vertex] (dm2) at (13.5,0) [shape=circle,draw=black,minimum size=2em] {$\Wm_2$};
    \node (dcdot) at (15,0) {$\cdots$};
    \node[vertex] (dmp-1) at (16.5,0) [shape=circle,draw=black,minimum size=2em] {\tiny$\Wm_{p\texttt{-}1}$};
    \node[vertex] (dNc) at (18,0) [shape=circle,draw=black,minimum size=2em] {$N_c'$};
    \node[vertex] (dNf1) at (20,0.8) [shape=rectangle,draw=black,minimum height=2em, minimum width=2em] {$N_f$};
    \node[vertex] (dNf2) at (20,-0.8) [shape=rectangle,draw=black,minimum height=2em, minimum width=2em] {$N_f$};

    \draw[-to, min distance=1cm]  (dm1) edge [out=120, in=60] node {} (dm1);
    \draw[-to, min distance=1cm]  (dm2) edge [out=120, in=60] node {} (dm2);
    \draw[->-=.5] ([yshift= 2pt] dm2.west) to ([yshift= 2pt] dm1.east);
    \draw[->-=.5] ([yshift= -2pt] dm1.east) to ([yshift= -2pt] dm2.west);
    \draw[->-=.5] ([yshift= 2pt] dcdot.west) to ([yshift= 2pt] dm2.east);
    \draw[->-=.5] ([yshift= -2pt] dm2.east) to ([yshift= -2pt] dcdot.west);
    \draw[->-=.5] ([yshift= 2pt] dmp-1.west) to ([yshift= 2pt] dcdot.east);
    \draw[->-=.5] ([yshift= -2pt] dcdot.east) to ([yshift= -2pt] dmp-1.west);
    \draw[-to, min distance=1cm]  (dmp-1) edge [out=120, in=60] node {} (dmp-1);
    \draw[->-=.5] ([yshift= 2pt] dNc.west) to ([yshift= 2pt] dmp-1.east);
    \draw[->-=.5] ([yshift= -2pt] dmp-1.east) to ([yshift= -2pt] dNc.west);
    \draw[->-=.5] ([yshift= 0pt] dNf1.west) to ([yshift= 2pt] dNc.east);
    \draw[->-=.5] ([yshift= -2pt] dNc.east) to ([yshift= -2pt] dNf2.west);
    \draw[->-=.7] ([xshift= -5pt] dNf2.north) to ([xshift= -5pt] dNf1.south);
    \draw[->-=.7] ([xshift= 5pt] dNf2.north) to ([xshift= 5pt] dNf1.south);

    \node at (-0.5+12,2) {$(B)$};
    
    \node at (12,1.3) {$\phi^{(p)}$};
    \node at (13.5,1.3) {$\phi^{(p\texttt{-}1)}$};
    \node at (16.5,1.3) {$\phi^{(2)}$};
    \node at (12.8,0.4) {\footnotesize$\Wr^{(p\texttt{-}1)}$};
    \node at (12.8,-0.4) {\footnotesize$r^{(p\texttt{-}1)}$};
    \node at (14.3,0.4) {\footnotesize$\Wr^{(p\texttt{-}2)}$};
    \node at (14.3,-0.4) {\footnotesize$r^{(p\texttt{-}2)}$};
    \node at (15.7,0.4) {\footnotesize$\Wr^{(2)}$};
    \node at (15.7,-0.4) {\footnotesize$r^{(2)}$};
    \node at (17.3,0.4) {\footnotesize$\Wr^{(1)}$};
    \node at (17.3,-0.4) {\footnotesize$r^{(1)}$};
    \node at (18.8,0.8) {$\Wq$};
    \node at (18.8,-0.8) {$q$};
    \node at (20-0.1,-0.1) {\tiny$\cdot$};
    \node at (20,-0.1) {\tiny$\cdot$};
    \node at (20+0.1,-0.1) {\tiny$\cdot$};
    \node at (21.3,0) {$M_{i=0,\cdots,p\texttt{-}1}$};
    \node at (15,-1.3) {$+\;V_{i=0,\cdots,p\texttt{-}1}^\pm$};

\end{tikzpicture}
\caption{\label{fig:dKP} The quiver representation of the deconfined Kim--Park duality. The gauge ranks of the tails are determined as explained in section \ref{sec:confinement}. Also, the rank of the rightmost gauge node on the RHS is given by $N_c' = p N_f-N_c$. $V_i^\pm$ on the dual side are gauge singlet fields coupled to the monopole operators of the dual theory.}
\end{figure}
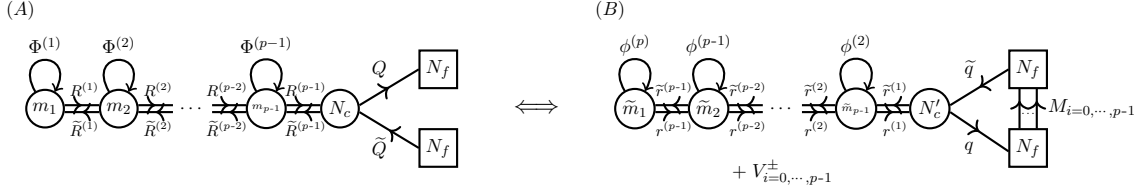

As we have discussed in section \ref{sec:confinement}, the adjoint field $X$ with the superpotential $W = \mathrm{Tr} X^{p+1}$, or $x$ on the dual side, can be deconfined as a linear tail if $N_c = \pm 1 \mod p$. Thus, we propose a deconfined version of the duality as shown in Figure \ref{fig:dKP} with the following superpotentials:
\begin{align}
W_A &= \eta^{p+1}+\eta \sum_{i=\a}^{p\texttt{-}1}\Tr \Phi^{(i)}
    +
    \sum_{i=\a}^{p\texttt{-}1} \hat{V}^{(i),\pm}
    +
     \hat{V}^{(\a,p\texttt{-}1),\mp}
     \sum_{i=\a}^{p-1}\hat{V}^{(i\texttt{-}\a\texttt{+}2,i),\mp}
     \,, \\
W_B &= \eta^{p+1}+\eta \sum_{i=2}^{p\texttt{-}\tilde{\a}\texttt{+}1}\Tr \phi^{(i)}
    +
    \sum_{i=2}^{p\texttt{-}\tilde{\a}\texttt{+}1} \hat{v}^{(i),\pm}
    +
    \hat{v}^{(2,p\texttt{-}\tilde{\a}\texttt{+}1),\mp}
    \sum_{i=1}^{p\texttt{-}\tilde{\a}\texttt{+}1}
    \hat{v}^{(i\texttt{-}\tilde{\a}\texttt{+}2,i),\mp}
    \nonumber \\
    &\quad +
    \sum_{i=0}^{p\texttt{-}\tilde{\a}}
    M_{p\texttt{-}i\texttt{-}1} 
    \tilde{q} \tilde{r}^{(1)}\cdots \tilde{r}^{(i)}
    r^{(i)}\cdots r^{(1)}q
     +\sum_{i=1}^{p\texttt{-}\tilde{\a}\texttt{+}1}
    V_{[-i]}^{\pm} \;\hat{v}^{(1,i),\pm}
    +
    V_{[N_c\texttt{+}i\texttt{-}1]}^{\mp} \;\hat{v}^{(1,i),\mp}
    \nonumber \\
    &\quad +
    \d_{\tilde{\a},2}
    \Tr(\tilde{r}^{(1)}r^{(1)})
    \Big(
    M_0\; \tilde{q} \tilde{r}^{(1)}\cdots \tilde{r}^{(p\texttt{-}2)}
    r^{(p\texttt{-}2)}\cdots r^{(1)}q
    +
    V_0^{\pm} \hat{v}^{(1,p\texttt{-}1),\pm}
    +
    V_0^{\mp} \hat{v}^{(1,p\texttt{-}1),\mp}
    \Big)
\end{align}
where $\hat{V}^{(i,j),\pm}$ and $\hat{v}^{(i,j),\pm}$ are monopole operators of Theory A and Theory B, respectively.
Note that the gauge nodes are labeled from the left for Theory A and from the right for Theory B such that an adjoint field of the $a$-th node and a bifundamental field between the $a$-th node and the $(a+1)$th node carry superscript $(a)$.

Most interestingly, one can prove this duality only assuming the Aharony duality \cite{Aharony:1997gp}, by sequentially applying the Aharony duality to each gauge node from the right, which will be discussed in the subsequent subsections.

\subsection{General derivation}
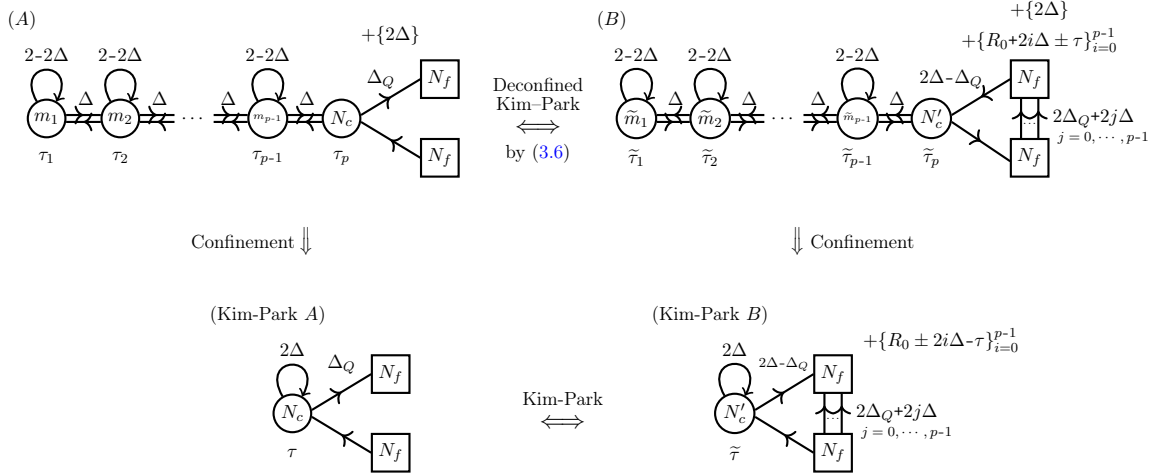
\begin{figure}[tbp]
\centering
\begin{tikzpicture}[thick,scale=0.65, every node/.style={scale=0.65}]
  \tikzset{vertex/.style={circle,fill=white!25,minimum size=12pt,inner sep=2pt}}
  \tikzset{every loop/.style={}}
    
    \node[vertex] (m1) at (0,0) [shape=circle,draw=black,minimum size=2em] {$m_1$};
    \node[vertex] (m2) at (1.5,0) [shape=circle,draw=black,minimum size=2em] {$m_2$};
    \node (ecdot) at (3,0) {$\cdots$};
    \node[vertex] (mp-1) at (4.5,0) [shape=circle,draw=black,minimum size=2em] {\tiny$m_{p\texttt{-}1}$};
    \node[vertex] (Nc) at (6,0) [shape=circle,draw=black,minimum size=2em] {$N_c$};
    \node[vertex] (eNf1) at (8,0.8) [shape=rectangle,draw=black,minimum height=2em, minimum width=2em] {$N_f$};
    \node[vertex] (eNf2) at (8,-0.8) [shape=rectangle,draw=black,minimum height=2em, minimum width=2em] {$N_f$};

    \draw[-to, min distance=1cm]  (m1) edge [out=120, in=60] node {} (m1);
    \draw[-to, min distance=1cm]  (m2) edge [out=120, in=60] node {} (m2);
    \draw[->-=.5] ([yshift= 2pt] m1.east) to ([yshift= 2pt] m2.west);
    \draw[->-=.5] ([yshift= -2pt] m2.west) to ([yshift= -2pt] m1.east);
    \draw[->-=.5] ([yshift= 2pt] m2.east) to ([yshift= 2pt] ecdot.west);
    \draw[->-=.5] ([yshift= -2pt] ecdot.west) to ([yshift= -2pt] m2.east);
    \draw[->-=.5] ([yshift= 2pt] ecdot.east) to ([yshift= 2pt] mp-1.west);
    \draw[->-=.5] ([yshift= -2pt] mp-1.west) to ([yshift= -2pt] ecdot.east);
    \draw[-to, min distance=1cm]  (mp-1) edge [out=120, in=60] node {} (mp-1);
    \draw[->-=.5] ([yshift= 2pt] mp-1.east) to ([yshift= 2pt] Nc.west);
    \draw[->-=.5] ([yshift= -2pt] Nc.west) to ([yshift= -2pt] mp-1.east);
    \draw[->-=.5] ([yshift= 2pt] Nc.east) to ([yshift= 0pt] eNf1.west);
    \draw[->-=.5] ([yshift= 0pt] eNf2.west) to ([yshift= -2pt] Nc.east);

    \node at (-0.5,2) {$(A)$};
    
    \node at (0,1.3) {$2\texttt{-}2\D$};
    \node at (1.5,1.3) {$2\texttt{-}2\D$};
    \node at (4.5,1.3) {$2\texttt{-}2\D$};
    \node at (0.8,0.4) {$\D$};
    \node at (2.3,0.4) {$\D$};
    \node at (3.7,0.4) {$\D$};
    \node at (5.3,0.4) {$\D$};
    \node at (6.8,0.8) {$\D_Q$};
    \node at (7,1.7) {$+\{2\D\}$};

    \node at (0,-0.8) {$\t_1$};
    \node at (1.5,-0.8) {$\t_2$};
    \node at (4.5,-0.8) {$\t_{p\texttt{-}1}$};
    \node at (6,-0.8) {$\t_p$};

    \node at (10,0.7) {Deconfined};
    \node at (10,0.3) {Kim--Park};
    \node at (10,-0.2) {\Large $\Longleftrightarrow$};
    \node at (10,-0.7) {by \eqref{eq: dKP derivation step}};
    
    \node[vertex] (dm1) at (12,0) [shape=circle,draw=black,minimum size=2em] {$\Wm_1$};
    \node[vertex] (dm2) at (13.5,0) [shape=circle,draw=black,minimum size=2em] {$\Wm_2$};
    \node (dcdot) at (15,0) {$\cdots$};
    \node[vertex] (dmp-1) at (16.5,0) [shape=circle,draw=black,minimum size=2em] {\tiny$\Wm_{p\texttt{-}1}$};
    \node[vertex] (dNc) at (18,0) [shape=circle,draw=black,minimum size=2em] {$N_c'$};
    \node[vertex] (dNf1) at (20,0.8) [shape=rectangle,draw=black,minimum height=2em, minimum width=2em] {$N_f$};
    \node[vertex] (dNf2) at (20,-0.8) [shape=rectangle,draw=black,minimum height=2em, minimum width=2em] {$N_f$};

    \draw[-to, min distance=1cm]  (dm1) edge [out=120, in=60] node {} (dm1);
    \draw[-to, min distance=1cm]  (dm2) edge [out=120, in=60] node {} (dm2);
    \draw[->-=.5] ([yshift= 2pt] dm2.west) to ([yshift= 2pt] dm1.east);
    \draw[->-=.5] ([yshift= -2pt] dm1.east) to ([yshift= -2pt] dm2.west);
    \draw[->-=.5] ([yshift= 2pt] dcdot.west) to ([yshift= 2pt] dm2.east);
    \draw[->-=.5] ([yshift= -2pt] dm2.east) to ([yshift= -2pt] dcdot.west);
    \draw[->-=.5] ([yshift= 2pt] dmp-1.west) to ([yshift= 2pt] dcdot.east);
    \draw[->-=.5] ([yshift= -2pt] dcdot.east) to ([yshift= -2pt] dmp-1.west);
    \draw[-to, min distance=1cm]  (dmp-1) edge [out=120, in=60] node {} (dmp-1);
    \draw[->-=.5] ([yshift= 2pt] dNc.west) to ([yshift= 2pt] dmp-1.east);
    \draw[->-=.5] ([yshift= -2pt] dmp-1.east) to ([yshift= -2pt] dNc.west);
    \draw[->-=.5] ([yshift= 0pt] dNf1.west) to ([yshift= 2pt] dNc.east);
    \draw[->-=.5] ([yshift= -2pt] dNc.east) to ([yshift= -2pt] dNf2.west);
    \draw[->-=.7] ([xshift= -5pt] dNf2.north) to ([xshift= -5pt] dNf1.south);
    \draw[->-=.7] ([xshift= 5pt] dNf2.north) to ([xshift= 5pt] dNf1.south);
    
    \node at (-0.5+12,2) {$(B)$};
    
    \node at (12,1.3) {$2\texttt{-}2\D$};
    \node at (13.5,1.3) {$2\texttt{-}2\D$};
    \node at (16.5,1.3) {$2\texttt{-}2\D$};
    \node at (12.8,0.4) {$\D$};
    \node at (14.3,0.4) {$\D$};
    \node at (15.7,0.4) {$\D$};
    \node at (17.3,0.4) {$\D$};
    \node at (18.4,0.8) {$2\D\texttt{-}\D_Q$};
    \node at (20-0.1,-0.1) {\tiny$\cdot$};
    \node at (20,-0.1) {\tiny$\cdot$};
    \node at (20+0.1,-0.1) {\tiny$\cdot$};
    \node at (21.3,0) {$2\D_Q \texttt{+} 2j\D$};
    \node at (21.5,-0.4) {\scriptsize $j=0,\cdots,p\texttt{-}1$};
    
    \node at (20.2,2.2) {$+\{2\D\}$};
    \node at (20.2,1.6) {$+\{R_0 \texttt{+} 2i\D \pm \t\}_{i=0}^{p\texttt{-}1}$};

    \node at (12,-0.8) {$\widetilde{\t}_1$};
    \node at (13.5,-0.8) {$\widetilde{\t}_2$};
    \node at (16.5,-0.8) {$\widetilde{\t}_{p\texttt{-}1}$};
    \node at (18,-0.8) {$\widetilde{\t}_p$};

    \node at (6.3-1,-2.5) 
    {\begin{tikzpicture}
      \node [rotate=90] {\LARGE\textcolor{black}{$\Longleftarrow$}};    
    \end{tikzpicture}
    };
    \node at (6.3-2.3,-2.5) {Confinement};

    \node at (15.3,-2.5) 
    {\begin{tikzpicture}
      \node [rotate=90] {\LARGE\textcolor{black}{$\Longleftarrow$}};    
    \end{tikzpicture}
    };
    \node at (15.3+1.3,-2.5) {Confinement};
    
    \node[vertex] (kNc) at (5,-6) [shape=circle,draw=black,minimum size=2em] {$N_c$};
    \node[vertex] (keNf1) at (7,0.8-6) [shape=rectangle,draw=black,minimum height=2em, minimum width=2em] {$N_f$};
    \node[vertex] (keNf2) at (7,-0.8-6) [shape=rectangle,draw=black,minimum height=2em, minimum width=2em] {$N_f$};

    \draw[-to, min distance=1cm]  (kNc) edge [out=120, in=60] node {} (kNc);
    \draw[->-=.5] ([yshift= 2pt] kNc.east) to ([yshift= 0pt] keNf1.west);
    \draw[->-=.5] ([yshift= 0pt] keNf2.west) to ([yshift= -2pt] kNc.east);

    \node at (5-0.5,2-6) {(Kim-Park $A$)};
    
    \node at (5,1.3-6) {$2\D$};
    \node at (5,-0.8-6) {$\t$};
    \node at (6,1-6) {$\D_Q$};

    \node[vertex] (kmNc) at (14,-6) [shape=circle,draw=black,minimum size=2em] {$N_c'$};
    \node[vertex] (kmNf1) at (16,0.8-6) [shape=rectangle,draw=black,minimum height=2em, minimum width=2em] {$N_f$};
    \node[vertex] (kmNf2) at (16,-0.8-6) [shape=rectangle,draw=black,minimum height=2em, minimum width=2em] {$N_f$};

    \draw[-to, min distance=1cm]  (kmNc) edge [out=120, in=60] node {} (kmNc);
    \draw[->-=.5] ([yshift= 2pt] kmNc.east) to ([yshift= 0pt] kmNf1.west);
    \draw[->-=.5] ([yshift= 0pt] kmNf2.west) to ([yshift= -2pt] kmNc.east);
    \draw[->-=.7] ([xshift= -5pt] kmNf2.north) to ([xshift= -5pt] kmNf1.south);
    \draw[->-=.7] ([xshift= 5pt] kmNf2.north) to ([xshift= 5pt] kmNf1.south);

    \node at (14-0.5,2-6) {(Kim-Park $B$)};
    
    \node at (16-0.1,-0.1-6) {\tiny$\cdot$};
    \node at (16,-0.1-6) {\tiny$\cdot$};
    \node at (16+0.1,-0.1-6) {\tiny$\cdot$};
    \node at (14,1.3-6) {$2\D$};
    \node at (15,1-6) {\scriptsize $2\D\texttt{-}\D_Q$};
    \node at (14,-0.8-6) {$\widetilde{\t}$};

    \node at (17.3,0-6) {$2\D_Q \texttt{+} 2j\D$};
    \node at (17.5,-0.4-6) {\scriptsize $j=0,\cdots,p\texttt{-}1$};

    \node at (18.2,2-6.5) {$+\{R_0 \pm 2i\D \texttt{-} \t\}_{i=0}^{p\texttt{-}1}$};

    \node at (10.5,0.3-6) {Kim-Park};
    \node at (10.5,-0.2-6) {\Large $\Longleftrightarrow$};

\end{tikzpicture}
\caption{\label{fig: dKP derivation} The normalized real masses and FI parameters of a deconfined Kim--Park dual pair and those of the corresponding original Kim--Park dual pair. The dual ranks $\tilde m_j$ are given by $\tilde m_j = j N_f+m_{p-j}-m_p$. For the singlets, $R_0 = N_f(1-\D_Q) - 2\D(N_c-1)$ where $\D=\frac{1}{p+1}$, which is fixed by the confining superpotential $\D W_\text{conf}^{(p),+}$ of the tail.
}
\end{figure}
Figure \ref{fig: dKP derivation} shows the normalized real masses and FI parameters for a deconfined Kim--Park dual pair, indicating the global charges of the elementary chiral fields and monopole operators. Note that the gauge ranks of the $\mathbb D_p[SU(N)]$ tails of the dual pair satisfy the following relation:
\begin{align}
\tilde m_j = j N_f+m_{p-j}-m_p \,,
\end{align}
where $m_j = \lfloor jN_c/p \rfloor$ and $\Wm_j = \lfloor jN_c'/p \rfloor$ for $j = 0, \dots, p$.\footnote{Although $m_j$ and $\tilde m_j$ are defined for $j = 1,\,\dots\,,\,p-1$, their definitions can naturally be extended to the $j = 0$ and $j = p$ cases as well.}
This relation naturally arises from the derivation of the deconfined KP duality using the Aharony duality. We apply the Aharony duality to each gauge node of Theory A from the right. In the $(j+1)$th step, we encounter the following quiver: 
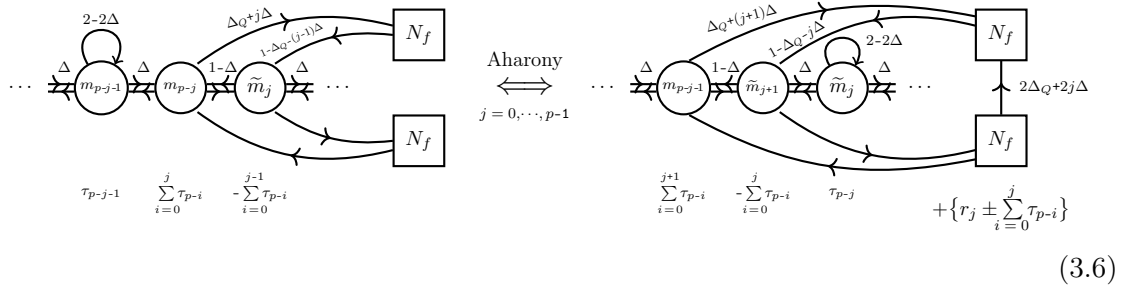
\begin{equation}
    \begin{tikzpicture}[thick,scale=0.7, every node/.style={scale=0.7}]
  \tikzset{vertex/.style={circle,fill=white!25,minimum size=2.5em,inner sep=2pt}}
  \tikzset{every loop/.style={}}

    \node[vertex] (A'n) at (0,0) [shape=circle,draw=black,minimum size=2.5em] {\scriptsize $m_{p\texttt{-}\!j\texttt{-}\!1}$};
    \node[vertex] (A'2n) at (1.5,0) [shape=circle,draw=black,minimum size=2.5em] {\scriptsize $m_{p\texttt{-}j}$};
    \node[vertex] (A'm) at (3,0) [shape=circle,draw=black,minimum size=2.5em] {$\Wm_j$};
    \node[vertex] (A'cd1) at (-1.5,0) {$\cdots$};
    \node[vertex] (A'cd2) at (4.5,0) {$\cdots$};
    \node[vertex] (A'F1) at (6,1) [shape=rectangle,draw=black,minimum height=2.5em, minimum width=2.5em] {$N_f$};
    \node[vertex] (A'F2) at (6,-1) [shape=rectangle,draw=black,minimum height=2.5em, minimum width=2.5em] {$N_f$};

    \draw[-to, min distance=1cm]  (A'n) edge [out=120, in=60] node {} (A'n);
    \draw[->-=.5] ([yshift= 2pt] A'cd1.east) to ([yshift= 2pt] A'n.west);
    \draw[->-=.5] ([yshift= -2pt] A'n.west) to ([yshift= -2pt] A'cd1.east);
    \draw[->-=.5] ([yshift= 2pt] A'n.east) to ([yshift= 2pt] A'2n.west);
    \draw[->-=.5] ([yshift= -2pt] A'2n.west) to ([yshift= -2pt] A'n.east);
    \draw[->-=.5] ([yshift= 2pt] A'm.west) to ([yshift= 2pt] A'2n.east);
    \draw[->-=.5] ([yshift= -2pt] A'2n.east) to ([yshift= -2pt] A'm.west);
    \draw[->-=.5] ([yshift= 2pt] A'm.east) to ([yshift= 2pt] A'cd2.west);
    \draw[->-=.5] ([yshift= -2pt] A'cd2.west) to ([yshift= -2pt] A'm.east);
    
    \draw[->-=.5] ([yshift= 0pt] A'F1.west) to [bend right=20] ([xshift= -6pt,yshift=13pt] A'm.east);
    \draw[->-=.5] ([xshift= -6pt,yshift=-13pt] A'm.east) to [bend right=20] ([yshift= 0pt] A'F2.west);
    
    \draw[->-=.5] ([xshift= -5pt,yshift=13pt] A'2n.east) to [bend left=25] ([yshift= 5pt] A'F1.west);
    \draw[->-=.5] ([yshift= -5pt] A'F2.west) to [bend left=25] ([xshift= -5pt,yshift=-13pt] A'2n.east);

    \node at (0,1.3) {\scriptsize $2\texttt{-}2\D$};
    \node at (-0.7,0.4) {\scriptsize $\D$};
    \node at (0.8,0.4) {\scriptsize $\D$};
    \node at (2.3,0.4) {\scriptsize $1\texttt{-}\D$};
    \node at (3.8,0.4) {\scriptsize $\D$};

    \node at (3.6,0.9)
    {\begin{tikzpicture}
      \node [rotate=25] {\scriptsize $1\texttt{-}\D_Q \texttt{-}( j \texttt{-} 1) \D$};    
    \end{tikzpicture}
    };
    \node at (2.8,1.3)
    {\begin{tikzpicture}
      \node [rotate=15] { $\D_Q \texttt{+} j\D$};    
    \end{tikzpicture}
    };

    \node at (0,-2) {\scriptsize $\t_{p\texttt{-}j\texttt{-}1}$};
    
    \node at (1.25,-1.7) {\tiny $j$};
    \node at (1.5,-2) {\scriptsize $\sum \t_{p\texttt{-}i}$};
    \node at (1.25,-2.3) {\tiny $i\!=\!0$};

    \node at (2.9,-1.7) {\tiny $j\texttt{-}1$};
    \node at (3,-2) {\scriptsize $\texttt{-}\sum \t_{p\texttt{-}i}$};
    \node at (2.9,-2.3) {\tiny $i\!=\!0$};

    \node at (8,0.5) {Aharony};
    \node at (8,0) {\LARGE $\Longleftrightarrow$};
    \node at (8,-0.5) {\scriptsize $j=0,\!\cdots\!,p\texttt{-1}$};

    \node[vertex] (B'n) at (11,0) [shape=circle,draw=black,minimum size=2.5em]{\scriptsize $m_{p\texttt{-}\!j\texttt{-}\!1}$};
    \node[vertex] (B'2m) at (12.5,0) [shape=circle,draw=black,minimum size=2.5em] {\scriptsize $\Wm_{j\texttt{+}1}$};
    \node[vertex] (B'm) at (14,0) [shape=circle,draw=black,minimum size=2.5em] {$\Wm_j$};
    \node[vertex] (B'cd1) at (9.5,0) {$\cdots$};
    \node[vertex] (B'cd2) at (15.5,0) {$\cdots$};
    \node[vertex] (B'F1) at (17,1) [shape=rectangle,draw=black,minimum height=2.5em, minimum width=2.5em] {$N_f$};
    \node[vertex] (B'F2) at (17,-1) [shape=rectangle,draw=black,minimum height=2.5em, minimum width=2.5em] {$N_f$};

    \draw[->-=.5] ([yshift= 2pt] B'cd1.east) to ([yshift= 2pt] B'n.west);
    \draw[->-=.5] ([yshift= -2pt] B'n.west) to ([yshift= -2pt] B'cd1.east);
    \draw[->-=.5] ([yshift= 2pt] B'2m.west) to ([yshift= 2pt] B'n.east);
    \draw[->-=.5] ([yshift= -2pt] B'n.east) to ([yshift= -2pt] B'2m.west);
    \draw[->-=.5] ([yshift= 2pt] B'2m.east) to ([yshift= 2pt] B'm.west);
    \draw[->-=.5] ([yshift= -2pt] B'm.west) to ([yshift= -2pt] B'2m.east);
    \draw[-to, min distance=0.8cm]  (B'm) edge [out=120, in=60] node {} (B'm);
    \draw[->-=.5] ([yshift= 2pt] B'm.east) to ([yshift= 2pt] B'cd2.west);
    \draw[->-=.5] ([yshift= -2pt] B'cd2.west) to ([yshift= -2pt] B'm.east);
    \draw[->-=.5] ([yshift= 5pt] B'F1.west) to [bend right=25] ([xshift= -5pt,yshift=13pt] B'2m.east);
    \draw[->-=.5] ([xshift= -5pt,yshift=-13pt] B'2m.east) to [bend right=25] ([yshift= -5pt] B'F2.west);
    \draw[->-=.5] ([xshift= 3pt,yshift=0pt] B'n.north) to [bend left=20] ([yshift= 10pt] B'F1.west);
    \draw[->-=.5] ([yshift= -10pt] B'F2.west) to [bend left=20] ([xshift= 3pt,yshift=0pt] B'n.south);
    
    \draw[->-=.6] ([xshift= 0pt] B'F2.north) to ([xshift= 0pt] B'F1.south);

    \node at (-0.7+11,0.4) {\scriptsize $\D$};
    \node at (0.8+11,0.4) {\scriptsize $1\texttt{-}\D$};
    \node at (2.3+11,0.4) {\scriptsize $\D$};
    \node at (3.8+11,0.4) {\scriptsize $\D$};
    \node at (3.8+11,0.9) {\scriptsize $2\texttt{-}2\D$};
    \node at (18,0) {\scriptsize $2\D_Q \texttt{+} 2j\D$};
    
    \node at (17+0.2,-2.3+0.35) {\scriptsize $j$};
    \node at (17,-2.3) {$+\big\{ r_j \pm \sum \t_{p\texttt{-}i} \big\}$};
    \node at (17+0.2,-2.3-0.3) {\scriptsize $i=0$};
    
    \node at (4.15+9,0.9)
    {\begin{tikzpicture}
      \node [rotate=29] { $1\texttt{-}\D_Q \texttt{-} j\D$};    
    \end{tikzpicture}
    };
    \node at (2.8+9.3,1.3)
    {\begin{tikzpicture}
      \node [rotate=15] { $\D_Q \texttt{+} (j\texttt{+}1)\D$};    
    \end{tikzpicture}
    };

    \node at (-0.25+11,-1.7) {\tiny $j\texttt{+}1$};
    \node at (0+11,-2) {\scriptsize $\sum \t_{p\texttt{-}i}$};
    \node at (-0.25+11,-2.3) {\tiny $i\!=\!0$};
    
    \node at (1.35+11,-1.7) {\tiny $j$};
    \node at (1.5+11,-2) {\scriptsize $\texttt{-}\sum \t_{p\texttt{-}i}$};
    \node at (1.35+11,-2.3) {\tiny $i\!=\!0$};

    \node at (3+11,-2) {\scriptsize $\t_{p\texttt{-}j}$};

\end{tikzpicture}
    \label{eq: dKP derivation step}
\end{equation}
where we then apply the Aharony duality on the $(p-j)$th node to obtain the RHS.
In the end, we get $\Wm_p = N_c' = p N_f - N_c$, the expected dual gauge rank of the Kim-Park duality. Note that the dual normalized FI parameter of the $U(\Wm_{j})$ gauge node is given by $\widetilde{\t}_j = \t_{p\texttt{-}j}$ for $j=1,\,\cdots\,,\,p-1$, which is indeed the value required by the confining superpotential deformation $\D W_\text{conf}^+$ on the dual side:\footnote{The normalized FI parameters $\t_{i}$ for $i=1,\cdots,p-1$ are fixed as $(m_{j\texttt{-}1} - 2 m_{j} + m_{j\texttt{+}1} + 2) (1-\D) + \t_j= 2$ by the monopole superpotential term conditions.}
\begin{align}
    (\Wm_{j\texttt{-}1} - 2 \Wm_{j} + \Wm_{j\texttt{+}1} + 2) (1-\D) + \widetilde{\t}_j= 2
    \,,\qquad
    \text{for}\quad j = 1,\, \dots\,, \,p-1.
\end{align}
Furthermore, in the second line of Figure \ref{fig: dKP derivation}, we present the corresponding Kim--Park dual pair with the normalized real masses and FI parameters obtained by confining the $\mathbb D_p[SU(N)]$ tails.
Confining the tail on the LHS, the R-symmetry is mixed with the $U(1)_T$ symmetry of the $U(N_c)$ gauge group such that the normalized FI parameters before and after the confinement, $\t_p$ and $\t$, are related as follows:
\begin{align}
    \t_p = \t + \D (1-[N_c]) \,.
\end{align}
Similarly, the dual FI parameter $\tilde \tau_p$, which is given by $\tilde \tau_p = -\sum_{i = 1}^p \tau_i$ according to \eqref{eq: dKP derivation step}, relates to  that after the confinement, $\tilde \tau$, as follows:
\begin{align}
     -\sum_{i=1}^p \t_i &= \widetilde{\t}_p = \widetilde{\t} + \D(1-[N_c']) \,,
\end{align}
which implies $\widetilde{\t} = -\t$.

We can also express the spectrum of the normalized real mass parameters of the singlets that arise from the repeated steps of the Aharony duality in terms of $\t$ and $\D$:
\begin{align}
    +\Big\{ r_j \pm &\sum_{i=p-j}^{p} \t_i \Big\}_{j=0}^{p-1}
    \;\;=\;\;
    +\big\{ R_0 +2i\D \pm \t\big\}_{i=0}^{p-1}
\end{align}
where $r_j = m_{p\texttt{-}j\texttt{-}1}(1-\D) + N_f(1-\D_Q - j\D) + \Wm_{j}\D - m_{p\texttt{-}j}+1$ and $R_0 = N_f(1-\D_Q) - 2\D(N_c-1)$. The RHS exactly matches that of the monopole-like singlets, which couple to the dressed monopole operators on the dual side of the Kim--Park dual pair.
In addition, the meson-like singlets in the bifundamental representation between the two $SU(N_f)$ flavor symmetries that are produced from \eqref{eq: dKP derivation step} also match those of the original Kim--Park duality. Therefore, the Kim--Park duality of the gauge rank $N_c = \pm 1 \mod p$ can be understood solely from the Aharony duality provided with the confinement of $\mathbb{D}_p[SU(N)]$, which also originates from the Aharony and BBP dualities as we have shown in section \ref{sec:confinement}. In the following subsections, we serve explicit examples for small $p$ cases.

\subsection{Example I: $p = 2$}

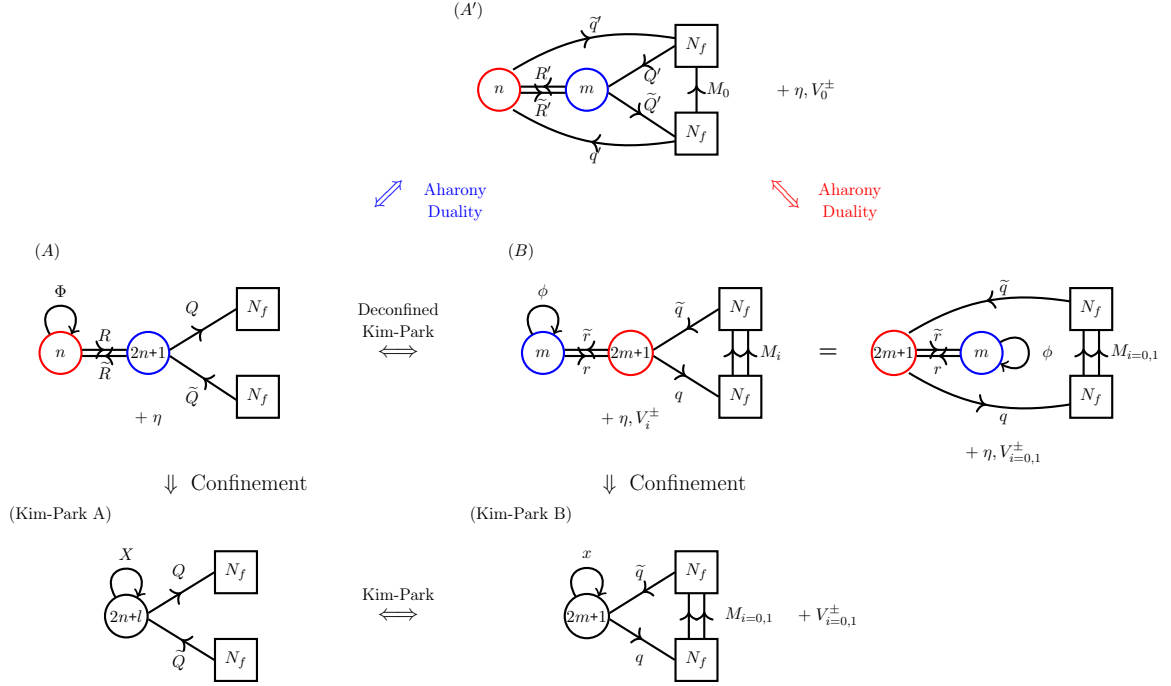
\begin{figure}[tbp]
\centering
\begin{tikzpicture}[thick,scale=0.58, every node/.style={scale=0.58}]
  \tikzset{vertex/.style={circle,fill=white!25,minimum size=2.5em,inner sep=2pt}}
  \tikzset{every loop/.style={}}
    
    \node[vertex] (An) at (0,0) [shape=circle,draw=red,minimum size=2.5em] {$n$};
    \node[vertex] (A2n) at (2,0) [shape=circle,draw=blue,minimum size=2.5em] {$\!2n \texttt{+}1\!$};
    \node[vertex] (AF1) at (4.5,1) [shape=rectangle,draw=black,minimum height=2.5em, minimum width=2.5em] {$N_f$};
    \node[vertex] (AF2) at (4.5,-1) [shape=rectangle,draw=black,minimum height=2.5em, minimum width=2.5em] {$N_f$};

    \draw[-to, min distance=1cm]  (An) edge [out=120, in=60] node {} (An);
    \draw[->-=.5] ([yshift= 2pt] An.east) to ([yshift= 2pt] A2n.west);
    \draw[->-=.5] ([yshift= -2pt] A2n.west) to ([yshift= -2pt] An.east);
    \draw[->-=.5] ([yshift= 2pt] A2n.east) to ([yshift= 0pt] AF1.west);
    \draw[->-=.5] ([yshift= -2pt] AF2.west) to ([yshift= -2pt] A2n.east);

    \node at (0,1.4) {$\Phi$};
    \node at (1,0.4) {$R$};
    \node at (1,-0.4) {$\WR$};
    \node at (3,1) {$Q$};
    \node at (3,-1) {$\WQ$};
    \node at (2,-1.5) {$+\;\eta$};

    \node[vertex] (Bn) at (11,0) [shape=circle,draw=blue,minimum size=2.5em] {$m$};
    \node[vertex] (B2n) at (13,0) [shape=circle,draw=red,minimum size=2.5em] {$\!2m \texttt{+} 1\! $};
    \node[vertex] (BF1) at (15.5,1) [shape=rectangle,draw=black,minimum height=2.5em, minimum width=2.5em] {$N_f$};
    \node[vertex] (BF2) at (15.5,-1) [shape=rectangle,draw=black,minimum height=2.5em, minimum width=2.5em] {$N_f$};

    \draw[-to, min distance=1cm]  (Bn) edge [out=120, in=60] node {} (Bn);
    \draw[->-=.5] ([yshift= 2pt] B2n.west) to ([yshift= 2pt] Bn.east);
    \draw[->-=.5] ([yshift= -2pt] Bn.east) to ([yshift= -2pt] B2n.west);
    \draw[->-=.5] ([yshift= 0pt] BF1.west) to ([yshift= 2pt] B2n.east);
    \draw[->-=.5] ([yshift= -2pt] B2n.east) to ([yshift= 0pt] BF2.west);
    \draw[->-=.6] ([xshift= -5pt] BF2.north) to ([xshift= -5pt] BF1.south);
    \draw[->-=.6] ([xshift= 5pt] BF2.north) to ([xshift= 5pt] BF1.south);

    \node at (11,1.4) {$\phi$};
    \node at (12,0.4) {$\Wr$};
    \node at (12,-0.4) {$r$};
    \node at (13.6+0.5,1) {$\Wq$};
    \node at (13.6+0.5,-1) {$q$};
    \node at (15.7+0.5,0) {$M_i$};
    \node at (13,-1.5) {$+\;\eta,V_i^\pm$};

    \node[vertex] (b2m) at (19,0) [shape=circle,draw=red,minimum size=2.5em] {$\!2m \texttt{+} 1\!$};
    \node (bm) at (21,0) [shape=circle,draw=blue,minimum size=2.5em] {$m$};
    \node[vertex] (bF1) at (23.5,1) [shape=rectangle,draw=black,minimum height=2.5em, minimum width=2.5em] {$N_f$};
    \node[vertex] (bF2) at (23.5,-1) [shape=rectangle,draw=black,minimum height=2.5em, minimum width=2.5em] {$N_f$};

    \draw[-to, min distance=1cm]  (bm) edge [out=30, in=-30] node {} (bm);
    \draw[->-=.5] ([yshift= 2pt] b2m.east) to ([yshift= 2pt] bm.west);
    \draw[->-=.5] ([yshift= -2pt] bm.west) to ([yshift= -2pt] b2m.east);
    \draw[->-=.5] ([yshift= 5pt] bF1.west) to [bend right=20] ([xshift=-5pt ,yshift= 13pt] b2m.east);
    \draw[->-=.5] ([xshift=-5pt ,yshift= -13pt] b2m.east) to [bend right=20] ([yshift= -5pt] bF2.west);
    \draw[->-=.6] ([xshift= -5pt] bF2.north) to ([xshift= -5pt] bF1.south);
    \draw[->-=.6] ([xshift= 5pt] bF2.north) to ([xshift= 5pt] bF1.south);

    \node at (22.5,0) {$\phi$};
    \node at (20,0.4) {$\Wr$};
    \node at (20,-0.4) {$r$};
    \node at (24.5,0) {$M_{i=0,1}$};
    \node at (21.5,1.5) {$\Wq$};
    \node at (21.5,-1.5) {$q$};
    \node at (21.5,-2.3) {$+\;\eta,V_{i=0,1}^\pm$};

    \node[vertex] (A'n) at (10,6) [shape=circle,draw=red,minimum size=2.5em] {$n$};
    \node[vertex] (A'm) at (12,6) [shape=circle,draw=blue,minimum size=2.5em] { $m$};
    \node[vertex] (A'F1) at (14.5,7) [shape=rectangle,draw=black,minimum height=2.5em, minimum width=2.5em] {$N_f$};
    \node[vertex] (A'F2) at (14.5,5) [shape=rectangle,draw=black,minimum height=2.5em, minimum width=2.5em] {$N_f$};

    \draw[->-=.5] ([yshift= 2pt] A'm.west) to ([yshift= 2pt] A'n.east);
    \draw[->-=.5] ([yshift= -2pt] A'n.east) to ([yshift= -2pt] A'm.west);
    \draw[->-=.5] ([yshift= 0pt] A'F1.west) to ([yshift= 2pt] A'm.east);
    \draw[->-=.5] ([yshift= -2pt] A'm.east) to ([yshift= -2pt] A'F2.west);
    \draw[->-=.6] ([yshift= 0pt] A'F2.north) to ([yshift= 0pt] A'F1.south);
    \draw[->-=.5] ([xshift= -5pt,yshift=13pt] A'n.east) to [bend left=20] ([yshift= 5pt] A'F1.west);
    \draw[->-=.5] ([yshift= -5pt] A'F2.west) to [bend left=20] ([xshift= -5pt,yshift=-13pt] A'n.east);

    \node at (11,6+0.4) {$R'$};
    \node at (11,6-0.4) {$\WR'$};
    \node at (13.5,6+0.3) {$Q'$};
    \node at (13.5,6-0.3) {$\WQ'$};
    \node at (12.2,6+1.5) {$\Wq'$};
    \node at (12.2,6-1.5) {$q'$};
    \node at (15,6) {$M_0$};
    \node at (17,6) {$+\;\eta,V_0^\pm$};

    \node[vertex] (e2n) at (1.5,-6) [shape=circle,draw=black,minimum size=2.5em] {$2n\texttt{+}l$};
    \node[vertex] (eF1) at (4,-6+1) [shape=rectangle,draw=black,minimum height=2.5em, minimum width=2.5em] {$N_f$};
    \node[vertex] (eF2) at (4,-6-1) [shape=rectangle,draw=black,minimum height=2.5em, minimum width=2.5em] {$N_f$};

    \draw[-to, min distance=1cm]  (e2n) edge [out=120, in=60] node {} (e2n);
    \draw[->-=.5] ([yshift= 2pt] e2n.east) to ([yshift= 0pt] eF1.west);
    \draw[->-=.5] ([yshift= 0pt] eF2.west) to ([yshift= -2pt] e2n.east);

    \node at (1.5,-6+1.4) {$X$};
    \node at (2.7,-6+1) {$Q$};
    \node at (2.7,-6-1) {$\WQ$};

    \node[vertex] (m2m) at (12,-6) [shape=circle,draw=black,minimum size=2.5em] {$\!2m\texttt{+}1\!$};
    \node[vertex] (mF1) at (14.5,-6+1) [shape=rectangle,draw=black,minimum height=2.5em, minimum width=2.5em] {$N_f$};
    \node[vertex] (mF2) at (14.5,-6-1) [shape=rectangle,draw=black,minimum height=2.5em, minimum width=2.5em] {$N_f$};

    \draw[-to, min distance=1cm]  (m2m) edge [out=120, in=60] node {} (m2m);
    \draw[->-=.5] ([yshift= 0pt] mF1.west) to ([yshift= 2pt] m2m.east);
    \draw[->-=.5] ([yshift= -2pt] m2m.east) to ([yshift= 0pt] mF2.west);
    \draw[->-=.6] ([xshift= -5pt] mF2.north) to ([xshift= -5pt] mF1.south);
    \draw[->-=.6] ([xshift= 5pt] mF2.north) to ([xshift= 5pt] mF1.south);

    \node at (12,-6+1.4) {$x$};
    \node at (13.2,-6+1) {$\Wq$};
    \node at (13.2,-6-1) {$q$};
    \node at (15.7,-6) {$M_{i=0,1}$};
    \node at (17.5,-6) {$+\;V_{i=0,1}^\pm$};

    \node at (9.3,7.8) {$(A')$};
    \node at (-0.3,2.3) {$(A)$};
    \node at (10.5,2.3) {$(B)$};
    \node at (4,-3) {\Large $\Downarrow\;$ Confinement};
    \node at (14,-3) {\Large $\Downarrow\;$ Confinement};
    \node at (7.7,1) {Deconfined};
    \node at (7.7,0.5) {Kim-Park};
    \node at (7.7,0) {\Large $\Longleftrightarrow$};
    \node at (17.5,0) {\Large $\boldsymbol{=}$};
    \node at (7.7,-6+0.5) {Kim-Park};
    \node at (7.7,-6) {\Large $\Longleftrightarrow$};
    \node at (0,-3.7) {(Kim-Park A)};
    \node at (10.5,-3.7) {(Kim-Park B)};

    \node at (7.5,3.6) 
    {\begin{tikzpicture}
      \node [rotate=45] {\Huge \textcolor{blue}{$\Longleftrightarrow$}};    
    \end{tikzpicture}
    };
    \node at (9,3.7) { \textcolor{blue}{Aharony}};
    \node at (9,3.2) { \textcolor{blue}{Duality}};
    \node at (16.5,3.6) 
    {\begin{tikzpicture}
      \node [rotate=-45] {\Huge \textcolor{red}{$\Longleftrightarrow$}};    
    \end{tikzpicture}
    };
    \node at (18,3.7) { \textcolor{red}{Aharony}};
    \node at (18,3.2) { \textcolor{red}{Duality}};
    
\end{tikzpicture}
\caption{\label{fig:dKP_p=2} The deconfined Kim--Park duality for $p = 2$ (the second line) and its derivation using the Aharony duality. The confinement of the $\mathbb D_p[SU(N)]$ tails of Theory A and Theory B leads to the original Kim--Park duality (the third line). Note that $m$ is defined by $m = N_f-n-1$. The extra gauge singlets not included in the quiver diagrams are indicated separately.}
\end{figure}

As the simplest example, let us consider the $p = 2$ case first. We assume $N_c \neq 1$ is odd so that $\gcd(p,N_c) = 1$. The corresponding deconfined Kim--Park duality is shown in the second line of Figure \ref{fig:dKP_p=2}.
The theory on each side of the duality has the superpotential as follows:
\begin{align}
W_A &= \mathrm{Tr} \, \tilde R \Phi R+\eta^{3}+\eta \, \mathrm{Tr} \, \Phi+\hat V^{(1),+}+\hat V^{(1),-} \,, \label{eq:p=2_A} \\
W_B &= \mathrm{Tr} \, \tilde r \phi r+\eta^{3}+\eta \, \mathrm{Tr} \, \phi+\hat v^{(2),+}+\hat v^{(2),-} \nonumber \\
&\quad +M_0 \, \tilde q \tilde r r q+M_1 \, \tilde q q+V_0^+ \, \hat v^{(1,2),+}+V_0^- \, \hat v^{(1,2),-}+V_1^+ \, \hat v^{(1),+}+V_1^- \, \hat v^{(1),-} \label{eq:p=2_B}
\end{align}
where the first terms of $W_A$ and $W_B$ are the cubic superpotential terms inherited from the $\mathcal N=4$ D-term of the $\mathbb D_p[SU(N)]$ tail, the next two terms are the superpotential terms triggering the confinement of $\mathbb D_p[SU(N)]$, and lastly the remaining terms of $W_B$ are the deconfined version of the dual superpotential terms of the Kim--Park duality, \eqref{eq:dual KP sup}, other than $\mathrm{Tr} \, X^{p+1}$. Due to these superpotentials, some of the monopole operators are lifted; namely, Theory A has $\hat V^{(2),\pm}$ and $\hat V^{(1,2),\pm}$ as chiral monopole operators, while $V^{(1),\pm}$ are lifted, and Theory B has no chiral monopole operator.
The charges of the elementary matter fields and the chiral monopole operators are given in Table \ref{tab:charge_p=2}.
\begin{table}[tbp]
\centering
\begin{tabular}{|c|c|c|c|c|c|}
\hline
 & $U(1)_R$ & $SU(N_f)_t$ & $SU(N_f)_u$ & $U(1)_A$ & $U(1)_T$ \\
\hline 
$Q$ & $\Delta_Q$ & $\overline{\mathbf{N_f}}$ & $\mathbf 1$ & $1$ & $0$ \\
$\tilde Q$ & $\Delta_Q$ & $\mathbf 1$ & $\mathbf{N_f}$ & $1$ & $0$ \\
$R$ & $1/3$ & $\mathbf 1$ & $\mathbf 1$ & $0$ & $0$ \\
$\tilde R$ & $1/3$ & $\mathbf 1$ & $\mathbf 1$ & $0$ & $0$ \\
$\Phi$ & $4/3$ & $\mathbf 1$ & $\mathbf 1$ & $0$ & $0$ \\
$\eta$ & $2/3$ & $\mathbf 1$ & $\mathbf 1$ & $0$ & $0$ \\
$\hat V^{(2),\pm}$ & $(1-\Delta_Q) N_f-4 n/3$ & $\mathbf 1$ & $\mathbf 1$ & $-N_f$ & $\pm1$ \\
$\hat V^{(1,2),\pm}$ & $(1-\Delta_Q) N_f-(4 n-2)/3$ & $\mathbf 1$ & $\mathbf 1$ & $-N_f$ & $\pm1$ \\
\hline
$q$ & $2/3-\Delta_Q$ & $\mathbf 1$ & $\overline{\mathbf{N_f}}$ & $1$ & $0$ \\
$\tilde q$ & $2/3-\Delta_Q$ & $\mathbf{N_f}$ & $\mathbf 1$ & $1$ & $0$ \\
$r$ & $1/3$ & $\mathbf 1$ & $\mathbf 1$ & $0$ & $0$ \\
$\tilde r$ & $1/3$ & $\mathbf 1$ & $\mathbf 1$ & $0$ & $0$ \\
$\phi$ & $4/3$ & $\mathbf 1$ & $\mathbf 1$ & $0$ & $0$ \\
$\eta$ & $2/3$ & $\mathbf 1$ & $\mathbf 1$ & $0$ & $0$ \\
$M_i$ & $2 \Delta_Q+2 i/3$ & $\overline{\mathbf{N_f}}$ & $\mathbf{N_f}$ & $2$ & $0$ \\
$V_i^\pm$ & $(1-\Delta_Q) N_f-(4 n-2 i)/3$ & $\mathbf 1$ & $\mathbf 1$ & $-N_f$ & $\pm1$ \\
\hline
\end{tabular}
\caption{\label{tab:charge_p=2} The charges of the elementary matter fields and the chiral monopole operators of Theory A (the upper box) and Theory B (the lower box) for $p = 2$. The hatted operators are monopoles, whereas the others are elementary matter fields. Note that the monopole operators of Theory B are not indicated in this table because they become Q-exact due to the superpotential. The subscript $i$ of $M_i$ and $V_i^\pm$ runs from 0 to 1.}
\end{table}

As shown in the figure, one can derive this duality by applying the Aharony duality on each gauge node sequentially from the right. Once we take the Aharony duality on the blue gauge node of Theory A, we get Theory A$'$ in the first line of Figure \ref{fig:dKP_p=2}, where the gauge rank changes from $2n+1$ to $N_f-n-1 \equiv m$. In addition, extra $N_f$ pairs of fundamental and anti-fundamental fields of the red node, denoted by $q'$ and $\tilde q'$ and gauge singlets $M_0$ and $V_0^\pm$ are generated, where the latter interact with the gauge sector via the following superpotential:
\begin{align}
W_{A'} = \eta^{3}+\eta \, \mathrm{Tr} \, R' \tilde R'+\hat v^{(1,2),\pm}+Q' R' \tilde q'+q' \tilde R' \tilde Q'+M_0 \, Q' \tilde Q'+V_0^\pm \, \hat{v}^{(2),\pm} \,.
\end{align}
There is also a new adjoint field of the red node, which however forms a mass term with $\Phi$, and both of them are integrated out.
The charges of the elementary matter fields and the monopole operators of Theory A$'$ are given in Table \ref{tab:charge_p=2_A'}.
\begin{table}[tbp]
\centering
\begin{tabular}{|c|c|c|c|c|c|}
\hline
 & $U(1)_R$ & $SU(N_f)_t$ & $SU(N_f)_u$ & $U(1)_A$ & $U(1)_T$ \\
\hline 
$Q'$ & $1-\Delta_Q$ & $\mathbf{N_f}$ & $\mathbf 1$ & $-1$ & $0$ \\
$\tilde Q'$ & $1-\Delta_Q$ & $\mathbf 1$ & $\overline{\mathbf{N_f}}$ & $-1$ & $0$ \\
$R'$ & $2/3$ & $\mathbf 1$ & $\mathbf 1$ & $0$ & $0$ \\
$\tilde R'$ & $2/3$ & $\mathbf 1$ & $\mathbf 1$ & $0$ & $0$ \\
$q'$ & $1/3+\Delta_Q$ & $\mathbf 1$ & $\mathbf{N_f}$ & $1$ & $0$ \\
$\tilde q'$ & $1/3+\Delta_Q$ & $\overline{\mathbf{N_f}}$ & $\mathbf 1$ & $1$ & $0$ \\
$\eta$ & $2/3$ & $\mathbf 1$ & $\mathbf 1$ & $0$ & $0$ \\
$M_0$ & $2 \Delta_Q$ & $\overline{\mathbf{N_f}}$ & $\mathbf{N_f}$ & $2$ & $0$ \\
$V_0^\pm$ & $(1-\Delta_Q) N_f-4 n/3$ & $\mathbf 1$ & $\mathbf 1$ & $-N_f$ & $\pm1$ \\
$\hat v^{(1),\pm}$ & $(1-\Delta_Q) N_f-(4 n-2)/3$ & $\mathbf 1$ & $\mathbf 1$ & $-N_f$ & $\pm1$ \\
\hline
\end{tabular}
\caption{\label{tab:charge_p=2_A'} The charges of the elementary matter fields and the chiral monopole operators of Theory A$'$ for $p = 2$. The hatted operators are monopoles, whereas the others are elementary matter fields.}
\end{table}

Next, we take another step of the Aharony duality on the red node of Theory A$'$, whose rank then changes from $n$ to $2 N_f-2n-1 = 2 m+1$. For the blue node, the adjoint field $\phi$ and extra fundamental and anti-fundamental fields are generated, but the latter form mass terms with $Q$ and $\tilde Q'$ and are integrated out. Also, gauge singlets $M_1$ and $V_1^\pm$ are generated and interact via the superpotential \eqref{eq:p=2_B}. The resulting theory is exactly the proposed dual theory of the deconfined Kim--Park duality for $p = 2$. The corresponding identity for the $S^3_b$ partition function is also given in appendix \ref{app: S_b^3 ptf}.

Moreover, as we argued in section \ref{sec:confinement}, the red node of Theory A with the superpotential \eqref{eq:p=2_A} confines and generates an adjoint field $X$ of the blue node with the superpotential term $\mathrm{Tr} \, X^3$; in the same manner, the blue node of Theory B also confines and generates an adjoint field $x$ with the superpotential term $\mathrm{Tr} \, x^3$, leading to the original Kim--Park duality for $p = 2$, shown in the third line of Figure \ref{fig:dKP_p=2}. We also provide the duality map of the chiral ring generators across this chain of dualities in Table \ref{tab:map_p=2}.
\begin{table}[tbp]
\centering
\begin{tabular}{|c|c|c|c|c|}
\hline
Kim--Park A & Theory A & Theory A$'$ & Theory B & Kim--Park B \\
\hline 
$\tilde Q Q$ & $\tilde Q Q$ & $M_0$ & $M_0$ & $M_0$ \\
$\tilde Q X Q$ & $\tilde Q \tilde R R Q$ & $q' \tilde q'$ & $M_1$ & $M_1$ \\
$\mathrm{Tr} \, X$ & $\eta \sim \mathrm{Tr} \, \tilde R R$ & $\eta$ & $\eta \sim \mathrm{Tr} \, \tilde r r$ & $\mathrm{Tr} \, x$ \\
$\hat V_0^\pm$ & $\hat V^{(2),\pm}$ & $V_0^\pm$ & $V_0^\pm$ & $V_0^\pm$ \\
$\hat V_1^\pm$ & $\hat V^{(1,2),\pm}$ & $\hat v^{(1),\pm}$ & $V_1^\pm$ & $V_1^\pm$ \\
\hline
\end{tabular}
\caption{\label{tab:map_p=2} The duality map of the chiral ring generators for $p = 2$.}
\end{table}

\subsection{Example II: $p = 3$}

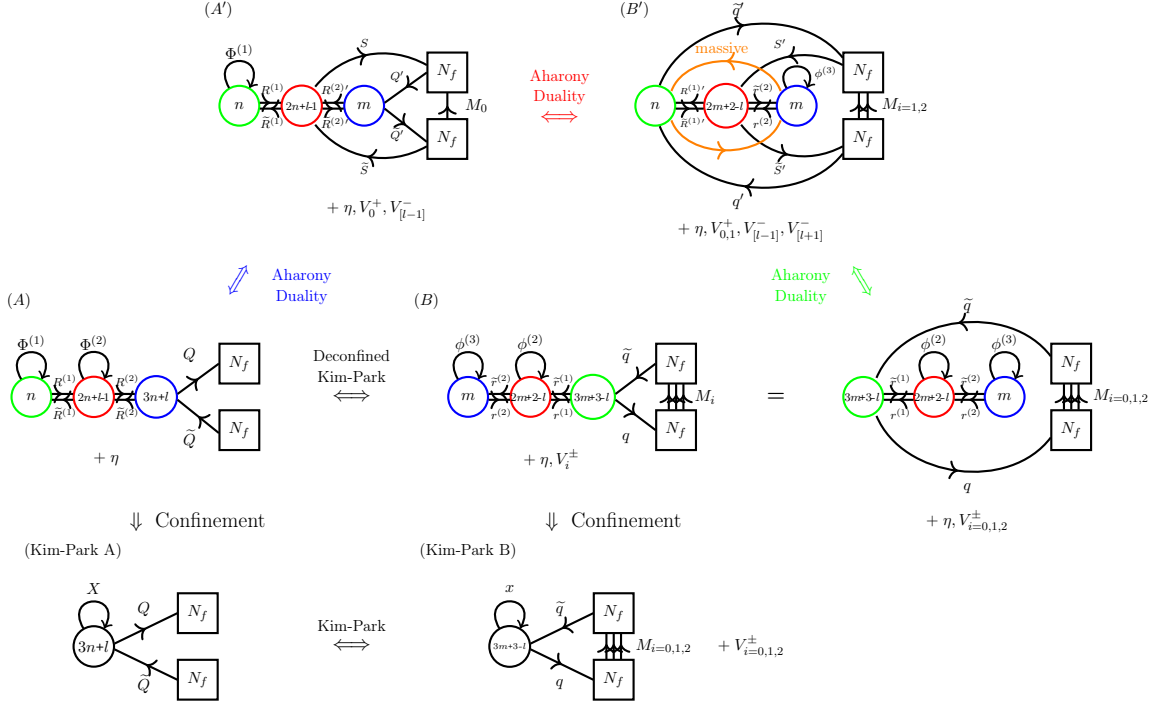
\begin{figure}[tbp]
\centering 
\begin{tikzpicture}[thick,scale=0.55, every node/.style={scale=0.55}]
  \tikzset{vertex/.style={circle,fill=white!25,minimum size=2.5em,inner sep=2pt}}
  \tikzset{every loop/.style={}}
    
    \node[vertex] (An) at (0,0) [shape=circle,draw=green,minimum size=2.5em] {$n$};
    \node[vertex] (A2n) at (1.5,0) [shape=circle,draw=red,minimum size=2.5em] {\scriptsize $2n \texttt{+} l\! \texttt{-}\!1$};
    \node (A3n) at (3,0) [shape=circle,draw=blue,minimum size=2.5em] {\footnotesize $3n \texttt{+} l$};
    \node[vertex] (AF1) at (5,0.8) [shape=rectangle,draw=black,minimum height=2.5em, minimum width=2.5em] {$N_f$};
    \node[vertex] (AF2) at (5,-0.8) [shape=rectangle,draw=black,minimum height=2.5em, minimum width=2.5em] {$N_f$};

    \draw[-to, min distance=1cm]  (An) edge [out=120, in=60] node {} (An);
    \draw[-to, min distance=1cm]  (A2n) edge [out=120, in=60] node {} (A2n);
    \draw[->-=.5] ([yshift= 2pt] An.east) to ([yshift= 2pt] A2n.west);
    \draw[->-=.5] ([yshift= -2pt] A2n.west) to ([yshift= -2pt] An.east);
    \draw[->-=.5] ([yshift= 2pt] A2n.east) to ([yshift= 2pt] A3n.west);
    \draw[->-=.5] ([yshift= -2pt] A3n.west) to ([yshift= -2pt] A2n.east);
    \draw[->-=.5] ([yshift= 2pt] A3n.east) to ([yshift= 0pt] AF1.west);
    \draw[->-=.5] ([yshift= -2pt] AF2.west) to ([yshift= -2pt] A3n.east);

    \node at (0,1.3) {$\Phi^{(1)}$};
    \node at (1.5,1.3) {$\Phi^{(2)}$};
    \node at (0.8,0.4) {\scriptsize$R^{(1)}$};
    \node at (0.8,-0.4) {\scriptsize$\WR^{(1)}$};
    \node at (2.3,0.4) {\scriptsize$R^{(2)}$};
    \node at (2.3,-0.4) {\scriptsize$\WR^{(2)}$};
    \node at (3.8,1) {$Q$};
    \node at (3.8,-1) {$\WQ$};
    \node at (1.8,-1.5) {$+\;\eta$};

    \node[vertex] (Bn) at (10+0.5,0) [shape=circle,draw=blue,minimum size=2.5em] {$m$};
    \node[vertex] (B2n) at (11.5+0.5,0) [shape=circle,draw=red,minimum size=2.5em] {\scriptsize $\!\!2m \!\texttt{+} 2\! \texttt{-}l$};
    \node (B3n) at (13+0.5,0) [shape=circle,draw=green,minimum size=2.5em] {\scriptsize $\!\!3m \!\texttt{+}3\!\texttt{-}l\!$};
    \node[vertex] (BF1) at (15+0.5,0.8) [shape=rectangle,draw=black,minimum height=2.5em, minimum width=2.5em] {$N_f$};
    \node[vertex] (BF2) at (15+0.5,-0.8) [shape=rectangle,draw=black,minimum height=2.5em, minimum width=2.5em] {$N_f$};

    \draw[-to, min distance=1cm]  (Bn) edge [out=120, in=60] node {} (Bn);
    \draw[-to, min distance=1cm]  (B2n) edge [out=120, in=60] node {} (B2n);
    \draw[->-=.5] ([yshift= 2pt] B2n.west) to ([yshift= 2pt] Bn.east);
    \draw[->-=.5] ([yshift= -2pt] Bn.east) to ([yshift= -2pt] B2n.west);
    \draw[->-=.5] ([yshift= 2pt] B3n.west) to ([yshift= 2pt] B2n.east);
    \draw[->-=.5] ([yshift= -2pt] B2n.east) to ([yshift= -2pt] B3n.west);
    \draw[->-=.5] ([yshift= 0pt] BF1.west) to ([yshift= 2pt] B3n.east);
    \draw[->-=.5] ([yshift= -2pt] B3n.east) to ([yshift= 0pt] BF2.west);
    \draw[->-=.6] ([xshift= -5pt] BF2.north) to ([xshift= -5pt] BF1.south);
    \draw[->-=.6] ([xshift= 0pt] BF2.north) to ([xshift= 0pt] BF1.south);
    \draw[->-=.6] ([xshift= 5pt] BF2.north) to ([xshift= 5pt] BF1.south);

    \node at (10+0.5,1.3) {$\phi^{(3)}$};
    \node at (11.5+0.5,1.3) {$\phi^{(2)}$};
    \node at (10.8+0.5,0.4) {\scriptsize$\Wr^{(2)}$};
    \node at (10.8+0.5,-0.4) {\scriptsize$r^{(2)}$};
    \node at (12.3+0.5,0.4) {\scriptsize$\Wr^{(1)}$};
    \node at (12.3+0.5,-0.4) {\scriptsize$r^{(1)}$};
    \node at (13.8+0.5,1) {$\Wq$};
    \node at (13.8+0.5,-1) {$q$};
    \node at (15.7+0.5,0) {$M_i$};
    \node at (12.5,-1.5) {$+\;\eta,V_i^\pm$};

    \node[vertex] (b3m) at (20,0) [shape=circle,draw=green,minimum size=2.5em] {\scriptsize $\!\!3m \!\texttt{+}3\!\texttt{-}l\!$};
    \node[vertex] (b2m) at (21.7,0) [shape=circle,draw=red,minimum size=2.5em] {\scriptsize $\!\!2m \!\texttt{+} 2\! \texttt{-}l$};
    \node (bm) at (23.4,0) [shape=circle,draw=blue,minimum size=2.5em] {$m$};
    \node[vertex] (bF1) at (25,0.8) [shape=rectangle,draw=black,minimum height=2.5em, minimum width=2.5em] {$N_f$};
    \node[vertex] (bF2) at (25,-0.8) [shape=rectangle,draw=black,minimum height=2.5em, minimum width=2.5em] {$N_f$};

    \draw[-to, min distance=1cm]  (b2m) edge [out=120, in=60] node {} (b2m);
    \draw[-to, min distance=1cm]  (bm) edge [out=120, in=60] node {} (bm);
    \draw[->-=.5] ([yshift= 2pt] b3m.east) to ([yshift= 2pt] b2m.west);
    \draw[->-=.5] ([yshift= -2pt] b2m.west) to ([yshift= -2pt] b3m.east);
    \draw[->-=.5] ([yshift= 2pt] b2m.east) to ([yshift= 2pt] bm.west);
    \draw[->-=.5] ([yshift= -2pt] bm.west) to ([yshift= -2pt] b2m.east);
    \draw[->-=.5] ([yshift= 5pt] bF1.west) to [bend right=60] ([xshift=-5pt ,yshift= 13pt] b3m.east);
    \draw[->-=.5] ([xshift=-5pt ,yshift= -13pt] b3m.east) to [bend right=60] ([yshift= -5pt] bF2.west);
    \draw[->-=.6] ([xshift= -5pt] bF2.north) to ([xshift= -5pt] bF1.south);
    \draw[->-=.6] ([xshift= 0pt] bF2.north) to ([xshift= 0pt] bF1.south);
    \draw[->-=.6] ([xshift= 5pt] bF2.north) to ([xshift= 5pt] bF1.south);

    \node at (21.7,1.3) {$\phi^{(2)}$};
    \node at (23.4,1.3) {$\phi^{(3)}$};
    \node at (20.9,0.4) {\scriptsize$\Wr^{(1)}$};
    \node at (20.9,-0.4) {\scriptsize$r^{(1)}$};
    \node at (22.6,0.4) {\scriptsize$\Wr^{(2)}$};
    \node at (22.6,-0.4) {\scriptsize$r^{(2)}$};
    \node at (26.2,0) {$M_{i=0,1,2}$};
    \node at (22.5,2.2) {$\Wq$};
    \node at (22.5,-2.2) {$q$};
    \node at (22.5,-3) {$+\;\eta,V_{i=0,1,2}^\pm$};

    \node[vertex] (A'n) at (5,7) [shape=circle,draw=green,minimum size=2.5em] {$n$};
    \node[vertex] (A'2n) at (6.5,7) [shape=circle,draw=red,minimum size=2.5em] {\scriptsize $2n \texttt{+} l\! \texttt{-}\!1$};
    \node[vertex] (A'm) at (8,7) [shape=circle,draw=blue,minimum size=2.5em] {$m$};
    \node[vertex] (A'F1) at (10,7.8) [shape=rectangle,draw=black,minimum height=2.5em, minimum width=2.5em] {$N_f$};
    \node[vertex] (A'F2) at (10,7-0.8) [shape=rectangle,draw=black,minimum height=2.5em, minimum width=2.5em] {$N_f$};

    \draw[-to, min distance=1cm]  (A'n) edge [out=120, in=60] node {} (A'n);
    \draw[->-=.5] ([yshift= 2pt] A'n.east) to ([yshift= 2pt] A'2n.west);
    \draw[->-=.5] ([yshift= -2pt] A'2n.west) to ([yshift= -2pt] A'n.east);
    \draw[->-=.5] ([yshift= 2pt] A'm.west) to ([yshift= 2pt] A'2n.east);
    \draw[->-=.5] ([yshift= -2pt] A'2n.east) to ([yshift= -2pt] A'm.west);
    \draw[->-=.5] ([yshift= 0pt] A'F1.west) to ([yshift= 2pt] A'm.east);
    \draw[->-=.5] ([yshift= -2pt] A'm.east) to ([yshift= -2pt] A'F2.west);
    \draw[->-=.5] ([yshift= 0pt] A'F2.north) to ([yshift= 0pt] A'F1.south);
    \draw[->-=.5] ([xshift= -5pt,yshift=13pt] A'2n.east) to [bend left=40] ([yshift= 5pt] A'F1.west);
    \draw[->-=.5] ([yshift= -5pt] A'F2.west) to [bend left=40] ([xshift= -5pt,yshift=-13pt] A'2n.east);

    \node at (5,8.3) {$\Phi^{(1)}$};
    \node at (5.8,7.4) {\scriptsize$R^{(1)}$};
    \node at (5.8,7-0.4) {\scriptsize$\WR^{(1)}$};
    \node at (7.3,7.4) {\scriptsize$R^{(2)}{}'$};
    \node at (7.3,7-0.4) {\scriptsize$\WR^{(2)}{}'$};
    \node at (8.8,7+0.7) {\scriptsize $Q'$};
    \node at (8.8,7-0.7) {\scriptsize $\WQ'$};
    \node at (8,7+1.5) {\scriptsize $S$};
    \node at (8,7-1.5) {\scriptsize $\WS$};
    \node at (10.7,7) {$M_0$};
    \node at (8.3,4.5) {$+\; \eta, V_0^+ , V_{[l-1]}^-$};
    
    \node[vertex] (B'n) at (15,7) [shape=circle,draw=green,minimum size=2.5em]{$n$};
    \node[vertex] (B'2m) at (16.7,7) [shape=circle,draw=red,minimum size=2.5em] {\scriptsize $\!2m \texttt{+} 2\! \texttt{-}l$};
    \node[vertex] (B'm) at (18.4,7) [shape=circle,draw=blue,minimum size=2.5em] {$m$};
    \node[vertex] (B'F1) at (20,7.8) [shape=rectangle,draw=black,minimum height=2.5em, minimum width=2.5em] {$N_f$};
    \node[vertex] (B'F2) at (20,7-0.8) [shape=rectangle,draw=black,minimum height=2.5em, minimum width=2.5em] {$N_f$};

    \draw[->-=.5] ([yshift= 2pt] B'2m.west) to ([yshift= 2pt] B'n.east);
    \draw[->-=.5] ([yshift= -2pt] B'n.east) to ([yshift= -2pt] B'2m.west);
    \draw[->-=.5] ([yshift= 2pt] B'2m.east) to ([yshift= 2pt] B'm.west);
    \draw[->-=.5] ([yshift= -2pt] B'm.west) to ([yshift= -2pt] B'2m.east);
    \draw[-to, min distance=0.8cm]  (B'm) edge [out=120, in=60] node {} (B'm);
    \draw[->-=.5] ([yshift= 5pt] B'F1.west) to [bend right=40] ([xshift= -5pt,yshift=13pt] B'2m.east);
    \draw[->-=.5] ([xshift= -5pt,yshift=-13pt] B'2m.east) to [bend right=40] ([yshift= -5pt] B'F2.west);
    \draw[->-=.5,orange] ([xshift=3pt,yshift= 10pt] B'm.west) to [bend right=70] ([xshift=-3pt,yshift=10pt] B'n.east);
    \draw[->-=.5,orange] ([xshift=-3pt,yshift=-10pt] B'n.east) to [bend right=70] ([xshift=3pt,yshift= -10pt] B'm.west);
    \draw[->-=.5] ([xshift= 3pt,yshift=0pt] B'n.north) to [bend left=60] ([yshift= 10pt] B'F1.west);
    \draw[->-=.5] ([yshift= -10pt] B'F2.west) to [bend left=60] ([xshift= 3pt,yshift=0pt] B'n.south);
    \draw[->-=.6] ([xshift= -3pt] B'F2.north) to ([xshift= -3pt] B'F1.south);
    \draw[->-=.6] ([xshift= 3pt] B'F2.north) to ([xshift= 3pt] B'F1.south);

    \node at (15.9,7+0.4) {\tiny$R^{(1)}{}'$};
    \node at (15.9,7-0.4) {\tiny$\WR^{(1)}{}'$};
    \node at (17.6,7+0.4) {\scriptsize$\Wr^{(2)}$};
    \node at (17.6,7-0.4) {\scriptsize$r^{(2)}$};
    \node at (19.1,7+0.8) {\scriptsize$\phi^{(3)}$};
    \node at (18,7+1.5) {\scriptsize$S'$};
    \node at (18,7-1.5) {\scriptsize$\WS'$};
    \node at (17,7+2.3) {$\Wq'$};
    \node at (17,7-2.3) {$q'$};
    \node at (21,7) {$M_{i=1,2}$};
    \node at (16.6,7+1.4) {\textcolor{orange}{massive}};
    \node at (17.3,4) {$+\; \eta, V_{0,1}^+ , V_{[l-1]}^-,V_{[l+1]}^-$};

    \node[vertex] (e3n) at (1.5,-6) [shape=circle,draw=black,minimum size=2.5em] {$3n\texttt{+}l$};
    \node[vertex] (eF1) at (4,-6+0.8) [shape=rectangle,draw=black,minimum height=2.5em, minimum width=2.5em] {$N_f$};
    \node[vertex] (eF2) at (4,-6-0.8) [shape=rectangle,draw=black,minimum height=2.5em, minimum width=2.5em] {$N_f$};

    \draw[-to, min distance=1cm]  (e3n) edge [out=120, in=60] node {} (e3n);
    \draw[->-=.5] ([yshift= 2pt] e3n.east) to ([yshift= 0pt] eF1.west);
    \draw[->-=.5] ([yshift= 0pt] eF2.west) to ([yshift= -2pt] e3n.east);

    \node at (1.5,-6+1.4) {$X$};
    \node at (2.7,-6+0.9) {$Q$};
    \node at (2.7,-6-0.9) {$\WQ$};

    \node[vertex] (m3m) at (11.5,-6) [shape=circle,draw=black,minimum size=2.5em] {\tiny$3m\texttt{+}3\texttt{-}l$};
    \node[vertex] (mF1) at (14,-6+0.8) [shape=rectangle,draw=black,minimum height=2.5em, minimum width=2.5em] {$N_f$};
    \node[vertex] (mF2) at (14,-6-0.8) [shape=rectangle,draw=black,minimum height=2.5em, minimum width=2.5em] {$N_f$};

    \draw[-to, min distance=1cm]  (m3m) edge [out=120, in=60] node {} (m3m);
    \draw[->-=.5] ([yshift= 0pt] mF1.west) to ([yshift= 2pt] m3m.east);
    \draw[->-=.5] ([yshift= -2pt] m3m.east) to ([yshift= 0pt] mF2.west);
    \draw[->-=.6] ([xshift= -5pt] mF2.north) to ([xshift= -5pt] mF1.south);
    \draw[->-=.6] ([xshift= 0pt] mF2.north) to ([xshift= 0pt] mF1.south);
    \draw[->-=.6] ([xshift= 5pt] mF2.north) to ([xshift= 5pt] mF1.south);

    \node at (11.5,-6+1.4) {$x$};
    \node at (12.7,-6+0.9) {$\Wq$};
    \node at (12.7,-6-0.9) {$q$};
    \node at (15.2,-6) {$M_{i=0,1,2}$};
    \node at (17.3,-6) {$+\;V_{i=0,1,2}^\pm$};

    \node at (4.5,9.3) {$(A')$};
    \node at (14.5,9.3) {$(B')$};
    \node at (-0.3,2.3) {$(A)$};
    \node at (9.5,2.3) {$(B)$};
    \node at (4,-3) {\Large $\Downarrow\;$ Confinement};
    \node at (14,-3) {\Large $\Downarrow\;$ Confinement};
    \node at (7.7,1) {Deconfined};
    \node at (7.7,0.5) {Kim-Park};
    \node at (7.7,0) {\Large $\Longleftrightarrow$};
    \node at (17.9,0) {\Large $\boldsymbol{=}$};
    \node at (12.7,7-0.3) {\Large \textcolor{red}{$\Longleftrightarrow$}};
    \node at (12.7,7+0.7) {\textcolor{red}{Aharony}};
    \node at (12.7,7+0.2) {\textcolor{red}{Duality}};
    \node at (7.7,-6+0.5) {Kim-Park};
    \node at (7.7,-6) {\Large $\Longleftrightarrow$};
    \node at (1,-3.7) {(Kim-Park A)};
    \node at (10.5,-3.7) {(Kim-Park B)};

    \node at (5,2.8) 
    {\begin{tikzpicture}
      \node [rotate=60] {\Huge \textcolor{blue}{$\Longleftrightarrow$}};    
    \end{tikzpicture}
    };
    \node at (6.5,2.9) { \textcolor{blue}{Aharony}};
    \node at (6.5,2.4) { \textcolor{blue}{Duality}};
    \node at (20,2.8) 
    {\begin{tikzpicture}
      \node [rotate=120] {\Huge \textcolor{green}{$\Longleftrightarrow$}};    
    \end{tikzpicture}
    };
    \node at (18.5,2.9) { \textcolor{green}{Aharony}};
    \node at (18.5,2.4) { \textcolor{green}{Duality}};

\end{tikzpicture}
\caption{\label{fig:dKP_p=3} The deconfined Kim--Park duality for $p = 3$ (the second line) and its derivation using the Aharony duality. The confinement of the $\mathbb D_p[SU(N)]$ tails of Theory A and Theory B leads to the original Kim--Park duality (the third line). Note that $m = N_f-n-1$ and $l = 1,\, 2$. The extra gauge singlets not included in the quiver diagrams are indicated separately.}
\end{figure}

Next, let us examine the $p = 3$ case, which can easily be generalized to higher $p$ as well. We assume $N_c$ is larger than and not divided by 3. The corresponding deconfined Kim--Park duality is shown in the second line of Figure \ref{fig:dKP_p=3}. 
The theory on each side of the duality has the superpotential as follows:
\begin{align}
W_A &= W_{\mathcal N=4}+\hat V^{(1),+}+\hat V^{(2),+}+\hat V^{(1,2),-}+\eta^{4}+\eta \left(\mathrm{Tr} \, \Phi^{(1)}+\mathrm{Tr} \, \Phi^{(2)}\right) \,, \label{eq:p=3_A} \\
W_B & = W_{\mathcal N=4}+\hat v^{(2),+}+\hat v^{(3),+}+\hat v^{(2,3),-}+\eta^{4}+\eta \left(\mathrm{Tr} \, \phi^{(2)}+\mathrm{Tr} \, \phi^{(3)}\right) \nonumber \\
&\quad +M_0 \, \tilde q \tilde r^{(1)} \tilde r^{(2)} r^{(2)} r^{(1)} q^{(1)}+M_1 \, \tilde q \tilde r^{(1)} r^{(1)} q+M_2 \, \tilde q q \nonumber \\
&\quad +V_0^+ \, \hat v^{(1,3),+}+V_{[l-1]}^- \, \hat v^{(1,3),-}+V_1^+ \, \hat v^{(1,2),+}+V_{[l+1]}^- \, \hat v^{(1,2),-}+V_2^+ \, \hat v^{(1),+}+V_{[l]}^- \, \hat v^{(1),-} \label{eq:p=3_B}
\end{align}
where we have defined $[n] = n \mod 3$.
The first lines of $W_A$ and $W_B$ are the superpotentials of confining $\mathbb D_p[SU(N)]$, where $W_{\mathcal N=4}$ includes the standard cubic superpotential terms from the $\mathcal N=4$ D-term. The second and third lines of $W_B$ are a deconfined version of the dual superpotential terms of the Kim--Park duality, \eqref{eq:dual KP sup}, other than $\mathrm{Tr} \, X^{p+1}$ as before. Due to these superpotentials, some of the monopole operators are lifted: Theory A has chiral monopole operators $V^{(3),\pm}$, $V^{(2,3),\pm}$, and $V^{(1,3),\pm}$, while Theory B has no chiral monopole operator. The charges of the elementary matter fields and the monopole operators are given in Table \ref{tab:charge_p=3}.
\begin{table}[tbp]
\centering
\begin{tabular}{|c|c|c|c|c|c|}
\hline
 & $U(1)_R$ & $SU(N_f)_t$ & $SU(N_f)_u$ & $U(1)_A$ & $U(1)_T$ \\
\hline 
$Q$ & $\Delta_Q$ & $\overline{\mathbf{N_f}}$ & $\mathbf 1$ & $1$ & $0$ \\
$\tilde Q$ & $\Delta_Q$ & $\mathbf 1$ & $\mathbf{N_f}$ & $1$ & $0$ \\
$R^{(a)}$ & $1/4$ & $\mathbf 1$ & $\mathbf 1$ & $0$ & $0$ \\
$\tilde R^{(a)}$ & $1/4$ & $\mathbf 1$ & $\mathbf 1$ & $0$ & $0$ \\
$\Phi^{(a)}$ & $3/2$ & $\mathbf 1$ & $\mathbf 1$ & $0$ & $0$ \\
$\eta$ & $1/2$ & $\mathbf 1$ & $\mathbf 1$ & $0$ & $0$ \\
$\hat V^{(3),+}$ & $(1-\Delta_Q) N_f-(3 n+l-1)/2$ & $\mathbf 1$ & $\mathbf 1$ & $-N_f$ & $1$ \\
$\hat V^{(3),-}$ & $(1-\Delta_Q) N_f-3 n/2$ & $\mathbf 1$ & $\mathbf 1$ & $-N_f$ & $-1$ \\
$\hat V^{(2,3),+}$ & $(1-\Delta_Q) N_f-(3 n+l-2)/2$ & $\mathbf 1$ & $\mathbf 1$ & $-N_f$ & $1$ \\
$\hat V^{(2,3),-}$ & $(1-\Delta_Q) N_f-(3 n+3 l-5)/2$ & $\mathbf 1$ & $\mathbf 1$ & $-N_f$ & $-1$ \\
$\hat V^{(1,3),+}$ & $(1-\Delta_Q) N_f-(3 n+l-3)/2$ & $\mathbf 1$ & $\mathbf 1$ & $-N_f$ & $1$ \\
$\hat V^{(1,3),-}$ & $(1-\Delta_Q) N_f-(3 n-1)/2$ & $\mathbf 1$ & $\mathbf 1$ & $-N_f$ & $-1$ \\
\hline
$q$ & $1/2-\Delta_Q$ & $\mathbf 1$ & $\overline{\mathbf{N_f}}$ & $-1$ & $0$ \\
$\tilde q$ & $1/2-\Delta_Q$ & $\mathbf{N_f}$ & $\mathbf 1$ & $-1$ & $0$ \\
$r^{(a)}$ & $1/4$ & $\mathbf 1$ & $\mathbf 1$ & $0$ & $0$ \\
$\tilde r^{(a)}$ & $1/4$ & $\mathbf 1$ & $\mathbf 1$ & $0$ & $0$ \\
$\phi^{(a+1)}$ & $3/2$ & $\mathbf 1$ & $\mathbf 1$ & $0$ & $0$ \\
$\eta$ & $1/2$ & $\mathbf 1$ & $\mathbf 1$ & $0$ & $0$ \\
$M_i$ & $2 \Delta_Q+i/2$ & $\overline{\mathbf{N_f}}$ & $\mathbf{N_f}$ & $2$ & $0$ \\
$V_i^\pm$ & $(1-\Delta_Q) N_f-(3 n+l-1-i)/2$ & $\mathbf 1$ & $\mathbf 1$ & $-N_f$ & $\pm1$ \\
\hline
\end{tabular}
\caption{\label{tab:charge_p=3} The charges of the elementary matter fields and the chiral monopole operators of Theory A (the upper box) and Theory B (the lower box) for $p = 3$. The hatted operators are monopoles, whereas the others are elementary matter fields. Note that the monopole operators of Theory B are not indicated in this table because they become Q-exact due to the superpotential. The superscript $a$ runs from 1 to 2, and the subscript $i$ runs from 0 to 2.}
\end{table}

Again, this duality can be derived from the Aharony duality, which we depict in Figure \ref{fig:dKP_p=3}. We first apply the Aharony duality to the blue gauge node of Theory A to obtain Theory A$'$, which changes the gauge rank from $3 n+l$ to $N_f-n-1 \equiv m$ and generate extra fields as shown in the figure, whose charges are given in Table \ref{tab:charge_p=3'}.
\begin{table}[tbp]
\centering
\begin{tabular}{|c|c|c|c|c|c|}
\hline
 & $U(1)_R$ & $SU(N_f)_t$ & $SU(N_f)_u$ & $U(1)_A$ & $U(1)_T$ \\
\hline 
$Q'$ & $1-\Delta_Q$ & $\mathbf{N_f}$ & $\mathbf 1$ & $-1$ & $0$ \\
$\tilde Q'$ & $1-\Delta_Q$ & $\mathbf 1$ & $\overline{\mathbf{N_f}}$ & $-1$ & $0$ \\
$R^{(1)}$ & $1/4$ & $\mathbf 1$ & $\mathbf 1$ & $0$ & $0$ \\
$\tilde R^{(1)}$ & $1/4$ & $\mathbf 1$ & $\mathbf 1$ & $0$ & $0$ \\
$R^{(2)}{}'$ & $3/4$ & $\mathbf 1$ & $\mathbf 1$ & $0$ & $0$ \\
$\tilde R^{(2)}{}'$ & $3/4$ & $\mathbf 1$ & $\mathbf 1$ & $0$ & $0$ \\
$S$ & $1/4+\Delta_Q$ & $\overline{\mathbf{N_f}}$ & $\mathbf 1$ & $1$ & $0$ \\
$\tilde S$ & $1/4+\Delta_Q$ & $\mathbf 1$ & $\mathbf{N_f}$ & $1$ & $0$ \\
$\Phi^{(1)}$ & $3/2$ & $\mathbf 1$ & $\mathbf 1$ & $0$ & $0$ \\
$\eta$ & $1/2$ & $\mathbf 1$ & $\mathbf 1$ & $0$ & $0$ \\
$M_0$ & $2 \Delta_Q$ & $\overline{\mathbf{N_f}}$ & $\mathbf{N_f}$ & $2$ & $0$ \\
$V_0^+$ & $(1-\Delta_Q) N_f-(3 n+l-1)/2$ & $\mathbf 1$ & $\mathbf 1$ & $-N_f$ & $1$ \\
$V_{[l-1]}^-$ & $(1-\Delta_Q) N_f-3 n/2$ & $\mathbf 1$ & $\mathbf 1$ & $-N_f$ & $-1$ \\
$\hat v^{(2),+}$ & $(1-\Delta_Q) N_f-(3 n+l-2)/2$ & $\mathbf 1$ & $\mathbf 1$ & $-N_f$ & $1$ \\
$\hat v^{(2),-}$ & $(1-\Delta_Q) N_f-(3 n+3 l-5)/2$ & $\mathbf 1$ & $\mathbf 1$ & $-N_f$ & $-1$ \\
$\hat v^{(1,2),+}$ & $(1-\Delta_Q) N_f-(3 n+l-3)/2$ & $\mathbf 1$ & $\mathbf 1$ & $-N_f$ & $1$ \\
$\hat v^{(1,2),-}$ & $(1-\Delta_Q) N_f-(3 n-1)/2$ & $\mathbf 1$ & $\mathbf 1$ & $-N_f$ & $-1$ \\
\hline
$R^{(1)}{}'$ & $3/4$ & $\mathbf 1$ & $\mathbf 1$ & $0$ & $0$ \\
$\tilde R^{(1)}{}'$ & $3/4$ & $\mathbf 1$ & $\mathbf 1$ & $0$ & $0$ \\
$r^{(2)}$ & $1/4$ & $\mathbf 1$ & $\mathbf 1$ & $0$ & $0$ \\
$\tilde r^{(2)}$ & $1/4$ & $\mathbf 1$ & $\mathbf 1$ & $0$ & $0$ \\
$S'$ & $3/4-\Delta_Q$ & $\mathbf{N_f}$ & $\mathbf 1$ & $-1$ & $0$ \\
$\tilde S'$ & $3/4-\Delta_Q$ & $\mathbf 1$ & $\overline{\mathbf{N_f}}$ & $-1$ & $0$ \\
$q'$ & $1/2+\Delta_Q$ & $\mathbf 1$ & $\mathbf{N_f}$ & $1$ & $0$ \\
$\tilde q'$ & $1/2+\Delta_Q$ & $\overline{\mathbf{N_f}}$ & $\mathbf 1$ & $1$ & $0$ \\
$\phi^{(3)}$ & $3/2$ & $\mathbf 1$ & $\mathbf 1$ & $0$ & $0$ \\
$\eta$ & $1/2$ & $\mathbf 1$ & $\mathbf 1$ & $0$ & $0$ \\
$M_i$ & $2 \Delta_Q+i/2$ & $\overline{\mathbf{N_f}}$ & $\mathbf{N_f}$ & $2$ & $0$ \\
$V_0^+$ & $(1-\Delta_Q) N_f-(3 n+l-1)/2$ & $\mathbf 1$ & $\mathbf 1$ & $-N_f$ & $1$ \\
$V_{[l-1]}^-$ & $(1-\Delta_Q) N_f-3 n/2$ & $\mathbf 1$ & $\mathbf 1$ & $-N_f$ & $-1$ \\
$V_1^+$ & $(1-\Delta_Q) N_f-(3 n+l-2)/2$ & $\mathbf 1$ & $\mathbf 1$ & $-N_f$ & $1$ \\
$V_{[l+1]}^-$ & $(1-\Delta_Q) N_f-(3 n+3 l-5)/2$ & $\mathbf 1$ & $\mathbf 1$ & $-N_f$ & $-1$ \\
$\hat v^{(1),+}$ & $(1-\Delta_Q) N_f-(3 n+l-3)/2$ & $\mathbf 1$ & $\mathbf 1$ & $-N_f$ & $1$ \\
$\hat v^{(1),-}$ & $(1-\Delta_Q) N_f-(3 n-1)/2$ & $\mathbf 1$ & $\mathbf 1$ & $-N_f$ & $-1$ \\
\hline
\end{tabular}
\caption{\label{tab:charge_p=3'} The charges of the elementary matter fields and the chiral monopole operators of Theory A$'$ (the upper box) and Theory B$'$ (the lower box) for $p = 3$. The hatted operators are monopoles, whereas the others are elementary matter fields. The subscript $i$ of $M_i$ runs from 0 to 1.}
\end{table}
In addition, the superpotential becomes
\begin{align}
W_{A'} &= \mathrm{Tr} \, \tilde R^{(1)} \Phi^{(1)} \, R^{(1)}+\mathrm{Tr} \, R^{(2)}{}' \tilde R^{(1)} R^{(1)} \tilde R^{(2)}{}'+\hat v^{(1),+}+\hat v^{(2,3),+}+\hat v^{(1,3),-} \nonumber \\
&\quad +\eta^{4}+\eta \left(\mathrm{Tr} \, \Phi^{(1)}+\mathrm{Tr} \, R^{(2)}{}' \tilde R^{(2)}{}'\right)+Q' R^{(2)}{}' S+\tilde S \tilde R^{(2)}{}' \tilde Q' \nonumber \\
&\quad +M_0 \, Q' \tilde Q'+V_0^+ \, \hat v^{(3),+}+V_{[l-1]}^- \, \hat v^{(3),-} \,.
\end{align}
Then we apply the Aharony duality to the red node of Theory A$'$ to obtain Theory B$'$, which changes the gauge rank from $2 n+l-1$ to $2 N_f-2 n-l=2 m+2-l$. Now, the superpotential is given by
\begin{align}
W_{B'} &= \mathrm{Tr} \, \tilde r^{(2)} \phi^{(3)} \, r^{(2)}+\hat{v}^{(1,2),+}+\hat{v}^{(3),+}+\hat{v}^{(1,3),-} \nonumber \\
&\quad +\eta^{4}+\eta \left(\mathrm{Tr} \, R^{(1)}{}' \tilde R^{(1)}{}'+\mathrm{Tr} \, \phi^{(3)}\right)+S' R^{(1)}{}' \tilde q'+q' \tilde R^{(1)}{}' \tilde S' + \tilde R^{(1)}{}' \tilde r^{(2)} r^{(2)} R^{(1)}{}' \nonumber \\
&\quad +M_0 \, S' \tilde r^{(2)} r^{(2)} \tilde S'+M_1 \, S' \tilde S'+V_0^+ \, \hat v^{(2,3),+}+V_{[l-1]}^- \, \hat v^{(2,3),-}+V_1^+ \, \hat v^{(2),+}+V_{[l+1]}^- \, \hat v^{(2),-} \,.
\end{align}
This step also generates extra fields as shown in the figure, among which the bifundamental fields between the green and blue gauge nodes, colored in orange in the figure, form a mass term by themselves and therefore can be integrated out. After integrating them out, we apply the last Aharony duality to the green gauge node of Theory B$'$ and obtain Theory B, which is exactly the proposed dual theory of the deconfined Kim--Park duality for $p = 3$.

As in the $p=2$ case, one can use the confinement of $\mathbb D_p[SU(N)]$ to show that the green and red nodes of Theory A confine, leaving a new adjoint field $X$ of the blue node with the superpotential $\mathrm{Tr} X^4$, and similarly for Theory B. Hence, we get the original Kim-Park duality for $p = 3$, shown in the third line of Figure \ref{fig:dKP_p=3}. We also provide the duality map of the chiral ring generators across this chain of dualities in Table \ref{tab:map_p=3}.

The same method can be applied to arbitrary $p$ as long as $N_c = \pm 1 \mod p$, which is the good theory condition for the $\mathbb D_p[SU(N)]$ tail, whereas the ugly cases are relegated to \cite{ugly}.\footnote{$\mathbb D_p[SU(N)]$ is either good or ugly.} Namely, under this condition, the Kim--Park duality can be proven only assuming the Aharony duality and the confinement of $\mathbb D_p[SU(N)]$, which also originates from the Aharony and BBP dualities as we have shown in section \ref{sec:confinement}.

\begin{table}[tbp]
\centering
\begin{tabular}{|c|c|c|c|c|c|}
\hline
KP A & Theory A & Theory A$'$ & Theory B$'$ & Theory B & KP B \\
\hline 
$\tilde Q Q$ & $\tilde Q Q$ & $M_0$ & $M_0$ & $M_0$ & $M_0$ \\
$\tilde Q X Q$ & $\tilde Q \tilde R^{(2)} R^{(2)} Q$ & $\tilde S S$ & $M_1$ & $M_1$ & $M_1$ \\
$\tilde Q X^2 Q$ & $\tilde Q \tilde R^{(2)} \tilde R^{(1)} R^{(1)} R^{(2)} Q$ & $\tilde S \tilde R^{(1)} R^{(1)} S$ & $q' \tilde q'$ & $M_2$ & $M_2$ \\
$\mathrm{Tr} \, X$ & $\eta \sim \mathrm{Tr} \, \tilde R^{(a)} R^{(a)}$ & $\eta \sim \mathrm{Tr} \, \tilde R^{(1)} R^{(1)}$ & $\eta \sim \mathrm{Tr} \, \tilde r^{(2)} r^{(2)}$ & $\eta \sim \mathrm{Tr} \, \tilde r^{(a)} r^{(a)}$ & $\mathrm{Tr} \, x$ \\
$\mathrm{Tr} \, X^2$ & $\mathrm{Tr} \, \tilde R^{(2)} \tilde R^{(1)} R^{(1)} R^{(2)}$ & $\Tr \tilde R^{(1)} R^{(1)} \tilde R^{(1)} R^{(1)}$ & $\Tr \tilde r^{(2)} r^{(2)} \tilde r^{(2)} r^{(2)}$ & $\mathrm{Tr} \, \tilde r^{(1)} \tilde r^{(2)} r^{(2)} r^{(1)}$ & $\mathrm{Tr} \, x^2$ \\
$\hat V_0^+$ & $\hat V^{(3),+}$ & $V_0^+$ & $V_0^+$ & $V_0^+$ & $V_0^+$ \\
$\hat V_{[l-1]}^-$ & $\hat V^{(3),-}$ & $V_{[l-1]}^-$ & $V_{[l-1]}^-$ & $V_{[l-1]}^-$ & $V_{[l-1]}^-$ \\
$\hat V_1^+$ & $\hat V^{(2,3),+}$ & $\hat v^{(2),+}$ & $V_1^+$ & $V_1^+$ & $V_1^+$ \\
$\hat V_{[l+1]}^-$ & $\hat V^{(2,3),-}$ & $\hat v^{(2),-}$ & $V_{[l+1]}^-$ & $V_{[l+1]}^-$ & $V_{[l+1]}^-$ \\
$\hat V_2^+$ & $\hat V^{(1,3),+}$ & $\hat v^{(1,2),+}$ & $\hat v^{(1),+}$ & $V_2^+$ & $V_2^+$ \\
$\hat V_{[l]}^-$ & $\hat V^{(1,3),-}$ & $\hat v^{(1,2),-}$ & $\hat v^{(1),-}$ & $V_{[l]}^-$ & $V_{[l]}^-$ \\
\hline
\end{tabular}
\caption{\label{tab:map_p=3} The duality map of the chiral ring generators for $p = 3$.}
\end{table}

\subsection{With a linear monopole superpotential}

One can also consider the deconfined version of the monopole-deformed Kim--Park duality. Specifically, here we discuss the case deformed by a single linear bare monopole superpotential term. We will show that this case can also be derived exactly in the same way but using the Benini--Benvenuti--Pasquetti (BBP) duality \cite{Benini:2017dud} instead of the Aharony duality.

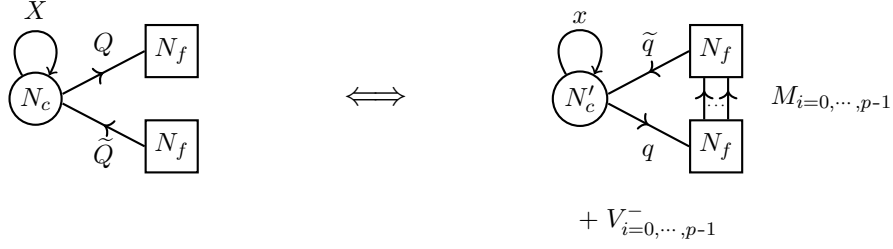
\begin{figure}[tbp]
\centering
\begin{tikzpicture}[thick,scale=0.9, every node/.style={scale=0.9}]
  \tikzset{vertex/.style={circle,fill=white!25,minimum size=12pt,inner sep=2pt}}
  \tikzset{every loop/.style={}}
    \node[vertex] (eNc) at (-1,0) [shape=circle,draw=black,minimum size=2em] {$N_c$};
    \node[vertex] (eNf1) at (1,0.7) [shape=rectangle,draw=black,minimum height=2em, minimum width=2em] {$N_f$};
    \node[vertex] (eNf2) at (1,-0.7) [shape=rectangle,draw=black,minimum height=2em, minimum width=2em] {$N_f$};

    \draw[-to, min distance=1cm]  (eNc) edge [out=120, in=60] node {} (eNc);
    \draw[->-=.5] ([yshift= 2pt] eNc.east) to ([yshift= 0pt] eNf1.west);
    \draw[->-=.5] ([yshift= 0pt] eNf2.west) to ([yshift= -2pt] eNc.east);

    \node at (-1,1.3) {$X$};
    \node at (0,0.8) {$Q$};
    \node at (0,-0.8) {$\WQ$};

    \node at (4,0) {\Large$\Longleftrightarrow$};

    \node[vertex] (mNc) at (7,0) [shape=circle,draw=black,minimum size=2em] {$N_c'$};
    \node[vertex] (mNf1) at (9,0.7) [shape=rectangle,draw=black,minimum height=2em, minimum width=2em] {$N_f$};
    \node[vertex] (mNf2) at (9,-0.7) [shape=rectangle,draw=black,minimum height=2em, minimum width=2em] {$N_f$};

    \draw[-to, min distance=1cm]  (mNc) edge [out=120, in=60] node {} (mNc);
    \draw[->-=.5] ([yshift= 0pt] mNf1.west) to ([yshift= 2pt] mNc.east);
    \draw[->-=.5] ([yshift= -2pt] mNc.east) to ([yshift= 0pt] mNf2.west);
    \draw[->-=.7] ( [xshift= -5pt] mNf2.north) to ( [xshift= -5pt] mNf1.south);
    \draw[->-=.7] ( [xshift= 5pt] mNf2.north) to ( [xshift= 5pt] mNf1.south);

    \node at (7,1.2) {$x$};
    \node at (8,0.8) {$\Wq$};
    \node at (8,-0.8) {$q$};
    \node at (10.7,0) {$M_{i=0,\cdots,p\texttt{-}1}$};
    \node at (9-0.1,-0.1) {\tiny $\cdot$};
    \node at (9,-0.1) {\tiny $\cdot$};
    \node at (9+0.1,-0.1) {\tiny $\cdot$};
    \node at (8,-1.8) {$+\;V_{i=0,\cdots,p\texttt{-}1}^-$};

\end{tikzpicture}
\caption{\label{fig:mKP} The quiver representation of the monopole-deformed Kim--Park duality. The dual gauge rank $N_c'$ is given by $N_c' = p N_f-N_c-p$. $V_i^-$ on the dual side are gauge singlet fields coupled to the monopole operators of the dual theory.}
\end{figure}

Firstly, we propose the monopole-deformed Kim--Park duality with a single linear superpotential term of a bare monopole operator, which has not been discussed in the literature as far as we are aware of. It is an IR duality between the following two theories.
\begin{itemize}
\item The 3d $\mathcal N=2$ $U(N_c)$ gauge theory with $N_f$ pairs of fundamental and antifundamental $Q$ and $\tilde Q$, one adjoint $X$, and the superpotential
\begin{align}
W = \mathrm{Tr} \, X^{p+1}+\hat V_0^+ \,, \label{eq:mKP_A}
\end{align}
where $\hat V_0^+$ is the bare monopole operator with $U(1)_T$ charge $+1$.
\item The 3d $\mathcal N=2$ $U(p N_f-N_c-p)$ gauge theory with $N_f$ pairs of fundamental and antifundamental $q$ and $\tilde q$, one adjoint $x$, $p N_f^2+p$ singlet $M_n$ and $V_n^-$, and the superpotential
\begin{align}
W = \mathrm{Tr} \, x^{p+1}+\sum_{n = 0}^{p-1} M_{p-1-n} \, \tilde q X^n q+\sum_{n=0}^{p-1} V_{p-1-n}^- \, \hat v_n^- + \hat v_0^+ \,, \label{eq:mKP_B}
\end{align}
where the $\hat v_n^\pm$ are the (dressed) monopole operators with $U(1)_T$ charges $\mp 1$.
\end{itemize}
The corresponding quiver diagram is shown in Figure \ref{fig:mKP}.
The charges of the elementary matter fields and the chiral monopole operators are given in Table \ref{tab:mcharge}. As a non-trivial test of the duality, we have checked the superconformal index matchings for several possible dual pairs with small gauge ranks and the number of flavors, whose details will not be presented in this paper. A more rigorous analysis as well as the generalization to deformation by dressed monopole operators will be discussed in \cite{monopole}.
\begin{table}[tbp]
\centering
\begin{tabular}{|c|c|c|c|c|}
\hline
 & $U(1)_R$ & $SU(N_f)_t$ & $SU(N_f)_u$ & $U(1)_A$  \\
\hline 
$Q$ & $\Delta_Q$ & $\overline{\mathbf{N_f}}$ & $\mathbf 1$ & $1$  \\
$\tilde Q$ & $\Delta_Q$ & $\mathbf 1$ & $\mathbf{N_f}$ & $1$  \\
$X$ & $2/(p+1)$ & $\mathbf 1$ & $\mathbf 1$ & $0$  \\
$\hat V_i^-$ & $2N_f(1-\Delta_Q) - 2 (2 N_c + p - i - 1)/(p+1)$ & $\mathbf 1$ & $\mathbf 1$ & $-2N_f$  \\
\hline
$q$ & $2/(p+1)-\Delta_Q$ & $\mathbf 1$ & $\overline{\mathbf{N_f}}$ & $-1$  \\
$\tilde q$ & $2/(p+1)-\Delta_Q$ & $\mathbf{N_f}$ & $\mathbf 1$ & $-1$  \\
$x$ & $2/(p+1)$ & $\mathbf 1$ & $\mathbf 1$ & $0$  \\
$M_i$ & $2 \Delta_Q+2 i/(p+1)$ & $\overline{\mathbf{N_f}}$ & $\mathbf{N_f}$ & $2$  \\
$V_i^-$ & $2N_f(1-\Delta_Q) - 2 (2 N_c + p - i - 1)/(p+1)$ & $\mathbf 1$ & $\mathbf 1$ & $-2 N_f$  \\
\hline
\end{tabular}
\caption{\label{tab:mcharge} The charges of the elementary matter fields and the chiral monopole operators of the Kim--Park dual pair with a single linear superpotential term of a bare monopole. The hatted operators are monopoles, whereas the others are elementary matter fields. Note that the monopole operators of the dual theory are not indicated in this table because they become Q-exact due to the superpotential. The subscript $i$ of $M_i$ and $V_i^\pm$ runs from 0 to 1.}
\end{table}

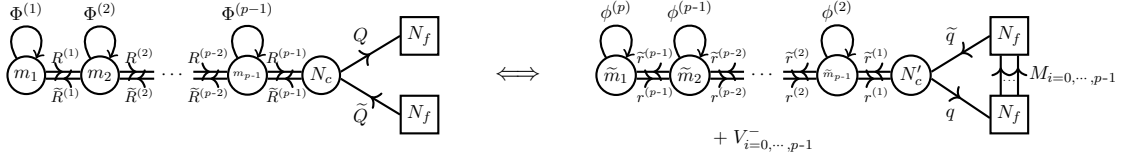
\begin{figure}[tbp]
\centering
\begin{tikzpicture}[thick,scale=0.65, every node/.style={scale=0.65}]
  \tikzset{vertex/.style={circle,fill=white!25,minimum size=12pt,inner sep=2pt}}
  \tikzset{every loop/.style={}}
    
    \node[vertex] (m1) at (0,0) [shape=circle,draw=black,minimum size=2em] {$m_1$};
    \node[vertex] (m2) at (1.5,0) [shape=circle,draw=black,minimum size=2em] {$m_2$};
    \node (ecdot) at (3,0) {$\cdots$};
    \node[vertex] (mp-1) at (4.5,0) [shape=circle,draw=black,minimum size=2em] {\tiny$m_{p\texttt{-}1}$};
    \node[vertex] (Nc) at (6,0) [shape=circle,draw=black,minimum size=2em] {$N_c$};
    \node[vertex] (eNf1) at (8,0.8) [shape=rectangle,draw=black,minimum height=2em, minimum width=2em] {$N_f$};
    \node[vertex] (eNf2) at (8,-0.8) [shape=rectangle,draw=black,minimum height=2em, minimum width=2em] {$N_f$};

    \draw[-to, min distance=1cm]  (m1) edge [out=120, in=60] node {} (m1);
    \draw[-to, min distance=1cm]  (m2) edge [out=120, in=60] node {} (m2);
    \draw[->-=.5] ([yshift= 2pt] m1.east) to ([yshift= 2pt] m2.west);
    \draw[->-=.5] ([yshift= -2pt] m2.west) to ([yshift= -2pt] m1.east);
    \draw[->-=.5] ([yshift= 2pt] m2.east) to ([yshift= 2pt] ecdot.west);
    \draw[->-=.5] ([yshift= -2pt] ecdot.west) to ([yshift= -2pt] m2.east);
    \draw[->-=.5] ([yshift= 2pt] ecdot.east) to ([yshift= 2pt] mp-1.west);
    \draw[->-=.5] ([yshift= -2pt] mp-1.west) to ([yshift= -2pt] ecdot.east);
    \draw[-to, min distance=1cm]  (mp-1) edge [out=120, in=60] node {} (mp-1);
    \draw[->-=.5] ([yshift= 2pt] mp-1.east) to ([yshift= 2pt] Nc.west);
    \draw[->-=.5] ([yshift= -2pt] Nc.west) to ([yshift= -2pt] mp-1.east);
    \draw[->-=.5] ([yshift= 2pt] Nc.east) to ([yshift= 0pt] eNf1.west);
    \draw[->-=.5] ([yshift= 0pt] eNf2.west) to ([yshift= -2pt] Nc.east);
    
    \node at (0,1.3) {$\Phi^{(1)}$};
    \node at (1.5,1.3) {$\Phi^{(2)}$};
    \node at (4.5,1.3) {$\Phi^{(p-1)}$};
    \node at (0.8,0.4) {\footnotesize$R^{(1)}$};
    \node at (0.8,-0.4) {\footnotesize$\WR^{(1)}$};
    \node at (2.3,0.4) {\footnotesize$R^{(2)}$};
    \node at (2.3,-0.4) {\footnotesize$\WR^{(2)}$};
    \node at (3.7,0.4) {\footnotesize$R^{(p\texttt{-}2)}$};
    \node at (3.7,-0.4) {\footnotesize$\WR^{(p\texttt{-}2)}$};
    \node at (5.3,0.4) {\footnotesize$R^{(p\texttt{-}1)}$};
    \node at (5.3,-0.4) {\footnotesize$\WR^{(p\texttt{-}1)}$};
    \node at (6.8,0.8) {$Q$};
    \node at (6.8,-0.8) {$\WQ$};

    \node at (10,0) {\Large $\Longleftrightarrow$};
    
    \node[vertex] (dm1) at (12,0) [shape=circle,draw=black,minimum size=2em] {$\Wm_1$};
    \node[vertex] (dm2) at (13.5,0) [shape=circle,draw=black,minimum size=2em] {$\Wm_2$};
    \node (dcdot) at (15,0) {$\cdots$};
    \node[vertex] (dmp-1) at (16.5,0) [shape=circle,draw=black,minimum size=2em] {\tiny$\Wm_{p\texttt{-}1}$};
    \node[vertex] (dNc) at (18,0) [shape=circle,draw=black,minimum size=2em] {$N_c'$};
    \node[vertex] (dNf1) at (20,0.8) [shape=rectangle,draw=black,minimum height=2em, minimum width=2em] {$N_f$};
    \node[vertex] (dNf2) at (20,-0.8) [shape=rectangle,draw=black,minimum height=2em, minimum width=2em] {$N_f$};

    \draw[-to, min distance=1cm]  (dm1) edge [out=120, in=60] node {} (dm1);
    \draw[-to, min distance=1cm]  (dm2) edge [out=120, in=60] node {} (dm2);
    \draw[->-=.5] ([yshift= 2pt] dm2.west) to ([yshift= 2pt] dm1.east);
    \draw[->-=.5] ([yshift= -2pt] dm1.east) to ([yshift= -2pt] dm2.west);
    \draw[->-=.5] ([yshift= 2pt] dcdot.west) to ([yshift= 2pt] dm2.east);
    \draw[->-=.5] ([yshift= -2pt] dm2.east) to ([yshift= -2pt] dcdot.west);
    \draw[->-=.5] ([yshift= 2pt] dmp-1.west) to ([yshift= 2pt] dcdot.east);
    \draw[->-=.5] ([yshift= -2pt] dcdot.east) to ([yshift= -2pt] dmp-1.west);
    \draw[-to, min distance=1cm]  (dmp-1) edge [out=120, in=60] node {} (dmp-1);
    \draw[->-=.5] ([yshift= 2pt] dNc.west) to ([yshift= 2pt] dmp-1.east);
    \draw[->-=.5] ([yshift= -2pt] dmp-1.east) to ([yshift= -2pt] dNc.west);
    \draw[->-=.5] ([yshift= 0pt] dNf1.west) to ([yshift= 2pt] dNc.east);
    \draw[->-=.5] ([yshift= -2pt] dNc.east) to ([yshift= -2pt] dNf2.west);
    \draw[->-=.7] ([xshift= -5pt] dNf2.north) to ([xshift= -5pt] dNf1.south);
    \draw[->-=.7] ([xshift= 5pt] dNf2.north) to ([xshift= 5pt] dNf1.south);

    \node at (12,1.3) {$\phi^{(p)}$};
    \node at (13.5,1.3) {$\phi^{(p\texttt{-}1)}$};
    \node at (16.5,1.3) {$\phi^{(2)}$};
    \node at (12.8,0.4) {\footnotesize$\Wr^{(p\texttt{-}1)}$};
    \node at (12.8,-0.4) {\footnotesize$r^{(p\texttt{-}1)}$};
    \node at (14.3,0.4) {\footnotesize$\Wr^{(p\texttt{-}2)}$};
    \node at (14.3,-0.4) {\footnotesize$r^{(p\texttt{-}2)}$};
    \node at (15.7,0.4) {\footnotesize$\Wr^{(2)}$};
    \node at (15.7,-0.4) {\footnotesize$r^{(2)}$};
    \node at (17.3,0.4) {\footnotesize$\Wr^{(1)}$};
    \node at (17.3,-0.4) {\footnotesize$r^{(1)}$};
    \node at (18.8,0.8) {$\Wq$};
    \node at (18.8,-0.8) {$q$};
    \node at (20-0.1,-0.1) {\tiny$\cdot$};
    \node at (20,-0.1) {\tiny$\cdot$};
    \node at (20+0.1,-0.1) {\tiny$\cdot$};
    \node at (21.3,0) {$M_{i=0,\cdots,p\texttt{-}1}$};
    \node at (15,-1.3) {$+\;V_{i=0,\cdots,p\texttt{-}1}^-$};

\end{tikzpicture}
\caption{\label{fig:dmKP} The quiver representation of the deconfined Kim--Park duality with monopole deformation. The gauge ranks of the tails are determined as explained in section \ref{sec:confinement}. Also, the rank of the rightmost gauge node on the RHS is given by $N_c' = p N_f-N_c-p$. $V_i^-$ on the dual side are gauge singlet fields coupled to the monopole operators of the dual theory.}
\end{figure}

Especially, if $N_c = \pm1 \mod p$, we can deconfine the adjoint fields into the $\mathbb D_p[SU(N)]$ tails as we did in the case without the monopole deformation, leading to a duality between two quiver gauge theories shown in Figure \ref{fig:dmKP} with the following superpotentials:
\begin{align}
W_A &= \hat{V}^{(p),+} + \eta^{p+1}+\eta \sum_{i=\a}^{p\texttt{-}1}\Tr \Phi^{(i)}
    +
    \sum_{i=\a}^{p\texttt{-}1} \hat{V}^{(i),+}
    +
     \hat{V}^{(\a,p\texttt{-}1),-}
     \sum_{i=\a}^{p-1}\hat{V}^{(i\texttt{-}\a\texttt{+}2,i),-}
     \,, \\
W_B &= \hat{v}^{(1),+} + \eta^{p+1}+\eta \sum_{i=2}^{p\texttt{-}\tilde{\a}\texttt{+}1}\Tr \phi^{(i)}
    +
    \sum_{i=2}^{p\texttt{-}\tilde{\a}\texttt{+}1} \hat{v}^{(i),+}
    +
    \hat{v}^{(2,p\texttt{-}\tilde{\a}\texttt{+}1),-}
    \sum_{i=1}^{p\texttt{-}\tilde{\a}\texttt{+}1}
    \hat{v}^{(i\texttt{-}\tilde{\a}\texttt{+}2,i),-}
    \nonumber \\
    &\quad +
    \sum_{i=0}^{p\texttt{-}\tilde{\a}}
    M_{p\texttt{-}i\texttt{-}1} 
    \tilde{q} \tilde{r}^{(1)}\cdots \tilde{r}^{(i)}
    r^{(i)}\cdots r^{(1)}q
     +\sum_{i=1}^{p\texttt{-}\tilde{\a}\texttt{+}1}
    V_{[N_c\texttt{+}i\texttt{-}1]}^{-} \;\hat{v}^{(1,i),-}
    \nonumber \\
    &\quad +
    \d_{\tilde{\a},2}
    \Tr(\tilde{r}^{(1)}r^{(1)})
    \Big(
    M_0\; \tilde{q} \tilde{r}^{(1)}\cdots \tilde{r}^{(p\texttt{-}2)}
    r^{(p\texttt{-}2)}\cdots r^{(1)}q
    +
    V_0^{-} \hat{v}^{(1,p\texttt{-}1),-}
    \Big)
\end{align}
which have extra monopole terms compared to the deconfined Kim--Park duality without the monopole deformation.
\paragraph{General derivation\\}
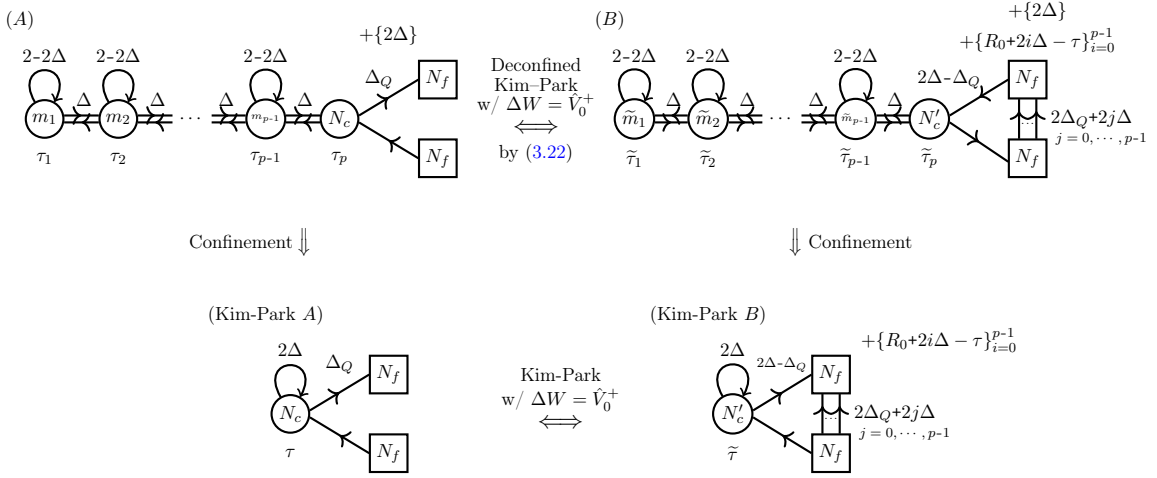
\begin{figure}[tbp]
\centering
\begin{tikzpicture}[thick,scale=0.65, every node/.style={scale=0.65}]
  \tikzset{vertex/.style={circle,fill=white!25,minimum size=12pt,inner sep=2pt}}
  \tikzset{every loop/.style={}}
    
    \node[vertex] (m1) at (0,0) [shape=circle,draw=black,minimum size=2em] {$m_1$};
    \node[vertex] (m2) at (1.5,0) [shape=circle,draw=black,minimum size=2em] {$m_2$};
    \node (ecdot) at (3,0) {$\cdots$};
    \node[vertex] (mp-1) at (4.5,0) [shape=circle,draw=black,minimum size=2em] {\tiny$m_{p\texttt{-}1}$};
    \node[vertex] (Nc) at (6,0) [shape=circle,draw=black,minimum size=2em] {$N_c$};
    \node[vertex] (eNf1) at (8,0.8) [shape=rectangle,draw=black,minimum height=2em, minimum width=2em] {$N_f$};
    \node[vertex] (eNf2) at (8,-0.8) [shape=rectangle,draw=black,minimum height=2em, minimum width=2em] {$N_f$};

    \draw[-to, min distance=1cm]  (m1) edge [out=120, in=60] node {} (m1);
    \draw[-to, min distance=1cm]  (m2) edge [out=120, in=60] node {} (m2);
    \draw[->-=.5] ([yshift= 2pt] m1.east) to ([yshift= 2pt] m2.west);
    \draw[->-=.5] ([yshift= -2pt] m2.west) to ([yshift= -2pt] m1.east);
    \draw[->-=.5] ([yshift= 2pt] m2.east) to ([yshift= 2pt] ecdot.west);
    \draw[->-=.5] ([yshift= -2pt] ecdot.west) to ([yshift= -2pt] m2.east);
    \draw[->-=.5] ([yshift= 2pt] ecdot.east) to ([yshift= 2pt] mp-1.west);
    \draw[->-=.5] ([yshift= -2pt] mp-1.west) to ([yshift= -2pt] ecdot.east);
    \draw[-to, min distance=1cm]  (mp-1) edge [out=120, in=60] node {} (mp-1);
    \draw[->-=.5] ([yshift= 2pt] mp-1.east) to ([yshift= 2pt] Nc.west);
    \draw[->-=.5] ([yshift= -2pt] Nc.west) to ([yshift= -2pt] mp-1.east);
    \draw[->-=.5] ([yshift= 2pt] Nc.east) to ([yshift= 0pt] eNf1.west);
    \draw[->-=.5] ([yshift= 0pt] eNf2.west) to ([yshift= -2pt] Nc.east);
    
    \node at (0,1.3) {$2\texttt{-}2\D$};
    \node at (1.5,1.3) {$2\texttt{-}2\D$};
    \node at (4.5,1.3) {$2\texttt{-}2\D$};
    \node at (0.8,0.4) {$\D$};
    \node at (2.3,0.4) {$\D$};
    \node at (3.7,0.4) {$\D$};
    \node at (5.3,0.4) {$\D$};
    \node at (6.8,0.8) {$\D_Q$};
    \node at (7,1.7) {$+\{2\D\}$};

    \node at (0,-0.8) {$\t_1$};
    \node at (1.5,-0.8) {$\t_2$};
    \node at (4.5,-0.8) {$\t_{p\texttt{-}1}$};
    \node at (6,-0.8) {$\t_p$};

    \node at (-0.5,2) {$(A)$};
    \node at (10,1.1) {Deconfined};
    \node at (10,0.7) {Kim--Park};
    \node at (10,0.3) {w/ $\D W = \hat{V}_0^+$};
    \node at (10,-0.2) {\Large $\Longleftrightarrow$};
    \node at (10,-0.7) {by \eqref{eq: dKP with V_0^+ derivation step}};
    
    \node[vertex] (dm1) at (12,0) [shape=circle,draw=black,minimum size=2em] {$\Wm_1$};
    \node[vertex] (dm2) at (13.5,0) [shape=circle,draw=black,minimum size=2em] {$\Wm_2$};
    \node (dcdot) at (15,0) {$\cdots$};
    \node[vertex] (dmp-1) at (16.5,0) [shape=circle,draw=black,minimum size=2em] {\tiny$\Wm_{p\texttt{-}1}$};
    \node[vertex] (dNc) at (18,0) [shape=circle,draw=black,minimum size=2em] {$N_c'$};
    \node[vertex] (dNf1) at (20,0.8) [shape=rectangle,draw=black,minimum height=2em, minimum width=2em] {$N_f$};
    \node[vertex] (dNf2) at (20,-0.8) [shape=rectangle,draw=black,minimum height=2em, minimum width=2em] {$N_f$};

    \draw[-to, min distance=1cm]  (dm1) edge [out=120, in=60] node {} (dm1);
    \draw[-to, min distance=1cm]  (dm2) edge [out=120, in=60] node {} (dm2);
    \draw[->-=.5] ([yshift= 2pt] dm2.west) to ([yshift= 2pt] dm1.east);
    \draw[->-=.5] ([yshift= -2pt] dm1.east) to ([yshift= -2pt] dm2.west);
    \draw[->-=.5] ([yshift= 2pt] dcdot.west) to ([yshift= 2pt] dm2.east);
    \draw[->-=.5] ([yshift= -2pt] dm2.east) to ([yshift= -2pt] dcdot.west);
    \draw[->-=.5] ([yshift= 2pt] dmp-1.west) to ([yshift= 2pt] dcdot.east);
    \draw[->-=.5] ([yshift= -2pt] dcdot.east) to ([yshift= -2pt] dmp-1.west);
    \draw[-to, min distance=1cm]  (dmp-1) edge [out=120, in=60] node {} (dmp-1);
    \draw[->-=.5] ([yshift= 2pt] dNc.west) to ([yshift= 2pt] dmp-1.east);
    \draw[->-=.5] ([yshift= -2pt] dmp-1.east) to ([yshift= -2pt] dNc.west);
    \draw[->-=.5] ([yshift= 0pt] dNf1.west) to ([yshift= 2pt] dNc.east);
    \draw[->-=.5] ([yshift= -2pt] dNc.east) to ([yshift= -2pt] dNf2.west);
    \draw[->-=.7] ([xshift= -5pt] dNf2.north) to ([xshift= -5pt] dNf1.south);
    \draw[->-=.7] ([xshift= 5pt] dNf2.north) to ([xshift= 5pt] dNf1.south);
    
    \node at (12-0.5,2) {$(B)$};
    \node at (12,1.3) {$2\texttt{-}2\D$};
    \node at (13.5,1.3) {$2\texttt{-}2\D$};
    \node at (16.5,1.3) {$2\texttt{-}2\D$};
    \node at (12.8,0.4) {$\D$};
    \node at (14.3,0.4) {$\D$};
    \node at (15.7,0.4) {$\D$};
    \node at (17.3,0.4) {$\D$};
    \node at (18.4,0.8) {$2\D\texttt{-}\D_Q$};
    \node at (20-0.1,-0.1) {\tiny$\cdot$};
    \node at (20,-0.1) {\tiny$\cdot$};
    \node at (20+0.1,-0.1) {\tiny$\cdot$};
    \node at (21.3,0) {$2\D_Q \texttt{+} 2j\D$};
    \node at (21.5,-0.4) {\scriptsize $j=0,\cdots,p\texttt{-}1$};
    
    \node at (20.2,2.2) {$+\{2\D\}$};
    \node at (20.2,1.6) {$+\{R_0 \texttt{+} 2i\D - \t\}_{i=0}^{p\texttt{-}1}$};

    \node at (12,-0.8) {$\widetilde{\t}_1$};
    \node at (13.5,-0.8) {$\widetilde{\t}_2$};
    \node at (16.5,-0.8) {$\widetilde{\t}_{p\texttt{-}1}$};
    \node at (18,-0.8) {$\widetilde{\t}_p$};

    \node at (6.3-1,-2.5) 
    {\begin{tikzpicture}
      \node [rotate=90] {\LARGE\textcolor{black}{$\Longleftarrow$}};    
    \end{tikzpicture}
    };
    \node at (6.3-2.3,-2.5) {Confinement};

    \node at (15.3,-2.5) 
    {\begin{tikzpicture}
      \node [rotate=90] {\LARGE\textcolor{black}{$\Longleftarrow$}};    
    \end{tikzpicture}
    };
    \node at (15.3+1.3,-2.5) {Confinement};
    
    \node[vertex] (kNc) at (5,-6) [shape=circle,draw=black,minimum size=2em] {$N_c$};
    \node[vertex] (keNf1) at (7,0.8-6) [shape=rectangle,draw=black,minimum height=2em, minimum width=2em] {$N_f$};
    \node[vertex] (keNf2) at (7,-0.8-6) [shape=rectangle,draw=black,minimum height=2em, minimum width=2em] {$N_f$};

    \draw[-to, min distance=1cm]  (kNc) edge [out=120, in=60] node {} (kNc);
    \draw[->-=.5] ([yshift= 2pt] kNc.east) to ([yshift= 0pt] keNf1.west);
    \draw[->-=.5] ([yshift= 0pt] keNf2.west) to ([yshift= -2pt] kNc.east);

    \node at (5-0.5,2-6) {(Kim-Park $A$)};
    \node at (5,1.3-6) {$2\D$};
    \node at (5,-0.8-6) {$\t$};
    \node at (6,1-6) {$\D_Q$};

    \node[vertex] (kmNc) at (14,-6) [shape=circle,draw=black,minimum size=2em] {$N_c'$};
    \node[vertex] (kmNf1) at (16,0.8-6) [shape=rectangle,draw=black,minimum height=2em, minimum width=2em] {$N_f$};
    \node[vertex] (kmNf2) at (16,-0.8-6) [shape=rectangle,draw=black,minimum height=2em, minimum width=2em] {$N_f$};

    \draw[-to, min distance=1cm]  (kmNc) edge [out=120, in=60] node {} (kmNc);
    \draw[->-=.5] ([yshift= 2pt] kmNc.east) to ([yshift= 0pt] kmNf1.west);
    \draw[->-=.5] ([yshift= 0pt] kmNf2.west) to ([yshift= -2pt] kmNc.east);
    \draw[->-=.7] ([xshift= -5pt] kmNf2.north) to ([xshift= -5pt] kmNf1.south);
    \draw[->-=.7] ([xshift= 5pt] kmNf2.north) to ([xshift= 5pt] kmNf1.south);

    \node at (16-0.1,-0.1-6) {\tiny$\cdot$};
    \node at (16,-0.1-6) {\tiny$\cdot$};
    \node at (16+0.1,-0.1-6) {\tiny$\cdot$};
    \node at (14-0.5,2-6) {(Kim-Park $B$)};
    \node at (14,1.3-6) {$2\D$};
    \node at (15,1-6) {\scriptsize $2\D\texttt{-}\D_Q$};
    \node at (14,-0.8-6) {$\widetilde{\t}$};

    \node at (17.3,0-6) {$2\D_Q \texttt{+} 2j\D$};
    \node at (17.5,-0.4-6) {\scriptsize $j=0,\cdots,p\texttt{-}1$};

    \node at (18.2,2-6.5) {$+\{R_0 \texttt{+} 2i\D -\t\}_{i=0}^{p\texttt{-}1}$};

    \node at (10.5,0.8-6) {Kim-Park};
    \node at (10.5,0.3-6) {w/ $\D W = \hat{V}_0^+$};
    \node at (10.5,-0.2-6) {\Large $\Longleftrightarrow$};

\end{tikzpicture}
\caption{\label{fig: dKP with bare monopole derivation} The normalized real masses and FI parameters of a deconfined Kim--Park dual pair with a single linear bare monopole superpotential term and those of the corresponding original Kim--Park dual pair. The dual ranks $\tilde m_j$ are given by $\tilde m_j = j N_f+m_{p-j}-m_p-j$. For the singlets, $R_0 = N_f(1-\D_Q) - 2\D(N_c-1)$ where $\D=\frac{1}{p+1}$, which is fixed by the confining superpotential $\D W_\text{conf}^{(p),+}$ of the tail.}
\end{figure}
Figure \ref{fig: dKP with bare monopole derivation} shows the normalized real masses and FI parameters for a deconfined Kim--Park dual pair with a single linear bare monopole superpotential term. The derivation is basically the same as that of the deconfined Kim-Park duality previously discussed. The only difference is that the intermediate steps are $\text{BBP}_1^+$ duality instead of the Aharony duality:
\begin{equation}
        \begin{tikzpicture}[thick,scale=0.7, every node/.style={scale=0.7}]
  \tikzset{vertex/.style={circle,fill=white!25,minimum size=2.5em,inner sep=2pt}}
  \tikzset{every loop/.style={}}

    \node[vertex] (A'n) at (0,0) [shape=circle,draw=black,minimum size=2.5em] {\scriptsize $m_{p\texttt{-}\!j\texttt{-}\!1}$};
    \node[vertex] (A'2n) at (1.5,0) [shape=circle,draw=black,minimum size=2.5em] {\scriptsize $m_{p\texttt{-}j}$};
    \node[vertex] (A'm) at (3,0) [shape=circle,draw=black,minimum size=2.5em] {$\Wm_j$};
    \node[vertex] (A'cd1) at (-1.5,0) {$\cdots$};
    \node[vertex] (A'cd2) at (4.5,0) {$\cdots$};
    \node[vertex] (A'F1) at (6,1) [shape=rectangle,draw=black,minimum height=2.5em, minimum width=2.5em] {$N_f$};
    \node[vertex] (A'F2) at (6,-1) [shape=rectangle,draw=black,minimum height=2.5em, minimum width=2.5em] {$N_f$};

    \draw[-to, min distance=1cm]  (A'n) edge [out=120, in=60] node {} (A'n);
    \draw[->-=.5] ([yshift= 2pt] A'cd1.east) to ([yshift= 2pt] A'n.west);
    \draw[->-=.5] ([yshift= -2pt] A'n.west) to ([yshift= -2pt] A'cd1.east);
    \draw[->-=.5] ([yshift= 2pt] A'n.east) to ([yshift= 2pt] A'2n.west);
    \draw[->-=.5] ([yshift= -2pt] A'2n.west) to ([yshift= -2pt] A'n.east);
    \draw[->-=.5] ([yshift= 2pt] A'm.west) to ([yshift= 2pt] A'2n.east);
    \draw[->-=.5] ([yshift= -2pt] A'2n.east) to ([yshift= -2pt] A'm.west);
    \draw[->-=.5] ([yshift= 2pt] A'm.east) to ([yshift= 2pt] A'cd2.west);
    \draw[->-=.5] ([yshift= -2pt] A'cd2.west) to ([yshift= -2pt] A'm.east);
    
    \draw[->-=.5] ([yshift= 0pt] A'F1.west) to [bend right=20] ([xshift= -6pt,yshift=13pt] A'm.east);
    \draw[->-=.5] ([xshift= -6pt,yshift=-13pt] A'm.east) to [bend right=20] ([yshift= 0pt] A'F2.west);
    
    \draw[->-=.5] ([xshift= -5pt,yshift=13pt] A'2n.east) to [bend left=25] ([yshift= 5pt] A'F1.west);
    \draw[->-=.5] ([yshift= -5pt] A'F2.west) to [bend left=25] ([xshift= -5pt,yshift=-13pt] A'2n.east);

    \node at (0,1.3) {\scriptsize $2\texttt{-}2\D$};
    \node at (-0.7,0.4) {\scriptsize $\D$};
    \node at (0.8,0.4) {\scriptsize $\D$};
    \node at (2.3,0.4) {\scriptsize $1\texttt{-}\D$};
    \node at (3.8,0.4) {\scriptsize $\D$};

    \node at (3.6,0.9)
    {\begin{tikzpicture}
      \node [rotate=25] {\scriptsize $1\texttt{-}\D_Q \texttt{-}( j \texttt{-} 1) \D$};    
    \end{tikzpicture}
    };
    \node at (2.8,1.3)
    {\begin{tikzpicture}
      \node [rotate=15] { $\D_Q \texttt{+} j\D$};    
    \end{tikzpicture}
    };

    \node at (0,-2) {\scriptsize $\t_{p\texttt{-}j\texttt{-}1}$};
    
    \node at (1.25,-1.7) {\tiny $j$};
    \node at (1.5,-2) {\scriptsize $\sum \t_{p\texttt{-}i}$};
    \node at (1.25,-2.3) {\tiny $i\!=\!0$};
    \node at (1.5,-2.6) {\scriptsize $\texttt{-}j\D$};

    \node at (2.9,-1.7) {\tiny $j\texttt{-}1$};
    \node at (3,-2) {\scriptsize $\texttt{-}\sum \t_{p\texttt{-}i}$};
    \node at (2.9,-2.3) {\tiny $i\!=\!0$};
    \node at (3.3,-2.6) {\scriptsize $\texttt{+}(p\texttt{+}j)\D$};

    \node at (8,0.5) {$\text{BBP}_1^+$};
    \node at (8,0) {\LARGE $\Longleftrightarrow$};
    \node at (8,-0.5) {\scriptsize $j=0,\!\cdots\!,p\texttt{-1}$};

    \node[vertex] (B'n) at (11,0) [shape=circle,draw=black,minimum size=2.5em]{\scriptsize $m_{p\texttt{-}\!j\texttt{-}\!1}$};
    \node[vertex] (B'2m) at (12.5,0) [shape=circle,draw=black,minimum size=2.5em] {\scriptsize $\Wm_{j\texttt{+}1}$};
    \node[vertex] (B'm) at (14,0) [shape=circle,draw=black,minimum size=2.5em] {$\Wm_j$};
    \node[vertex] (B'cd1) at (9.5,0) {$\cdots$};
    \node[vertex] (B'cd2) at (15.5,0) {$\cdots$};
    \node[vertex] (B'F1) at (17,1) [shape=rectangle,draw=black,minimum height=2.5em, minimum width=2.5em] {$N_f$};
    \node[vertex] (B'F2) at (17,-1) [shape=rectangle,draw=black,minimum height=2.5em, minimum width=2.5em] {$N_f$};

    \draw[->-=.5] ([yshift= 2pt] B'cd1.east) to ([yshift= 2pt] B'n.west);
    \draw[->-=.5] ([yshift= -2pt] B'n.west) to ([yshift= -2pt] B'cd1.east);
    \draw[->-=.5] ([yshift= 2pt] B'2m.west) to ([yshift= 2pt] B'n.east);
    \draw[->-=.5] ([yshift= -2pt] B'n.east) to ([yshift= -2pt] B'2m.west);
    \draw[->-=.5] ([yshift= 2pt] B'2m.east) to ([yshift= 2pt] B'm.west);
    \draw[->-=.5] ([yshift= -2pt] B'm.west) to ([yshift= -2pt] B'2m.east);
    \draw[-to, min distance=0.8cm]  (B'm) edge [out=120, in=60] node {} (B'm);
    \draw[->-=.5] ([yshift= 2pt] B'm.east) to ([yshift= 2pt] B'cd2.west);
    \draw[->-=.5] ([yshift= -2pt] B'cd2.west) to ([yshift= -2pt] B'm.east);
    \draw[->-=.5] ([yshift= 5pt] B'F1.west) to [bend right=25] ([xshift= -5pt,yshift=13pt] B'2m.east);
    \draw[->-=.5] ([xshift= -5pt,yshift=-13pt] B'2m.east) to [bend right=25] ([yshift= -5pt] B'F2.west);
    \draw[->-=.5] ([xshift= 3pt,yshift=0pt] B'n.north) to [bend left=20] ([yshift= 10pt] B'F1.west);
    \draw[->-=.5] ([yshift= -10pt] B'F2.west) to [bend left=20] ([xshift= 3pt,yshift=0pt] B'n.south);
    
    \draw[->-=.6] ([xshift= 0pt] B'F2.north) to ([xshift= 0pt] B'F1.south);

    \node at (-0.7+11,0.4) {\scriptsize $\D$};
    \node at (0.8+11,0.4) {\scriptsize $1\texttt{-}\D$};
    \node at (2.3+11,0.4) {\scriptsize $\D$};
    \node at (3.8+11,0.4) {\scriptsize $\D$};
    \node at (3.8+11,0.9) {\scriptsize $2\texttt{-}2\D$};
    \node at (18,0) {\scriptsize $2\D_Q \texttt{+} 2j\D$};
    
    \node at (17+0.2,-2.3+0.35) {\scriptsize $j$};
    \node at (17,-2.3) {$+\big\{ r_j - \sum \t_{p\texttt{-}i} \big\}$};
    \node at (17+0.2,-2.3-0.3) {\scriptsize $i=0$};
    
    \node at (4.15+9,0.9)
    {\begin{tikzpicture}
      \node [rotate=29] { $1\texttt{-}\D_Q \texttt{-} j\D$};    
    \end{tikzpicture}
    };
    \node at (2.8+9.3,1.3)
    {\begin{tikzpicture}
      \node [rotate=15] { $\D_Q \texttt{+} (j\texttt{+}1)\D$};    
    \end{tikzpicture}
    };

    \node at (-0.25+11,-1.7) {\tiny $j\texttt{+}1$};
    \node at (0+11,-2) {\scriptsize $\sum \t_{p\texttt{-}i}$};
    \node at (-0.25+11,-2.3) {\tiny $i\!=\!0$};
    \node at (0+11,-2.6) {\scriptsize $\texttt{-}(j\texttt{+}1)\D$};
    
    \node at (1.35+11,-1.7) {\tiny $j$};
    \node at (1.5+11,-2) {\scriptsize $\texttt{-}\sum \t_{p\texttt{-}i}$};
    \node at (1.35+11,-2.3) {\tiny $i\!=\!0$};
    \node at (1.7+11,-2.6) {\scriptsize $\texttt{+}(p\texttt{+}j\texttt{+}1)\D$};

    \node at (3+11,-2) {\scriptsize $\t_{p\texttt{-}j}$};

\end{tikzpicture}
    \label{eq: dKP with V_0^+ derivation step}
\end{equation}
where the dual gauge ranks and the dual normalized FI parameters are given by
\begin{align}
    \Wm_j &= j N_f + m_{p\texttt{-}j} - m_p - j
    \,,\qquad \text{for}\quad j=0,\cdots,p
    \nonumber\\
    \widetilde{\t}_j &= \t_{p\texttt{-}j}
    \,,\qquad\qquad\qquad\qquad\quad \text{for}\quad j=1,\cdots,p-1
\end{align}
which are equivalent to the condition for the confinement of $\mathbb{D}_p[SU(N_c')]$ with confining superpotential deformation $\D W_\text{conf}^{(p),+}$ on the dual side. Especially, $\Wm_p$ is given by $\Wm_p = N_c' = p N_f - N_c - p$, which is the desired dual gauge rank for the Kim-Park duality deformed by a linear bare monopole term. Moreover, the bare monopole operators $\hat{V}_0^+$ and $\hat{v}_0^+$ of Theory A and Theory B, respectively, are mapped to the monopole operators $\hat{V}^{(p),+}$ and $\hat{v}^{(p),+}$ under the deconfinement of the adjoint fields, which demands that the FI parameters of the deconfined dual pair, $\t_p$ and $\widetilde{\t}_p=- \sum_{i=1}^p \t_i + 2p\D$, should satisfy the superpotential conditions
\begin{align}
    &N_f(1-\D_Q) + m_{p\texttt{-}1}(1-\D) - N_c + 1 + \t_p \overset{!}{=} 2 \,,
    \nonumber\\
    &N_f(1-2\D+\D_Q) + \Wm_{p\texttt{-}1}(1-\D) - N_c' + 1 + \widetilde{\t}_p \overset{!}{=} 2 \,.
\end{align}
Then, together with the shift \eqref{eq: mixing under gauging}, the FI parameters of the confined dual pair are related as
\begin{align}
    \widetilde{\t} = -\t + 2p\D \,,
\end{align}
which indeed saturates the condition from the single linear bare monopole superpotential term as it should:
\begin{align}
    &N_f(1-\D_Q) - 2\D(N_c-1) + \t = 2 \,,
    \nonumber\\
    &N_f(1- 2\D + \D_Q) - 2\D(N_c'-1) + \widetilde{\t} = 2 \,.
\end{align}
We can also check the monopole-like singlet contributions arising from the sequential $\text{BBP}_1^+$ procedure are exactly the same as that of the one bare monopole deformed Kim--Park duality as follows:
\begin{align}
    +\Big\{ r_j - \sum_{i=p-j}^{p} \t_i \Big\}_{j=0}^{p-1}
    \;\;=\;\;
    +\big\{ R_0 +2i\D - \t\big\}_{i=0}^{p-1}
\end{align}
where $r_j =  m_{p\texttt{-}j\texttt{-}1}(1-\D) + \Wm_j\D + N_f(1-\D_Q-j\D) - m_{p\texttt{-}j} + 1$ and $R_0 = N_f(1-\D_Q) - 2\D(N_c-1)$. Similarly, the meson-like singlets from the sequential $\text{BBP}_1^+$ procedure match those of the one bare monopole deformed Kim--Park duality. Thus, our proposal of the monopole-deformed Kim--Park duality can be understood as a consequence of the $\text{BBP}_1^+$ duality if $N = \pm 1 \mod p$.

\begin{figure}[tbp]
\centering
\begin{tikzpicture}[thick,scale=0.58, every node/.style={scale=0.58}]
  \tikzset{vertex/.style={circle,fill=white!25,minimum size=2.5em,inner sep=2pt}}
  \tikzset{every loop/.style={}}
    
    \node[vertex] (An) at (0,0) [shape=circle,draw=red,minimum size=2.5em] {$n$};
    \node[vertex] (A2n) at (2,0) [shape=circle,draw=blue,minimum size=2.5em] {$\!2n \texttt{+}1\!$};
    \node[vertex] (AF1) at (4.5,1) [shape=rectangle,draw=black,minimum height=2.5em, minimum width=2.5em] {$N_f$};
    \node[vertex] (AF2) at (4.5,-1) [shape=rectangle,draw=black,minimum height=2.5em, minimum width=2.5em] {$N_f$};

    \draw[-to, min distance=1cm]  (An) edge [out=120, in=60] node {} (An);
    \draw[->-=.5] ([yshift= 2pt] An.east) to ([yshift= 2pt] A2n.west);
    \draw[->-=.5] ([yshift= -2pt] A2n.west) to ([yshift= -2pt] An.east);
    \draw[->-=.5] ([yshift= 2pt] A2n.east) to ([yshift= 0pt] AF1.west);
    \draw[->-=.5] ([yshift= -2pt] AF2.west) to ([yshift= -2pt] A2n.east);

    \node at (0,1.4) {$\Phi$};
    \node at (1,0.4) {$R$};
    \node at (1,-0.4) {$\WR$};
    \node at (3,1) {$Q$};
    \node at (3,-1) {$\WQ$};
    \node at (2,-1.5) {$+\;\eta$};

    \node[vertex] (Bn) at (11,0) [shape=circle,draw=black,minimum size=2.5em] {$m$};
    \node[vertex] (B2n) at (13,0) [shape=circle,draw=black,minimum size=2.5em] {$\!2m \texttt{+} 1\! $};
    \node[vertex] (BF1) at (15.5,1) [shape=rectangle,draw=black,minimum height=2.5em, minimum width=2.5em] {$N_f$};
    \node[vertex] (BF2) at (15.5,-1) [shape=rectangle,draw=black,minimum height=2.5em, minimum width=2.5em] {$N_f$};

    \draw[-to, min distance=1cm]  (Bn) edge [out=120, in=60] node {} (Bn);
    \draw[->-=.5] ([yshift= 2pt] B2n.west) to ([yshift= 2pt] Bn.east);
    \draw[->-=.5] ([yshift= -2pt] Bn.east) to ([yshift= -2pt] B2n.west);
    \draw[->-=.5] ([yshift= 0pt] BF1.west) to ([yshift= 2pt] B2n.east);
    \draw[->-=.5] ([yshift= -2pt] B2n.east) to ([yshift= 0pt] BF2.west);
    \draw[->-=.6] ([xshift= -5pt] BF2.north) to ([xshift= -5pt] BF1.south);
    \draw[->-=.6] ([xshift= 5pt] BF2.north) to ([xshift= 5pt] BF1.south);

    \node at (11,1.4) {$\phi$};
    \node at (12,0.4) {$\Wr$};
    \node at (12,-0.4) {$r$};
    \node at (13.6+0.5,1) {$\Wq$};
    \node at (13.6+0.5,-1) {$q$};
    \node at (15.7+0.5,0) {$M_i$};
    \node at (13,-1.5) {$+\;\eta\;,V_i^-$};

    \node[vertex] (b2m) at (19,0) [shape=circle,draw=red,minimum size=2.5em] {$\!2m \texttt{+} 1\!$};
    \node (bm) at (21,0) [shape=circle,draw=blue,minimum size=2.5em] {$m$};
    \node[vertex] (bF1) at (23.5,1) [shape=rectangle,draw=black,minimum height=2.5em, minimum width=2.5em] {$N_f$};
    \node[vertex] (bF2) at (23.5,-1) [shape=rectangle,draw=black,minimum height=2.5em, minimum width=2.5em] {$N_f$};

    \draw[-to, min distance=1cm]  (bm) edge [out=30, in=-30] node {} (bm);
    \draw[->-=.5] ([yshift= 2pt] b2m.east) to ([yshift= 2pt] bm.west);
    \draw[->-=.5] ([yshift= -2pt] bm.west) to ([yshift= -2pt] b2m.east);
    \draw[->-=.5] ([yshift= 5pt] bF1.west) to [bend right=20] ([xshift=-5pt ,yshift= 13pt] b2m.east);
    \draw[->-=.5] ([xshift=-5pt ,yshift= -13pt] b2m.east) to [bend right=20] ([yshift= -5pt] bF2.west);
    \draw[->-=.6] ([xshift= -5pt] bF2.north) to ([xshift= -5pt] bF1.south);
    \draw[->-=.6] ([xshift= 5pt] bF2.north) to ([xshift= 5pt] bF1.south);

    \node at (22.5,0) {$\phi$};
    \node at (20,0.4) {$\Wr$};
    \node at (20,-0.4) {$r$};
    \node at (24.5,0) {$M_{i=0,1}$};
    \node at (21.5,1.5) {$\Wq$};
    \node at (21.5,-1.5) {$q$};
    \node at (21.5,-2.3) {$+\;\eta\;,V_{i=0,1}^-$};

    \node[vertex] (A'n) at (10,6) [shape=circle,draw=red,minimum size=2.5em] {$n$};
    \node[vertex] (A'm) at (12,6) [shape=circle,draw=blue,minimum size=2.5em] { $m$};
    \node[vertex] (A'F1) at (14.5,7) [shape=rectangle,draw=black,minimum height=2.5em, minimum width=2.5em] {$N_f$};
    \node[vertex] (A'F2) at (14.5,5) [shape=rectangle,draw=black,minimum height=2.5em, minimum width=2.5em] {$N_f$};

    \draw[->-=.5] ([yshift= 2pt] A'm.west) to ([yshift= 2pt] A'n.east);
    \draw[->-=.5] ([yshift= -2pt] A'n.east) to ([yshift= -2pt] A'm.west);
    \draw[->-=.5] ([yshift= 0pt] A'F1.west) to ([yshift= 2pt] A'm.east);
    \draw[->-=.5] ([yshift= -2pt] A'm.east) to ([yshift= -2pt] A'F2.west);
    \draw[->-=.6] ([yshift= 0pt] A'F2.north) to ([yshift= 0pt] A'F1.south);
    \draw[->-=.5] ([xshift= -5pt,yshift=13pt] A'n.east) to [bend left=20] ([yshift= 5pt] A'F1.west);
    \draw[->-=.5] ([yshift= -5pt] A'F2.west) to [bend left=20] ([xshift= -5pt,yshift=-13pt] A'n.east);

    \node at (11,6+0.4) {$R'$};
    \node at (11,6-0.4) {$\WR'$};
    \node at (13.5,6+0.3) {$Q'$};
    \node at (13.5,6-0.3) {$\WQ'$};
    \node at (12.2,6+1.5) {$\Wq'$};
    \node at (12.2,6-1.5) {$q'$};
    \node at (15,6) {$M_0$};
    \node at (17,6) {$+\;\eta\;,V_0^-$};

    \node[vertex] (e2n) at (1.5,-6) [shape=circle,draw=black,minimum size=2.5em] {$2n\texttt{+}1$};
    \node[vertex] (eF1) at (4,-6+1) [shape=rectangle,draw=black,minimum height=2.5em, minimum width=2.5em] {$N_f$};
    \node[vertex] (eF2) at (4,-6-1) [shape=rectangle,draw=black,minimum height=2.5em, minimum width=2.5em] {$N_f$};

    \draw[-to, min distance=1cm]  (e2n) edge [out=120, in=60] node {} (e2n);
    \draw[->-=.5] ([yshift= 2pt] e2n.east) to ([yshift= 0pt] eF1.west);
    \draw[->-=.5] ([yshift= 0pt] eF2.west) to ([yshift= -2pt] e2n.east);

    \node at (1.5,-6+1.4) {$X$};
    \node at (2.7,-6+1) {$Q$};
    \node at (2.7,-6-1) {$\WQ$};

    \node[vertex] (m2m) at (12,-6) [shape=circle,draw=black,minimum size=2.5em] {$\!2m\texttt{+}1\!$};
    \node[vertex] (mF1) at (14.5,-6+1) [shape=rectangle,draw=black,minimum height=2.5em, minimum width=2.5em] {$N_f$};
    \node[vertex] (mF2) at (14.5,-6-1) [shape=rectangle,draw=black,minimum height=2.5em, minimum width=2.5em] {$N_f$};

    \draw[-to, min distance=1cm]  (m2m) edge [out=120, in=60] node {} (m2m);
    \draw[->-=.5] ([yshift= 0pt] mF1.west) to ([yshift= 2pt] m2m.east);
    \draw[->-=.5] ([yshift= -2pt] m2m.east) to ([yshift= 0pt] mF2.west);
    \draw[->-=.6] ([xshift= -5pt] mF2.north) to ([xshift= -5pt] mF1.south);
    \draw[->-=.6] ([xshift= 5pt] mF2.north) to ([xshift= 5pt] mF1.south);

    \node at (12,-6+1.4) {$x$};
    \node at (13.2,-6+1) {$\Wq$};
    \node at (13.2,-6-1) {$q$};
    \node at (15.7,-6) {$M_{i=0,1}$};
    \node at (17.5,-6) {$+\;V_{i=0,1}^-$};

    \node at (9.3,7.8) {$(A')$};
    \node at (-0.3,2.3) {$(A)$};
    \node at (10.5,2.3) {$(B)$};
    \node at (4,-3) {\Large $\Downarrow\;$ Confinement};
    \node at (14,-3) {\Large $\Downarrow\;$ Confinement};
    \node at (7.7,1) {Deconfined};
    \node at (7.7,0.5) {Kim-Park};
    \node at (7.7,0) {\Large $\Longleftrightarrow$};
    \node at (17.5,0) {\Large $\boldsymbol{=}$};
    \node at (7.7,-6+0.5) {Kim-Park};
    \node at (7.7,-6) {\Large $\Longleftrightarrow$};
    \node at (0,-3.7) {(Kim-Park A)};
    \node at (10.5,-3.7) {(Kim-Park B)};

    \node at (7.5,3.6) 
    {\begin{tikzpicture}
      \node [rotate=45] {\Huge \textcolor{blue}{$\Longleftrightarrow$}};    
    \end{tikzpicture}
    };
    \node at (9,3.7) { \textcolor{blue}{BBP}};
    \node at (9,3.2) { \textcolor{blue}{Duality}};
    \node at (16.5,3.6) 
    {\begin{tikzpicture}
      \node [rotate=-45] {\Huge \textcolor{red}{$\Longleftrightarrow$}};    
    \end{tikzpicture}
    };
    \node at (18,3.7) { \textcolor{red}{BBP}};
    \node at (18,3.2) { \textcolor{red}{Duality}};

\end{tikzpicture}
\caption{\label{fig:dmKP_p=2} The deconfined Kim--Park duality for $p = 2$ with monopole deformation (the second line) and its derivation using the BBP duality. The gauge ranks of the dual theory are specified by $m = N_f-n-2$. $V_i^-$ on the dual side are gauge singlet fields coupled to the monopole operators of the dual theory. The confinement of the $\mathbb D_p[SU(N)]$ tails of Theory A and Theory B leads to the monopole-deformed Kim--Park duality (the third line).}
\end{figure}
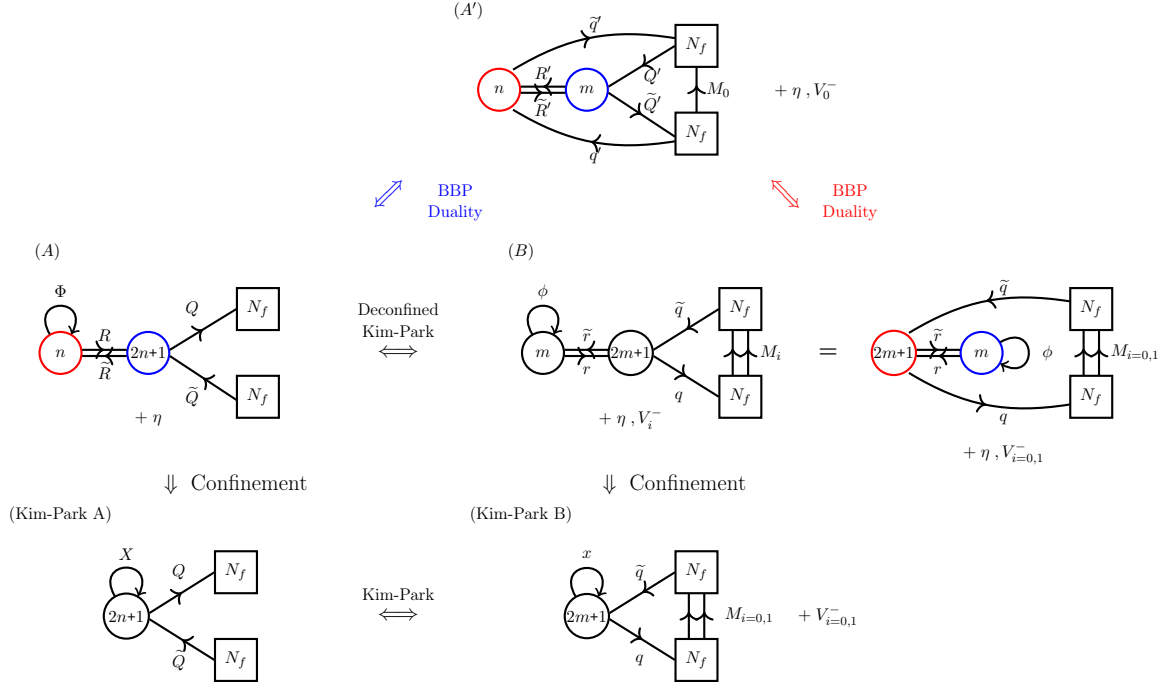

\paragraph{Example for $p=2$\\}
Let us consider an example. For $p = 2$, the deconfined dual pair are shown in the second line of Figure \ref{fig:dmKP_p=2}.
Each theory has the superpotential as follows:
\begin{align}
W_A &= \mathrm{Tr} \, \tilde R \Phi R+\eta^{3}+\eta \, \mathrm{Tr} \, \Phi+\hat V^{(1),+}+\hat V^{(1),-}+\hat V^{(2),+} \,, \\
W_B & = \tilde r \phi r+\eta^{3}+\eta \, \mathrm{Tr} \, \phi+\hat v^{(2),+}+\hat v^{(2),-}+M_0 \, \tilde q \tilde r r q+M_1 \, \tilde q q+V_0^- \, \hat v^{(1,2),-}+V_1^- \, \hat v^{(1),-}+\hat v^{(1),+}
\end{align}
where the last terms of $W_A$ and $W_B$ correspond to $\hat V_0^+$ in \eqref{eq:mKP_A} and $\hat v_0^+$ in \eqref{eq:mKP_B}, respectively.
Note that $\hat V^{(2),+}$ and $\hat V^{(1,2),+}$ of Theory A, which are nontrivial chiral monopole operators when the monopole deformation is absent, now become Q-exact due to this last term. Since this term breaks one of the $U(1)$ global symmetries, the corresponding current multiplet, whose fermion component is given by a combination of $\mathrm{Tr} \, Q \psi_Q^\dagger$ and $\mathrm{Tr} \, \psi_{\tilde Q}^\dagger \tilde Q$, can combine with $\hat V^{(2),+}$ and become a long multiplet. In a similar manner, $\hat V^{(1,2),+}$ would combine with a short multiplet including one combination of $\mathrm{Tr} \, \tilde R R Q \psi_Q^\dagger$ and $\mathrm{Tr} \, \psi_{\tilde Q}^\dagger \tilde Q \tilde R R$ and become a long multiplet. Thus, the nontrivial chiral monopole operators are only $\hat V^{(2),-}$ and $\hat V^{(1,2),-}$. Similarly, the last term of $W_B$ makes $\hat v^{(1),+}$ and $\hat v^{(1,2),+}$ of Theory B Q-exact.
The charges of the elementary matter fields and the chiral monopole operators are given in Table \ref{tab:mcharge_p=2}.
\begin{table}[tbp]
\centering
\begin{tabular}{|c|c|c|c|c|}
\hline
 & $U(1)_R$ & $SU(N_f)_t$ & $SU(N_f)_u$ & $U(1)_A$ \\
\hline 
$Q$ & $\D_Q$ & $\overline{\mathbf{N_f}}$ & $\mathbf 1$ & $1$ \\
$\tilde Q$ & $\Delta_Q$ & $\mathbf 1$ & $\mathbf{N_f}$ & $1$ \\
$R$ & $1/3$ & $\mathbf 1$ & $\mathbf 1$ & $0$ \\
$\tilde R$ & $1/3$ & $\mathbf 1$ & $\mathbf 1$ & $0$ \\
$\Phi$ & $4/3$ & $\mathbf 1$ & $\mathbf 1$ & $0$ \\
$\eta$ & $2/3$ & $\mathbf 1$ & $\mathbf 1$ & $0$ \\
$\hat V^{(2),-}$ & $2 (1-\Delta_Q) N_f-(8 n+6)/3$ & $\mathbf 1$ & $\mathbf 1$ & $-2 N_f$ \\
$\hat V^{(1,2),-}$ & $2 (1-\Delta_Q) N_f-(8 n+4)/3$ & $\mathbf 1$ & $\mathbf 1$ & $-2 N_f$ \\
\hline
$q$ & $2/3-\D_Q$ & $\mathbf 1$ & $\overline{\mathbf{N_f}}$ & $-1$ \\
$\tilde q$ & $2/3-\Delta_Q$ & $\mathbf{N_f}$ & $\mathbf 1$ & $-1$ \\
$r$ & $1/3$ & $\mathbf 1$ & $\mathbf 1$ & $0$ \\
$\tilde r$ & $1/3$ & $\mathbf 1$ & $\mathbf 1$ & $0$ \\
$\phi$ & $4/3$ & $\mathbf 1$ & $\mathbf 1$ & $0$ \\
$\eta$ & $2/3$ & $\mathbf 1$ & $\mathbf 1$ & $0$ \\
$M_i$ & $2 \Delta_Q+2 i/3$ & $\overline{\mathbf{N_f}}$ & $\mathbf{N_f}$ & $2$ \\
$V_i^-$ & $2 (1-\Delta_Q) N_f-(8 n+6-2 i)/3$ & $\mathbf 1$ & $\mathbf 1$ & $-2 N_f$ \\
\hline
\end{tabular}
\caption{\label{tab:mcharge_p=2} The charges of the elementary matter fields and the chiral monopole operators of Theory A (the upper box) and Theory B (the lower box) for $p = 2$ with a single linear superpotential term of a bare monopole operator. The hatted operators are monopoles, whereas the others are elementary matter fields. Note that the monopole operators of Theory B are not indicated in this table because they become Q-exact due to the superpotential. The subscript $i$ of $M_i$ and $V_i^-$ runs from 0 to 1.}
\end{table}

Now one can derive this duality exactly in the same way as in the case without the monopole deformation but using the BBP duality instead of the Aharony duality. As shown in the figure, we obtain the proposed dual theory by applying the BBP duality on each gauge node sequentially from the right. Theory A$'$ in the intermediate step has the following superpotential:
\begin{align}
W_{A'} = \eta^{3}+\eta \, \mathrm{Tr} \, R' \tilde R'+\hat v^{(1),+}+\hat v^{(1,2),-}+Q' R' \tilde q'+q' \tilde R' \tilde Q'+M_0 \, Q' \tilde Q'+V_0^- \, \hat{v}^{(2),-}+\hat v^{(2),+} \,.
\end{align}
The charges of the elementary matter fields and the monopole operators are given in Table \ref{tab:mcharge_p=2_A'}.
\begin{table}[tbp]
\centering
\begin{tabular}{|c|c|c|c|c|c|}
\hline
 & $U(1)_R$ & $SU(N_f)_t$ & $SU(N_f)_u$ & $U(1)_A$ \\
\hline 
$Q'$ & $1-\Delta_Q$ & $\mathbf{N_f}$ & $\mathbf 1$ & $-1$ \\
$\tilde Q'$ & $1-\Delta_Q$ & $\mathbf 1$ & $\overline{\mathbf{N_f}}$ & $-1$ \\
$R'$ & $2/3$ & $\mathbf 1$ & $\mathbf 1$ & $0$ \\
$\tilde R'$ & $2/3$ & $\mathbf 1$ & $\mathbf 1$ & $0$ \\
$q'$ & $1/3+\Delta_Q$ & $\mathbf 1$ & $\mathbf{N_f}$ & $1$ \\
$\tilde q'$ & $1/3+\Delta_Q$ & $\overline{\mathbf{N_f}}$ & $\mathbf 1$ & $1$ \\
$\eta$ & $2/3$ & $\mathbf 1$ & $\mathbf 1$ & $0$ \\
$M_0$ & $2 \Delta_Q$ & $\overline{\mathbf{N_f}}$ & $\mathbf{N_f}$ & $2$ \\
$V_0^-$ & $2 (1-\Delta_Q) N_f-(8 n+6)/3$ & $\mathbf 1$ & $\mathbf 1$ & $-2 N_f$ \\
$\hat v^{(1),-}$ & $2 (1-\Delta_Q) N_f-(8 n+4)/3$ & $\mathbf 1$ & $\mathbf 1$ & $-2 N_f$ \\
\hline
\end{tabular}
\caption{\label{tab:mcharge_p=2_A'} The charges of the elementary matter fields and the chiral monopole operators of Theory A$'$ for $p = 2$ with a single linear superpotential term of a bare monopole operator. The hatted operators are monopoles, whereas the others are elementary matter fields.}
\end{table}
The duality map of the chiral ring generators is the same as the case without the monopole deformation given in Table \ref{tab:map_p=2} except the fact that now $\hat V^{(2),+}$ and $\hat V^{(1,2),+}$ of Theory A and their counterparts on the other dual sides are Q-exact as we explained.

As before, the same method can be applied to arbitrary $p$ as long as $N_c = \pm 1 \mod p$, the good theory condition for the $\mathbb D_p[SU(N)]$ tail.
\\

\section{Conclusion}
\label{sec:conclusion}
In this paper, we propose confining deformation of a class of 3d reduced Agyres--Douglas theories, $\mathbb{D}_p[SU(N)]$, satisfying the good theory condition, $N = \pm1 \mod p$. The $\mathbb{D}_p[SU(N)]$ theories are 3d $\CN=4\,$ linear quiver theories with unitary gauge groups, which flow to the Wess--Zumino theories of a single $SU(N)$ flavor adjoint chiral field $M$ with the $A_p$-type superpotential $W = \Tr M^{p+1}$ once we turn on confining superpotential deformation $\D W_\text{conf}^{(p),\pm}$ given in \eqref{eq: confining deformation}. We have proved this confinement of deformed $\mathbb D_p[SU(N)]$ by using two basic dualities, the Benini--Benvenuti--Pasquetti duality and the Aharony duality, which can be obtained from the former by mass deformation.
By utilizing this confinement of the linear quiver theory $\mathbb{D}_p[SU(N)]$ into an adjoint field, we provide a new understanding of the 3d $\CN=2$ $U(N)$ adjoint SQCD with the $A_p$-type superpotential as the SQCD attached to the linear quiver tail. Surprisingly, this new perspective on the adjoint SQCD supplies a novel derivation of the Kim--Park duality by only assuming the Aharony duality. Furthermore, we also suggest a new Kim--Park-like duality deformed by a single linear bare monopole superpotential term that can be similarly derived by using the one-monopole BBP duality.

Let us address some possible future directions.
\paragraph{Extension to general $N$}
In the present paper, we have discussed the confinement of \emph{good} $\mathbb{D}_p[SU(N)]$ theories in the sense of \cite{Gaiotto:2008ak}, which satisfy the condition $N = \pm 1 \mod p$ as we show in appendix \ref{app: D_p[G] review}. On the other hand, the construction of the $\mathbb{D}_p[SU(N)]$ theories is provided for general $N$ \cite{Closset:2020afy, Giacomelli:2020ryy}, which generically results in \emph{ugly} quivers, and one may ask whether it is possible to extend our discussion of confining phenomena for $N = \pm 1 \mod p$ to arbitrary positive integer $N$. Indeed, we have checked that the same confining phenomena occur for ugly $\mathbb D_p[SU(N)]$ theories as well, where the underlying idea is essentially the same but the deformation is slightly modified.  A more detailed analysis of such generic cases will appear in the subsequent work \cite{ugly}.
\paragraph{Kim--Park-like dualities}
We have provided a novel way of understanding the Kim--Park duality, a 3d $\mathcal N=2$ IR duality for adjoint SQCDs with the $A_p$-type superpotential, using the confinement of $\mathbb D_p[SU(N)]$ and the sequential Aharony duality. We have also worked out the Kim--Park duality deformed by a single linear bare monopole term using the sequential one-monopole BBP duality in a similar way. One may wonder how far such a deconfined version of understanding would bring us to more general Kim--Park-like dualities. There are already some known Kim--Park-like dualities, e.g., in \cite{Hwang:2015wna,Amariti:2019rhc,Hwang:2022jjs} including chiral-like generalization and monopole deformation. We believe those can also be understood from our deconfined picture, although there are some subtleties. For example, the Kim--Park duality with two linear bare monopole terms can easily be checked for $p = 2$ using the sequential two-monopole BBP duality. However, the current method doesn't work for higher $p$ cases in a straightforward manner, which will be interesting to figure out.
\paragraph{Deconfinement for other types of adjoint superpotentials}
As stated in the preceding 4d story \cite{Maruyoshi:2023mnv}, it is very interesting to examine whether it is possible to obtain the other classified fixed points of 4d adjoint SQCDs labeled by the ADE-type superpotentials \cite{Intriligator:2003mi} from the $D_p[SU(N)]$ theories. Though the detailed dynamics of the 3d version may be different from the 4d one, the structures of the Seiberg-like dualities are quite parallel since they can directly be obtained from the circular compactification \cite{Aharony:2013dha,Kutasov:1995ve,Brodie:1996vx,Kim:2013cma,Kutasov:2014yqa,Hwang:2018uyj,Amariti:2018wht,Benvenuti:2018bav,Amariti:2019rhc,Nii:2020eui,Hwang:2020wpd,Amariti:2020xqm,Benvenuti:2020gvy,Okazaki:2021pnc,Okazaki:2021gkk,Hwang:2022jjs,Comi:2022aqo,Giacomelli:2023zkk,Amariti:2022lbw}. Thus, one may address the same question in 3d as well: is it possible to obtain the other ADE-type confinement from the 3d Argyres--Douglas theories $\mathbb{D}_p[SU(N)]$? Two obvious cases are the $E_6$ and $E_8$ types: $W_{E_6} = \Tr X^3 + \Tr Y^4$, $W_{E_8} = \Tr X^3 + \Tr Y^5$, in which cases, two adjoints in each WZ theory are decoupled. Once we gauge the WZ theory, we can obtain an SQCD with two adjoint fields, which can be deconfined straightforwardly into two linear quiver tails of $\mathbb{D}_p[SU(N)]$ and $\mathbb{D}_q[SU(N)]$ with $(p,q)=(2,3)$ for $E_6$ and $(p,q) = (2,4)$ for $E_8$. One advantage of the 3d version is that the $\mathbb{D}_p[SU(N)]$ theories are Lagrangian theories, which are more controllable than the non-Lagrangian 4d $D_p[SU(N)]$ theories. Hence, it would be interesting to explore the $E_6$ and $E_8$ type Seiberg-like dualities in 3d using our method, which are yet known anywhere including 4d.

\acknowledgments
CH is supported by the National Natural Science Foundation of China under Grant No.~12247103. SK is supported by a KIAS Individual Grant PG092101 at Korea Institute for Advanced Study.

\newpage
\appendix

\section{Review of the $D_p[SU(N)]$ Argyres--Douglas theories and their 3d reduction}
\label{app: D_p[G] review}
The Argyres--Douglas(AD) theories are a class of interacting $4d$ $\CN=2$ superconformal field theories characterized by dimensions of the Coulomb branch operators being fractional numbers \cite{Argyres:1995jj}. A collection of AD theories can be constructed by wrapping $N$ of M5-branes on a sphere with a particular irregular puncture, marked by two distinct types \cite{Xie:2012hs, Xie:2013jc}. Further refinements are possible by introducing an additional regular puncture onto the sphere \cite{Wang:2015mra}. Here we are interested in the latter refined version with a Type I irregular puncture which is labeled by $(A_{N-1}, A_{p-1})$ and a full regular puncture. This theory is typically denoted by $D_p[SU(N)]$ \cite{Cecotti:2013lda}; it has an $SU(N)$ flavor symmetry at least and notably lacks a Lagrangian description.
In the presence of a Coulomb branch operator of dimension two, one can deform the theory with it preserving the $\CN=2$ supersymmetry. This marginal deformation is usually associated with gauge couplings and one can find a cusp where the theory is described by weakly gauged distinct SCFTs with vector multiplets. In the context of the $D_p[SU(N)]$ theories, the specific procedure is outlined as follows \cite{Closset:2020afy, Giacomelli:2020ryy} :
\begin{align}
    D_p[SU(N)] = D_{p-q}[SU(N-n)] \leftarrow SU(N-n) \rightarrow \CD_q(N-n,N)
    \label{eq: decomposing D_p[SU(N)]}
\end{align}
where $n=N/\text{gcd}(N,p)$ and $q=p/\text{gcd}(N,p)$. The middle $\leftarrow SU(m) \rightarrow$ denotes a diagonal $SU(m)$ gauging of the theories on both sides.
One of those theories whose $SU(N-n)$ symmetry is gauged is again the AD theory associated with a full regular puncture, the theory theory is the AD theory associated with a partially closed regular puncture labeled by a partition, $[(q-1)^N,1^{N-n}]$, which can be written as
\begin{align}
\CD_q(N-n, N) = D_q\left[SU(q N-n), [(q-1)^N,1^{N-n}]\right] \,,
\end{align}
following the notation in \cite{Beem:2023ofp}.

One important observation for $\CD_q(N-n, N)$ is that its dimensional reduction to 3d, $\CD_q(N-n, N)_{3d}$, is described by a $3d$ $\CN=4$ linear quiver gauge theory with (special) unitary nodes:
\begin{center}
\begin{equation}
\begin{tikzpicture}
  \tikzset{vertex/.style={circle,fill=white!25,minimum size=12pt,inner sep=2pt}}
  \tikzset{every loop/.style={}}
    \node at (-2.5,0) {$\CD_q(N-n,N)_{3d}\quad:\qquad$ };    
    \node[vertex] (N) at (0,0) [shape=rectangle,draw=black,minimum height=2em, minimum width=2em] {\scriptsize $N\texttt{-}n$};
    \node[vertex] (m1) at (1.4,0) [shape=circle,draw=black] {$m_1$};
    \node[vertex] (m2) at (2.8,0) [shape=circle,draw=black] {$m_2$};
    \node[draw=none,fill=none] (ndots) at (4.2,0)  {$\cdots$};
    \node[vertex] (mq1) at (5.6,0) [shape=circle,draw=black] {\scriptsize $m_{q\!-\!1}$};
    \node[vertex] (Nn) at (7,0) [shape=rectangle,draw=black,minimum height=2em, minimum width=2em] {$N$};
    
      \draw  (N) -- (m1);
      \draw  (m1) -- (m2);
      \draw  (m2) -- (ndots);
      \draw  (ndots) -- (mq1);
      \draw (mq1) -- (Nn);
\end{tikzpicture}
\label{eq: 3d CD theory}
\end{equation}
\end{center}
where the circles denote $U(m_j)$ gauge nodes with $m_j = N-n+ \lfloor \frac{j n}{q} \rfloor$ and the lines are for the hypermultiplets in the bi-fundamental representations. Thus, we can get the $3d$ reduction of the $D_p[SU(N)]$ theories, denoted by $\mathbb{D}_p[SU(N)]$, by recursively decomposing it with a bunch of $\CD_q(M, L)$ blocks whose dimensional reductions are given by linear quivers \eqref{eq: 3d CD theory}. Let us provide some simple examples.
\paragraph{Example\\}
Consider $D_6[SU(4)]$ where $n=2$ and $q=3$ so that the decomposition \eqref{eq: decomposing D_p[SU(N)]} is given by
\begin{align}
    D_6[SU(4)] &= D_3[SU(2)] \leftarrow SU(2) \rightarrow \CD_3(2,4)
    \nonumber\\
    &= D_0[SU(0)] \leftarrow SU(0) \rightarrow \CD_3(0,2) \leftarrow SU(2) \rightarrow \CD_3(2,4)  
    \nonumber\\
    &=\CD_3(0,2) \leftarrow SU(2) \rightarrow \CD_3(2,4) \,.
    \label{eq: D_6[SU(4)] decomposition}
\end{align}
Since the 3d reductions of $\CD_3(0,2)$ and $\CD_3(2,4)$ are given by the following quivers:
\begin{center}
\begin{tikzpicture}
  \tikzset{vertex/.style={circle,fill=white!25,minimum size=12pt,inner sep=2pt}}
  \tikzset{every loop/.style={}}
    \node at (-2.5,-1) {$\CD_3(2,4)_{3d}\quad:$ };    
    \node[vertex] (4) at (0,-1) [shape=rectangle,draw=black] {$2$};
    \node[vertex] (3) at (1,-1) [shape=circle,draw=black] {$2$};
    \node[vertex] (2) at (2,-1) [shape=circle,draw=black] {$3$};
    \node[vertex] (2f) at (3,-1) [shape=rectangle,draw=black] {$4$};

    \node at (-2.5,0) {$\CD_3(0,2)_{3d}\quad:$ };    
    \node[vertex] (f2) at (0,0) [shape=rectangle,draw=black] {$0$};
    \node[vertex] (1) at (1,0) [shape=circle,draw=black] {$0$};
    \node[vertex] (0) at (2,0) [shape=circle,draw=black] {$1$};
    \node[vertex] (0f) at (3,0) [shape=rectangle,draw=black] {$2$};

    \node[vertex] (t2) at (5,0) [shape=circle,draw=black] {$1$};
    \node[vertex] (t1) at (6,0) [shape=rectangle,draw=black] {$2$};

    \node at (4,0) {$=$};
    
      \draw  (4) -- (3);
      \draw  (3) -- (2);
      \draw  (2) -- (2f);
      
      \draw  (f2) -- (1);
      \draw  (1) -- (0);
      \draw  (0) -- (0f);

      \draw  (t2) -- (t1);
\end{tikzpicture}
\end{center}
the $3d$ version of $D_6[SU(4)]$ is as follows:
\begin{center}
\begin{tikzpicture}
  \tikzset{vertex/.style={circle,fill=white!25,minimum size=12pt,inner sep=2pt}}
  \tikzset{every loop/.style={}}
    \node at (-2.5,0) {$\mathbb{D}_6[SU(4)]\quad:$ };    
    \node[vertex] (4) at (0,0) [shape=circle,draw=black] {$1$};
    \node[vertex] (3) at (1,0) [shape=circle,draw=blue] {$2$};
    \node[vertex] (2) at (2,0) [shape=circle,draw=black] {$2$};
    \node[vertex] (su2) at (3,0) [shape=circle,draw=black] {$3$};
    \node[vertex] (1) at (4,0) [shape=rectangle,draw=black] {$4$};
    
      \draw  (4) -- (3);
      \draw  (3) -- (2);
      \draw  (2) -- (su2);
      \draw  (su2) -- (1);
\end{tikzpicture}
\end{center}
where the blue circle denotes the $SU(2)$ gauge node.
\paragraph{No bad $\mathbb{D}_p[SU(N)]$ for coprime $(p,N)$\\}
We can check that any $\mathbb{D}_p[SU(N)]$ theory with a coprime integer pair $(p,N)$, which only contains unitary gauge nodes, is not a bad theory. Let $N \cdot j= k\;p + \a$ for some non-negative integer $k$ and $\a = N\cdot j \mod p$, then
\begin{align}
    m_{j-1} + m_{j+1} - 2 m_j &= 
    \Big\lfloor \frac{N(j-1)}{p} \Big\rfloor 
    + \Big\lfloor \frac{N(j+1)}{p} \Big\rfloor 
    - 2 \Big\lfloor \frac{N j}{p} \Big\rfloor
    \nonumber\\
    &=
    \Big\lfloor k - \frac{N}{p} + \frac{\a}{p} \Big\rfloor 
    + \Big\lfloor k + \frac{N}{p} + \frac{\a}{p} \Big\rfloor 
    - 2 \Big\lfloor k + \frac{\a}{p} \Big\rfloor
    \nonumber\\
    &= 
    \Big\lfloor \frac{\a}{p} - \frac{N}{p} \Big\rfloor 
    +\Big\lfloor \frac{\a}{p} + \frac{N}{p} \Big\rfloor.
    \label{eq: not bad condition}
\end{align}
Setting $\a/p \equiv \eta < 1$ and $N/p \equiv n + \xi$ for some non-negative integer $n$ such that $0\leq \xi < 1$, \eqref{eq: not bad condition} can be written
\begin{align}
     m_{j-1} + m_{j+1} - 2 m_j &= \lfloor \eta - \xi \rfloor + \lfloor \eta + \xi \rfloor
     \nonumber\\
     &=
     -1 \quad \text{or}\quad 0 \quad\text{or}\quad 1 \,,
\end{align}
in other words,
\begin{align}
   -1 \leq m_{j-1} + m_{j+1} - 2 m_j \leq 1 \,,
   \label{eq: m bound}
\end{align}
which ensures $\mathbb{D}_p[SU(N)]$ with coprime $(p,N)$ is not a bad theory. A natural question is when $\mathbb{D}_p[SU(N)]$ becomes good. To answer it, let us use a property,
\begin{align}
    m_j + m_{p-j} &= \Big\lfloor \frac{N j}{p} \Big\rfloor + \Big\lfloor \frac{N (p-j)}{p} \Big\rfloor
    \nonumber\\
    &= N + \Big\lfloor \frac{N j}{p} \Big\rfloor + \Big\lfloor - \frac{N j}{p} \Big\rfloor
    = N-1 + \d_{j,0} + \d_{j,p}
\end{align}
and suppose there exists an integer $0<\b<p$ such that
\begin{align}
\label{eq: m}
    m_{\b-1} + m_{\b+1} - 2m_\b = s
\end{align}
where $s$ can have its value as $-1$ or $0$ or $1$ according to \eqref{eq: m bound}. Then we have
\begin{align}
    m_{p-\b-1} + m_{p-\b+1} - 2m_{p-\b} &=
    (N-1 +\d_{p-\b-1,0} + \d_{p-\b-1,p} - m_{\b-1}) 
    \nonumber\\
    &\quad + (N-1 + \d_{p-\b+1,0} + \d_{p-\b+1,p} - m_{\b+1}) 
    \nonumber\\
    &\quad - 2(N-1 + \d_{p-\b,0} + \d_{p-\b,p} - m_\b)
    \nonumber\\
    &= -s + \d_{\b,1} + \d_{\b,p-1} \,.
    \label{eq: conj m}
\end{align}
\eqref{eq: m} and \eqref{eq: conj m} tell us that the first and $(p-1)$th nodes are always good due to the bound \eqref{eq: m bound}. Thus, let us focus on $1<\b<p-1$, in which case, if the $\b$-th node is good having $s = 1$, the $(p-\b)$th node is ugly having $s=-1$. Therefore, all the intermediate nodes of a good theory have no way but to satisfy $s=0$. Namely, for an integer $\s$ that can have values $0$ or $1$, we get
\begin{align}
    m_2-2m_1 &= \s
    \nonumber\\
    m_{j-1} + m_{j+1} &= 2m_j
    \, , \qquad 1 <j< p-1 \, , \nonumber \\
    m_{p-2}-2 m_{p-1} &= 1-\s \,,
\end{align}
which only happens for $N = \pm 1 \mod p$.

\section{Review of the confinement of $D_p[SU(N)]$ and the Maruyoshi--Nardoni--Song duality}
\label{app: MNS duality}
An s-confining supersymmetric gauge theory can be described by a Wess--Zumino (WZ) model of chiral multiplets with smooth confining superpotential.\footnote{A notion of \emph{smooth} confinement was first introduced in \cite{Csaki:1996sm}.} In that case, the chiral multiplets of WZ correspond to gauge invariant composite or solitonic particles of the gauge theory. More interestingly, such a confining phenomenon is not necessarily restricted to Lagrangian theories. 
For example, there have been several observations of the confinement-like behavior of the $4d$ $\CN=2$ Agyres--Douglas theories \cite{Bolognesi:2015wta, Xie:2021omd, Kang:2023dsa, Bajeot:2023gyl,Maruyoshi:2023mnv} with suitable deformations.
The example relevant to our study is $D_p[SU(N)]$ with $\gcd(p,N)=1$; a properly deformed $D_p[SU(N)]$ theory behaves as a WZ theory of a chiral multiplet in the adjoint representation of the $SU(N)$ symmetry.
Recently, a new duality between a $4d$ $\CN=1$ adjoint SQCD with the gauge group $SU(N)$ and the $A_p$-type superpotential of an adjoint chiral field $X$:
\begin{align}
    W = \Tr X^{p+1}
\end{align}
and an $\CN=1$ $SU(N)$ SQCD coupled to the $D_p[SU(N)]$ theory with a proper Coulomb branch superpotential deformation has been proposed in \cite{Maruyoshi:2023mnv}, which we call the {\it Maruyoshi--Nardoni--Song} (MNS) duality. The setup is, schematically, as follows:
\begin{center}
\begin{equation}
\begin{tikzpicture}
  \tikzset{vertex/.style={circle,fill=white!25,minimum size=12pt,inner sep=2pt}}
  \tikzset{every loop/.style={}}
    \node[vertex] (D) at (-1,0) [shape=ellipse,draw=black,minimum height=2em, minimum width=2em] {\scriptsize$D_p[SU(N)]$};
    \node[vertex] (N) at (2,0) [shape=circle,minimum size=2em,draw=black] {$N$};
    \node[vertex] (nf1) at (4,0.6) [shape=rectangle,draw=black,minimum height=2em, minimum width=2em] {$N_f$};
    \node[vertex] (nf2) at (4,-0.6) [shape=rectangle,draw=black,minimum height=2em, minimum width=2em] {$N_f$};
    \node at (1,0.3) {$\CN\!=\!1$};
    \node at (3,0.7) {$Q$};
    \node at (3,-0.7) {$\WQ$};
    
      \draw  (D) -- (N);
      \draw[->-=.5] ([yshift= 2pt] N.east) to ([yshift= 0pt] nf1.west);
      \draw[->-=.5] ([yshift= 0pt] nf2.west) to ([yshift= -2pt] N.east);
\end{tikzpicture}
\end{equation}
\end{center}
together with a superpotential deformation
\begin{align}
    \D W = u_0
    \label{eq: MNS dual deformation}
\end{align}
where $u_0$ is the Coulomb branch multiplet in the $D_p[SU(N)]$ theory having the lowest dimension. Also, ``$\CN=1$'' in the diagram denotes a diagonal gauging with an $\CN=1$ vector multiplet. It is assumed in \cite{Maruyoshi:2023mnv} that $\gcd(p, N)=1$ and $1 < p < N$ where the former condition implies $D_p[SU(N)]$ does not have a Lagrangian dual description.
The gauge invariant operators of the dual pair are mapped as follows:
\begin{align}
    \Tr X^i
    \quad&\leftrightarrow\quad
    v_{i-2} \;\;,\qquad i = 2,\cdots, p
    \nonumber\\
    M_i\equiv \WQ X^{i-1}Q
    \quad&\leftrightarrow\quad
    Q\m^{i-1}\WQ\;\;,\quad i=1,\cdots, p
    \nonumber\\
    B^{(n_1,\cdots,n_p)}
    \quad&\leftrightarrow\quad
    B^{(n_1,\cdots,n_p)}
    \label{eq: chiral ring map of MNS dual}
\end{align}
where $v_j$ are the $\CN=1$ descendant scalar chiral operators from the Coulomb branch operators $u_i$, $\m$ is the moment map for the $SU(N)$ group, and $B^{(n_1,\cdots,n_p)}$ are the generalized baryons:
\begin{align}
    B^{(n_1,\cdots,n_p)} = Q^{n_1} (XQ)^{n_2}\cdots(X^{j-1}Q)^{n_j} 
\end{align}
with $\{n_j\}$ is a partition of $N$ satisfying $n_j \leq N_f$.

\section{Monopole maps under basic dualities in linear quivers}
\label{app: monopole map}
In this appendix, we examine how the monopole operators in linear quivers are mapped under the basic $3d$ $\CN=2$ dualities we are interested in: the Aharony duality and the Benini--Benvenuti--Pasquetti (BBP) duality. Firstly, the Aharony duality \cite{Aharony:1997gp}, 3d reminiscent of the Seiberg duality \cite{Seiberg:1994pq}, is an IR duality between the following 3d $\mathcal N=2$ theories.
\begin{itemize}
    \item The 3d $\CN=2$ $U(N_c)$ gauge theory with $N_f$ pairs of fundamental and anti-fundamental $Q$ and $\WQ$ without superpotential,
    \begin{align}
        W_A = 0 \,.
    \end{align}

    \item The 3d $\CN=2$ $U(N_f - N_c)$ gauge theory with $N_f$ pairs of fundamental and anti-fundamental $q$ and $\Wq$ together with $N_f^2 + 2$ singlets $M$, $V^\pm$, and superpotential,
    \begin{align}
        W_B = M \Wq q + V^+ \hat{v}^+ + V^- \hat{v}^-
    \end{align}
    where $\hat{v}^\pm$ are the monopole operators. $M$ and $V^\pm$ correspond to meson $\WQ Q$ and monopole operators $\hat{V}^\pm$ of the original theory. 
\end{itemize}
The charges of elementary fields are given in Table \ref{tab: aharony charge}.
\begin{table}[tbp]
\centering
\begin{tabular}{|c|c|c|c|c|c|}
\hline
 & $U(1)_R$ & $SU(N_f)_t$ & $SU(N_f)_u$ & $U(1)_A$ & $U(1)_T$ \\
\hline 
$Q$ & $\Delta_Q$ & $\mathbf{N_f}$ & $\mathbf 1$ & $1$ & $0$ \\
$\WQ$ & $\Delta_Q$ & $\mathbf 1$ & $\mathbf{N_f}$ & $1$ & $0$ \\
$\hat{V}^\pm$ & $N_f (1-\Delta_Q) - N_c+1$ & $\mathbf 1$ & $\mathbf 1$ & $-N_f$ & $\pm1$ \\
\hline
$q$ & $1-\Delta_Q$ & $\mathbf 1$ & $\overline{\mathbf{N_f}}$ & $1$ & $0$ \\
$\Wq$ & $1-\Delta_Q$ & $\overline{\mathbf{N_f}}$ & $\mathbf 1$ & $1$ & $0$ \\
$M$ & $2 \Delta_Q$ & $\mathbf{N_f}$ & $\mathbf{N_f}$ & $2$ & $0$ \\
$V^\pm$ & $N_f (1-\Delta_Q)-N_c + 1$ & $\mathbf 1$ & $\mathbf 1$ & $-N_f$ & $\pm1$ \\
$\hat{v}^\pm$ & $-N_f (1-\Delta_Q) + N_c + 1$ & $\mathbf 1$ & $\mathbf 1$ & $N_f$ & $\mp 1$ \\
\hline
\end{tabular}
\caption{\label{tab: aharony charge} The charges of the elementary matter fields and the chiral monopole operators of the Aharony dual pair.}
\end{table}
The BBP duality, another similar 3d $\mathcal N=2$ IR duality, involves monopole superpotential terms \cite{Benini:2017dud}. The version with a single linear monopole superpotential term:
\begin{itemize}
    \item The 3d $\CN=2$ $U(N_c)$ gauge theory with $N_f$ pairs of fundamental and anti-fundamental $Q$ and $\WQ$ with one-monopole superpotential,
    \begin{align}
        W_A = \hat V^\pm
    \end{align}

    \item The 3d $\CN=2$ $U(N_f - N_c - 1)$ gauge theory with $N_f$ pairs of fundamental and anti-fundamental $q$ and $\Wq$ together with $N_f^2 + 1$ singlets $M$, $V^\mp$ and superpotential,
    \begin{align}
        W_B = \hat{v}^\pm  +  M \Wq q  + V^\mp \hat{v}^\mp
    \end{align}
    where $\hat{v}^\pm$ being the monopole operators. $M$ and $V^\mp$ correspond to meson $\WQ Q$ and monopole operator $\hat{V}^\mp$ of the original theory. 
\end{itemize}
We call this duality $\text{BBP}_1^\pm$ in short. The charges of the elementary fields for the one-monopole BBP duality are in Table \ref{tab: BBP1 charge}.
\begin{table}[tbp]
\centering
\begin{tabular}{|c|c|c|c|c|}
\hline
 & $U(1)_R$ & $SU(N_f)_t$ & $SU(N_f)_u$ & $U(1)_A$ \\
\hline 
$Q$ & $\Delta_Q$ & $\mathbf{N_f}$ & $\mathbf 1$ & $1$ \\
$\WQ$ & $\Delta_Q$ & $\mathbf 1$ & $\mathbf{N_f}$ & $1$ \\
$\hat{V}^\mp$ & $2N_f (1-\Delta_Q) - 2N_c$ & $\mathbf 1$ & $\mathbf 1$ & $-2N_f$ \\
\hline
$q$ & $1-\Delta_Q$ & $\mathbf 1$ & $\overline{\mathbf{N_f}}$ & $1$ \\
$\Wq$ & $1-\Delta_Q$ & $\overline{\mathbf{N_f}}$ & $\mathbf 1$ & $1$ \\
$M$ & $2 \Delta_Q$ & $\mathbf{N_f}$ & $\mathbf{N_f}$ & $2$ \\
$V^\mp$ & $2N_f (1-\Delta_Q)-2N_c $ & $\mathbf 1$ & $\mathbf 1$ & $-2N_f$ \\
$\hat{v}^\mp$ & $-2N_f (1-\Delta_Q) +2 N_c + 2$ & $\mathbf 1$ & $\mathbf 1$ & $2N_f$ \\
\hline
\end{tabular}
\caption{\label{tab: BBP1 charge} The charges of the elementary matter fields and the chiral monopole operators of the one-monopole BBP dual pair.}
\end{table}

\paragraph{In linear quivers\\}
Now, let us check how the monopole operators in linear quivers are mapped under those two basic dualities by matching their quantum numbers. Consider linear quivers in \eqref{eq: linear quiver} whose gauge and flavor nodes are unitary and special unitary groups respectively.
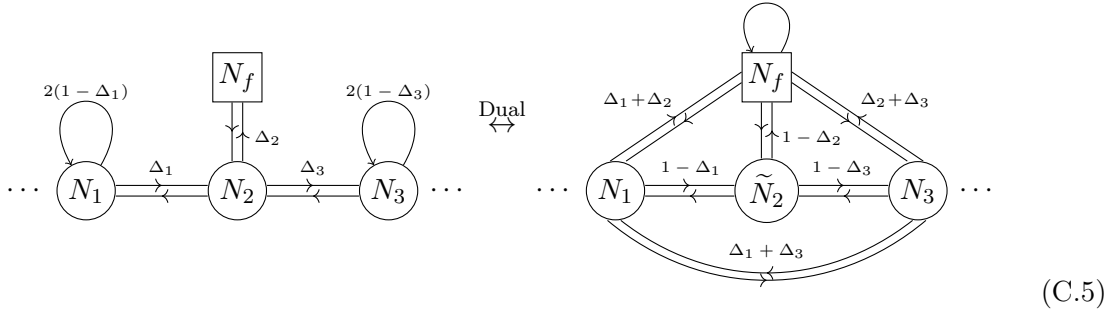
\begin{center}
\begin{equation}
\begin{tikzpicture}
  \tikzset{vertex/.style={circle,fill=white!25,minimum size=12pt,inner sep=2pt}}
  \tikzset{every loop/.style={}}
  
    \node[vertex] (N1) at (0,0) [shape=circle,draw=black] {$N_1$};
    \node[vertex] (N2) at (2,0) [shape=circle,draw=black] {$N_2$};
    \node[vertex] (Nf) at (2,1.5) [shape=rectangle,draw=black,minimum height=1.7em, minimum width=1.7em] {$N_f$};
    \node[vertex] (N3) at (4,0) [shape=circle,draw=black] {$N_3$};

    \draw[-to, min distance=1.3cm]  (N1) edge [out=60, in=120] node {} (N1);
    \draw[-to, min distance=1.3cm]  (N3) edge [out=60, in=120] node {} (N3);
    \draw[->-=.5] ([xshift= 2pt] N2.north) to ([xshift= 2pt] Nf.south);
    \draw[->-=.5] ([xshift= -2pt] Nf.south) to ([xshift= -2pt] N2.north);
    \draw[->-=.5] ([yshift= 2pt] N1.east) to ([yshift= 2pt] N2.west);
    \draw[->-=.5] ([yshift= -2pt] N2.west) to ([yshift= -2pt] N1.east);
    \draw[->-=.5] ([yshift= 2pt] N2.east) to ([yshift= 2pt] N3.west);
    \draw[->-=.5] ([yshift= -2pt] N3.west) to ([yshift= -2pt] N2.east);

    \node at (-0.8,0) {$\cdots$};
    \node at (0,1.3) {\tiny $2(1-\D_1)$};
    \node at (1,0.3) {\tiny $\D_1$};
    \node at (2+0.4,0.7) {\tiny $\D_2$};
    \node at (3,0.3) {\tiny $\D_3$};
    \node at (4,1.3) {\tiny $2(1-\D_3)$};
    \node at (4.8,0) {$\cdots$};

    \node at (5.5,1) {$\overset{\text{Dual}}{\leftrightarrow}$};

    \node[vertex] (N1d) at (7,0) [shape=circle,draw=black] {$N_1$};
    \node[vertex] (N2d) at (9,0) [shape=circle,draw=black] {$\WN_2$};
    \node[vertex] (Nfd) at (9,1.5) [shape=rectangle,draw=black,minimum height=1.7em, minimum width=1.7em] {$N_f$};
    \node[vertex] (N3d) at (11,0) [shape=circle,draw=black] {$N_3$};

    \node at (6.2,0) {$\cdots$};
    \draw[-to, min distance=1cm]  (Nfd) edge [out=60, in=120] node {} (Nfd);
    \draw[->-=.5] ([xshift= 2pt] N2d.north) to ([xshift= 2pt] Nfd.south);
    \draw[->-=.5] ([xshift= -2pt] Nfd.south) to ([xshift= -2pt] N2d.north);
    \draw[->-=.5] ([yshift= 2pt] N1d.east) to ([yshift= 2pt] N2d.west);
    \draw[->-=.5] ([yshift= -2pt] N2d.west) to ([yshift= -2pt] N1d.east);
    \draw[->-=.5] ([yshift= 2pt] N2d.east) to ([yshift= 2pt] N3d.west);
    \draw[->-=.5] ([yshift= -2pt] N3d.west) to ([yshift= -2pt] N2d.east);
    \draw[->-=.5] ([xshift= -2pt] N1d.north) to ([yshift= 2pt] Nfd.west);
    \draw[->-=.5] ([yshift= -2pt] Nfd.west) to ([xshift= 4pt] N1d.north);
    \draw[->-=.5] ([xshift= 2pt] N3d.north) to ([yshift= 2pt] Nfd.east);
    \draw[->-=.5] ([yshift= -2pt] Nfd.east) to ([xshift= -4pt] N3d.north);
    \draw[->-=.5] ([xshift= -3pt] N1d.south) to [bend right=40] ([xshift= 3pt] N3d.south);
    \draw[->-=.5] ([xshift= -2pt] N3d.south) to [bend left=38] ([xshift= 2pt] N1d.south);
    \node at (11.8,0) {$\cdots$};

    \node at (8,0.3) {\tiny $1-\D_1$} ;
    \node at (9+0.6,0.7) {\tiny $1-\D_2$};
    \node at (7.3,1.2) {\tiny $\D_1 \!+\! \D_2$} ;
    \node at (9,-0.8) {\tiny $\D_1+\D_3$};
    \node at (10.7,1.2) {\tiny $\D_2 \!+\! \D_3$} ;
    \node at (10,0.3) {\tiny $1-\D_3$} ;

\end{tikzpicture}
\label{eq: linear quiver}
\end{equation}
\end{center}
While we have depicted only three gauge nodes labeled as the 1st, 2nd, and the 3rd nodes, there are also other nodes on the left labeled as the 0th node, the $(-1)$th node, and so on and those on the right labeled as the 4th node, the 5th node, and so on. We have also written normalized real masses of the chiral multiplets around the arrows. The monopole operators of the LHS that are charged positively or negatively under the $U(1)_T$ symmetries from the $i$-th gauge node to $j$-th gauge node are denoted by $\hat v^{(i,j),\pm}$ respectively. Then, their R-charges are given by
\begin{align}
    &R[\hat{v}^{(l,1),\pm}] = r_L + (-2 N_1 + N_2 + 2 )(1-\D_1)
    \nonumber\\
    &R[\hat{v}^{(2),\pm}] = N_1 (1-\D_1) + N_f (1-\D_2) + N_3 (1-\D_3) - N_2 + 1
    \nonumber\\
    &R[\hat{v}^{(3,r),\pm}] = (N_2 - 2 N_3 + 2 ) (1-\D_3) + r_R
    \nonumber\\
    &R[\hat{v}^{(l,2),\pm}] = r_L + (N_2-N_1)(1-\D_1) + N_f (1-\D_2) + N_3 (1-\D_3) - N_2 + 1
    \nonumber\\
    &R[\hat{v}^{(2,r),\pm}] = N_1(1-\D_1) + N_f(1-\D_2) + (N_2 - N_3)(1-\D_3) - N_2 + 1 + r_R
    \nonumber\\
    &R[\hat{v}^{(l,r),\pm}] = r_L + (N_2-N_1)(1-\D_1) + N_f(1-\D_2) + (N_2 - N_3)(1-\D_3) - N_2 + 1 + r_R.
\end{align}
where $l < 2$ and $r > 2$. Here $r_L$ and $r_R$ are the R-charge contributions from the left side of the $U(N_1)$ node and the right side of the $U(N_3)$ node respectively, which are invariant under the local duality action acting on the $U(N_2)$ node. We have omitted the monopole operators charged only under the nodes on the left side of the $U(N_1)$ node or on the right side of the $U(N_3)$ node because they are irrelevant to our discussion. The R-charges of the monopole operators of the RHS quiver, $\hat u^{(i,j),\pm}$, are similarly given by
\begin{align}
    &R[\hat{u}^{(l,1),\pm}] = r_L + (N_f - \WN_2 + N_3)(1-\D_1) + N_f(1-\D_2) + N_3(1-\D_3) + \WN_2 - N_1 - N_f - N_3 + 1
    \nonumber\\
    &R[\hat{u}^{(2),\pm}] = -N_1 (1-\D_1) - N_f (1-\D_2) - N_3 (1-\D_3) - \WN_2 + N_1 + N_f + N_3 + 1
    \nonumber\\
    &R[\hat{u}^{(3,r),\pm}] = N_1(1-\D_1) + N_f(1-\D_2) + (N_1 - \WN_2 + N_f)(1-\D_3) + \WN_2 - N_1 - N_f - N_3 + 1 + r_R
    \nonumber\\
    &R[\hat{u}^{(l,2),\pm}] = r_L + (N_f - N_1 - \WN_2 + N_3 + 2 )(1-\D_1)
    \nonumber\\
    &R[\hat{u}^{(2,r),\pm}] = (N_1 + N_f - \WN_2 - N_3 + 2)(1-\D_3) + r_R
    \nonumber\\
    &R[\hat{u}^{(l,r),\pm}] = r_L + (N_f - \WN_2 + N_3)(1-\D_1) + N_f(1-\D_2) 
    \nonumber\\
    &\qquad\qquad\qquad\qquad\qquad\qquad\qquad\qquad
    + (N_1 - \WN_2 + N_f)(1-\D_3)  -N_1 + \WN_2 - N_f - N_3 + 1 + r_R
\end{align}

\paragraph{Aharony duality:}
Suppose we take the Aharony duality on the $U(N_2)$ node, whose dual rank is then given by $\WN_2 = N_1 + N_f + N_3 - N_2$. We find relations between the two sets of the monopole R-charges as follows:
\begin{align}
    &R[ \hat{v}^{(l,1),\pm}] = R[\hat{u}^{(l,2),\pm}] \,,
    \nonumber\\
    &R[ \hat{v}^{(l,2),\pm} ] = R[\hat{u}^{(l,1),\pm}] \,,
    \nonumber\\
    &R[ \hat{v}^{(3,r),\pm} ] = R[\hat{u}^{(2,r),\pm}] \,,
    \nonumber\\
    &R[ \hat{v}^{(2,r),\pm} ] = R[\hat{u}^{(3,r),\pm}] \,,
    \nonumber\\
    &R[ \hat{v}^{(l,r),\pm} ] = R[ \hat{u}^{(l,r),\pm} ] \,,
    \nonumber\\
    & R[\hat{v}^{(2),\pm}] + R[\hat{u}^{(2),\pm}] = 2 \,,
\end{align}
which suggest the following map of monopole operators:
\begin{empheq}[box=\widefbox]{align}    
    \hat{v}^{(\cdots,1),\pm}
    \qquad&\leftrightarrow\qquad
    \hat{u}^{(\cdots,1,2),\pm}
    \nonumber\\
    \hat{v}^{(\cdots,1,2),\pm}
    \qquad&\leftrightarrow\qquad
    \hat{u}^{(\cdots,1),\pm}
    \nonumber\\
    \hat{v}^{(2),\pm}
    \qquad&\leftrightarrow\qquad
    v^{(2),\pm}
    \nonumber\\
    \hat{v}^{(\cdots,3),\pm}
    \qquad&\leftrightarrow\qquad
    \hat{u}^{(\cdots,2,3),\pm}
    \nonumber\\
    \hat{v}^{(\cdots,2,3),\pm}
    \qquad&\leftrightarrow\qquad
    \hat{u}^{(\cdots,3),\pm}
    \label{eq: local Aharony monopole map}
\end{empheq}
where $v^{(2),\pm}$ are the singlets coupled to the monopole operators $\hat{u}^{(2),\pm}$ in the dual theory. The $U(1)$ topological charges of the $i$-th gauge node for both sides, $T_i$ and $\WT_i$, are mapped as
\begin{align}
    T_i  = \WT_i + \d_{2,i} \; (\WT_{i-1} - 2 \WT_i + \WT_{i+1} ).
\end{align}

\paragraph{One-monopole BBP:}
Suppose we take the $\text{BBP}_1^+$ duality on the $U(N_2)$ node, whose dual rank is given by $\WN_2 = N_1 + N_f + N_3 - N_2 - 1$. The one-monopole superpotential term $\hat{v}^{(2),+}$ breaks the $U(1)_A \times U(1)_{T_2}$ into a combination of them. Since the mixing of the $U(1)_A$ symmetry to the R-symmetry is already taken into account as a shift of $\Delta_i$, it is enough to introduce the mixing coefficients $\t_{i}$ of the $U(1)_{T_i}$ symmetries with the R-symmetry for $i=1,2,3$ to find the correct combination. Then R-charges of the monopole operators of the LHS quiver are shifted as follows:
\begin{align}
    R[\hat{v}^{(l,1),\pm}] &\rightarrow R[\hat{v}^{(l,1),\pm}] \pm \t_1
    \nonumber\\
    R[\hat{v}^{(2),+}] &\rightarrow R[\hat{v}^{(2),+}] +\t_2 \overset{!}{=}2
    \nonumber\\
    R[\hat{v}^{(2),-}] &\rightarrow R[\hat{v}^{(2),-}] -\t_2
    \nonumber\\
    R[\hat{v}^{(3,r),\pm}] &\rightarrow R[\hat{v}^{(3,r),\pm}] \pm \t_3
    \nonumber\\
    R[\hat{v}^{(l,2),\pm}] &\rightarrow R[\hat{v}^{(l,2),\pm}] \pm \t_1 \pm \t_2
    \nonumber\\
    R[\hat{v}^{(2,r),\pm}] &\rightarrow R[\hat{v}^{(2,r),\pm}] \pm \t_2 \pm \t_3
    \nonumber\\
    R[\hat{v}^{(l,r),\pm}] &\rightarrow R[\hat{v}^{(l,r),\pm}] \pm \t_1\pm\t_2\pm\t_3
\end{align}
and also similarly for the RHS monopoles: 
\begin{align}
    R[\hat{u}^{(l,1),\pm}] &\rightarrow R[\hat{u}^{(l,1),\pm}] \pm \t'_1
    \nonumber\\
    R[\hat{u}^{(2),+}] &\rightarrow R[\hat{u}^{(2),+}] \pm \t'_2
    \nonumber\\
    R[\hat{u}^{(3,r),\pm}] &\rightarrow R[\hat{u}^{(3,r),\pm}] \pm \t'_3
    \nonumber\\
    R[\hat{u}^{(l,2),\pm}] &\rightarrow R[\hat{u}^{(l,2),\pm}] \pm \t'_1 \pm \t'_2
    \nonumber\\
    R[\hat{u}^{(2,r),\pm}] &\rightarrow R[\hat{u}^{(2,r),\pm}] \pm \t'_2 \pm \t'_3
    \nonumber\\
    R[\hat{u}^{(l,r),\pm}] &\rightarrow R[\hat{u}^{(l,r),\pm}] \pm \t'_1\pm\t'_2\pm\t'_3
\end{align}
where we impose $\t_2 + \t'_2 = 1$ from the dual monopole superpotential term $\hat{u}^{(2),+}$. By considering all the $\t_i$ and $\t'_i$ independent combinations of the R-charges together with the global symmetry charges, a unique way of mapping the monopole operators is given by
\begin{empheq}[box=\widefbox]{align}    
    \hat{v}^{(l,1),+}
    \qquad&\leftrightarrow\qquad
    \hat{u}^{(l,1),+}
    \nonumber\\
    \hat{v}^{(3,r),+}
    \qquad&\leftrightarrow\qquad
    \hat{u}^{(3,r),+}
    \nonumber\\
    \hat{v}^{(l,1),-}
    \qquad&\leftrightarrow\qquad
    \hat{u}^{(l,2),-}
    \nonumber\\
    \hat{v}^{(l,2),-}
    \qquad&\leftrightarrow\qquad
    \hat{u}^{(l,1),-}
    \nonumber\\
    \hat{v}^{(2),-}
    \qquad&\leftrightarrow\qquad
    v^{(2),-}
    \nonumber\\
    \hat{v}^{(3,r),-}
    \qquad&\leftrightarrow\qquad
    \hat{u}^{(2,r),-}
    \nonumber\\
    \hat{v}^{(2,r),-}
    \qquad&\leftrightarrow\qquad
    \hat{u}^{(3,r),-}
    \label{eq: local BBP1 monopole map}
\end{empheq}
if $(1-\D_1)+\t_1 = \t'_1 + \t'_2$ and $(1-\D_3)+\t_3 = \t'_2 + \t'_3$, where the positively charged monopole operators are trivially mapped, while the negatively charged ones are mapped as Aharony duality.
\paragraph{Monopole maps for section \ref{sec: p=2 derivation example}}
Let us also provide some specific monopole maps used in the section \ref{sec: p=2 derivation example}. The first one comes from the one-monopole BBP duality where one can check that
\begin{align}
    R[\hat u^{(1,2),+}] = R[\hat v^{(1),+}] + 2(1-\D_1) \,,
\end{align}
which suggests a natural monopole map $\hat v_1^{(1),+} \leftrightarrow \hat u^{(1,2),+}$ with $\hat v_1^{(1),+}$ being a positive monopole operator dressed once by an adjoint field of the $U(N_1)$ gauge node of the LHS theory in \eqref{eq: linear quiver}. Secondly, we also have
\begin{align}
    R[\hat v^{(1,2),+}] = R[\hat u^{(1,2),+}] + 4\D_1-2 \,.
\end{align}
When $\D_1 = 2/3$, which is the case for the $p=2$ case, we can modify the above relation as 
\begin{align}
    R[\hat v^{(1,2),+}] = R[\hat u^{(1,2),+}] + 2-2\D_1 \,,
\end{align}
which suggests a monopole map $\hat v^{(1,2),+} \leftrightarrow \text{tr}\big( \Wr r\, \hat u^{(1,2),+} \big)$ where $r$ and $\Wr$ denote the bi-fundamental chirals between $U(N_1)$ and $U(\WN_2)$ gauge nodes of the RHS theory in \eqref{eq: linear quiver}. Lastly, consider deconfining an adjoint via $\text{BBP}_1^+$ duality:
\begin{equation}
\begin{tikzpicture}
  \tikzset{vertex/.style={circle,fill=white!25,minimum size=12pt,inner sep=2pt}}
  \tikzset{every loop/.style={}}
  
    \node[vertex] (AN) at (0.7,0) [shape=circle,draw=black,minimum size=2em] {$N$};
    \node[vertex] (AF) at (2.3,0) [shape=rectangle,draw=black,minimum height=2em, minimum width=2em] {$F$};

    \draw[-to, min distance=1cm]  (AN) edge [out=120, in=60] node {} (AN);
    \draw[->-=.5] ([yshift= 2pt] AN.east) to ([yshift= 2pt] AF.west);
    \draw[->-=.5] ([yshift= -2pt] AF.west) to ([yshift= -2pt] AN.east);

    \node at (0.7,1.2) {\scriptsize $2\D_1$};
    \node at (1.5,0.4) {\scriptsize $\D_2$};

    \node at (4.4,0.2) {\scriptsize $\text{BBP}_1^+$};
    \node at (4.4,-0.2) {\scriptsize $\longleftrightarrow$};

    \node[vertex] (BN1) at (0+6,0) [shape=circle,draw=black,minimum size=2em] {\scriptsize $\! N\texttt{-}1\!$};
    \node[vertex] (BN) at (1.5+6,0) [shape=circle,draw=black,minimum height=2em, minimum width=2em] {$N$};
    \node[vertex] (BF) at (3+6,0) [shape=rectangle,draw=black,minimum height=2em, minimum width=2em] {$F$};

    \draw[->-=.5] ([yshift= 2pt] BN1.east) to ([yshift= 2pt] BN.west);
    \draw[->-=.5] ([yshift= -2pt] BN.west) to ([yshift= -2pt] BN1.east);
    \draw[->-=.5] ([yshift= 2pt] BN.east) to ([yshift= 2pt] BF.west);
    \draw[->-=.5] ([yshift= -2pt] BF.west) to ([yshift= -2pt] BN.east);

    \node at (0.75+6,+0.4) {\scriptsize $\D_1$};
    \node at (2.2+6,+0.4) {\scriptsize $\D_2$};
\end{tikzpicture} \,.
\label{eq: adjoint deconf}
\end{equation}
One can check that
\begin{align}
    R[V_0^-] &= 
    F(1-\D_2) - 3\D_1 (N-1) - \t
    =
    R[\hat v^{(1,2),-}] \,,
    \nonumber\\
    R[\hat V_{N-1}^-] &=
    R(1-\D_2) - \D_1(N-1) - \t
    =
    R[\hat v^{(2),-}]
\end{align}
where $\t$ is a possible R-symmetry mixing coefficient of the $U(1)$ topological symmetry of the $U(N)$ gauge group in the RHS quiver. Hence, these relations suggest monopole map $\hat V_0^- \leftrightarrow \hat v^{(1,2),-}$ and $\hat V_{N-1}^- \leftrightarrow \hat v^{(2),-}$.
\section{Monopole map under the deconfinement of an adjoint into a $\mathbb D_p[SU(N)]$ tail}
\label{app: monopole map under gauging}
Let us discuss how the dressed monopole operators in the $A_p$-type adjoint SQCD with the $U(N)$ gauge group are mapped under deconfining an adjoint chiral into a $\mathbb{D}_p[SU(N)]$ tail:
\begin{center}
\begin{equation}
\begin{tikzpicture}
  \tikzset{vertex/.style={circle,fill=white!25,minimum size=12pt,inner sep=2pt}}
  \tikzset{every loop/.style={}}
    \node[vertex] (N) at (-2.5,0) [shape=circle,draw=black,minimum size=2em] {$N$};
    \node[vertex] (Nf) at (-1,0) [shape=rectangle,draw=black,minimum height=2em, minimum width=2em] {$N_f$};

    \draw[-to, min distance=1cm]  (N) edge [out=60, in=120] node {} (N);
    \draw[->-=.5] ([yshift= 2pt] N.east) to ([yshift= 2pt] Nf.west);
    \draw[->-=.5] ([yshift= -2pt] Nf.west) to ([yshift= -2pt] N.east);

    \node at (-2.5,1.2) {\tiny $X$};
    \node at (-1.8,0.3) {\tiny$Q$};
    \node at (-1.8,-0.3) {\tiny$\WQ$};
    \node at (-1.7,-1) {\scriptsize$W = \Tr X^{p+1}$};

    \node at (1.5,0.1) {$\xrightarrow{\text{Deconfine}}$};

    \node[vertex] (m1) at (4,0) [shape=circle,draw=black,minimum size=2em] {$m_1$};
    \node (cdot) at (5.2,0) {$\cdots$};
    \node[vertex] (mpm1) at (6.5,0) [shape=circle,draw=black,minimum size=2em] {\tiny$m_{p\!-\!1}$};
    \node[vertex] (NN) at (8,0) [shape=circle,draw=black,minimum size=2em] {$N$};
    \node[vertex] (NfNf) at (9.5,0) [shape=rectangle,draw=black,minimum height=2em, minimum width=2em] {$N_f$};

    \draw[-to, min distance=1cm]  (m1) edge [out=60, in=120] node {} (m1);
    \draw[-to, min distance=1cm]  (mpm1) edge [out=60, in=120] node {} (mpm1);
    \draw[->-=.5] ([yshift= 2pt] m1.east) to ([yshift= 2pt] cdot.west);
    \draw[->-=.5] ([yshift= -2pt] cdot.west) to ([yshift= -2pt] m1.east);
    \draw[->-=.5] ([yshift= 2pt] cdot.east) to ([yshift= 2pt] mpm1.west);
    \draw[->-=.5] ([yshift= -2pt] mpm1.west) to ([yshift= -2pt] cdot.east);
    \draw[->-=.5] ([yshift= 2pt] mpm1.east) to ([yshift= 2pt] NN.west);
    \draw[->-=.5] ([yshift= -2pt] NN.west) to ([yshift= -2pt] mpm1.east);
    \draw[->-=.5] ([yshift= 2pt] NN.east) to ([yshift= 2pt] NfNf.west);
    \draw[->-=.5] ([yshift= -2pt] NfNf.west) to ([yshift= -2pt] NN.east);

    \node at (4.6,0.3) {\tiny$Q^{(1)}$};
    \node at (4.6,-0.3) {\tiny$\WQ^{(1)}$};
    \node at (5.7,0.3) {\tiny$Q^{(p\!-\!2)}$};
    \node at (5.7,-0.3) {\tiny$\WQ^{(p\!-\!2)}$};
    \node at (7.3,0.3) {\tiny$Q^{(p\!-\!1)}$};
    \node at (7.3,-0.3) {\tiny$\WQ^{(p\!-\!1)}$};
    \node at (8.7,0.3) {\tiny$Q$};
    \node at (8.7,-0.3) {\tiny$\WQ$};
    \node at (4,1.2) {\tiny $\Phi^{(1)}$};
    \node at (6.5,1.2) {\tiny $\Phi^{(p\!-\!1)}$};

    \node at (7,-1) {\scriptsize$W = \sum_{i=1}^{p-1} \WQ^{(i)} \Phi^{(i)} Q^{(i)} + \sum_{i=2}^{p-1} Q^{(i-1)}\Phi^{(i)} \WQ^{(i-1)} + \D W_{\text{conf}}^{(p),\pm}$};
    
\end{tikzpicture} .
\end{equation}
\end{center}
The Coulomb branch of the LHS theory is described by $2p$ dressed monopole operators, $\hat{V}_{i=0,\cdots,p-1}^\pm$ whose R-charge is given as,
\begin{align}
    R[\hat{V}_i^{\pm}] = N_f(1-\D_Q) + \frac{2}{p+1} (1-N+i)\;,
    \quad\text{with}\;\; \D_Q = R[Q] = R[\WQ].
    \label{eq: dressed R}
\end{align}
Suppose the $U(1)_T$ topological symmetry of the adjoint SQCD on the LHS is linearly mapped to that of the right-most gauge node of the deconfined theory on the RHS. Since all the monopole operators in the LHS theory are charged $\pm 1$ under the $U(1)_T$, the corresponding Coulomb branch operators in the deconfined theory should also be charged under the topological symmetry of the right-most gauge node. We claim such $2p$ candidates are $\hat{v}^{(j,p),\pm}$, $j=1,\cdots,p$ where we set the right-most gauge node as $p$-th gauge node. The R-charges of them can be computed as,
\begin{align}
    R[\hat{v}^{(j,p),\pm}] = N_f(1-\D_Q) + \frac{p}{p+1}( m_{j-1} - m_j + N) - N + 1
    \pm \sum_{i=j}^{p} \t_i.
    \label{eq: v^(...p) R-charge}
\end{align}
with $m_0\equiv 0$, $m_p \equiv N$, and $\t_i$'s are the R-charge mixing coefficient of the $U(1)_{T_i}$ topological symmetries. The coefficients $\t_{i<p}$ are fixed by the constraints from the confining superpotential deformation $W_{\text{conf}}^{(p),\pm}$,
\begin{align}
    &(m_{i-1} - 2 m_i + m_{i+1} +2)(1-\frac{1}{p+1}) \pm \t_i \overset{!}{=} 2
    \qquad \text{for} \quad i=1,\,\cdots\,,\,p-1 \,,
\end{align}
which fix the coefficients as
\begin{align}
    \t_j = \pm \frac{1}{p+1} \big(
    2-p\; \d_{[N\texttt{+}j],0}
    \big)
    \,,\qquad
    j=0,\,\cdots\,,\,p-1 \,,
\end{align}
where $[n]:= n \,\text{mod}\, p$. In addition, $\t_p$ can be fixed by a consistency condition that the R-charge spectrum of the positively and negatively charged monopole operators should be the same, which is the case on the LHS; namely,
\begin{align}
    \sum_{j=1}^{p} R[\hat{v}^{(j,p),+}] \overset{!}{=} \sum_{j=1}^{p} R[\hat{v}^{(j,p),-}]
    \quad \to \quad
    \t_p = - \frac{1}{p} \sum_{j=1}^{p-1} j\;\t_j
    =
    \pm \frac{1-[N]}{p+1}.
    \label{eq: tau_p construction}
\end{align}
With those fixings, one can neatly match the quantum numbers from \eqref{eq: dressed R} and \eqref{eq: v^(...p) R-charge}, which results in the following map of the Coulomb branch operators under the deconfinement.
\begin{empheq}[box=\widefbox]{align} 
    &\text{For}\;\; W_\text{conf}^{(p),+} \; :\;
    \Big\{
    \begin{array}{ccc}
       \hat{V}_{[-j]}^{+} & \quad\leftrightarrow\quad & \hat{v}^{(j,p),+} \\
        \hat{V}_{[j\texttt{-}1+N]}^{-} & \quad\leftrightarrow\quad &\hat{v}^{(j,p),-} 
    \end{array}
    \nonumber\\
    &\text{For}\;\; W_\text{conf}^{(p),-} \; :\;
    \Big\{
    \begin{array}{ccc}
       \hat{V}_{[j\texttt{-}1+N]}^{+} & \quad\leftrightarrow\quad &\hat{v}^{(j,p),+}  \\
        \hat{V}_{[-j]}^{-}& \quad\leftrightarrow\quad & \hat{v}^{(j,p),-}
    \end{array}
\end{empheq}
\section{$S^3_b$ partition function identities and quiver diagrams}
\label{app: S_b^3 ptf}
In this appendix, we review the localization formula of the squashed three-sphere partition function \cite{Hama:2011ea} and its identities from the basic $3d$ $\CN=2$ IR dualities. We also explain how to read the partition function from a given quiver diagram.
The squashed three-sphere partition function of a $3d$ $\CN=2$ gauge theory with gauge group $G$ and a set of chiral multiplets, collectively denoted by $\Phi$, having $R$-charge $r_\Phi$ and charged under the gauge and global symmetries with weights $\r_\Phi$ and $\phi_\Phi$ respectively, with real mass parameter $M$, is given by an integral over the Coulomb branch parameter $\s$ as follows:
\begin{align}
    Z_{S_b^3} = \frac{1}{|W|} 
    \int \prod_{i=1}^{\text{rank}(G)} d \s_j
    e^{2\pi i \xi \Tr \s}
    \frac{
    \prod_{\Phi} s_b (\frac{i Q}{2}(1-r_\Phi) - \rho_\Phi(\s) - \phi_\Phi(M) )
    }
    {
    \prod_{\a} s_b(\a(\s) + \frac{i Q}{2})
    }
    \label{eq: S_b^3 ptf integral formula}
\end{align}
where $W$ is the Weyl group of $G$, $Q\equiv b + b^{-1}$ parametrizes the squashing parameter $b$, $\a$ runs over the roots of the gauge group $G$, and $\xi$ is the Fayet-Illiopoulos parameter for the $U(1)$ factors in $G$. The double-sine function $s_b(\frac{iQ}{2}(1-r)-m)$ for a chiral multiplet is defined by
\begin{align}
    s_b(x) := \prod_{l,n\geq 0}
    \frac{l b + n b^{-1} + Q/2 - i x}{l b + n b^{-1} + Q/2 + i x} \,,
\end{align}
which satisfies a property $s_b(x) s_b(-x) = 1$, reflecting a cancellation of the contributions of two coupled massive chiral fields. Let us write down the integral identities from the 3d $\CN=2$ basic dualities where we follow the conventions of \cite{Benini:2017dud}:
\paragraph{Aharony duality:\\}
\begin{align}
    &\frac{1}{N_c !} \int \prod_{j=1}^{N_c} dx_j
    e^{\pi i \xi \sum_{j=1}^{N_c} x_j}
    \frac{\prod_{j=1}^{N_c} \prod_{a=1}^{N_f}s_b( x_j - m_a + i Q/2 )\prod_{b=1}^{N_f} s_b(-x_j - \Wm_b + i Q/2) }
    {\prod_{i\neq j}^{N_c} s_b(x_i - x_j + i Q/2)}
    \nonumber\\
    &=s_b\Big( \frac{iQ}{2}(1-N_f+N_c-1)+\frac{\sum_{a=1}^{N_f} m_a +\sum_{b=1}^{N_f} \Wm_b}{2} \pm \frac{\xi}{2} \Big)
    \prod_{a,b=1}^{N_f} s_b (-m_a - \Wm_b + iQ/2)
    \nonumber\\
    &\quad\times
    \frac{e^{\frac{\pi i}{2}\xi \sum_{a=1}^{N_f}( m_a - \Wm_b )}}{(N_f - N_c)!} \int \prod_{j=1}^{N_f - N_c}dx_j
    e^{\pi i (-\xi) \sum_{j=1}^{N_f-N_c} x_j}
    \frac{\prod_{j=1}^{N_f-N_c} \prod_{a=1}^{N_f}s_b( -x_j + m_a)\prod_{b=1}^{N_f} s_b(x_j + \Wm_b) }
    {\prod_{i\neq j}^{N_f-N_c} s_b(x_i - x_j + i Q/2)}
\end{align}
where we have used a short-hand notation $s_b(x \pm y )\equiv s_b(x+y)s_b(x-y)$. Note that we explicitly keep the overall phase factor in the last line though it vanishes by a usual condition, $\sum_{a=1}^{N_f} m_a = \sum_{b=1}^{N_f} \Wm_b$. We keep this factor because it plays an important role in a quiver, where $U(N)$ rather than $SU(N)$ is gauged, to properly transport the FI-parameter to the adjacent gauge nodes.

For the Aharony duality, in addition to the $S^3_b$ partition function, the identity for the superconformal index has also been proven in \cite{Hwang:2017kmk} using wall-crossing phenomena in vortex quantum mechanics and the factorization technique of the index \cite{Hwang:2012jh,Hwang:2015wna}, given the vanishing background flux for the flavor symmetry.

\paragraph{One-monopole BBP duality:\\}
\begin{align}
    &\frac{1}{N_c !} \int \prod_{j=1}^{N_c} dx_j
    e^{\pi i \xi \sum_{j=1}^{N_c} x_j}
    \frac{\prod_{j=1}^{N_c} \prod_{a=1}^{N_f}s_b( x_j - m_a + i Q/2 )\prod_{b=1}^{N_f} s_b(-x_j - \Wm_b + i Q/2) }
    {\prod_{i\neq j}^{N_c} s_b(x_i - x_j + i Q/2)}
    \nonumber\\
    &=s_b\Big( \frac{iQ}{2}(1-N_f+N_c-1)+\frac{\sum_{a=1}^{N_f} m_a +\sum_{b=1}^{N_f} \Wm_b}{2} + \frac{\xi}{2} \Big)
    \prod_{a,b=1}^{N_f} s_b (-m_a - \Wm_b + iQ/2)
    \nonumber\\
    &\quad\times
    \frac{e^{\frac{\pi i}{2}\sum_{a=1}^{N_f}\Big( \xi (m_a - \Wm_a) - (m_a^2 - \Wm_a^2) \Big) }}{N_c'!} \int \prod_{j=1}^{N_c'}dx_j
    e^{\pi i (-\xi+iQ) \sum_{j=1}^{N_c'} x_j}
    \frac{\prod_{j=1}^{N_c'} \prod_{a=1}^{N_f}s_b( -x_j + m_a)\prod_{b=1}^{N_f} s_b(x_j + \Wm_b) }
    {\prod_{i\neq j}^{N_c'} s_b(x_i - x_j + i Q/2)}
\end{align}
where $N_c' = N_f - N_c - 1$ and the superpotential condition is given as
\begin{align}
    \sum_{a=1}^{N_f} m_a + \sum_{b=1}^{N_f}\Wm_b = \xi + iQ(N_f - N_c - 1).
    \label{eq: BBP1 condition}
\end{align}

\paragraph{Partition function from a quiver diagram\\}
Using the localization formula \eqref{eq: S_b^3 ptf integral formula}, one can easily read the partition function from a given quiver diagram in the following way.
Let us focus on the cases where the theory only contains matter fields in the bi-fundamental or adjoint representations of $U(N)$ or $SU(N)$ groups. Each component of a quiver diagram is associated with a squashed three-sphere partition function contribution as follows:
\begin{equation}
    \begin{tikzpicture}[baseline={([yshift=-.5ex]current bounding box.center)}]
  \tikzset{vertex/.style={circle,fill=white!25,minimum size=12pt,inner sep=2pt}}
  \tikzset{every loop/.style={}}
    \node[vertex] (N) at (0,0) [shape=circle,draw=black,minimum size=2em] {$N$};
    \node at (0,-0.6) {\scriptsize $\eta$};
    \end{tikzpicture}
    \;\;\Longleftrightarrow \;\;
    \frac{1}{N!} \int \prod_{i=1}^{N} dx_i \; e^{\pi i\; \frac{iQ}{2}\eta \sum_{i=1}^{N} x_i} \frac{1}{\prod_{i\neq j}^{N} s_b ( x_i - x_j + \frac{iQ}{2} )} \,,
    \label{eq: circular node}
\end{equation}
\begin{equation}
\begin{tikzpicture}[baseline={([yshift=-.5ex]current bounding box.center)}]
  \tikzset{vertex/.style={circle,fill=white!25,minimum size=12pt,inner sep=2pt}}
  \tikzset{every loop/.style={}}
    \node[vertex] (N) at (0,0) {$N$};
    \node[vertex] (M) at (1.5,0) {$M$};
    \node at (0.7,0.3) {$\D$};
    \draw[->-=.55] ([yshift= 0pt] N.east) to ([yshift= 0pt] M.west);
    \end{tikzpicture}
    \;\;\Longleftrightarrow\;\;
    \prod_{i=1}^{N} \prod_{j=1}^{M} 
    s_b \Big( y_j - x_i + \frac{iQ}{2}(1-\D) \Big) \,,
    \qquad\qquad
    \label{eq: arrow}
\end{equation}
\begin{equation}
    +\big\{\D\big\}
    \;\;\Longleftrightarrow\;\;
    s_b \Big(\frac{iQ}{2}(1-\D) \Big) \,.
    \qquad\qquad\qquad
    \label{eq: singlet}
\end{equation}
The circular node \eqref{eq: circular node} implies the $U(N)$ gauge group, and we associate it with an integration over the gauge variables with a vector multiplet contribution and normalized FI parameter $\eta$ presented under the node.\footnote{The original FI parameter $\xi$ in \eqref{eq: S_b^3 ptf integral formula} is related to the normalized one, $\eta$, as $\xi = \frac{iQ}{2}\eta$. We will ignore the FI parameter for the flavor nodes in the paper since they will always be the special unitary groups.} The arrow in \eqref{eq: arrow} between two nodes, either circular (gauge) or square (flavor), represents a chiral multiplet contribution in the (anti-)fundamental representation of the $U(N)$ ($U(M)$) node whose fugacities are $x_i$ $(y_j)$ with the normalized real mass parameter $\D$.\footnote{Similarly to the FI, the normalized real mass is defined by $\D = \frac{2}{iQ} m+r$ where $m$ is an original real mass parameter appearing in \eqref{eq: S_b^3 ptf integral formula} and $r$ is the R-charge of the chiral multiplet.} We also put a list of normalized real masses with curly brackets as in \eqref{eq: singlet} to represent extra singlet contributions. For simplicity, we omit the overall phase corresponding to the background BF coupling in the quiver diagram. The only place in the paper where such a phase arises is in \eqref{eq: dKP with V_0^+ derivation step} and each step will give $e^{\frac{\pi i}{2}\sum_{a=1}^{N_f}(m_a^2 - \Wm_a^2) }$ where $m_a$ and $\Wm_a$ are the real masses of the two flavor nodes.
Let us draw how the Aharony and one-monopole BBP dualities are realized in a linear quiver theory:
\begin{equation}
\begin{tikzpicture}
  \tikzset{vertex/.style={circle,fill=white!25,minimum size=12pt,inner sep=2pt}}
  \tikzset{every loop/.style={}}
    \node[vertex] (n1) at (0-1.5,0) [shape=circle,draw=black,minimum size=2em] {$N_1$};
    \node[vertex] (n2) at (1.5-1.5,0) [shape=circle,draw=black,minimum height=2em, minimum width=2em] {$N_2$};
    \node[vertex] (n3) at (3-1.5,0) [shape=circle,draw=black,minimum height=2em, minimum width=2em] {$N_3$};

    \draw[->-=.5] ([yshift= 2pt] n1.east) to ([yshift= 2pt] n2.west);
    \draw[->-=.5] ([yshift= -2pt] n2.west) to ([yshift= -2pt] n1.east);
    \draw[->-=.5] ([yshift= 2pt] n2.east) to ([yshift= 2pt] n3.west);
    \draw[->-=.5] ([yshift= -2pt] n3.west) to ([yshift= -2pt] n2.east);

    \node at (-0.7-1.5,0) {$\cdots$};
    \node at (3.7-1.5,0) {$\cdots$};
    \node at (0.75-1.5,0.4) {\scriptsize $\D_1$};
    \node at (2.25-1.5,0.4) {\scriptsize $\D_2$};
    \node at (0-1.5,-0.7) {\scriptsize $\th$};
    \node at (1.5-1.5,-0.7) {\scriptsize $\xi$};
    \node at (3-1.5,-0.7) {\scriptsize $\l$};
    
    \node[vertex] (N1) at (0+5,0+3) [shape=circle,draw=black,minimum size=2em] {$N_1$};
    \node[vertex] (N2) at (1.5+5,0+3) [shape=circle,draw=black,minimum height=2em, minimum width=2em] {\scriptsize $N^{(0)}\!$};
    \node[vertex] (N3) at (3+5,0+3) [shape=circle,draw=black,minimum height=2em, minimum width=2em] {$N_3$};

    \draw[-to, min distance=1cm]  (N1) edge [out=120, in=60] node {} (N1);
    \draw[-to, min distance=1cm]  (N3) edge [out=120, in=60] node {} (N3);
    \draw[->-=.5] ([yshift= 2pt] N1.east) to ([yshift= 2pt] N2.west);
    \draw[->-=.5] ([yshift= -2pt] N2.west) to ([yshift= -2pt] N1.east);
    \draw[->-=.5] ([yshift= 2pt] N2.east) to ([yshift= 2pt] N3.west);
    \draw[->-=.5] ([yshift= -2pt] N3.west) to ([yshift= -2pt] N2.east);
    \draw[->-=.5] ([xshift= 6pt,yshift= -2pt] N1.north) to [bend right=-60] ([xshift= -6pt,yshift= -2pt] N3.north);
    \draw[->-=.5] ([xshift= -9pt,yshift= -4pt] N3.north) to [bend right=60] ([xshift= 9pt,yshift= -4pt] N1.north);

    \node at (-0.7+5,0+3) {$\cdots$};
    \node at (3.7+5,0+3) {$\cdots$};
    \node at (0+5,1.2+3) {\scriptsize $2\D_1$};
    \node at (3+5,1.2+3) {\scriptsize $2\D_2$};
    \node at (0.75+5,0.25+3) {\scriptsize $1\texttt{-}\D_1$};
    \node at (2.25+5,0.25+3) {\scriptsize $1\texttt{-}\D_2$};
    \node at (1.5+5,1.2+3) {\scriptsize $\D_1 \texttt{+} \D_2$};
    \node at (0+5,-0.7+3) {\scriptsize $\th+\xi$};
    \node at (1.5+5,-0.7+3) {\scriptsize $-\xi$};
    \node at (3+5,-0.7+3) {\scriptsize $\l+\xi$};

    \node at (6.3,1.5) {\scriptsize $+\{N_1(1\texttt{-}\D_1) \texttt{+} N_3 (1\texttt{-}\D_2) \texttt{-}N_2 \texttt{+} 1 \pm \xi\}$};

    \node[vertex] (b1) at (5,0-3) [shape=circle,draw=black,minimum size=2em] {$N_1$};
    \node[vertex] (b2) at (6.5,0-3) [shape=circle,draw=black,minimum height=2em, minimum width=2em] {\scriptsize $N^{(1)}\!$};
    \node[vertex] (b3) at (8,0-3) [shape=circle,draw=black,minimum height=2em, minimum width=2em] {$N_3$};

    \draw[-to, min distance=1cm]  (b1) edge [out=120, in=60] node {} (b1);
    \draw[-to, min distance=1cm]  (b3) edge [out=120, in=60] node {} (b3);
    \draw[->-=.5] ([yshift= 2pt] b1.east) to ([yshift= 2pt] b2.west);
    \draw[->-=.5] ([yshift= -2pt] b2.west) to ([yshift= -2pt] b1.east);
    \draw[->-=.5] ([yshift= 2pt] b2.east) to ([yshift= 2pt] b3.west);
    \draw[->-=.5] ([yshift= -2pt] b3.west) to ([yshift= -2pt] b2.east);
    \draw[->-=.5] ([xshift= 6pt,yshift= -2pt] b1.north) to [bend right=-60] ([xshift= -6pt,yshift= -2pt] b3.north);
    \draw[->-=.5] ([xshift= -9pt,yshift= -4pt] b3.north) to [bend right=60] ([xshift= 9pt,yshift= -4pt] b1.north);

    \node at (4.3,0-3) {$\cdots$};
    \node at (8.7,0-3) {$\cdots$};
    \node at (5,1.2-3) {\scriptsize $2\D_1$};
    \node at (8,1.2-3) {\scriptsize $2\D_2$};
    \node at (5.75,0.25-3) {\scriptsize $1\texttt{-}\D_1$};
    \node at (7.25,0.25-3) {\scriptsize $1\texttt{-}\D_2$};
    \node at (1.5+5,1.2-3) {\scriptsize $\D_1 \texttt{+} \D_2$};
    \node at (5,-0.7-3) {\scriptsize $\th\texttt{+}\xi \mp \D_1$};
    \node at (6.5,-0.7-3) {\scriptsize $-\xi \pm 1$};
    \node at (8,-0.7-3) {\scriptsize $\l\texttt{+}\xi \mp \D_2$};

    \node at (9,0.8-3) {\scriptsize $+\{2\mp 2\xi\}$};

    
    \node at (3,1.5) 
    {\begin{tikzpicture}
      \node [rotate=45] {\textcolor{black}{$\Longleftrightarrow$}};    
    \end{tikzpicture}
    };
    \node at (3,2) {\scriptsize Aharony};
    
    \node at (3,-1.5) 
    {\begin{tikzpicture}
      \node [rotate=135] {\textcolor{black}{$\Longleftrightarrow$}};    
    \end{tikzpicture}
    };
    \node at (3,-1) {\scriptsize $\text{BBP}_1^\pm$};
    \node at (2.9,-2) {\scriptsize if \eqref{eq: BBP1 real mass condition}};

    \node at (-0.5,3.5) {$N^{(\a)} \equiv N_1+N_3 - N_2 - \a$};

\end{tikzpicture}
\label{eq: basic duality in quiver}
\end{equation}
Here $\text{BBP}_1^\pm$ denotes the BBP duality with a single positive/negative linear monopole superpotential, whose normalized real mass parameters should satisfy the constraint from the monopole superpotential term as follows:
\begin{align}
    \text{BBP}_1^\pm
    \; &: \;
    N_1(1-\D_1) + N_3(1-\D_2) - N_2 + 1 \pm \xi = 2.
    \label{eq: BBP1 real mass condition}
\end{align}
\paragraph{Example: $p=2$ deconfined Kim--Park duality\\}
As an example, let us demonstrate the derivation of the partition function identity for the Kim--Park duality via the deconfined Kim--Park duality for $p=2$ case. Since we are considering $N_c = \pm 1 \mod p$ cases, the generic gauge rank of the original theory is $N_c = 2n+1$ for some non-negative integer $n$. The squashed three sphere partition function of $U(2n+1)$ adjoint SQCD with superpotential $W = \Tr X^{2+1}$ is given by
\begin{align}
    Z=&\frac{1}{(2n+1)!} \int \prod_{j=1}^{2n+1} dx_j
    e^{\pi i \xi \sum_{j=1}^{2n+1}x_j}
    \prod_{i,j=1}^{2n+1} s_b\Big(x_i - x_j + \frac{iQ}{2}(1-\d)\Big)
    \nonumber\\
    &\qquad\times
    \frac
    { \prod_{j=1}^{2n+1} \prod_{a=1}^{N_f} s_b(x_j - m_a + \frac{iQ}{2})
    \prod_{b=1}^{N_f} s_b(-x_j -\Wm_b + \frac{iQ}{2})}
    {\prod_{i\neq j}^{2n+1} s_b(x_i - x_j + \frac{iQ}{2})}
    \label{eq: squashed ptf of adjoint SQCD}
\end{align}
where $\d=\frac{2}{1+p} = \frac23$ is the R-charge of the adjoint, whereas $m_a$ and $\Wm_b$ are the real mass parameters of the $N_f$ flavors, not the normalized ones. By deconfining the adjoint for $p=2$, we have
\begingroup\allowdisplaybreaks
\begin{align}
    Z = & 
    s_b\Big(\frac{iQ}{2}(1-\d)\Big)
    \frac{1}{n!} \int \prod_{i=1}^{n} dy_i
    \frac{\prod_{i,j=1}^{n} s_b\big(y_i - y_j + \frac{iQ}{2}(1-(2-\d))\big) }
    {\prod_{i\neq j}^{n} s_b(y_i - y_j + \frac{iQ}{2})}
    \nonumber\\
    &\times
    \frac{1}{(2n+1)!} \int \prod_{j=1}^{2n+1} dx_j
    e^{\pi i \xi \sum_{j=1}^{2n+1}x_j}
    \frac{1}{\prod_{i\neq j}^{2n+1} s_b(x_i - x_j + \frac{iQ}{2}) }
    \nonumber\\
    &\times 
    \prod_{j=1}^{2n+1}
    \Bigg(
    \prod_{a=1}^{N_f}s_b\Big(x_j-m_a+\frac{iQ}{2}\Big)
    \prod_{i=1}^{n} s_b\Big(x_j - y_i + \frac{iQ}{2}\big(1-\frac{\d}{2}\big)\Big)
    \Bigg)
    \nonumber\\
    &\times
    \prod_{j=1}^{2n+1}
    \Bigg(
    \prod_{b=1}^{N_f} s_b\Big(-x_j - \Wm_b + \frac{iQ}{2}\Big)
    \prod_{i=1}^{n} s_b\Big(-x_j+y_i + \frac{iQ}{2}\big(1-\frac{\d}{2}\big)\Big)
    \Bigg)
    \nonumber\\
    \overset{\text{Aharony}}{=} &
    s_b\Big(\frac{iQ}{2}(1-\d)\Big)
    s_b\Big( \frac{iQ}{2}\big(1-N_f+\frac{4}{3}n\big) +\frac{\sum_{a=1}^{N_f}m_a + \sum_{b=1}^{N_f}\Wm_b}{2} \pm \frac{\xi}{2} \Big)
    \prod_{a,b=1}^{N_f} s_b\Big(-m_a - \Wm_b + \frac{iQ}{2} \Big)
    \nonumber\\
    &\times
    \frac{1}{(N_f - n - 1)!} \int \prod_{j=1}^{N_f - n - 1} dx_j
    e^{\pi i (-\xi) \sum_{j=1}^{N_f - n - 1} x_j}
    \frac{\prod_{j=1}^{N_f - n - 1}\prod_{a=1}^{N_f} s_b(-x_j + m_a)\prod_{b=1}^{N_f} s_b(x_j + \Wm_b)}
    {\prod_{i\neq j}^{N_f - n - 1} s_b(x_i - x_j + \frac{iQ}{2})}
    \nonumber\\
    &\times
    \frac{1}{n!}\int \prod_{i=1}^{n} dy_i
    e^{\pi i \xi \sum_{i=1}^{n}y_i}
    \frac{1}{\prod_{i\neq j}^{n} s_b(y_i - y_j + \frac{iQ}{2})}
    \nonumber\\
    &\times
    \prod_{i=1}^{n}
    \Bigg(
    \prod_{a=1}^{N_f} s_b\Big(y_i-m_a+\frac{iQ}{2}\big(1-\frac{\d}{2}\big)\Big)
    \prod_{j=1}^{N_f-n-1} s_b\Big(y_i - x_j + \frac{iQ}{2}\frac{\d}{2}\Big)
    \Bigg)
    \nonumber\\
    &\times
    \prod_{i=1}^{n}
    \Bigg(
    \prod_{b=1}^{N_f} s_b\Big(-y_i-\Wm_b + \frac{iQ}{2}\big(1-\frac{\d}{2}\big)\Big)
    \prod_{j=1}^{N_f-n-1} s_b\Big(-y_i + x_j + \frac{iQ}{2}\frac{\d}{2}\Big)
    \Bigg)
    \nonumber\\
    \overset{\text{Aharony}}{=} &
    \prod_{k=0}^{1}
    s_b\Big( \frac{iQ}{2}\big(1-N_f+\frac{4}{3}n-k\d\big) +\frac{\sum_{a=1}^{N_f}m_a + \sum_{b=1}^{N_f}\Wm_b}{2} \pm \frac{\xi}{2} \Big)
    \prod_{a,b=1}^{N_f}
    s_b\Big( -m_a - \Wm_b + \frac{iQ}{2}(1-k \d) \Big)
    \nonumber\\
    &\times
    \frac{1}{(2N_f-2n-1)!}\int \prod_{i=1}^{2N_f - 2n - 1 }dy_i
    e^{\pi i (-\xi) \sum_{i=1}^{2N_f-2n-1}y_i}
    \nonumber\\
    &\times
    \Bigg\{
    s_b\Big(\frac{iQ}{2}(1-\d)\Big)
    \frac{1}{(N_f-n-1)!}
    \int\prod_{j=1}^{N_f-n-1}dx_j
    \prod_{i,j=1}^{N_f - n - 1}
    s_b\Big( x_i - x_j + \frac{iQ}{2}\big(1-(2-\d)\big) \Big)
    \nonumber\\
    &\times
    \frac{
    \prod_{j=1}^{N_f-n-1}\prod_{i=1}^{2N_f-2n-1}
    s_b\Big( x_j - y_i + \frac{iQ}{2}\big(1-\frac{\d}{2}\big) \Big)
    s_b\Big( -x_j + y_i + \frac{iQ}{2}\big(1-\frac{\d}{2}\big) \Big)
    }
    {\prod_{i\neq j}^{N_f - n - 1} s_b\big( x_i - x_j + \frac{iQ}{2} \big)}
    \Bigg\}
    \nonumber\\
    &\times
    \frac{
    \prod_{i=1}^{2N_f-2n-1}
    \prod_{a=1}^{N_f}
    s_b\Big( - y_i + m_a + \frac{iQ}{2}\frac{\d}{2} \Big)
    \prod_{b=1}^{N_f}
    s_b\Big(  y_i + \Wm_b + \frac{iQ}{2}\frac{\d}{2} \Big)
    }
    {\prod_{i\neq j}^{2N_f - 2n - 1} s_b\big( y_i - y_j + \frac{iQ}{2} \big)}.
    \label{eq: KP ptf derivation}
\end{align}
\endgroup
Note that the curly bracket in the last step of \eqref{eq: KP ptf derivation} is the partition function of the confining $D_{p=2}[SU(2n+1)]$ theory. Hence, by switching it with an adjoint chiral contribution, we finally get
\begin{align}
    Z =& \prod_{k=0}^{1}
    s_b\Big( \frac{iQ}{2}\big(1-N_f+\frac{4}{3}n-k\d\big) +\frac{\sum_{a=1}^{N_f}m_a + \sum_{b=1}^{N_f}\Wm_b}{2} \pm \frac{\xi}{2} \Big)
    \prod_{a,b=1}^{N_f}
    s_b\Big( -m_a - \Wm_b + \frac{iQ}{2}(1-k \d) \Big)
    \nonumber\\
    &\times
    \frac{1}{(2N_f-2n-1)!}\int \prod_{i=1}^{2N_f - 2n - 1 }dy_i
    e^{\pi i (-\xi) \sum_{i=1}^{2N_f-2n-1}y_i}
    \prod_{i,j=1}^{2N_f-2n-1}
    s_b\Big( y_i - y_j +\frac{iQ}{2}(1-\d) \Big)
    \nonumber\\
    &\times
    \frac{
    \prod_{i=1}^{2N_f-2n-1}
    \prod_{a=1}^{N_f}
    s_b\Big( - y_i + m_a + \frac{iQ}{2}\frac{\d}{2} \Big)
    \prod_{b=1}^{N_f}
    s_b\Big(  y_i + \Wm_b + \frac{iQ}{2}\frac{\d}{2} \Big)
    }
    {\prod_{i\neq j}^{2N_f - 2n - 1} s_b\big( y_i - y_j + \frac{iQ}{2} \big)} \,,
    \label{eq: p=2 dual KP ptf}
\end{align}
which is nothing but the squashed three-sphere partition function of the dual $U(2N_f - 2n - 1)$ $=$ $U(2 N_f - N_c)$ adjoint SQCD with $W=\Tr x^{2+1}$ superpotential. One can check the charges of the monopole and meson operators of the dual theory are precisely those consistent with the singlet terms in the superpotential.
\begin{figure}[tbp]
\centering
\begin{tikzpicture}[thick,scale=0.58, every node/.style={scale=0.58}]
  \tikzset{vertex/.style={circle,fill=white!25,minimum size=2.5em,inner sep=2pt}}
  \tikzset{every loop/.style={}}

    \node[vertex] (e2n) at (1.5,0) [shape=circle,draw=black,minimum size=2.5em] {$2n\texttt{+}1$};
    \node[vertex] (eF1) at (4,+1) [shape=rectangle,draw=black,minimum height=2.5em, minimum width=2.5em] {$N_f$};
    \node[vertex] (eF2) at (4,-1) [shape=rectangle,draw=black,minimum height=2.5em, minimum width=2.5em] {$N_f$};

    \draw[-to, min distance=1cm]  (e2n) edge [out=120, in=60] node {} (e2n);
    \draw[->-=.5] ([yshift= 2pt] e2n.east) to ([yshift= 0pt] eF1.west);
    \draw[->-=.5] ([yshift= 0pt] eF2.west) to ([yshift= -2pt] e2n.east);

    \node at (1.5,+1.4) {$\d$};
    \node at (2.7,+1) {$\D_Q$};
    \node at (1.5,-1) {$\eta$};

    \node[vertex] (m2m) at (12,0) [shape=circle,draw=black,minimum size=2.5em] {$\!2m\texttt{+}1\!$};
    \node[vertex] (mF1) at (14.5,+1) [shape=rectangle,draw=black,minimum height=2.5em, minimum width=2.5em] {$N_f$};
    \node[vertex] (mF2) at (14.5,-1) [shape=rectangle,draw=black,minimum height=2.5em, minimum width=2.5em] {$N_f$};

    \draw[-to, min distance=1cm]  (m2m) edge [out=120, in=60] node {} (m2m);
    \draw[->-=.5] ([yshift= 0pt] mF1.west) to ([yshift= 2pt] m2m.east);
    \draw[->-=.5] ([yshift= -2pt] m2m.east) to ([yshift= 0pt] mF2.west);
    \draw[->-=.6] ([xshift= -5pt] mF2.north) to ([xshift= -5pt] mF1.south);
    \draw[->-=.6] ([xshift= 5pt] mF2.north) to ([xshift= 5pt] mF1.south);

    \node at (12,+1.4) {$\d$};
    \node at (13,+1) {$\d\texttt{-}\D_Q$};
    \node at (15.7,0.2) {$2\D_Q $};
    \node at (15.7,-0.2) {$2\D_Q \texttt{+} \d$};
    \node at (18,1) {$+\{ N_f(1\texttt{-}\D_Q) \texttt{-} \frac{4}{3}n \texttt{+} j\d \}_{j=0}^{1}$};
    \node at (12,-1) {$-\eta$};

    \node[vertex] (An) at (0-4,-6) [shape=circle,draw=black,minimum size=2.5em] {$n$};
    \node[vertex] (A2n) at (2-4,0-6) [shape=circle,draw=black,minimum size=2.5em] {$\!2n \texttt{+}1\!$};
    \node[vertex] (AF1) at (4.5-4,1-6) [shape=rectangle,draw=black,minimum height=2.5em, minimum width=2.5em] {$N_f$};
    \node[vertex] (AF2) at (4.5-4,-1-6) [shape=rectangle,draw=black,minimum height=2.5em, minimum width=2.5em] {$N_f$};

    \draw[-to, min distance=1cm]  (An) edge [out=120, in=60] node {} (An);
    \draw[->-=.5] ([yshift= 2pt] An.east) to ([yshift= 2pt] A2n.west);
    \draw[->-=.5] ([yshift= -2pt] A2n.west) to ([yshift= -2pt] An.east);
    \draw[->-=.5] ([yshift= 2pt] A2n.east) to ([yshift= 0pt] AF1.west);
    \draw[->-=.5] ([yshift= -2pt] AF2.west) to ([yshift= -2pt] A2n.east);

    \node at (0-4,1.4-6) {$2\texttt{-}\d$};
    \node at (1-4,0.4-6) {$\d/2$};
    \node at (3-4,1-6) {$\D_Q$};
    \node at (2-4,1.7-6) {$+\{\d\}$};
    \node at (0-4,-1-6) {$0$};
    \node at (2-4,-1-6) {$\eta$};

    \node[vertex] (A'n) at (5,6-12) [shape=circle,draw=black,minimum size=2.5em] {$n$};
    \node[vertex] (A'm) at (7,6-12) [shape=circle,draw=black,minimum size=2.5em] { $m$};
    \node[vertex] (A'F1) at (9.5,7-12) [shape=rectangle,draw=black,minimum height=2.5em, minimum width=2.5em] {$N_f$};
    \node[vertex] (A'F2) at (9.5,5-12) [shape=rectangle,draw=black,minimum height=2.5em, minimum width=2.5em] {$N_f$};

    \draw[->-=.5] ([yshift= 2pt] A'm.west) to ([yshift= 2pt] A'n.east);
    \draw[->-=.5] ([yshift= -2pt] A'n.east) to ([yshift= -2pt] A'm.west);
    \draw[->-=.5] ([yshift= 0pt] A'F1.west) to ([yshift= 2pt] A'm.east);
    \draw[->-=.5] ([yshift= -2pt] A'm.east) to ([yshift= -2pt] A'F2.west);
    \draw[->-=.6] ([yshift= 0pt] A'F2.north) to ([yshift= 0pt] A'F1.south);
    \draw[->-=.5] ([xshift= -5pt,yshift=13pt] A'n.east) to [bend left=20] ([yshift= 5pt] A'F1.west);
    \draw[->-=.5] ([yshift= -5pt] A'F2.west) to [bend left=20] ([xshift= -5pt,yshift=-13pt] A'n.east);

    \node at (6.1,6+0.4-12) {$1\texttt{-}\d/2$};
    \node at (7.7,6+0.7-12) {$1\texttt{-}\D_Q$};
    \node at (7.2,6+1.6-12) {$\d/2\texttt{+}\D_Q$};
    \node at (10.3,6-12) {$2\D_Q$};
    \node at (7.7,8.5-12) {$+\{ \d,N_f(1\texttt{-}N_f)\texttt{-}\frac{4}{3}n\pm\eta \}$};
    \node at (5,-0.7-6) {$\eta$};
    \node at (7,-0.7-6) {$-\eta$};

    \node[vertex] (b2m) at (15,-6) [shape=circle,draw=black,minimum size=2.5em] {$\!2m \texttt{+} 1\!$};
    \node (bm) at (17,-6) [shape=circle,draw=black,minimum size=2.5em] {$m$};
    \node[vertex] (bF1) at (19.5,1-6) [shape=rectangle,draw=black,minimum height=2.5em, minimum width=2.5em] {$N_f$};
    \node[vertex] (bF2) at (19.5,-1-6) [shape=rectangle,draw=black,minimum height=2.5em, minimum width=2.5em] {$N_f$};

    \draw[-to, min distance=1cm]  (bm) edge [out=30, in=-30] node {} (bm);
    \draw[->-=.5] ([yshift= 2pt] b2m.east) to ([yshift= 2pt] bm.west);
    \draw[->-=.5] ([yshift= -2pt] bm.west) to ([yshift= -2pt] b2m.east);
    \draw[->-=.5] ([yshift= 5pt] bF1.west) to [bend right=20] ([xshift=-5pt ,yshift= 13pt] b2m.east);
    \draw[->-=.5] ([xshift=-5pt ,yshift= -13pt] b2m.east) to [bend right=20] ([yshift= -5pt] bF2.west);
    \draw[->-=.6] ([xshift= -5pt] bF2.north) to ([xshift= -5pt] bF1.south);
    \draw[->-=.6] ([xshift= 5pt] bF2.north) to ([xshift= 5pt] bF1.south);

    \node at (18-0.2,0.7-6) {$2\texttt{-}\d$};
    \node at (16,0.4-6) {$\d/2$};
    \node at (20.5,0.2-6) {$2\D_Q$};
    \node at (20.5,-0.2-6) {$2\D_Q\texttt{+}\d$};
    \node at (17.5,1.5-6) {$\d\texttt{-}\D_Q$};
    \node at (18.5,2.5-6) {$+\{\d, N_f(1\texttt{-}\D_Q) \texttt{-} \frac{4}{3}n \texttt{+} j\d \}_{j=0}^{1}$};
    \node at (15,-0.7-6) {$-\eta$};
    \node at (17,-0.7-6) {$0$};


    \node at (0,-2.5) 
    {\begin{tikzpicture}
      \node [rotate=60] {\Huge $\Longrightarrow$};    
    \end{tikzpicture}
    };
    \node at (-2,-2.5) {Confinement};

    \node at (3,-6) 
    {\begin{tikzpicture}
      \node [rotate=0] {\Huge $\Longleftrightarrow$};    
    \end{tikzpicture}
    };
    \node at (3,-5.5) {Aharony};

    \node at (12.5,-6) 
    {\begin{tikzpicture}
      \node [rotate=0] {\Huge $\Longleftrightarrow$};    
    \end{tikzpicture}
    };
    \node at (12.5,-5.5) {Aharony};

    \node at (13.5,-2.5) 
    {\begin{tikzpicture}
      \node [rotate=-60] {\Huge $\Longleftarrow$};    
    \end{tikzpicture}
    };
    \node at (15,-2.5) {Confinement};

    \node at (8,0) 
    {\begin{tikzpicture}
      \node [rotate=0] {\Huge $\Longleftrightarrow$};    
    \end{tikzpicture}
    };
    \node at (8,0.5) {Kim--Park};

\end{tikzpicture}
\caption{\label{fig: p=2 KP example} The quiver expression for the $p=2$ Kim--Park duality derivation. The dual gauge rank is given by $2 m+1$ where $m=N_f-n-1$.}
\end{figure}
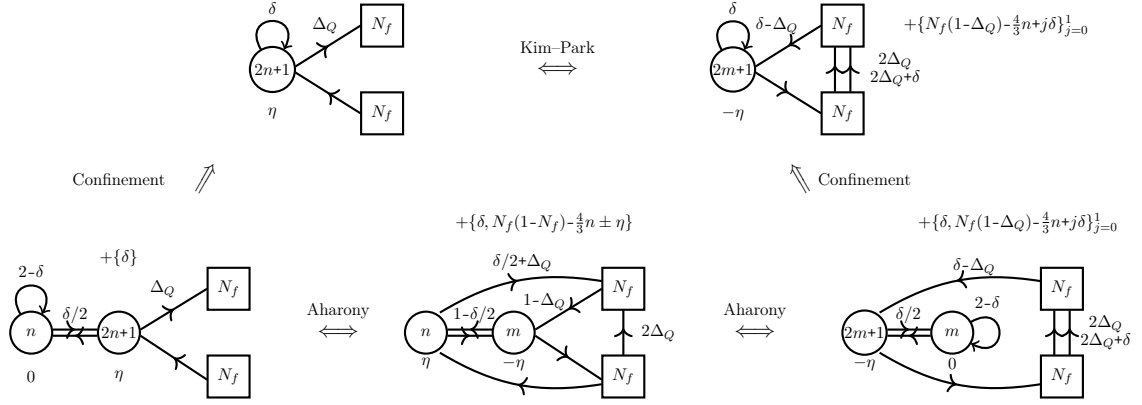
This derivation can be expressed more efficiently by using quiver diagrams. See Figure \ref{fig: p=2 KP example}, where the R-charge of the flavors are singled out as $\D_Q$ from the real mass parameters of the $SU(N_f)$ flavor nodes by defining $m_a = \frac{iQ}{2}\D_Q + \m_a$ and $\Wm_b = \frac{iQ}{2}\D_Q + \widetilde{\m}_b$ with the condition $\sum_{a}^{N_f}\m_a = \sum_{b}^{N_f}\widetilde{\m}_b = 0$.

\section{Rank--reducing duality}
\label{app: rank reducing}
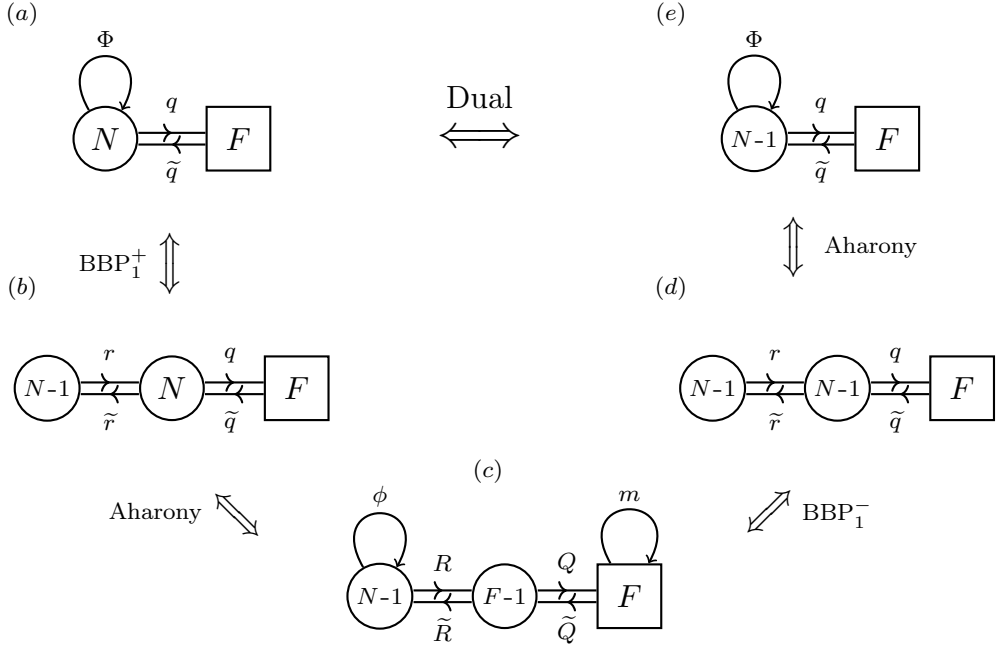
\begin{figure}[tbp]
\centering
\begin{tikzpicture}[thick,scale=1.1, every node/.style={scale=1.1}]
\tikzset{vertex/.style={circle,fill=white!25,minimum size=12pt,inner sep=2pt}}
  \tikzset{every loop/.style={}}
    \node at (-0.3,1.5) {\scriptsize $(a)$};
    \node[vertex] (AN) at (0.7,0) [shape=circle,draw=black,minimum size=2em] {$N$};
    \node[vertex] (AF) at (2.3,0) [shape=rectangle,draw=black,minimum height=2em, minimum width=2em] {$F$};

    \draw[-to, min distance=1cm]  (AN) edge [out=120, in=60] node {} (AN);
    \draw[->-=.5] ([yshift= 2pt] AN.east) to ([yshift= 2pt] AF.west);
    \draw[->-=.5] ([yshift= -2pt] AF.west) to ([yshift= -2pt] AN.east);

    \node at (0.7,1.2) {\scriptsize $\Phi$};
    \node at (1.5,0.4) {\scriptsize $q$};
    \node at (1.5,-0.4) {\scriptsize $\Wq$};

    \node at (-0.3,1.2-3) {\scriptsize $(b)$};
    \node[vertex] (BN1) at (0,-3) [shape=circle,draw=black,minimum size=2em] {\scriptsize $N\texttt{-}1$};
    \node[vertex] (BN) at (1.5,-3) [shape=circle,draw=black,minimum height=2em, minimum width=2em] {$N$};
    \node[vertex] (BF) at (3,-3) [shape=rectangle,draw=black,minimum height=2em, minimum width=2em] {$F$};

    \draw[->-=.5] ([yshift= 2pt] BN1.east) to ([yshift= 2pt] BN.west);
    \draw[->-=.5] ([yshift= -2pt] BN.west) to ([yshift= -2pt] BN1.east);
    \draw[->-=.5] ([yshift= 2pt] BN.east) to ([yshift= 2pt] BF.west);
    \draw[->-=.5] ([yshift= -2pt] BF.west) to ([yshift= -2pt] BN.east);

    \node at (0.75,-3+0.4) {\scriptsize $r$};
    \node at (0.75,-3-0.4) {\scriptsize $\Wr$};
    \node at (2.2,-3+0.4) {\scriptsize $q$};
    \node at (2.2,-3-0.4) {\scriptsize $\Wq$};
    
    \node at (5.3,1.5-5.5) {\scriptsize $(c)$};
    \node[vertex] (CN1) at (4,-5.5) [shape=circle,draw=black,minimum size=2em] {\scriptsize $N\texttt{-}1$};
    \node[vertex] (CF1) at (5.5,-5.5) [shape=circle,draw=black,minimum height=2em, minimum width=2em] {\scriptsize $\! F\texttt{-}1 \!$};
    \node[vertex] (CF) at (7,-5.5) [shape=rectangle,draw=black,minimum height=2em, minimum width=2em] {$F$};

    \draw[-to, min distance=1cm]  (CN1) edge [out=120, in=60] node {} (CN1);
    \draw[->-=.5] ([yshift= 2pt] CN1.east) to ([yshift= 2pt] CF1.west);
    \draw[->-=.5] ([yshift= -2pt] CF1.west) to ([yshift= -2pt] CN1.east);
    \draw[->-=.5] ([yshift= 2pt] CF1.east) to ([yshift= 2pt] CF.west);
    \draw[->-=.5] ([yshift= -2pt] CF.west) to ([yshift= -2pt] CF1.east);
    \draw[-to, min distance=1cm]  (CF) edge [out=120, in=60] node {} (CF);

    \node at (4,-5.5+1.2) {\scriptsize $\phi$};
    \node at (4.75,-5.5+0.4) {\scriptsize $R$};
    \node at (4.75,-5.5-0.4) {\scriptsize $\WR$};
    \node at (6.25,-5.5+0.4) {\scriptsize $Q$};
    \node at (6.25,-5.5-0.4) {\scriptsize $\WQ$};
    \node at (7,-5.5+1.2) {\scriptsize $m$};

    \node at (7.5,1.2-3) {\scriptsize $(d)$};
    \node[vertex] (DN) at (8,-3) [shape=circle,draw=black,minimum size=2em] {\scriptsize $N\texttt{-}1$};
    \node[vertex] (DN1) at (9.5,-3) [shape=circle,draw=black,minimum height=2em, minimum width=2em] {\scriptsize $N\texttt{-}1$};
    \node[vertex] (DF) at (11,-3) [shape=rectangle,draw=black,minimum height=2em, minimum width=2em] {$F$};

    \draw[->-=.5] ([yshift= 2pt] DN.east) to ([yshift= 2pt] DN1.west);
    \draw[->-=.5] ([yshift= -2pt] DN1.west) to ([yshift= -2pt] DN.east);
    \draw[->-=.5] ([yshift= 2pt] DN1.east) to ([yshift= 2pt] DF.west);
    \draw[->-=.5] ([yshift= -2pt] DF.west) to ([yshift= -2pt] DN1.east);

    \node at (8.75,-3+0.4) {\scriptsize $r$};
    \node at (8.75,-3-0.4) {\scriptsize $\Wr$};
    \node at (10.2,-3+0.4) {\scriptsize $q$};
    \node at (10.2,-3-0.4) {\scriptsize $\Wq$};

    \node at (7.5,1.5) {\scriptsize $(e)$};
    \node[vertex] (EN) at (8.5,0) [shape=circle,draw=black,minimum size=2em] {\scriptsize $N\texttt{-}1$};
    \node[vertex] (EF) at (10.1,0) [shape=rectangle,draw=black,minimum height=2em, minimum width=2em] {$F$};

    \draw[-to, min distance=1cm]  (EN) edge [out=120, in=60] node {} (EN);
    \draw[->-=.5] ([yshift= 2pt] EN.east) to ([yshift= 2pt] EF.west);
    \draw[->-=.5] ([yshift= -2pt] EF.west) to ([yshift= -2pt] EN.east);

    \node at (8.5,1.2) {\scriptsize $\Phi$};
    \node at (9.3,0.4) {\scriptsize $q$};
    \node at (9.3,-0.4) {\scriptsize $\Wq$};

    \node at (1.5,-1.5) 
    {\begin{tikzpicture}
      \node [rotate=90] {\textcolor{black}{$\Longleftrightarrow$}};    
    \end{tikzpicture}
    };
    \node at (0.8,-1.5) {\scriptsize $\text{BBP}_1^+$};

    \node at (2.3,-4.5) 
    {\begin{tikzpicture}
      \node [rotate=135] {\textcolor{black}{$\Longleftrightarrow$}};    
    \end{tikzpicture}
    };
    \node at (1.3,-4.5) {\scriptsize $\text{Aharony}$};

    \node at (8.7,-4.5) 
    {\begin{tikzpicture}
      \node [rotate=45] {\textcolor{black}{$\Longleftrightarrow$}};    
    \end{tikzpicture}
    };
    \node at (9.5,-4.5) {\scriptsize $\text{BBP}_1^-$};

    \node at (9,-1.3) 
    {\begin{tikzpicture}
      \node [rotate=90] {\textcolor{black}{$\Longleftrightarrow$}};    
    \end{tikzpicture}
    };
    \node at (9.9,-1.3) {\scriptsize $\text{Aharony}$};

    \node at (5.2,0) 
    {\begin{tikzpicture}
      \node [rotate=0] {\Large \textcolor{black}{$\Longleftrightarrow$}};    
    \end{tikzpicture}
    };
    \node at (5.2,0.5) { $\text{Dual}$};
    
\end{tikzpicture}
\caption{\label{fig: p=2 rank reducing} Rank--reducing procedure for $p=2$, starting with $U(N)$ adjoint SQCD (theory $(a)$) to get $U(N-1)$ SQCD (theory $(e)$). Repeatedly applying it eventually gives a WZ theory without a gauge group.}
\end{figure}
In this appendix, we will show the rank--reducing procedure for $p=2$ case as described in Figure \ref{fig: p=2 rank reducing}. Consider theory (a), the $U(N)$ adjoint SQCD theory with superpotential
\begin{align}
    W_{(a)} = u + u \hat{V}_{N-1}^{-} + u \,v\, \hat{V}_{N-1}^{-} + \Wq \Phi q + \a\, \hat{V}_0^{-} + \b\, \hat{V}_1^{-} \,,
    \label{eq: rank reducing Wa}
\end{align}
where $\hat{V}_{j}^\pm$ are the $j$-th dressed monopole operators and $\a$, $\b$ and $u$ are singlets. Deconfining the adjoint $\Phi$ into $U(N-1)$ tail gives theory $(b)$ with superpotential,
\begin{align}
    W_{(b)} = u + u\, \hat{v}^{(2),-} + \Wq \Wr r q + \a\, \hat{v}^{(1,2),-} + \b\, \text{tr}(r\Wr\, \hat{v}^{(1,2),-}) + u\,v\,\hat{v}^{(2),-} + \hat{v}^{(1),+} + \g\, \hat{v}^{(1),-}
\end{align}
where $\Phi$ is mapped to $\Wr r$ and $\g$ is a singlet corresponding to $\det \Phi$. The monopole operators are mapped as $\hat{V}_{N-1}^{-} \to \hat{v}^{(2),-}$, $\hat{V}_0^- \to \hat{v}^{(1,2),-}$. Applying the Aharony duality to the second gauge node produces theory $(c)$ with superpotential,
\begin{align}
    W_{(c)} = \hat{v}^{(2),-} + \WR \phi R + m\,\WQ Q + \hat{v}^{(1,2),+} + \a\,\hat{v}^{(1),-} + \b\,\hat{v}_1^{(1),-} + \g\, \hat{v}^{(1,2),-} + v^+ \, \hat{v}^{(2),+} + \Wv\, \hat{v}^{(2),-}
\end{align}
with monopole operators mapped as $\hat{v}^{(1,2),-} \to \hat{v}^{(1),-}$, $\hat{v}^{(1),\pm}\to \hat{v}^{(1,2),\pm}$. The singlets $v^+$ and $\Wv$ are from the Aharony duality. Crucially, since $\Wv$ has R-charge 0, we can redefine $\Wv \to \Wv+1$, which results in a linear monopole term $\hat{v}^{(2),-}$. Therefore, by further applying the BBP duality to the second gauge node gives theory $(d)$ with superpotential,
\begin{align}
    W_{(d)} = \hat{v}^{(2),-}  + \Wq \Wr r q  + v^{(1),+} + \a \, \hat{v}^{(1),-} + \b\, \hat{v}^{(1,2),-} + \g\,\text{tr}(\Wr r \, \hat{v}^{(1,2),-}) + \Wv\,\hat v^{(2),-} \,,
\end{align}
where the monopole operators are mapped as $\hat{v}^{(1,2),+}\to \hat{v}^{(1),+}$, $\hat{v}^{(1),-}\to\hat{v}^{(1),-}$, $\hat{v}_1^{(1),-}\to \hat{v}^{(1,2),-}$, $\hat{v}^{(1,2),-}\to \text{tr}(\Wr r \,\hat{v}^{(1,2),-})$. Note that $\hat{v}^{(2),\pm}$ here get quantum relations,
\begin{align}
    \hat{v}^{(2),+} 
    \; &\sim \; \hat{v}^{(1),-} \text{tr}\big((\Wr r)^{N-2} \hat{v}^{(1,2),+}\big) , \\
    \hat{v}^{(2),-} 
    \; &\sim \; \hat{v}^{(1),+} \text{tr}\big((\Wr r)^{N-2} \hat{v}^{(1,2),-}\big)
    \label{eq: qrforv2}
\end{align}
as discussed in the main text.
Finally, confining the first gauge node into an adjoint chiral $\Phi$ via the Aharony duality gives theory $(e)$ with superpotential,
\begin{align}
    W_{(e)} = \Wu + \Wu \, \hat{V}_{N-2}^{-} + \Wu\,\Wv\,\hat{V}_{N-2}^{-}
    + \Wq \Phi q + \b \, \hat{V}_0^{-} + \g\, \hat{V}_1^{-}
    \label{eq: rank reducing N}
\end{align}
where the monopole operators $\hat{V}_{j}^-$ are mapped from $\text{tr}((\Wr r)^{j}\hat{v}^{(1,2),-})$, the singlet $u$ is mapped from the monopole operator $\hat{v}^{(1),+}$, and massive singlets are integrated out. One should notice that \eqref{eq: rank reducing N} is the same as \eqref{eq: rank reducing Wa}, the superpotential of the original theory $(a)$, up to the gauge rank of the theory reduced by one. Thus, one can repeatedly reduce the rank of the gauge group down to $N=3$ in this way.

For $N=3$, we again apply the same chain of dualities to the theory. The only difference is that we expect the second relation \eqref{eq: qrforv2} is deformed by the singlet $\gamma$ for $N=3$ because now the singlet $\gamma$ has exactly the charges required by the relation.\footnote{As discussed in the main text, the relation can also be deformed by $\hat v^{(1),+}$, which we omit for simplicity since it does not affect the final result.} Modulo this, we follow the same steps as we have done and end up with $U(2)$ adjoint SQCD with $F$ flavor chirals and superpotential,
\begin{align}
    W_{(a)} = u + u \hat{V}_1^- + u\,v\,\hat{V}_1^- + \Wq \Phi q  + \b\, \hat{V}_0^- + \g \hat{V}_1^- + \g + \g\,v 
\end{align}
where we have removed tildes from $\tilde u$ and $\tilde v$. Note that the last two terms originate from the mixing of $\gamma$ into the quantum relation \eqref{eq: qrforv2} we explained. Since $\gamma$ and $v$ are massive due to the last term, we integrate them out and obtain 
\begin{align}
    W_{(a)} = \b\, \hat{V}_0^- + u + \big( \hat{V}_1^- \big)^2
    + \Wq \Phi q \,.
\end{align}
As before, deconfining the adjoint by BBP gives,
\begin{align}
    W_{(b)} = \b \, \hat{v}^{(1,2),-} + u + u \big( \hat{v}^{(2),-} \big)^2 + \hat{v}^{(1),+} + \xi \, \hat{v}^{(1),-}
    + \Wq \Wr r q \,.
\end{align}
Since the process is the same as above, we do not repeatedly showcase the operator map for simplicity, except $\xi$ which comes from $\det\Phi$. By applying Aharony we get,
\begin{align}
    W_{(c)} = \b \, \hat{v}^{(1),-} + \hat{v}^{(1,2),+} + \xi \, \hat{v}^{(1,2),-} + v^+ \hat{v}^{(2),+} + \hat{v}^{(2),-} + \WR \phi R + m\,\WQ Q \,,
\end{align}
and from the linear monopole superpotential term $\hat v^{(2),-}$, we can apply BBP duality to get
\begin{align}
    W_{(d)} = \b \, \hat{v}^{(1),-} + \hat{v}^{(1),+} + \xi \,\text{tr}(\Wr r) \hat{v}^{(1,2),-} + \hat{v}^{(2),-} + \Wq \Wr r q
\end{align}
where we have integrated out massive chiral fields. Applying the Aharony duality to confine the $U(1)$ gauge node, we get an SQED with $F$ flavors and superpotential
\begin{align}
    W_{(e)} = \Wv^+ + \xi \, \Phi \, \hat{V}^- + \Wv^+\,\hat{V}^- + \Wq q\, \Phi
\end{align}
where the singlet $\Wv^+$ is mapped from the monopole operator $\hat{v}^{(1),+}$, and we also integrated out massive singlet $\b$. Note that the adjoint $\Phi$ is now a singlet since the gauge group is $U(1)$. Thus, we can apply Aharony and obtain the $U(F-1)$ gauge theory with superpotential,
\begin{align}
    W = \Wv^+ + \xi \, \Phi \, V^- + \Wv^+ \, V^- + V^+ \hat{v}^+ + V^- \hat{v}^- 
\end{align}
where monopole operators are mapped as $\hat{V}^\pm \to V^\pm$. By integrating out massive singlets $\Wv^+$ and $V^-$, we have
\begin{align}
    W = \xi \, \Phi + \hat v^- + V^+\, \hat v^+ \,.
\end{align}
Again we get massive singlets $\xi$ and $\Phi$, which can be integrated out. Finally, applying the BBP duality with a linear monopole term $\hat v^-$ brings us a Wess--Zumino theory with superpotential,
\begin{align}
    W_{\text{WZ}} = V^+ \,\Wv + \Wv\,\det M
\end{align}
where the monopole operator $\hat v^+$ is mapped to a singlet $\Wv$ that couples to $\det M$. Integrating out the massive singlets $V^+$ and $\Wv$ washes out all the superpotential terms, completing the proof.
\begin{align}
    W_{\text{WZ}} = 0\,.
\end{align}

\section{Comparison of the vacuum moduli spaces}
\label{app:moduli}

The confining $\mathbb D_p[SU(N)]$ model considered in section \ref{sec:confinement} has a set of marginal operators spanning the conformal manifold of the theory, which may lead to the ambiguity of superpotential for the proposed duality. In this appendix, we aim to determine the superpotential of the model such that its moduli space is consistent with that of the dual WZ.

In section \ref{sec:confinement}, we have proposed a duality between the $\mathbb{D}_p[SU(N)]$ theory deformed by the superpotential
\begin{align}
\label{eq:Dp}
\eta^{p+1}+\eta \sum_{i=\alpha}^{p-1}\mathrm{tr}_i \, \Phi^{(i)}
+
\sum_{i=\alpha}^{p-1} \hat{v}^{(i),\pm}
+
\hat{v}^{(\alpha,p-1),\mp}\;
\sum_{i=\alpha}^{p-1}
\hat{v}^{(i-\alpha+2,i),\mp}
\end{align}
and the WZ theory of an $N \times N$ matrix-valued chiral $X$ with the superpotential
\begin{align}
\label{eq:WZ}
	\mathrm{Tr} \, X^{p+1} \,,
\end{align}
where we distinguish $\mathrm{tr}_i$ and $\mathrm{Tr}$ for clarity, which respectively denote the trace over the $i$th gauge node of $\mathbb D_p[SU(N)]$ and the trace over the common global $SU(N)$ symmetry. Note that all the fields are regarded as matrix valued, appropriate for their representations.
\\

Let us first consider the $p=2$ case. For $p=2$, one side of the duality is an adjoint SQCD with the superpotential
\begin{align}
\label{eq:p=2}
	\mathrm{tr} \, \Phi Q \tilde Q + 2 \, \eta \, \mathrm{tr} \, \Phi + \eta^{3} + V_0^+ + V_0^- \,,
\end{align}
where the coefficients are tuned such that the moduli space is consistent with the proposed duality.\footnote{The first term usually comes with factor $\sqrt 2$, which is omitted here because the overall coefficient as well as those of the monopole terms do not affect the subsequent discussion and will be ignored  for simplicity.}
Note that the additional singlet $\eta$ is inserted to be mapped to the trace part of $X$ on the dual WZ side, which has the superpotential
\begin{align}
\label{eq:WZ p=2}
	\mathrm{Tr} \, X^3 \,.
\end{align}

If $\eta$ is absent, this reduces to the known duality proposed by Bajeot--Benvenuti--Sacchi (BBS) in \cite{Bajeot:2023gyl} between the adjoint SQCD with the superpotential
\begin{align}
\label{eq:BBS}
	\mathrm{tr} \, \Phi Q \tilde Q + V_0^+ + V_0^-
\end{align}
and the WZ with the superpotential
\begin{align}
\label{eq:BBS WZ}
	\mathrm{Tr} \, X|_\text{traceless}{}^3 = \mathrm{Tr} \, X^3-\xi \, \mathrm{Tr} \, X \,,
\end{align}
where the second term is introduced to ensure the tracelessness of $X$. On the WZ side, one can easily recover the superpotential \eqref{eq:WZ p=2} without the traceless condition by coupling an extra field $\eta$ using the relation
\begin{align}
\label{eq:relation}
	\mathrm{Tr} \, X^3 = \mathrm{Tr} \, X|_\text{traceless}{}^3 + \frac3d \eta \, \mathrm{Tr} \, X|_\text{traceless}{}^2 + \frac{1}{d^2} \eta^3
\end{align}
where $d$ is the dimension of $X$ that is shifted by the identity matrix $\mathbb I_d$ of the same dimension as follows:
\begin{align}
	X = X|_\text{traceless}+\frac1d \eta \, \mathbb I_d \,.
\end{align}
The trace of $X$ is now given by $\eta$,
\begin{align}
	\mathrm{Tr} \, X = \eta \,.
\end{align}
Under the BBS duality, $\mathrm{Tr} \, X|_\text{traceless}^2$ of the WZ is mapped to $\mathrm{tr} \, \Phi$ of the adjoint SQCD because it is the only operator consistent with the symmetry charges. Hence, the last two terms of \eqref{eq:relation} involving $\eta$ are mapped to
\begin{align}
\alpha \, \eta \, \mathrm{tr} \, \Phi + \beta \, \eta^3
\end{align}
on the adjoint SQCD side, which are exactly the interaction terms of $\eta$ in \eqref{eq:p=2}. $\alpha$ and $\beta$ are undetermined coefficients depending on the detailed duality map, which in \eqref{eq:p=2} was tuned to be consistent with the moduli space of the dual pair, as we explain shortly.

Now let us turn to the moduli space of supersymmetric vacua, which can be parameterized by VEVs of gauge invariant chiral operators subject to some relations. As shown in \eqref{eq:p=2}, we aim to fix the values of $\alpha$ and $\beta$ to give the same moduli space as the dual WZ. First of all, we notice that the last two monopole terms of the superpotential \eqref{eq:p=2} imply that all the monopole operators $\hat V_i^\pm$ are lifted; i.e., the Coulomb branch, and also mixed branches, of the moduli space are empty \cite{Hwang:2022jjs}. Thus, we only need to consider the Higgs branch.

At first, rather than considering both $\eta \, \mathrm{tr} \, \Phi$ and $\eta^3$ terms, we would like to turn on $\eta \, \mathrm{tr} \, \Phi$ only. In that case, the F-term conditions are given as
\begin{gather}
	Q \tilde Q+2 \eta \, \mathbb I_n = 0 \,, \\
	\Phi Q = \tilde Q \Phi = 0 \,, \\
	\mathrm{tr} \, \Phi = 0 \,,
\end{gather}
where the first two conditions imply that
\begin{gather}
	\mathrm{tr} \, Q \tilde Q = \mathrm{Tr} \, \tilde Q Q = -2n \eta \,, \\
	\eta \Phi = 0 \,.
\end{gather}
For $\eta \neq 0$, the only gauge invariant operators having nonzero VEVs are $\tilde Q Q$ and $\eta$, where the latter is actually identified with the trace part of $\tilde Q Q$. Therefore, once we define
\begin{align}
	X = \tilde Q Q+\eta \, \mathbb I_{2 n+1} \,,
\end{align}
we find that
\begin{align}
	X^2 &= \tilde Q Q \tilde Q Q+2 \eta \tilde Q Q+\eta^2 \, \mathbb I_{2 n+1} = \eta^2 \, \mathbb I_{2n+1} \,, \label{eq:X^2} \\
	\mathrm{Tr} X &= \mathrm{Tr} \, \tilde Q Q+(2n+1) \eta = \eta \,, \label{eq:TrX}
\end{align}
which characterize the Higgs branch of the theory. While vanishing $\eta$ seems to allow nontrivial VEVs of $\mathrm{tr} \, \Phi^{k > 1}$, the superconformal index analysis indicates these operators are all lifted due to the monopole superpotential, which is not surprising because $\Phi$ also parameterizes the Coulomb branch in the original $\mathcal N=4$ theory. Thus, the Higgs branch is parameterized by $X$ only.

We notice that this Higgs branch can be realized on the WZ side by employing a slightly deformed superpotential:
\begin{align}
	W = \mathrm{Tr} \, X^3-\xi (\mathrm{Tr} \, X-\eta)-\eta^3 \,,
\end{align}
leading to a set of F-term conditions as follows:
\begin{gather}
	3 X^2-\xi \, \mathbb I_{2 n+1} = 0 \,, \\
	\mathrm{Tr} \, X-\eta = 0 \,, \\
	\xi-3 \eta^2 = 0 \,,
\end{gather}
which are exactly equivalent to the Higgs branch relations \eqref{eq:X^2} and \eqref{eq:TrX} on the adjoint SQCD side. If we integrate out massive $\xi$ and $\eta$, we get
\begin{align}
	W = \mathrm{Tr} \, X^3-(\mathrm{Tr} X)^3,
\end{align}
which differs from the original WZ by $-(\mathrm{Tr} X)^3$. Therefore, we can recover the original dual pair by simply deforming back the deformed dual pair by $\eta^3 = (\mathrm{Tr} \, X)^3$, resulting in $\alpha = 2$ and $\beta = 1$ as shown in \eqref{eq:p=2}. One might wonder if these values would be renormalized along the RG flow. However, as is well known, the superpotential is only affected by the wavefuntion renormalization, at least perturbatively, which doesn't change the argument except suitable field rescaling in the equations, maintaining the values of $\alpha$ and $\beta$.

As a consistency check, one can give mass to $\eta$ by coupling it to an extra singlet $\xi$ and see if our duality reduces to the expected BBS duality. The superpotential is now given as
\begin{align}
	\mathrm{tr} \, \Phi Q \tilde Q + 2 \eta \, \mathrm{tr} \, \Phi + \eta^{3} - \xi \eta + V_0^+ + V_0^- \,,
\end{align}
where both $\eta$ and $\xi$ are massive and can be integrated out, leading to \eqref{eq:BBS} of the BBS dual pair upon the equations of motion:
\begin{gather}
\xi = 2 \, \mathrm{tr} \, \Phi \,, \label{eq:xi} \\
\eta = 0 \,.
\end{gather}
On the WZ side, the extra coupling corresponds to $-\xi \, \mathrm{Tr} X$ since $\eta = \mathrm{Tr} \, X$, which exactly gives the dual superpotential \eqref{eq:BBS WZ}. Moreover, \eqref{eq:xi} validates the duality map between $\mathrm{tr} \, \Phi$ and $\xi = \mathrm{Tr} X^2$ previously claimed.
\\

Such comparison of the moduli spaces can be generalized to higher $p$ as well. As before, first we consider the superpotential without $\eta^{p+1}$:
\begin{align}
	W_{\mathcal N=4} + \eta \sum_{i = 1}^{p-1} \alpha_i \, \mathrm{tr}_i \, \Phi^{(i)} + \sum_{i = 1}^{p-1} v^{(i),\pm}+v^{(1,p-1),\mp} \,,
\end{align}
where the first term collectively denotes the standard $\mathcal N=4$ superpotential terms. Again, the monopole terms are expected to lift the Coulomb branch and mixed branches completely, which is consistent with the index analysis. Thus, let us focus on the Higgs branch of the moduli space. The relevant F-term conditions are as follows:
\begin{gather}
	\tilde Q^{(j-1)} Q^{(j-1)} + Q^{(j)} \tilde Q^{(j)}+ \alpha_j \eta \, \mathbb I_{m_j} = 0 \,, \\
	\Phi^{(j)} Q^{(j)}+Q^{(j)} \Phi^{(j+1)} = \tilde Q^{(j)} \Phi^{(j)}+\Phi^{(j+1)} Q^{(j)} = 0 \,, \\
	\sum_{i = 1}^{p-1} \alpha_i \mathrm{tr}_i \Phi^{(i)} = 0
\end{gather}
for $j = 1, \dots, p-1$ with $Q^{(0)} = \tilde Q^{(0)} = \Phi^{(p)} = 0$. $m_j$ is the rank of the $j$th gauge node given as $m_j = \lfloor j N / p \rfloor$. Defining 
\begin{align}
X = \tilde Q^{(p-1)} Q^{(p-1)} + \eta \, \mathbb I_{N}
\end{align}
we find that
\begin{align}
X^p &= \left(\tilde Q^{(p-1)} Q^{(p-1)} + \eta \mathbb I_N\right)^p \\
&= f_1 (\alpha_1,\dots,\alpha_{p-1}) \, \eta \tilde Q^{(p-1)} \dots \tilde Q^{(1)} Q^{(1)} \dots Q^{(p-1)} \nonumber \\
&\quad +f_2 (\alpha_2,\dots,\alpha_{p-1}) \, \eta^2 \tilde Q^{(p-1)} \dots \tilde Q^{(2)} Q^{(2)} \dots Q^{(p-1)} \nonumber \\
&\qquad \vdots \nonumber \\
&\quad +f_{p-1} (\alpha_{p-1}) \eta^{p-1} \, \tilde Q^{(p-1)} Q^{(p-1)} \nonumber \\
&\quad +\eta^p \, \mathbb I_N \,, \\
\mathrm{Tr} \, X &= \beta \eta \label{eq:TrX 2}
\end{align}
where $f_j$ is a polynomial in terms of coefficients $\alpha_j, \dots, \alpha_{p-1}$ and $\beta$ is given as
\begin{align}
\label{eq:beta}
\beta = \sum_{i = 1}^p (-1)^{p-i} \alpha_i m_i
\end{align}
with $\alpha_p = 1$. As in the $p=2$ case, $\alpha_i$ can be tuned such that $f_1 = \dots = f_{p-1} = 0$, implying that
\begin{align}
\label{eq:X^p}
X^p = \eta^p \, \mathbb I_N \,.
\end{align}
Together with \eqref{eq:TrX 2}, this characterizes the Higgs branch for higher $p$ parameterized by $X$.

This Higgs branch can be realized on the WZ side by the following superpotential:
\begin{align}
	\mathrm{Tr} \, X^{p+1} - \xi (\mathrm{Tr} \, X-\beta \eta)-\beta \eta^{p+1} \,,
\end{align}
whose F-term conditions imply that
\begin{align}
	X^p &= \eta^p \, \mathbb I_N \,, \\
	\mathrm{Tr} \, X &= \beta \eta \,,
\end{align}
which exactly agrees with the relations \eqref{eq:X^p} and \eqref{eq:TrX 2} on the $\mathbb D_p[SU(N)]$ side.
Once we integrate out massive $\xi$ and $\eta$, the superpotential reduces to
\begin{align}
	\mathrm{Tr} \, X^{p+1}-\beta^{-p} (\mathrm{Tr} \, X)^{p+1} \,.
\end{align}
which differs from the original WZ by the second term. Hence, we can recover the original dual pair by deforming back this deformed dual pair by $\beta^{-p} \eta^{p+1} = \beta^{-p} (\mathrm{Tr} \, X)^{p+1}$ with $\beta$ defined by \eqref{eq:beta}, resulting in the proposed duality between the $\mathbb{D}_p[SU(N)]$ theory deformed by the superpotential \eqref{eq:Dp} and the WZ theory with the superpotential \eqref{eq:WZ}.

\newpage
\bibliographystyle{JHEP}
\bibliography{ref}

\end{document}